\RequirePackage{etex}
\documentclass[12pt,a4paper]{article} 
\pdfoutput=1
\usepackage[backref=page, colorlinks=true, urlcolor=blue, anchorcolor=blue, citecolor=blue, filecolor=blue, linkcolor=blue, menucolor=blue, linktocpage=true, pdfproducer=medialab,pdfa=true]{hyperref}
\usepackage{datetime} 
\usepackage{graphicx}
\usepackage{etex}
\usepackage{jheppub}
\usepackage{epstopdf}
\usepackage{epsfig}
\usepackage{dcolumn}  
\usepackage{bm}    
\usepackage{amssymb} 
\usepackage{amsmath}
\usepackage[Bourbaki-arrow]{dynkin-diagrams}
\usepackage{amsfonts} 
\usepackage{slashed}
\usepackage{ytableau}
\usepackage[mathscr]{euscript}
\usepackage{epsfig}
\usepackage[utf8]{inputenc}
\usepackage{tikz}
\usetikzlibrary{tikzmark,calc,,arrows,shapes,decorations.pathreplacing}
\usepackage{xcolor}
\usepackage{mathrsfs}
\usepackage{verbatim}
\usepackage{cases}
\usepackage{bbding}
\usepackage{pifont}
\usepackage{ulem}
\usepackage{dcolumn}
\usepackage{pgf}
\usepackage{pstricks}
\usepackage{pst-node}
\usepackage[figuresright]{rotating}
\usepackage{mathtools}
\usepackage{braket}
\usepackage{titlesec}
\usepackage[most]{tcolorbox}
\tcbuselibrary{breakable} 
\usepackage{stmaryrd}
\usepackage{cancel}
\usepackage{textcomp}
\usepackage{extarrows}
\usepackage{accents}
\usepackage{ulem}
\usepackage{color}
\usepackage{amsthm}
\usepackage{caption}
\usepackage{subcaption}
\usepackage{stackengine}
\usepackage{arydshln}
\usepackage{bm}
\usepackage[all,cmtip]{xy}
\usepackage{soul}
\usepackage{shuffle}
\usepackage{tikz}
\tikzstyle{vertex}=[circle, draw, minimum size=0pt]
\usetikzlibrary{automata,arrows,positioning,calc}

\newcommand{\nn}{\mathfrak{n}}

\usetikzlibrary{backgrounds,calc,positioning}

\usetikzlibrary{arrows.meta}

\def\lm{\limits}

\makeatletter
\DeclareFontFamily{OMX}{MnSymbolE}{}
\DeclareSymbolFont{MnLargeSymbols}{OMX}{MnSymbolE}{m}{n}
\SetSymbolFont{MnLargeSymbols}{bold}{OMX}{MnSymbolE}{b}{n}
\DeclareFontShape{OMX}{MnSymbolE}{m}{n}{
	<-6>  MnSymbolE5
	<6-7>  MnSymbolE6
	<7-8>  MnSymbolE7
	<8-9>  MnSymbolE8
	<9-10> MnSymbolE9
	<10-12> MnSymbolE10
	<12->   MnSymbolE12
}{}
\DeclareFontShape{OMX}{MnSymbolE}{b}{n}{
	<-6>  MnSymbolE-Bold5
	<6-7>  MnSymbolE-Bold6
	<7-8>  MnSymbolE-Bold7
	<8-9>  MnSymbolE-Bold8
	<9-10> MnSymbolE-Bold9
	<10-12> MnSymbolE-Bold10
	<12->   MnSymbolE-Bold12
}{}

\let\llangle\@undefined
\let\rrangle\@undefined
\DeclareMathDelimiter{\llangle}{\mathopen}%
{MnLargeSymbols}{'164}{MnLargeSymbols}{'164}
\DeclareMathDelimiter{\rrangle}{\mathclose}%
{MnLargeSymbols}{'171}{MnLargeSymbols}{'171}
\makeatother

\makeatletter
\newsavebox{\mstrut}
\newcommand{\bbra}[1]{%
	\sbox{\mstrut}{\(#1\)}%
	\mathinner{\left\langle\kern-0.5\ht\mstrut\left\langle{#1}\right|\mkern-2mu\right|}%
}
\newcommand{\kett}[1]{%
	\sbox{\mstrut}{\(#1\)}%
	\mathinner{\left|\mkern-2mu\left|{#1}\right\rangle\kern-0.5\ht\mstrut\right\rangle}%
}
\makeatother

\titleclass{\subsubsubsection}{straight}[\subsection]

\newcounter{subsubsubsection}[subsubsection]
\renewcommand\thesubsubsubsection{\thesubsubsection.\arabic{subsubsubsection}}

\titleformat{\subsubsubsection}
{\normalfont\normalsize\bfseries}{\thesubsubsubsection}{1em}{}
\titlespacing*{\subsubsubsection}
{0pt}{3.25ex plus 1ex minus .2ex}{1.5ex plus .2ex}

\makeatletter
\renewcommand\paragraph{\@startsection{paragraph}{5}{\z@}%
	{3.25ex \@plus1ex \@minus.2ex}%
	{-1em}%
	{\normalfont\normalsize\bfseries}}
\renewcommand\subparagraph{\@startsection{subparagraph}{6}{\parindent}%
	{3.25ex \@plus1ex \@minus .2ex}%
	{-1em}%
	{\normalfont\normalsize\bfseries}}
\def\toclevel@subsubsubsection{4}
\def\toclevel@paragraph{5}
\def\toclevel@paragraph{6}
\def\l@subsubsubsection{\@dottedtocline{4}{7em}{4em}}
\def\l@paragraph{\@dottedtocline{5}{10em}{5em}}
\def\l@subparagraph{\@dottedtocline{6}{14em}{6em}}
\makeatother

\usepackage{young}
\usepackage{graphicx,rotating,booktabs}

\usepackage{varwidth}

\newdimen\tableauside\tableauside=1.0ex
\newdimen\tableaurule\tableaurule=0.4pt
\newdimen\tableaustep
\def\phantomhrule#1{\hbox{\vbox to0pt{\hrule height\tableaurule width#1\vss}}}
\def\phantomvrule#1{\vbox{\hbox to0pt{\vrule width\tableaurule height#1\hss}}}
\def\sqr{\vbox{%
		\phantomhrule\tableaustep
		\hbox{\phantomvrule\tableaustep\kern\tableaustep\phantomvrule\tableaustep}%
		\hbox{\vbox{\phantomhrule\tableauside}\kern-\tableaurule}}}
\def\squares#1{\hbox{\count0=#1\noindent\loop\sqr
		\advance\count0 by-1 \ifnum\count0>0\repeat}}
\def\tableau#1{\vcenter{\offinterlineskip
		\tableaustep=\tableauside\advance\tableaustep by-\tableaurule
		\kern\normallineskip\hbox
		{\kern\normallineskip\vbox
			{\gettableau#1 0 }%
			\kern\normallineskip\kern\tableaurule}%
		\kern\normallineskip\kern\tableaurule}}
\def\gettableau#1 {\ifnum#1=0\let\next=\null\else
	{{\tiny\yng(1)}}s{#1}\let\next=\gettableau\fi\next}

\renewcommand{\(}{\left(}
\renewcommand{\)}{\right)}

\newcommand{\includeCroppedPdf}[2][]{%
	\IfFileExists{./#2-crop.pdf}{}{%
		\immediate\write18{pdfcrop #2 #2-crop.pdf}}%
	\includegraphics[#1]{#2-crop.pdf}}        

\newcommand{\be}{ \begin{equation}}
	\newcommand{\ee}{\end{equation}}
\newcommand{\bea}[1]{\begin{eqnarray}\label{#1} }
	\newcommand{\eea}{\end{eqnarray}}

\def\ZZZ{{\hskip-3pt\hbox{ Z\kern-1.6mm Z}}}
\def\zzz{{\hskip-3pt\hbox{ z\kern-1mm z}}}

\newcommand{\eps}{\epsilon}

\newcommand*\circled[1]{\tikz[baseline=(char.base)]{
		\node[shape=circle,draw,inner sep=2pt] (char) {#1};}}

\def\bal#1\eal{\begin{align}#1\end{align}}

\renewcommand{\(}{\left(}
\renewcommand{\)}{\right)}

\newcommand{\Res}{\textrm{Res}}

\def\myequiv{:= }

\def\mye{e}

\newcommand{\longsquiggly}{\xymatrix{{}\ar@{~>}[r]&{}}}

\def\mye{e}

\newcommand\sqbox[1]{{
		\setbox0=\hbox{\scalebox{1}{\mbox{$\Box$}}}
		\setbox1=\hbox{\mbox{\raisebox{0.35ex}{\scriptsize #1}}}
		\mbox{\raisebox{-0.2ex}{\rlap{\hbox to \wd0{\hss{\box1}\hss}}\box0}}
}}

\newcommand\sqboxs[1]{{
		\setbox0=\hbox{\scalebox{1.3}{\mbox{$\Box$}}}
		\setbox1=\hbox{\mbox{\raisebox{0.5ex}{\scriptsize #1}}}
		\mbox{\raisebox{-0.2ex}{\rlap{\hbox to \wd0{\hss{\box1}\hss}}\box0}}
}}

\newcommand\sqboxss[1]{{
		\setbox0=\hbox{\scalebox{2}{\mbox{$\Box$}}}
		\setbox1=\hbox{\mbox{\raisebox{0.6ex}{\scriptsize #1}}}
		\mbox{\raisebox{-0.2ex}{\rlap{\hbox to \wd0{\hss{\box1}\hss}}\box0}}
}}

\title{Quiver Yangians as Coulomb branch algebras}

\author[a,b]{Tiantai Chen}
\author{and}
\author[a]{Wei Li}

\affiliation[a]{Institute of Theoretical Physics, Chinese Academy of Sciences,\\
	\hspace*{0.3cm}Zhongguancun East Road 55, Beijing 100190, China}
\affiliation[b]{School of Physical Sciences, University of Chinese Academy of Sciences,\\
\hspace*{0.3cm}Yuquan Road 19, Beijing 100049, China}

\emailAdd{chentiantai@itp.ac.cn}
\emailAdd{weili@mail.itp.ac.cn}

\date{\today}

\abstract{For a 3D $\mathcal{N}=4$ gauge theory, turning on the $\Omega$-background in $\mathbb{R}\times \mathbb{R}^2_{\eps}$ deforms the Coulomb branch chiral ring into the quantum Coulomb branch algebra, generated by the $\frac{1}{2}$-BPS monopoles together with the complex scalar in the vector-multiplet.
We conjecture that for a 3D $\mathcal{N}=4$ quiver gauge theory with unitary gauge group, the quantum Coulomb branch algebra can be formulated as the truncated shifted quiver Yangian Y$(\widehat{Q},\widehat{W})$ based on the triple quiver $\widehat{Q}$  of the original quiver $\mathrm{Q}$ with canonical potential $\widehat{W}$.
We check this conjecture explicitly for general tree-type quivers $\mathrm{Q}$ by considering the action of monopoles on the $\frac{1}{2}$-BPS vortex configurations.
The Hilbert spaces of vortices approaching different vacua at  spatial infinity furnish different representations of the shifted quiver Yangian, and all the charge functions have only simple poles.
For quivers beyond tree-type, 
our conjecture is consistent with known results on special examples.}

\begin{document}
\maketitle

\section{Introduction and Summary}
\label{sec:Intro}

Supersymmetric quantum field theories with $8$ supercharges occupy an optimal middle ground, offering sufficient symmetry to yield exact results while retaining rich non-trivial dynamics. 
3D $\mathcal{N}=4$ gauge theories exemplify this balance and allow for the convergence of beautiful physics and mathematics. 
In particular, the moduli space generically factorizes into a Higgs branch and a Coulomb branch, and the extended supersymmetry ensures that both branches are hyper-K\"ahler manifolds \cite{Seiberg:1996nz}. 
Furthermore, 3D mirror symmetry interchanges the Higgs  branch of one  3D $\mathcal{N}=4$ gauge theory with the Coulomb branch of its dual theory \cite{Intriligator:1996ex,Hanany_1997,Aharony_1997}, and the symplectic duality relates the Higgs branch and the Coulomb branch of the same theory as a pair of symplectic varieties (see e.g.\ \cite{braden2022quantizationsconicalsymplecticresolutions,Bullimore:2016nji,kamnitzer2022symplecticresolutionssymplecticduality}).

\medskip

In this paper, we focus on the Coulomb branches of 3D $\mathcal{N}=4$ gauge theories, which are harder to study than the Higgs branches since, unlike the Higgs branch, they are not protected against  quantum corrections.\footnote{The Higgs branch can be constructed as a hyper-K\"ahler quotient and the Higgs branch chiral ring arises via holomorphic symplectic reduction.} 
For a 3D $\mathcal{N}=4$ gauge theory with gauge group $G$, the classical Coulomb branch $\mathcal{M}_{C}$ is parametrized by the VEV's of the triplet of scalars in the $\mathcal{N}=4$ vector multiplet together with the (periodic) dual photon, namely $\mathcal{M}^{\textrm{classical}}_C=(\mathbb{R}^3\times S^1)^{\textrm{rank}(G)}/\textrm{Weyl}(G)$. 
The quantum-corrected Coulomb branch $\mathcal{M}_C$ is generated by $\frac{1}{2}$-BPS monopole operators dressed by gauge-invariant\footnote{The gauge invariance is w.r.t.\ the subgroup of $G$ preserved by the monopole singularities.} polynomials of the vector-multiplet scalars.\footnote{As an algebraic variety, the quantum-corrected Coulomb branch is birationally equivalent to the classical one.}

\medskip

In this paper, we are interested in the \textit{algebraic structure} of the quantum Coulomb branch $\mathcal{M}_{C}$ of a 3D $\mathcal{N}=4$ quiver gauge theory.
One way of understanding the ``Coulomb branch algebra" is as follows \cite{bullimore2015coulomb}. 
One first rewrites the $\mathcal{N}=4$ theory in terms of the $\mathcal{N}=2$ language.
Choosing an $\mathcal{N}=2$ subalgebra of the $\mathcal{N}=4$ supersymmetry corresponds to selecting one complex structure out of the triplet of complex structures in the hyper-K\"ahler manifold, thus viewing the hyper-K\"ahler manifold $\mathcal{M}_C$ as a complex symplectic manifold, with the complex symplectic form given by the linear combination of the hyper-K\"ahler forms that matches with the choice of the  $\mathcal{N}=2$ subalgebra.

The dressed monopole operators that respect this $\mathcal{N}=2$ subalgebra generate an $\mathcal{N}=2$ chiral ring $\mathbb{C}[\mathcal{M}_C]$ (of holomorphic functions on $\mathcal{M}_C$), which is independent of the gauge couplings\footnote{In a 3D $\mathcal{N}=4$ theory, the gauge couplings are real parameters of the theory and cannot be complexified.} and whose character is captured by the monopole formula of \cite{Cremonesi_2014}.
The complex symplectic form induces the Poisson brackets of the holomorphic functions, which can be determined via e.g.\ the abelianization map of \cite{bullimore2015coulomb}, giving rise to the  Coulomb branch Poisson algebra $\mathcal{A}$.
Once the Poisson algebra $\mathcal{A}$  is determined, one can (canonically) quantize it to obtain the quantum Coulomb branch algebra $\mathcal{A}_{\hbar}$.\footnote{One can then unify all the complex structures and recover the hyper-K\"ahler structure in twistor space, thereby obtaining the Coulomb branch algebra for general complex structures \cite{bullimore2015coulomb}.}

\medskip

Mathematically, the Coulomb branch is an important object for geometric quantization theory, which aims to understand a geometric object via its (quantum) ring of functions and vice versa: for the Coulomb branch, the geometric object is $\mathcal{M}_C$ and the ring of functions is $\mathbb{C}[\mathcal{M}_C]$ with its quantum version being $\mathcal{A}_{\hbar}$. 
The Coulomb branch is important because it provides many interesting examples of \textit{non-commutative} algebras, whereas in the traditional geometric representation theory the ring of functions is normally commutative. 
For a 3D $\mathcal{N}=4$ gauge theory, the non-commutative Coulomb branch algebra was constructed via deformation quantization in a series of papers \cite{Nakajima:2015txa,Braverman:2016wma,Braverman:2016pwk,braverman2024ringobjectsequivariantderived}.
(See Sec.~\ref{sssec:DeformationQuantization} for a more detailed review.)

\medskip

One can also bypass the step of obtaining the Poisson algebra and instead directly put the theory on the $\Omega$-background $\mathbb{R}\times \mathbb{R}^2_{\eps}$, which has the effect of quantizing the Coulomb branch chiral ring into the  quantum Coulomb branch algebra $\mathbb{C}_\eps[\mathcal{M}_C]$.
In the $\Omega$-deformed theory, by studying how the $\frac{1}{2}$-BPS monopole operators and the complex scalar act on the vortex configurations, one can deduce the algebraic relations among these Coulomb branch chiral ring generators \cite{Bullimore:2016hdc}.
The vortex Hilbert space forms a Verma module of the quantum Coulomb branch algebra.

\medskip

The mathematical construction of the quantum Coulomb branch algebra in \cite{Braverman:2016pwk, Braverman:2016wma} applies to general 3D $\mathcal{N}=4$ gauge theories with compact gauge group $G$ and matter in symplectic representation $(R,\bar{R})$. 
However, often it is useful to express the algebra in terms of generators and relations, and this has been done in a class by class manner. 
For general quantization parameter $\hbar$,  this has been done for a 3D $\mathcal{N}=4$ quiver gauge theory with finite ADE quiver in \cite{Braverman:2016pwk} and the quantum Coulomb branch algebra $\mathcal{A}_{\hbar}$ (called quantized Coulomb branch there) was shown to be the truncated shifted Yangian of ADE type.
This has since been generalized to the BCFG quiver in \cite{Nakajima:2019olw} and the Jordan quiver in \cite{Kodera:2016faj_jordan_quiver}, where $\mathcal{A}_{\hbar}$ was shown to be the truncated shifted Yangian of BCFG type and the truncated shifted affine Yangian of $\mathfrak{gl}_1$, respectively.\footnote{
For special quantization parameter $\hbar$, this has been studied for more general classes of quivers, such as the simple quivers in \cite{Kamnitzer:2019mjs,weekes2019generatorscoulombbranchesquiver} and quivers with symmetrizable adjacency matrices  \cite{Nakajima:2019olw}.
}

\medskip

In this paper, we propose that for a 3D $\mathcal{N}=4$ quiver gauge theory based on the quiver $\mathrm{Q}$ and with unitary gauge group factors $U(N^{(a)})$, the quantum Coulomb branch algebra can be formulated as certain truncated shifted quiver Yangian 
based on the \textit{triple quiver} $\widehat{Q}$ of the original quiver $\mathrm{Q}$:
\begin{equation}\label{eq:MainConjecture}
\mathcal{A}_{\hbar}(\mathrm{Q},\{N^{(a)}\})
\ \simeq \
\textrm{truncated shifted }
\mathrm{Y}_{\hbar}(\widehat{Q},\widehat{W}) \,.
\end{equation}
Here the triple quiver $\widehat{Q}$ is defined as follows: apart from all the nodes and arrows in the original quiver $\mathrm{Q}$, for each arrow $I^{a\rightarrow b}\in \mathrm{Q}_1$, the triple quiver $\widehat{Q}$ includes its opposite arrow $I^{b \rightarrow a}$, and for each node $a\in \mathrm{Q}_0$, it also includes a self-loop $I^{a \rightarrow a}$.

Quiver Yangians are a class of Yangian-type algebras that generalize the original Yangian based on a Lie algebra $\mathfrak{g}$.
They were originally defined for quivers from toric Calabi-Yau threefolds \cite{Li:2020rij,Yamazaki:2022cdg} and were later generalized to cover general quivers (with potentials) \cite{Li:2023zub}.
The shifted quiver Yangians were introduced to study general representations \cite{Galakhov:2021xum,Li:2023zub}. For recent studies on the quiver Yangian, see \cite{Galakhov:2020vyb,Galakhov:2021vbo,Bao:2022fpk,Bao:2022jhy,Bao:2023ece,Galakhov:2023mpc,Galakhov:2024bzs}.

\medskip

Our conjecture is based on three pieces of evidence.
\begin{enumerate}
\item The conjecture holds for all the known results above, namely the cases of the ABCDEFG quiver and the affine $\mathfrak{gl}_1$ quiver from \cite{Braverman:2016pwk,Kodera:2016faj_jordan_quiver,Nakajima:2019olw}.
\item We check the conjecture explicitly for all the 3D $\mathcal{N}=4$ \textit{tree-type} quiver gauge theories with unitary gauge groups.
\item For general quivers, we support our conjecture by a chain of  mathematical arguments based on some older theorems, conjectures, and expectations.
\end{enumerate}

In the explicit check for the tree-type quivers, instead of directly translating the mathematical definition in \cite{Braverman:2016wma,Braverman:2016pwk} into generators and relations, as done in \cite{Braverman:2016pwk,Kodera:2016faj_jordan_quiver, Nakajima:2019olw}, we will adopt the more physical method of \cite{Bullimore:2016hdc}, namely deducing the relations among the monopole operators and vector-multiplet scalars by studying how they act on the vortex configurations.
More specifically:\footnote{Note that this method of defining the quantum Coulomb branch algebra by first determining their action on the vortices is reminiscent of how the quiver Yangian was first constructed as  the BPS algebra for the $\frac{1}{2}$-BPS D-brane bound states system in type IIA string theory on a generic toric Calabi-Yau threefolds, by first fixing how the  generators of the BPS algebra act on the BPS states (using their description in terms of certain 3D colored crystals \cite{Ooguri:2008yb})
	\cite{Li:2020rij}. 
}
\begin{enumerate}
\item 
We generalize the results  of  \cite{Bullimore:2016hdc} on A-type quivers to 3D $\mathcal{N}=4$ quiver gauge theories with general tree-type quivers $\mathrm{Q}$: we derive the $\frac{1}{2}$-BPS vortex configurations and the vortex Hilbert space, and then by studying how the  $\frac{1}{2}$-BPS (first fundamental) minuscule monopoles and vector-multiplet complex scalars act on the vortex Hilbert space, we obtain the quantum Coulomb branch algebra $\mathbb{C}_{\eps}[\mathcal{M}_C]$ (for $\epsilon \neq 0$) explicitly in terms of relations among the monopole operators and the vector-multiplet scalars.\footnote{Note that this method works because for the quiver gauge theories with $U(N)$ factors, which are the focus of this paper, the minuscule monopole operators (together with vector-multiplet scalars) generate the entire chiral ring $\mathbb{C}[\mathcal{M}_C]$ \cite{bullimore2015coulomb,Braverman:2016wma}, see later; moreover, $\mathbb{C}_{\eps}[\mathcal{M}_C]$ with $\epsilon \neq 0$ can be generated using only the first fundamental minuscule monopoles (together with vector-multiplet scalars) \cite{weekes2019generatorscoulombbranchesquiver}.} 
\item We then determine the field redefinitions to rewrite the quantum Coulomb branch algebra $\mathbb{C}_{\eps}[\mathcal{M}_C]$ with $\eps \neq 0$ in terms of the (shifted) quiver Yangian 
Y$_{\hbar}(\widehat{Q},\widehat{W})$  
of the triple quiver\footnote{Physically, the triple quiver appears because the vortex quantum mechanics (the effective theory on the $\frac{1}{2}$-BPS vortex world-line) is an $\mathcal{N}=4$ quiver quantum mechanics whose quiver is the triple quiver of the original quiver that defines the 3D $\mathcal{N}=4$ quiver gauge theory.} $\widehat{Q}$ of the original tree-type quiver $\mathrm{Q}$, with $\widehat{W}$ being the canonical potential for the triple quiver \cite{Ginzburg:2006fu} and the parameter identification $\eps=\hbar \neq 0$.\footnote{For tree-type theories, after imposing the superpotential constraints and performing appropriate spectral shifts, there is only one parameter for the quiver Yangian, denoted by $\hbar$, which precisely matches the $\Omega$-deformation parameter $\epsilon$. 
For more general quivers, the quiver Yangian can have more than one parameter, in which case one of them would match $\epsilon$ while the remaining ones are the parameters for the classical Coulomb branch algebra.} 
In this translation, the vector-multiplet scalars are mapped to the Cartan generators of the quiver Yangian, and the $\frac{1}{2}$-BPS first fundamental minuscule monopole operators are mapped to the raising and lowering operators of the quiver Yangian (in the Chevalley basis).
\end{enumerate}

An advantage of translating the algebra into the (shifted) quiver Yangian language is due to its elegant representation theory. 
The representations of a (shifted) quiver Yangian are specified by the framings of the quiver and can be expressed in terms of the finite-codimensional ideals of the Jacobian algebra $J(Q,W)$. 
The action of the quiver Yangian generators on these ideals has simple and uniform expressions \cite{Li:2020rij,Li:2023zub}, and in particular, the information of the representation is entirely captured by a rational function (of the equivariant weights of the framing arrows) that can be read off easily from the framing.

With this translation, we can then use the quiver Yangian representations to characterize the vortices. 
For a given quiver gauge theory (with given quiver $\mathrm{Q}$ and gauge and flavor ranks), the Hilbert spaces of vortices approaching different vacua at spatial infinity furnish different  representations of the same shifted quiver Yangian Y$(\widehat{Q},\widehat{W})$.
The information on the vacuum $\nu$, which contains the information on ranks of the gauge and flavor groups ($\{N^{(a)}\}$ and $N_{\mathtt{f}}$), is translated into the framing of the quiver $\widehat{Q}$, and the action of the quiver Yangian generators on the states in the corresponding representation $\mathcal{R}_{\nu}$ has rather uniform expression for different states and  different $\nu$.

\medskip

Note that for the triple quiver $\widehat{Q}$ of a general  quiver $\mathrm{Q}$, a general framing would give rise to higher-order poles, see \cite[App.~C]{Li:2023zub}.
It was proposed there that one should consider framings with enough arrows going back to the framing node in order to truncate the representations in such a way so as to avoid the potential higher-order poles. 
We will see that the quiver Yangian representations $\mathcal{R}_{\nu}$ considered in this paper all automatically satisfy the property that the charge functions only have simple poles --- the representations are automatically truncated (due to the presence of certain framing arrows to the framing node) to the extent that the potential states with higher-order poles simply will not enter the representations.

\medskip

Due to this truncation, the representation $\mathcal{R}_{\nu}$ is not a Verma module w.r.t.\ the shifted quiver Yangian.
Since by construction $\mathcal{R}_{\nu}$ is a Verma module of the Coulomb branch algebra generated by monopoles and vector-multiplet scalars (for generic masses), we can define the truncated shifted quiver Yangian as the pre-image of the map to the latter algebra.
(It is reassuring that in our computations, although both the truncation and the shift a priori might depend on the information of the vacuum $\nu$, in the end only the dependence on the gauge and flavor ranks remains.) 
In summary, in the main conjecture \eqref{eq:MainConjecture},  Y$_{\hbar}(\widehat{Q},\widehat{W})$ only depends on the quiver $\mathrm{Q}$, whereas the 
information of the truncation and shift is contained in the ranks of the gauge and flavor groups.

\medskip

The paper is organized as follows. 
In Sec.~\ref{sec:Review} we review the 3D $\mathcal{N}=4$ quiver gauge theory.
Sec.~\ref{sec:vortex_tree_typeQGT} studies the general 3D $\mathcal{N}=4$ tree-type quiver gauge theory and describes its vortex Hilbert space and the Coulomb branch algebra.
Sec.~\ref{sec:QYasMonopoleAlgebra} presents the explicit map between the Coulomb branch algebra and the quiver Yangian based on the triple quiver, and reproduces the vortex Hilbert space $\mathcal{H}_{\nu}$ as the representation $\mathcal{R}_{\nu}$ of the quiver Yangian. 
We present the ADE examples  in Sec.~\ref{ssec:Dtype_theory} and App.~\ref{appsec:ADE_examples}, and in particular an example in Sec.~\ref{sec:kleaf_example} to demonstrate how the potential higher-order pole problem is avoided automatically.
In Sec.~\ref{sec:Conjecture} we present our main conjecture for general quivers.
In Sec.~\ref{sec:BeyondSimple} we study the non-simply-laced quivers and quivers with edge-loops.
We end in Sec.~\ref{sec:Discussion} with discussion and a list of future problems. 
Finally App.~\ref{appsec:Rev-3D-N=4} contains a review on 3D $\mathcal{N}=4$ gauge theories, App.~\ref{appsec:details} provides more details on the computations in Sec.~\ref{sec:vortex_tree_typeQGT}, and App.~\ref{appsec:ADE_examples} 
shows examples on ADE-type quivers.

\section{Review}
\label{sec:Review}
\subsection{\texorpdfstring{3D $\mathcal{N}=4$ quiver gauge theories}{3D N=4 quiver gauge theories}}
\label{ssec:3DN=4QGT}

We briefly review the 3D $\mathcal{N}=4$ quiver gauge theories, in particular the various constructions of the quantum Coulomb branch algebra; for more details on 3D $\mathcal{N}=4$ gauge theories see App.~\ref{appsec:Rev-3D-N=4}.

\medskip

The 3D $\mathcal{N}=4$ supersymmetry has 8 supercharges $\mathcal{Q}_\alpha^{a\dot{a}}$, with $a,\dot{a}=1,2$ the indices for the R-symmetry $SU(2)_H\times SU(2)_C$, and $\alpha=\pm$ the index for the Lorentz symmetry $SU(2)_E$. 
The superalgebra is:
\begin{equation}
\{\mathcal{Q}_\alpha^{a\dot{a}},\mathcal{Q}_\beta^{b\dot{b}}\}
=-2\epsilon^{ab}\epsilon^{\dot{a}\dot{b}}\sigma^\mu_{\alpha\beta}P_\mu+2\epsilon_{\alpha\beta}(\epsilon^{ab}Z^{\dot{a}\dot{b}}+\epsilon^{\dot{a}\dot{b}}Z^{ab})\,,
\end{equation}
where $(\sigma^{\mu})^{\alpha}_{\ \beta}$ are the Pauli matrices, $\epsilon^{12}=\epsilon_{21}=1$, and $Z^{ab},Z^{\dot{a}\dot{b}}$ are the central charges.

\subsubsection{Quiver and framed quiver}
We consider 3D $\mathcal{N}=4$  quiver gauge theories, for which the compact gauge group $G$ and the field content are captured by a quiver $\textrm{Q}$:
\begin{equation}
\textrm{Q}=\{\textrm{Q}_0,\textrm{Q}_1;s,t:\textrm{Q}_1\to \textrm{Q}_0\}\,,
\end{equation}
where $\textrm{Q}_0$ is the set of nodes and $\textrm{Q}_1$ the set of arrows; the functions $s:\textrm{Q}_1\to \textrm{Q}_0$ and $t:\textrm{Q}_1\to \textrm{Q}_0$ map an arrow to its source and its target, respectively. 
For a 3D $\mathcal{N}=4$ unitary quiver gauge theory:
\begin{itemize}
\item each node $a\in \textrm{Q}_0$ corresponds to a gauge factor $U(N^{(a)})$
with an associated $\mathcal{N}=4$ vector-multiplet; (We will call them ``gauge nodes" to distinguish them from the flavor nodes to be introduced later.)
\item each  arrow $I^{a\rightarrow b}\in \textrm{Q}_1$ corresponds to an $\mathcal{N}=4$ hypermultiplet, in a unitary bi-fundamental representation $R_{I^{a\rightarrow b}}$ of $U(N^{(b)})\times U(N^{(a)})$.
\end{itemize}

\medskip

The theory has global symmetries $G_H\times G_C$, where the subscripts $H$ and $C$ refer to the Higgs branch and the Coulomb branch, respectively. 
The symmetry $G_H$ is the  flavor symmetry, and for quiver gauge theories, $G_H$ can be specified as follows. 
In addition to the quiver $\mathrm{Q}$ that specifies the gauge content of the theory, we introduce the flavor nodes and additional arrows between the flavor nodes and the gauge nodes of $a\in \mathrm{Q}_0$: 
\begin{itemize}
\item each flavor node $\mathtt{f}$ corresponds to a $U(N_{\mathtt{f}})$ factor of the flavor symmetry $G_H$;
\item each newly introduced arrow between the flavor node $\mathtt{f}$ and the gauge node $a\in \mathrm{Q}_0$ corresponds to a hypermultiplet that transforms in the bi-fundamental representation under $U(N^{(a)})\times U(N_{\mathtt{f}})$.
\end{itemize}
We see that the interpretation of the thus extended quiver by the flavor nodes is similar to the original quiver $\mathrm{Q}$, with the only difference being that the flavor nodes are non-dynamical: they have no associated vector-multiplet. 
In drawing the extended quiver, we will use circle nodes (resp.\ square nodes) to represent  the gauge nodes (resp.\ the flavor nodes).

\medskip

The symmetry $G_C$ acts as the large gauge symmetry due to the periodicity of the dual photon\footnote{In 3D the on-shell d.o.f of $A_\mu$ is a scalar, and one can thus introduce the dual photon $\phi$ through $\mathrm{d}\phi=\ast\mathrm{d}A$; it is periodic: $\phi\sim\phi+2\pi g^2$, with $g$ the gauge coupling.} $\phi$ and is identified with the Pontryagin dual: $G_C=\mathrm{Hom}(\pi_1(G),U(1))$ \cite{Bullimore:2016hdc}. 
For quiver gauge theories, 
$G_C=U(1)^{|\mathrm{Q}_0|}$.

\subsubsection{\texorpdfstring{Decomposing into $\mathcal{N}=2$ multiplets}{Decomposing into N=2 multiplets}}
\label{sssec:IntoN2}
Throughout the paper, we will describe the $\mathcal{N}=4$ multiplets using the $\mathcal{N}=2$ language, where only $U(1)_H\times U(1)_C$ is preserved (see App.~\ref{ssec:3D_N=4_SUSY}).
\begin{itemize}
\item For each $a\in \mathrm{Q}_0$, the associated 3D $\mathcal{N}=4$ vector-multiplet  is decomposed into 
\begin{equation}
\begin{aligned}
&\textrm{3D $\mathcal{N}=2$ v.m.} \quad (\sigma^{(a)}\,,\, \lambda^{(a)}\, ,\, A^{(a)}_{\mu}) \quad
\oplus \quad \textrm{3D $\mathcal{N}=2$ c.m.} \quad (\varphi^{(a)}\, , \eta^{(a)})\,,
\end{aligned}    
\end{equation}
where $\sigma^{(a)}$ and $\varphi^{(a)}$ are real and complex scalars, with $(\sigma^{(a)}, \varphi^{(a)}, \varphi^{(a)\dag})$ transforming as a triplet of $SU(2)_C$, and 
$(\lambda^{(a)}\, ,\, \eta^{(a)})$ are complex spinors transforming as a doublet under $SU(2)_H$.
The 3D $\mathcal{N}=2$  chiral multiplet $(\varphi^{(a)},\eta^{(a)})$ lives in the adjoint representation of the gauge factor $U(N^{(a)})$.

\item For each arrow $I^{a\rightarrow b}\in \mathrm{Q}_1$, the associated 3D $\mathcal{N}=4$ hypermultiplet is decomposed into:
\begin{equation}
\begin{aligned}
&\textrm{3D $\mathcal{N}=2$ c.m.} \quad (X_{I}, \psi^{X}_I )
\quad
\oplus \quad 
\textrm{3D $\mathcal{N}=2$ a.c.m.} \quad (\bar{Y}_I, \bar{\psi}^{Y}_I) \,,
\end{aligned}    
\end{equation}
where $(X_I,\bar{Y}_I)$ transform as a doublet under $SU(2)_H$, and the fermions $(\psi^X_I,\bar{\psi}^{Y}_I)$ transform as a doublet under $SU(2)_C$. 
The chiral multiplets $(X_I,\psi^X_I)$ (resp.\ $(Y_I,\psi^Y_I)$) live in a unitary representation $R$ of $U(N^{(b)})\times \overline{U(N^{(a)})}$ (resp.\ $U(N^{(a)})\times \overline{U(N^{(b)})}$ ).
Throughout this paper, we will choose $R$ to be the bi-fundamental representation.
\end{itemize}

As a result of this decomposition, the quiver encoding the 3D $\mathcal{N}=4$ gauge and matter content maps to an associated triple quiver $\widehat{Q}=(\widehat{Q}_0,\widehat{Q}_1; s,t)$, defined as follows:
\begin{equation}\label{eq:tripleQuiver}
\widehat{Q}_0:=\mathrm{Q}_0\,, \qquad \widehat{Q}_1:=\mathrm{Q}_1\cup \,  {}^{\textrm{op}}\mathrm{Q}_1\cup
	\{\mathrm{self\ loops}\}\,,
\end{equation}
where
\begin{equation}\label{eq:tripleQ1}
	{}^{\textrm{op}}\mathrm{Q}_1:=\{I^{b\rightarrow a}\, |\, I^{a\rightarrow b} \in \mathrm{Q}_1\}
	\qquad\textrm{and} \qquad
	\{\mathrm{self\ loops}\}:=\{I^{a\rightarrow a}\,|\, a\in \mathrm{Q}_0\};
\end{equation}
The triple quiver $\widehat{Q}$ encodes the 3D $\mathcal{N}=2$ multiplets as follows.
\begin{itemize}
\item Each node $a\in \widehat{Q}_0$ corresponds to a gauge factor $U(N^{(a)})$ together with a 3D $\mathcal{N}=2$ vector-multiplet $(\sigma^{(a)}, \lambda^{(a)}, A^{(a)}_{\mu})$.
	\item Each self-loop $I^{a\rightarrow a}$ corresponds to a 3D $\mathcal{N}=2$ chiral multiplet $(\varphi^{(a)},\eta^{(a)})$, living in the adjoint representation of $U(N^{(a)})$.
	\item Each pair of arrows $\{I^{a\rightarrow b},I^{b\rightarrow a}\}$ corresponds to a pair of 3D $\mathcal{N}=2$ chiral multiplets $\{(X_I,\psi^X_I),(Y_I,\psi^Y_I)\}$, living in the bi-fundamental representations of $U(N^{(b)})\times \overline{U(N^{(a)})}$ and $U(N^{(a)})\times \overline{U(N^{(b)})}$, respectively.
\end{itemize}

\subsubsection{Masses and FI-parameters}

In App.~\ref{appssec:Lagrangian} we present the Lagrangian of a general 3D $\mathcal{N}=4 $ gauge theory, together with its deformations in terms of masses and FI parameters. 
The real and complex masses, $m_{\mathbb{R}}$ and $m_{\mathbb{C}}$, can be viewed as the background values of the vector-multiplet scalars $\sigma$ and $\varphi$, respectively, associated to the Higgs branch symmetry $G_H$ (the flavor symmetry). 
Similarly, the real and complex FI parameters, $t_{\mathbb{R}}$ and $t_{\mathbb{C}}$, can be introduced as the background values of the scalars in the twisted $\mathcal{N}=4$ vector-multiplet, associated to the Coulomb branch  symmetry $G_C$.
For more details see App.~\ref{appssec:Lagrangian}.

\medskip

In this paper we will 
set $t_{\mathbb{C}}=0$ to preserve the $U(1)_H$ symmetry, and let $t_{\mathbb{R}}$ be non-zero to resolve the Higgs branch (see later). Then turning on mass parameters will localize the vacua to isolated fixed points. 
(We set $m_{\mathbb{R}}=0$ for simplicity, such that $\sigma$ is trivially solved by $\sigma=0$; and it is enough to turn on generic $m_{\mathbb{C}}\neq 0$ to remove all the flat directions.)
To summarize, for a quiver gauge theory, the masses and FI-parameters that we turn on are:
\begin{equation}
m_{\mathbb{C}} =\{m_i|i=1,2,\dots,N_{\mathtt{f}}\} \in \mathfrak{t}^{(H)}_{\mathbb{C}} 
\qquad\textrm{and}\qquad
t_{\mathbb{R}}=\{t^{(a)}_{\mathbb{R}}|a\in \textrm{Q}_0\}  \in \mathfrak{t}^{(C)}\,. 
\end{equation}

\subsection{\texorpdfstring{BPS states in the 3D $\mathcal{N}=4$ quiver gauge theories}{BPS states in the 3D N=4 quiver gauge theories}}\label{ssec:BPS_in_3D_N4}
One focus of this paper is the algebra of monopole operators, for which we will study various BPS objects in the 3D $\mathcal{N}=4$ theory: the vacua, the vortices, and the monopoles. They preserve the following supersymmetries \cite{Bullimore:2016hdc}:
\begin{equation}\label{eq:BPS_property}
\begin{aligned}
\text{Vacua:}&\quad \{\mathcal{Q}^{1\dot{1}}_{\pm},\mathcal{Q}^{1\dot{2}}_{\pm},\mathcal{Q}^{2\dot{1}}_{\pm},\mathcal{Q}^{2\dot{2}}_{\pm}\}\,,\\
\text{Vortices:}&\quad \{\mathcal{Q}^{1\dot{1}}_{-},\mathcal{Q}^{1\dot{2}}_{-},\mathcal{Q}^{2\dot{1}}_{+},\mathcal{Q}^{2\dot{2}}_{+}\}\,,\\
\text{Monopoles:}&\quad \{\mathcal{Q}^{1\dot{1}}_{-},\mathcal{Q}^{2\dot{1}}_{-},\mathcal{Q}^{1\dot{1}}_{+},\mathcal{Q}^{2\dot{1}}_{+}\}\,.\\
\end{aligned}
\end{equation}
For convenience we will use complex coordinates:
$t:=x^3$, $z=x^1+ix^2$ and $\bar{z}=x^1-ix^2$, 
and we will also compactify the $z$-plane to $\mathbb{CP}^1$.

\subsubsection{The vacua}

Now we review the supersymmetric vacua for a general 3D $\mathcal{N}=4$ gauge theory, and leave the specialization to the quiver gauge theories to Sec.~\ref{sec:vortex_tree_typeQGT}.
The supersymmetric vacua preserve all the eight supercharges $\mathcal{Q}^{a\dot{a}}_{\alpha}$; they can be obtained by requiring all the non-scalar fields to vanish, and the scalars to take certain constant values such that the total energy vanishes. 
These conditions can be translated into the following equations:
\begin{equation}\label{eq:SUSY-vacua}
\begin{aligned}
&\mu_{\mathbb{R}}(T)+t_{\mathbb{R}}=0\,, \qquad \mu_{\mathbb{C}}(T)+t_{\mathbb{C}}=0\,,\qquad [\varphi,\varphi^\dag]=[\varphi,\sigma]=0\,,\\
&\varphi X+X m_{\mathbb{C}}=\sigma X+X m_{\mathbb{R}} =0\,, \qquad Y\varphi+m_{\mathbb{C}}Y=Y\sigma+m_{\mathbb{R}}Y =0\,,
\end{aligned}    
\end{equation}
whose solutions, modulo the gauge group $G$, give the vacua of the theory. 
By $\mathcal{N}=4$ supersymmetry, the moduli space of vacua $\mathcal{M}$ is hyper-K\"{a}hler \cite{Alvarez-Gaume:1981exv,Crew:2021ipc}.
The moduli space contains the Higgs branch and the Coulomb branch. 
\begin{itemize}
\item Classically, the Higgs branch $\mathcal{M}_H$ is the branch with $m_{\mathbb{R}}=m_{\mathbb{C}}=0$ and $t_{\mathbb{R}},t_{\mathbb{C}}\neq 0$. Hence the equations involving $\varphi,\sigma$ are trivially solved by $\varphi=\sigma=0$, giving
\begin{equation}\label{eq:MHiggsDef}
\mathcal{M}_H=\{\mu_{\mathbb{R}}(T)+t_{\mathbb{R}}=\mu_{\mathbb{C}}(T)+t_{\mathbb{C}}=0\}/ G\subseteq \mathbb{C}^{2N}\,,
\end{equation}
where $N$ is the total number of hypermultiplets, and $\mu_{\mathbb{R}},\mu_{\mathbb{C}}\in \mathfrak{g}^*$ are the real and complex momentum maps of the $G$-action on the hypermultiplet scalars, given by:
\begin{equation}
\mu_{\mathbb{R}}(T)=\bar{X}TX-YT\bar{Y}\,,\qquad \mu_{\mathbb{C}}(T)= YT X\,,\qquad T\in \mathfrak{g}\,.
\end{equation}

Since the Higgs branch receives no quantum correction \cite{Seiberg:1994aj,Argyres:1996eh}, the classical description \eqref{eq:MHiggsDef} is actually exact.
There is another, equivalent, definition \cite{Bullimore:2016hdc}:
\begin{equation}
\mathcal{M}_H=\{\mu_{\mathbb{C}}(T)+t_{\mathbb{C}}=0\}^{\textrm{stab}}/ G_{\mathbb{C}}\,,
\end{equation}
where $G_{\mathbb{C}}$ is the complexified gauge group and a stability condition is imposed: $X\neq 0$ (resp.\ $Y\neq 0$) when $t_{\mathbb{R}} < 0$ (resp.\ $t_{\mathbb{R}}> 0$). 
The coordinate ring of the Higgs branch is the $\frac{1}{2}$-BPS chiral ring $\mathbb{C}[\mathcal{M}_H]$, annihilated by the supercharges with $SU(2)_H$ charge $+\frac{1}{2}$, namely $\mathcal{Q}^{1\dot{a}}_{\alpha}$
\cite{Bullimore:2018gva}. 
For quiver gauge theories, which we consider in this paper, the Higgs branch is identified with the corresponding Nakajima quiver variety \cite{Nakajima:1994nid}.

\item The (classical) Coulomb branch $\mathcal{M}_C^{\text{classical}}$ is the branch where the $m_{\mathbb{R},\mathbb{C}}$ are generic but $t_{\mathbb{R}}=t_{\mathbb{C}}=0$, and it is parameterized by the VEV's of the vector-multiplet scalars. They take their values in the Cartan subalgebra of the gauge algebra $\mathfrak{g}$.
However, this description is not yet complete since the dual photon is missing. 
This problem can be resolved after we take into account the monopole operators, see below.
In addition, unlike the Higgs branch, the classical Coulomb branch will be corrected upon quantization.\footnote{For explicit computations of the corrections, see e.g.\ \cite{Intriligator:1996ex,Seiberg:1996nz}.} 
Taking both into account, the quantized Coulomb branch $\mathcal{M}_C$ is the fibration \cite{bullimore2015coulomb,Nakajima:2015txa,Braverman:2016wma}:
\begin{equation}
\begin{aligned}
T^{\vee}_{\mathbb{C}}\rightarrow{}&\mathcal{M}_C\\[-1ex]
&\ \hspace{-.2ex}\downarrow\\[-.5ex]
&\mathfrak{t}_{\mathbb{C}}/W
\end{aligned}
\end{equation}
where $T^{\vee}_{\mathbb{C}}$ is the complexified dual of the maximal torus of $G$, and in the base $\mathfrak{t}_{\mathbb{C}}/W$, $\mathfrak{t}_{\mathbb{C}}$ is the complexified Cartan subalgebra of $\mathfrak{g}$ and $W$ is the Weyl group of $G$. 
The base space can be parameterized by the gauge-invariant $\varphi$-polynomials and the fibers are parametrized by the VEV's of the monopole operators. 
Together they generate the $\frac{1}{2}$-BPS chiral ring $\mathbb{C}[\mathcal{M}_C]$, which is annihilated by the supercharges with $SU(2)_C$ charge $+\frac{1}{2}$, namely $\mathcal{Q}^{a\dot{1}}_{\alpha}$. 
\end{itemize}

\medskip
To summarize, the Higgs branch is obtained as the solution of the vacuum equations \eqref{eq:SUSY-vacua} with $m_{\mathbb{R}}=m_{\mathbb{C}}=0$ and $t_{\mathbb{R}},t_{\mathbb{C}}\neq 0$, while the Coulomb branch is obtained as the solution of \eqref{eq:SUSY-vacua} with $m_{\mathbb{R}},m_{\mathbb{C}}$ generic but $t_{\mathbb{R}}=t_{\mathbb{C}}= 0$. 
In this paper we will consider the \textit{mixed branch}, in which
\begin{equation}\label{eq:Mixed}
m_{\mathbb{C}}\,, t_{\mathbb{R}}\neq 0
\qquad\textrm{and}\qquad
m_{\mathbb{R}}=t_{\mathbb{C}}=0\,.
\end{equation}
As a result, the vacua are localized to isolated points. This mixed branch can be viewed as the fixed points on the Higgs branch under the infinitesimal $G\times G_H$ transformations with parameters $(\varphi,m_{\mathbb{C}})$ and $(\sigma,m_{\mathbb{R}})$, see the $2^{\text{nd}}$ line of \eqref{eq:SUSY-vacua}.

\subsubsection{Vortices and the vortex moduli space}\label{sssec:vortex_moduli}

The $\frac{1}{2}$-BPS vortices preserve half of the supersymmetry, $\mathcal{Q}^{1\dot{a}}_{-}$ and $\mathcal{Q}^{2\dot{a}}_{+}$. 
They are invariant under a diagonal $U(1)_H\times U(1)_E$ transformation.
The corresponding BPS equations are
\begin{align}
&\partial_t(X,Y,\varphi,\sigma,A_\mu)=0\,,\quad \label{eq:vortex-t-indep}\\
&D_z\varphi=D_{\bar{z}}\varphi=D_z\sigma=D_{\bar{z}}\sigma=0\,,\quad [\varphi,\varphi^\dagger]=[\varphi,\sigma]=0\,,\label{eq:vortex_constant}\\
&D_{\bar{z}}X=D_{\bar{z}}Y=0\,,\quad 4[D_z,D_{\bar{z}}]=\mu_{\mathbb{R}}(T)+t_{\mathbb{R}}\,,\quad \mu_{\mathbb{C}}(T)+t_{\mathbb{C}}
=0\,,\label{eq:vortex_moduli}\\
&(\varphi+m_{\mathbb{C}})X=(\sigma+m_{\mathbb{R}}) X=0\,, \quad Y(\varphi+m_{\mathbb{C}})=Y(\sigma+m_{\mathbb{R}}) =0\,,\label{eq:vortex_constraint}
\end{align}
\noindent where we have adopted the temporal gauge $A_t=0$. 
(One can derive these by combining the $\frac{1}{4}$-BPS conditions imposed by $(\mathcal{Q}^{1\dot{1}}_{-}, \mathcal{Q}^{2\dot{1}}_{+})$, see \eqref{eq:14BSPhyper} -- \eqref{eq:14BPSremaining}, and those by $(\mathcal{Q}^{1\dot{2}}_{-}, \mathcal{Q}^{2\dot{2}}_{+})$, which are related to the previous ones by a $SU(2)_C$ rotation.)
The vortices are \textit{time-independent} solutions (see \eqref{eq:vortex-t-indep}), behaving like strings aligned parallel to the $t$-axis, and are topologically classified by their vortex numbers. 
(We will study them more carefully below, as they form a natural representation of the Coulomb branch chiral ring $\mathbb{C}[\mathcal{M}_C]$.)

\medskip

Let us unpack this set of equations. 
\begin{enumerate}
\item \eqref{eq:vortex-t-indep} merely says that the vortices are time-independent.
\item Together with \eqref{eq:vortex-t-indep},  \eqref{eq:vortex_constant} states that the scalars $\varphi,\sigma$ are constant and live in the Cartan subalgebra of $\mathfrak{g}$.

\item After taking into account \eqref{eq:vortex-t-indep}, the solution of \eqref{eq:vortex_moduli} gives the  vortex moduli space. 
We will consider vortices approaching supersymmetric vacuum $\nu$ at spatial infinity in the $x^{1,2}$-plane, i.e.\ $|z|\to\infty$, and the moduli space of such configurations, $\mathcal{M}_\nu$, is given by:
\begin{equation}\label{eq:vor_mod1}
\mathcal{M}_\nu=\left\{
\begin{array}{lcc}
D_{\bar{z}}X=0\, \quad D_{\bar{z}}Y=0\,,& &  \\
4[D_z,D_{\bar{z}}]=\mu_{\mathbb{R}}(T)+t_{\mathbb{R}}\,,&\qquad\mathrm{with}&X,Y\overset{|z|\to\infty}{\longrightarrow}G\cdot\nu \\
\mu_{\mathbb{C}}(T)=0\,,& &
\end{array}
\right\}/\mathcal{G}\,,
\end{equation}
where $(X,Y)$ are time-independent, $G\cdot \nu$ is the $G$-orbit of $\nu$, and
the group $\mathcal{G}$ consists of all gauge transformations that are constant at infinity, see also \cite{Bullimore:2016hdc}.
Each point on $\mathcal{M}_\nu$ specifies a principal $G$-bundle over $\mathbb{CP}^1$ 
with holomorphic sections $(X,Y)$ of the associated bundles in the symplectic representation $(R,\bar{R})$ of $G$, satisfying the defining equations in \eqref{eq:vor_mod1}.
	
\medskip
	
There is a mathematically equivalent definition of the vortex moduli space, in which the gauge group is complexified and as a result the equation involving the real momentum map becomes a gauge-fixing condition and can be dropped from the definition:
\begin{equation}\label{eq:MnuC}
\mathcal{M}_\nu\simeq \left\{
\begin{array}{lcc}
D_{\bar{z}}X=0\,, \quad
D_{\bar{z}}Y=0\,,& &  \\
\mu_{\mathbb{C}}(T)=0\,,&\qquad\mathrm{with}&X,Y\overset{|z|\to\infty}{\longrightarrow}G_{\mathbb{C}}\cdot\nu 
\end{array}
\right\}/\mathcal{G}_{\mathbb{C}}\,.
\end{equation}
For convenience, we will choose the following gauge. 
First, by a complex gauge transformation we set $D_{\bar{z}}= \partial_{\bar{z}}$, and thus $X$ and $Y$ become holomorphic polynomials. 
After this, there is still a gauge redundancy that consists of holomorphic polynomial-valued gauge group elements $g(z)$, and the moduli space is defined modulo this redundancy. 
Each point in $\mathcal{M}_\nu$ specifies a $G_{\mathbb{C}}$-bundle with the holomorphic sections $(X,Y)$ of the associated bundles in the representation $(R,\bar{R})$, satisfying $\mu_{\mathbb{C}}(T)=0$ and approaching the vacuum $\nu$ at infinity.

The vortices are topologically classified by the vortex numbers $\nn$, defined in terms of the first Chern class of the principal $G_{\mathbb{C}}$-bundles: 
\begin{equation}
\nn=\frac{1}{2\pi}\int_{\mathbb{CP}^1} \mathrm{Tr}(F)
\quad \in \quad
\pi_{1}(G)\,,
\end{equation}
and $\mathcal{M}_\nu$ can be decomposed into different components labeled by their vortex numbers:
\begin{equation}
\mathcal{M}_\nu=\bigcup_\nn \mathcal{M}_\nu^\nn \,.
\end{equation}

\item Finally, eq.~\eqref{eq:vortex_constraint}
should be interpreted as the constraints, which will localize the vortex configurations to certain fixed points on the vortex moduli space.   
The combination $\varphi+m_{\mathbb{C}}$ can be interpreted as an infinitesimal $T_H$ transformation, compensated by a gauge transformation. 
The Hilbert space of the $\frac{1}{2}$-BPS vortices with boundary condition $\nu$ is denoted by $\mathcal{H}_\nu$, and comprises the fixed points of the $T_{H}$-action on $\mathcal{M}_\nu$.
\end{enumerate}

\subsubsection{Monopoles}\label{sec:monopole}

As noted before, in order to characterize the Coulomb branch of a 3D $\mathcal{N}=4$ gauge theory, besides the local operators (namely the  vector-multiplet scalars $\varphi$ and $\sigma$), we also need to include the dual photon $\phi$, which can be captured by the $\frac{1}{2}$-BPS monopole operators.

The BPS equations for the monopoles are obtained by requiring the invariance under $\mathcal{Q}^{a\dot{1}}_\alpha$, see \eqref{eq:Q11VM}, \eqref{eq:Q11HM} and \eqref{eq:Q21VM}, \eqref{eq:Q21HM}. 
First, with vanishing FI parameters, the equations involving hypermultiplet scalars are trivially solved, i.e.\ by $X=Y=0$. The non-trivial part gives the monopole equations:\footnote{Or in components, 
$[D_{\bar{z}},\mathcal{D}_t]=0$, $[D_{z},\mathcal{D}^{\dag}_t]=0$, and $4[D_z,D_{\bar{z}}]=[\mathcal{D}_t,\mathcal{D}_t^{\dag}]$.
}
\begin{equation}
F=\ast D\sigma\,,
\end{equation}
\noindent
together with the constraints $D\varphi=[\varphi,\varphi^\dag]=[\varphi,\sigma]=0$.
The equations with non-zero FI parameters are similar but more complicated and we omit them here.
(The derivation is similar to the vortex case: one combines the $\frac{1}{4}$-BPS condition imposed by $(\mathcal{Q}^{1\dot{1}}_{-}, \mathcal{Q}^{2\dot{1}}_{+})$, see \eqref{eq:14BSPhyper} -- \eqref{eq:14BPSremaining}, and those by $(\mathcal{Q}^{2\dot{1}}_{-}, \mathcal{Q}^{1\dot{1}}_{+})$, which are related to the previous ones by a $SU(2)_H$ rotation.) 

The monopoles are characterized by the magnetic charge $A\in \Lambda_{\text{cocharacter}}:=\mathrm{Hom}(U(1),G)$ (modulo the Weyl group). 
On a small $S^2$ of radius $r$ surrounding a monopole with charge $A$, we have
\begin{equation}
F\sim A\, \mathrm{sin}\theta\ \mathrm{d}\theta\wedge \mathrm{d}\phi
\qquad\textrm{and}\qquad \sigma\sim \frac{A}{r}\,,
\end{equation}
where $(\theta,\phi)$ are the coordinates on the sphere. 
The monopole operator $\hat{v}_A$ is defined as the operator that creates such a monopole configuration. 
Particularly important to us are the monopole operators with $A$ being minuscule cocharacters,\footnote{Recall that a minuscule cocharacter is defined as the cocharacter whose pairing with any root equals $0,\pm 1$; it can be viewed as the ``smallest" cocharacter. 
In particular, for GL$_N$,  any dominant cocharacter
can be written as a sum of minuscule ones.} and they are called ``minuscule monopole operators". 
For the quiver gauge theories with $U(N)$ factors, which are the focus of this paper, the minuscule monopole operators (together with vector-multiplet scalars) generate the entire chiral ring \cite{bullimore2015coulomb,Braverman:2016wma}.
We will focus on minuscule monopole operators in most of the paper and will drop ``minuscule" when there is no risk of confusion. 

\medskip

The monopole operators and the vector-multiplet scalars together generate the $\frac{1}{2}$-BPS Coulomb branch  chiral ring $\mathbb{C}[\mathcal{M}_C]$.
The Hilbert series of $\mathbb{C}[\mathcal{M}_C]$, which counts the chiral operators (i.e.\ the gauge-invariant monopole operators dressed by the vector-multiplet scalars) graded by their conformal dimensions and quantum numbers w.r.t.\ the global symmetries, is computed by the monopole formula of \cite{Cremonesi_2014}.
We will be interested in the algebraic relations among these operators.

For each minuscule cocharacter $A\in \mathrm{Hom}(U(1),G)$, there exists a pair of minuscule monopole operators with opposite charges $\pm A$. Explicitly, they are given by
\begin{equation}
\hat{v}^{\pm}_A=e^{\pm\frac{1}{g_A^2}(\sigma_A+i\phi_A)}\,,
\end{equation}
where $g_A,\sigma_A$, and $\phi_A$ are the gauge coupling, the real vector-multiplet scalar, and the dual photon associated to the $U(1)_A$ factor of $G$, respectively. 
Later, we will study the action of the (first fundamental) minuscule monopole operators on the $\frac{1}{2}$-BPS vortex configurations, and derive the quantum Coulomb branch algebra from their action (see Sec.~\ref{ssec:monopole_operator}).

\subsection{\texorpdfstring{$\Omega$-deformation and $\mathcal{N}=4$ quantum mechanics}{Ω-deformation and N=4 quantum mechanics}}\label{ssec:review_vortex_QM}

One way to canonically quantize the Poisson algebra of the Coulomb branch chiral ring is to place the 3D theory on the $\Omega$-background $\mathbb{R}\times \mathbb{R}^2_{\epsilon}$ \cite{Nekrasov:2002qd,Nekrasov_2010}, which effectively compactifies the 3D theory into an $\mathcal{N}=4$ quantum mechanics \cite{Bullimore:2016hdc}.

\subsubsection{\texorpdfstring{$\Omega$-deformation}{Ω-deformation}}\label{sssec:omega_defor}

Consider the vector that rotates the $x^{1,2}$-plane:
\begin{equation}\label{eq:omega}
V=x^1\partial_2-x^2\partial_1=i(z\partial_z-\bar{z}\partial_{\bar{z}})\,.
\end{equation}
We will turn on the $\Omega$-deformation associated to this vector (with the equivariant parameter $\epsilon$), which effectively localizes onto the fields invariant under the rotation \eqref{eq:omega} and compactifies the theory to the origin of the $x^{1,2}$-plane. 
As a result, the vortices become line-segments along the $t$-axis (at $x^{1,2}=0$), the monopoles sit at $x^{1,2}=0$ and some points in $t$.
Different vortices are separated by the monopoles. 

Therefore, turning on the $\Omega$-deformation allows one to define a non-commutative algebra of monopole operators, by considering their time-ordered action on the vortices. 

\subsubsection{\texorpdfstring{$\mathcal{N}=4$ quantum mechanics}{N=4 quantum mechanics}}

The four supercharges of the vortex (see \eqref{eq:BPS_property}) give rise to the four super-coordinates of the quantum mechanics $\{\theta^\pm,\bar{\theta}^\pm\}$:
\begin{equation}
\mathcal{Q}^{1\dot{1}}_{-}\rightarrow \bar{\theta}^+\,,\quad
\mathcal{Q}^{1\dot{2}}_-\rightarrow\bar{\theta}^-\,,\quad
\mathcal{Q}^{2\dot{1}}_+\rightarrow\theta^-\,,\quad
\mathcal{Q}^{2\dot{2}}_+\rightarrow \theta^+\,,
\end{equation}
where we have adopted the $\mathcal{N}=(2,2)$ convention. 
For convenience we list the $ U(1)_H \times U(1)_C\times U(1)_E$ charges of these super-coordinates:
\begin{equation}
\begin{array}{c|cccc}
&\theta^+ &\bar{\theta}^+&\theta^- &\bar{\theta}^-\\
\hline  
U(1)_H&\frac{1}{2}&-\frac{1}{2}&\frac{1}{2}&-\frac{1}{2}\\
U(1)_C&\frac{1}{2}&-\frac{1}{2}&-\frac{1}{2}&\frac{1}{2}\\
U(1)_E &-\frac{1}{2}&\frac{1}{2}&-\frac{1}{2}&\frac{1}{2}
\end{array}\,.
\end{equation}
The corresponding super-derivatives are:
\begin{equation}
\mathtt{D}_\pm=\frac{\partial}{\partial\theta^\pm}+i\bar{\theta}^\pm \partial_t\,,\qquad \bar{\mathtt{D}}_\pm=\frac{\partial}{\partial\bar{\theta}^\pm}+i\theta^\pm\partial_t\,.
\end{equation}

The 3D scalar fields $X,Y$ together with the spinor components $\psi^X_+,\bar{\psi}^X_+,\psi^Y_+,\bar{\psi}^Y_+$ now sit inside the 1D $\mathcal{N}=4$ chiral superfields:
\begin{equation}\label{eq:vqm_XY}
\begin{aligned}
\Phi_X&=X+\theta^+\psi^X_+ +\theta^-\bar{\psi}^X_++\dots+\theta^+\theta^-F_X\,,\\
\Phi_Y&=Y+\theta^+\psi^Y_+ +\theta^-\bar{\psi}^Y_++\dots+\theta^+\theta^-F_Y\,,
\end{aligned}
\end{equation}
where $F_X,F_Y$ are the auxiliary fields. 
The 3D gauge field $A_{\bar{z}}$, together with the spinor components $\bar{\lambda}_-,\eta_-$ are also re-packaged into a 1D chiral superfield:
\begin{equation}\label{eq:vqm_Azb}
\Phi_{A_{\bar{z}}}=A_{\bar{z}}+\theta^+\bar{\lambda}_- +\theta^-\eta_-+\dots+\theta^+\theta^-F_{A_{\bar{z}}}\,,
\end{equation}
with $F_{A_{\bar{z}}}$ the auxiliary field. 
The 1D superpotential is given by
\begin{equation}\label{eq:vqm_superpotential}
W=\int \mathrm{d}^2z\ \Phi_Y D_{\bar{z}}\Phi_X\,,
\end{equation}
where $D_{\bar{z}}=\partial_{\bar{z}}-i\Phi_{A_{\bar{z}}}$. Finally, the remaining 3D fields are contained in the 1D vector-multiplet
\begin{equation}\label{eq:vqm_vector}
\begin{aligned}
V= \theta^-\bar{\theta}^+\varphi&+\theta^+\bar{\theta}^-\varphi^\dag+ \theta^-\bar{\theta}^+(\bar{\theta}^- \lambda_- +\theta^+\eta_+)-\theta^+\bar{\theta}^-(\theta^- \bar{\lambda}_+ +\bar{\theta}^+\bar{\eta}_-)\\
&+\theta^+\bar{\theta}^+(A_t+i\sigma)+\theta^-\bar{\theta}^-(A_t-i\sigma)+\theta^4D\,,
 \end{aligned}
\end{equation}
where we adopt the Wess-Zumino gauge, with $A_t$ the gauge field in 1D and $D$ the auxiliary field. 
From $V$ one can define a twisted chiral multiplet $\Sigma=\bar{\mathtt{D}}_+\mathtt{D}_-V$:
\begin{equation}
\Sigma=\varphi+\bar{\theta}^-\lambda_- +\theta^+\eta_+ +\theta^+\bar{\theta}^-(D+[\mathcal{D}_t,\mathcal{D}_t^\dag])+\dots\,,
\end{equation}
and the K\"{a}hler potential is given by
\begin{equation}\label{eq:vqm_kahler}
K=\int\mathrm{d}^2z\left(\mathrm{Tr}\Sigma\Sigma^\dag +|e^{\frac{V}{2}}\Phi_X|^2+|e^{-\frac{V}{2}}\Phi_Y|^2+\mathrm{Tr}|e^{\frac{V}{2}}D_{\bar{z}}e^{-\frac{V}{2}}|^2\right)\,.
\end{equation}

\smallskip

In 3D, the complex mass $m_{\mathbb{C}}$ is the background value of $\varphi$, associated to the flavor symmetry $G_H$; this remains true in 1D.
Recall that in 3D, the real FI parameter $t_{\mathbb{R}}$ is introduced via  \eqref{eq:coupleTV-V} where $\tilde{\Sigma}$ is the $\mathcal{N}=2$ vector multiplet contained in the $\mathcal{N}=4$ twisted vector-multiplet; in 1D, the real FI parameter $t_{\mathbb{R}}$ is introduced via the twisted superpotential:
\begin{equation}
\widetilde{W}=\int\mathrm{d}^2 z \, t_{\mathbb{R}}\Sigma\,.
\end{equation}

\smallskip

Finally, one can check that the ground states of the $\mathcal{N}=4$ quantum mechanics describe the vortex configurations \eqref{eq:vor_mod1}:
$D_{\bar{z}}X=D_{\bar{z}}Y=0$ and $YTX=:\mu_{\mathbb{C}}(T)=0$ are the $F$-term equations, derived from the superpotential $W$; $4[D_{z},D_{\bar{z}}]=\mu_{\mathbb{R}}(T)+t_{\mathbb{R}}$ is the $D$-term equation; 
the other constraints such as $D_{\bar{z}}\varphi=(\varphi+m_{\mathbb{C}})X=
Y(\varphi+m_{\mathbb{C}})=0$ arise as mass terms from the K\"{a}hler potential. 
Namely, the vortex configurations in the 3D theory correspond to the supersymmetric vacua of the $\mathcal{N}=4$ quantum mechanics.

\medskip

This quantum mechanics theory can be decomposed into different components:
\begin{equation}
\bigcup_{\nn} \textrm{QM}(\nn,\nu),
\end{equation}
where each component is labeled by the vortex number $\nn$ and the boundary condition $\nu$ of the ground state, which corresponds to a vortex configuration of the 3D theory. 
We will call each QM$(\nn,\nu)$ a \textit{vortex quantum mechanics}, with vortex number $\nn$ and boundary condition $\nu$.
The Higgs branch of $\textrm{QM}(\nn,\nu)$ reproduces the vortex moduli space $\mathcal{M}_\nu^\nn$, and the monopoles can be formulated as interfaces between different quantum mechanics theories $\textrm{QM}(\nn,\nu)$ and $\textrm{QM}(\nn',\nu)$ \cite{Bullimore:2016hdc}.

\medskip

\subsection{Hilbert space of the vortex states}\label{ssec:vortex_Hilbert}
	
We now describe the Hilbert space, $\mathcal{H}_\nu$, of the vortex states asymptotically approaching the vacuum $\nu$.

The moduli space $\mathcal{M}_\nu$ of such configurations is given in \eqref{eq:vor_mod1}, but we still need to impose the constraints in \eqref{eq:vortex_constant} and \eqref{eq:vortex_constraint}. 
The first set of equations means that 
the vector-multiplet scalars $\varphi$ and $\sigma$ are constants, and they live in the (complexified) Cartan subalgebra of $\mathfrak{g}$:
$\varphi\in \mathfrak{t}_{\mathbb{C}}$, $\sigma \in \mathfrak{t}$.
For the second set of equations, since $m_{\mathbb{R}}=0$, the constraints $\sigma X=Y\sigma=0$ are trivially solved by $\sigma=0$.\footnote{In Sec.~\ref{sssec:vortex_Hilbert_space} we will see that the trivial solution $\sigma=0$ is in fact the only solution.}
Finally, we are left with the following constraints
\begin{equation}\label{eq:vor_non_defor}
(\varphi+m_{\mathbb{C}}) X=0\,,\qquad Y(\varphi+m_{\mathbb{C}}) =0\,.
\end{equation}
Turning on the $\Omega$-deformation will deform these two conditions into \cite{Bullimore:2018gva}:
\begin{equation}\label{eq:vor_defor}
\begin{aligned}
\
&(\varphi+m_{\mathbb{C}}+\epsilon z\partial_z+\frac{\eps}{2})X=0
\quad \textrm{and} \quad
Y(\varphi+m_{\mathbb{C}}+\epsilon z\overleftarrow{\partial_z}+\frac{\eps}{2})=0\,.
\end{aligned}
\end{equation}
In the vortex quantum mechanics, $\epsilon$ can be understood as the background value\footnote{The background value $\epsilon$ can be introduced by the replacement
$\varphi+m_{\mathbb{C}}\Longrightarrow \varphi+m_{\mathbb{C}}+\epsilon (z\partial_z-\bar{z}\partial_{\bar{z}})+\epsilon r_H$.
Since $X,Y$ are both holomorphic and have $U(1)_H$ charge $+\frac{1}{2}$, this explains the equations \eqref{eq:vor_defor}.} of the vector-multiplet scalar $\varphi$ associated to a new flavor symmetry, $U(1)_\epsilon$, which is a diagonal combination of $U(1)_H$ and $U(1)_E$:
\begin{equation}
U(1)_\epsilon=\{e^{i\epsilon r_H} \times e^{i\epsilon r_E}|\epsilon\in\mathbb{R}\}\,,
\end{equation}
where $r_H$ and $r_E$ are the generators of the  $U(1)_H$ and the $U(1)_E$, respectively. 
Under $U(1)_\eps$ the four supercharges $\mathcal{Q}^{1\dot{a}}_-$ and $\mathcal{Q}^{2\dot{a}}_+$ are invariant.

Through the relation between the supersymmetric vacua and the cohomologies in supersymmetric quantum mechanics \cite{Witten:1982im}, one can identify the vortex Hilbert space $\mathcal{H}_\nu$ with the equivariant cohomology of the moduli space $\mathcal{M}_\nu$ \cite{Bullimore:2016hdc}:
\begin{equation}
\mathcal{H}_\nu=H^*_{T_{H}\times U(1)_\epsilon}(\mathcal{M}_\nu,\mathbb{C})\,,
\end{equation}
which has a basis consisting of the fixed points of the ${T_{H}\times U(1)_\epsilon}$ action on $\mathcal{M}_\nu$, given by \eqref{eq:vor_defor}. 
Finally, let us also fix the normalization of states. For two states $|p),|p')\in \mathcal{H}_\nu$, the inner product is given by:
\begin{equation}\label{eq:vortex_normal}
( p'|p)=\frac{1}{\omega_p(m_{\mathbb{C}},\epsilon)}\delta_{pp'}\,,
\end{equation}
where $\omega_p(m_{\mathbb{C}},\epsilon)$ is the equivariant weight of the tangent space of $\mathcal{M}_\nu$ at the fixed point $p$.

\subsection{Quantum Coulomb branch algebra}
\label{ssec:CBalgebra}

The Coulomb branch chiral ring $\mathbb{C}[\mathcal{M}_C]$, as the coordinate ring  of the Coulomb branch $\mathcal{M}_C$, is generated by the holomorphic functions of the VEV's of the monopole operators and the complex scalars $\varphi$ from the vector-multiplets. 
Below we will briefly review various approaches to quantize $\mathbb{C}[\mathcal{M}_C]$ into a non-commutative algebra, namely the quantum Coulomb branch algebra $\mathbb{C}_\eps[\mathcal{M}_C]$.

\subsubsection{Quantization of Coulomb branch chiral ring}
\label{sssec:DeformationQuantization}

One approach to construct the Coulomb branch algebra is to endow the commutative ring $\mathbb{C}[\mathcal{M}_C]$ with a Poisson bracket and then canonically quantize the resulting Poisson algebra into a non-commutative algebra $\mathbb{C}_\eps[\mathcal{M}_C]$ \cite{bullimore2015coulomb}.
Physically, this can be realized by placing the theory in the $\Omega$-background $\mathbb{R}\times \mathbb{R}^2_{\epsilon}$.

Recall that by writing the $\mathcal{N}=4$ supersymmetry in $\mathcal{N}=2$ language, we have chosen one complex structure out of the triplet of complex structures in the hyper-K\"ahler manifold $\mathcal{M}_C$, thus viewing $\mathcal{M}_C$ as a complex symplectic manifold.

For abelian theories, the Coulomb branch receives no non-perturbative correction (due to the absence of dynamical monopoles). Furthermore it only has 1-loop corrections, and hence its geometry can be obtained explicitly from perturbative computations \cite{Seiberg_1996,Seiberg:1996nz,Intriligator:1996ex}. The Poisson brackets of the Coulomb branch operators can then be deduced from the complex symplectic form.

For non-abelian theories, the geometry of the Coulomb branch is difficult to study due to non-perturbative corrections from monopoles. 
However, the Poisson brackets of the Coulomb branch operators can be obtained via the ``abelianization map" of \cite{bullimore2015coulomb}, namely by embedding the Coulomb branch chiral ring $\mathbb{C}[\mathcal{M}_C]$ into a larger Poisson algebra, $\mathbb{C}[\mathcal{M}^{\text{abel}}_C]$, which consists of holomorphic functions on the ``abelian patch" of the Coulomb branch, i.e.\ the loci where the non-abelian gauge symmetry is broken to abelian factors \cite{bullimore2015coulomb}. 
The Poisson algebra structure of $\mathbb{C}[\mathcal{M}^{\text{abel}}_C]$ can be readily computed for the abelian theory and it induces the Poisson algebra structure of the chiral ring $\mathbb{C}[\mathcal{M}_C]$, whose canonical quantization then gives rise to the quantum Coulomb branch algebra $\mathbb{C}_\eps[\mathcal{M}_C]$.

\medskip

Apart from the field-theoretic construction above, there is an algebro-geometric formulation of the Coulomb branch and the algebra, developed in \cite{Nakajima:2015txa,Braverman:2016wma,Braverman:2016pwk}. 
The key idea is to consider the equivariant Borel-Moore homology over a certain moduli space of the triple $\mathcal{R}_{G,R}$, associated to the gauge group $G$ and the symplectic
representation $(R,\bar{R})$. 
Then one defines the appropriate convolution algebra over the equivariant homology space, and finally quantizes it via a non-commutative deformation.
Let us sketch the construction here.  

First, let us denote $\mathcal{O}:=\mathbb{C}[[z]]$ (the formal power series of $z$) and  $D:=\text{Spec}[\mathcal{O}]$ (the formal disk), and  denote $\mathcal{K}:=\mathbb{C}((z))$  (the field of fractions of $\mathcal{O}$)  and $D^*:=\text{Spec}[\mathcal{K}]=D-\{0\}$ (the punctured disk). 
Let us denote the $\mathcal{O}$- and $\mathcal{K}$-valued versions of $G_{\mathbb{C}}$ (the complexification of $G$) by $G_{\mathbb{C}}^{\mathcal{O}}$ and $G_{\mathbb{C}}^{\mathcal{K}}$, respectively. 
The affine Grassmannian $\text{Gr}_{G_{\mathbb{C}}}:=G_{\mathbb{C}}^{\mathcal{K}}/G_{\mathbb{C}}^{\mathcal{O}}$ can be viewed as the moduli space of the pair $(\mathscr{P}, \varphi)$ where $\mathscr{P}$ is an algebraic $G$-principal bundle over $D$ and $\varphi$ is a trivialization of $\mathscr{P}$ over $D^{*}$.

The moduli space of the triple $(\mathscr{P},\varphi,s)$, denoted as $\mathcal{R}_{G,R}$, can be considered as the generalization of $\text{Gr}_{G_{\mathbb{C}}}$ in the presence of matter.
The additional information $s$ is a section of the associated vector bundle in the representation $R$ on $D^{*}$ such that $s$ extends to a regular section $\mathscr{P}_{\textrm{triv}}\times_G(R,\bar{R})$ on $D$, which, under $\varphi$, is extended to a regular section $\mathscr{P}\times_G(R,\bar{R})$ on $D$. 
The moduli space is defined on the raviolo $\tilde{D}:=D_1\cup_{D^*}D_2$ (two formal disks $D_{1,2}$ glued over a formal punctured disk $D^*$). 
The reason to  pass from $\mathbb{P}^1$ to the raviolo $\tilde{D}$ is to consider the convolution product (defined as the pullback-pushforward in the correspondence space), and in this process the graded dimension of the cohomology space does not change.

\medskip

There is a natural action of $G^{\mathcal{O}}_{\mathbb{C}}$ on $\mathcal{R}_{G,R}$  given by varying the trivialization $\varphi$. 
In \cite{Nakajima:2015txa} it is proposed that the space of holomorphic functions on the Coulomb branch $\mathcal{M}_C$ is the $G^{\mathcal{O}}_{\mathbb{C}}$-equivariant Borel-Moore homology of $\mathcal{R}_{G,R}$, denoted by $H^{G_{\mathbb{C}}^{\mathcal{O}}}_*(\mathcal{R}_{G,R})$. 
In \cite{Braverman:2016wma} it is further proven that one can define a convolution product
$\star$ on this space, under which it is promoted to a commutative algebra:
\begin{equation}
	\mathcal{A}=(H^{G_{\mathbb{C}}^{\mathcal{O}}}_*(\mathcal{R}_{G,R}),\star)\,.
\end{equation}
For the rigorous definition on the convolution product $\star$ see  \cite[Sec.~3]{Braverman:2016wma}. 
Furthermore, $\mathcal{A}$ is a graded algebra, and the graded dimensions agree perfectly with the monopole formula \cite{Nakajima:2015txa}. 
The Coulomb branch algebra of the 3D $\mathcal{N}=4$ theory, with gauge group $G$ and matter in the representation $(R,\bar{R})$, is defined as the algebra $\mathcal{A}$, and the Coulomb branch $\mathcal{M}_C$ is obtained as its spectrum
\begin{equation}
	\mathcal{M}_C=\text{Spec}[\mathcal{A}]\,.
\end{equation}

To quantize the commutative algebra $\mathcal{A}$, we need to include the loop rotation group $C^{\times}$, 
which acts on the disk $D$ by 
$z\to tz$ with $t\in C^{\times}$,
and consider the $(G_{\mathbb{C}}^{\mathcal{O}}\rtimes C^{\times})$-equivariant Borel-Moore homology $H^{G_{\mathbb{C}}^{\mathcal{O}}\rtimes C^{\times}}_*(\mathcal{R}_{G,R})$. 
The quantum Coulomb branch algebra $\mathcal{A}_{\hbar}$ is defined as \cite{Braverman:2016wma}
\begin{equation}
\mathcal{A}_{\hbar}:=(H^{G_{\mathbb{C}}^{\mathcal{O}}\rtimes C^{\times}}_*(\mathcal{R}_{G,R}),\star)\,,
\end{equation}
which is a graded $\mathbb{C}[\hbar]$-algebra, with $\hbar$ the deformation parameter.

The construction above applies to general 3D $\mathcal{N}=4$ gauge theories, with arbitrary compact gauge group $G$ and matter in the symplectic representation $(R,\bar{R})$. 
One can then extract the explicit algebraic relations in terms of generators from this definition.
For generic quantum parameters, and for ADE-type quivers, BCFG-type quivers, and the Jordan quiver, it was shown in \cite{Braverman:2016pwk}, \cite{Nakajima:2019olw}, and \cite{Kodera:2016faj_jordan_quiver}, that the quantum Coulomb branch algebras are given by the truncated shifted Yangians of ADE type, of BCFG type, and of affine $\mathfrak{gl}_1$, respectively.  
We would like to generalize these results to general quivers. 

\subsubsection{Coulomb branch algebra from vortices}\label{sssec:algebra_from_action}

An alternative approach \cite{Bullimore:2016hdc} to obtain the quantum Coulomb branch algebra $\mathbb{C}_\eps[\mathcal{M}_C]$ is to study the action of the monopole operators and the complex scalars $\varphi$ on the vortex Hilbert space $\mathcal{H}_\nu$ of the 3D $\mathcal{N}=4$ theory. 
For generic mass parameters, the vortex configurations furnish a faithful representation of the Coulomb branch algebra, and hence we can bootstrap the algebraic relations of the operators from their action on $\mathcal{H}_\nu$.
Turning on the $\Omega$-background will modify the action and the algebraic relations of the Coulomb branch operators, leading to the non-commutative algebra $\mathbb{C}_\eps[\mathcal{M}_C]$. 
The vortex Hilbert space $\mathcal{H}_\nu$ furnishes a Verma module under the action of the quantum Coulomb branch algebra \cite{Braverman:2004vv,Braverman:2010ef,Bullimore:2016hdc}. 
We will follow this approach to explicitly compute the quantum Coulomb branch algebra for the tree-type theories.

\medskip

This construction can be  compared to the mathematical definition \cite{Nakajima:2015txa,Braverman:2016wma, Braverman:2016pwk}.
Consider the correspondence space  $\mathcal{M}^{\nn,\nn'}_\nu$, which is the moduli space of two vortex configurations (with vortex numbers $\nn$ and $\nn'$, respectively). 
$\mathcal{M}^{\nn,\nn'}_\nu$ can be projected onto the two moduli spaces $\mathcal{M}^\nn_\nu$ and $\mathcal{M}^{\nn'}_\nu$, which are related by the monopole operator of charge $\nn'-\nn$:
\begin{equation}\label{eq:correspondence_space}
\xymatrix{
  & \mathcal{M}^{\nn,\nn'}_\nu \ar[dr]^{\pi'} \ar@{->}[dl]_-{\pi} \\
  \mathcal{M}^{\nn}_\nu \ar[rr]^{\hat{v}_{\nn'-\nn}} & &\mathcal{M}^{\nn'}_\nu
}\,.
\end{equation}
Let $\boldsymbol{\mathcal{A}}_\nu$ denote the sum of the $T_H\times U(1)_\epsilon$-equivariant cohomologies of the correspondence spaces 
\begin{equation}
\boldsymbol{\mathcal{A}}_\nu=\bigoplus_{\nn,\nn'\in \pi_1(G)} H^*_{T_{H}\times U(1)_\epsilon}(\mathcal{M}^{\nn,\nn'}_\nu )\,.
\end{equation}
It was proposed in \cite{Bullimore:2016hdc} that one can define a convolution product over $\boldsymbol{\mathcal{A}}_\nu$ via the standard pullback-pushforward and it  realizes the OPEs of the monopole operators, and the resulting non-commutative algebra is equivalent to the algebra generated by the Coulomb branch operators and the projectors $\{e_{\nn}\}$ (from $\mathcal{H}_\nu$ to the subsector of $\mathcal{H}_{\nu}$ with vortex number $\nn$).

\section{Vortices in tree-type quiver gauge theory}
\label{sec:vortex_tree_typeQGT}

In this section we study the vortices in the 3D  $\mathcal{N}=4$ unitary quiver gauge theories whose underlying quivers are of tree type. They are important since they furnish a representation of the Coulomb branch algebra. 
Moreover, finding the correspondence between the (shifted) quiver Yangian and the quantum Coulomb branch algebra is also based on the fact that the shifted quiver Yangian has a representation isomorphic to the vortex Hilbert space.

We will solve the $\frac{1}{2}$-BPS vortex configurations for the 3D $\mathcal{N}=4$ theories, and then describe the vortex quantum mechanics. 
The simplest class of examples of the tree-type quiver gauge theories are the linear quiver gauge theories, whose quiver is given by the A-type Dynkin diagram, studied in \cite{Bullimore:2016hdc}, and this section is a generalization of the corresponding results there.

\subsection{Tree-type quiver gauge theories}\label{ssec:tree-type_quiver}

In the following three sections, we will focus on the  3D $\mathcal{N}=4$ quiver gauge theories whose underlying quivers are of \textit{tree type}, with the flavor group $U(N_{\mathtt{f}})$.
The underlying quiver $\mathrm{Q}=(\mathrm{Q}_0,\mathrm{Q}_1)$ and the external flavor node (denoted by $\mathtt{f}$)   satisfy the following conditions:
\begin{enumerate}
\item The quiver $\mathrm{Q}$ is of tree type: namely it is connected and contains no cycle. 
\item The flavor node $\mathtt{f}$ promotes $\mathrm{Q}$ to a rooted tree: there is one particular node $\mathtt{r}\in \mathrm{Q}_0$ which is linked to $\mathtt{f}$, and the node $\mathtt{r}$ serves as the root of the tree. 
(In the example in Fig.~\ref{fig:tree-qui}, $\mathtt{r}=7$.)

\end{enumerate}

For the 3D $\mathcal{N}=4$ quiver gauge theory, the orientations of the arrows in $\mathrm{Q}$ are immaterial. 
However, to facilitate later computations we assign a convenient choice of the orientations as follows.
First, the arrow between $\mathtt{f}$ and the root $\mathtt{r}$ is chosen to be pointing from $\mathtt{f}$ to $\mathtt{r}$.
Then we choose the orientations of  arrows in $\mathrm{Q}$ such that all of them  point away from the root.
An immediate consequence of the tree structure and our choice of the arrow directions is that for any node $a\in \mathrm{Q}_0$, there exists a unique path from $\mathtt{f}$ to $a$. 
See Fig.~\ref{fig:tree-qui} for an example of the tree-type quiver with this choice of the orientations of arrows.
\begin{figure}[h!]
\centering
\begin{tikzpicture}
\node (N) at (0,0)[rectangle,draw] {$\mathtt{f}$};
\node (v0) at (-1.5,0)[circle,draw] {7};
\node (v1) at (-3,1)[circle,draw] {6};
\node (v2) at (-3,0)[circle,draw] {5};
\node (v4) at (-4.5,0)[circle,draw] {2};
\node (v3) at (-4.5,1)[circle,draw] {4};
\node (v5) at (-4.5,-1)[circle,draw] {3};
\node (v6) at (-6,0)[circle,draw] {1};
			
\draw[->] (N) to (v0);
\draw[->] (v0) to (v1);
\draw[->] (v0) to (v2);
\draw[->] (v2) to (v3);
\draw[->] (v2) to (v4);
\draw[->] (v2) to (v5);
\draw[->] (v4) to (v6);
			
\end{tikzpicture}
\caption{A tree-type quiver, with the flavor node $\mathtt{f}$ and the root $\mathtt{r}=7$. All the arrows in $\mathrm{Q}$ point away from the root, and there is a unique path from $\mathtt{f}$ to any $a\in \mathrm{Q}_0$.  }
\label{fig:tree-qui}
\end{figure}

\medskip

We define the \textbf{precursor} and the \textbf{successor} of a node as follows.
For $a\in \mathrm{Q}_0$ and
$ a\neq \mathtt{r}$, its \textbf{precursor} $\mathfrak{p}(a)$ is the node right before $a$ on the (unique) path from the root $\mathtt{r}$ to $a$. 
In addition, the precursor of the root $\mathtt{r}$ is defined to be the flavor node $\mathtt{f}$.
For $a\in \mathrm{Q}_0$, its \textbf{set of successors} $\mathfrak{s}(a)$ is the set of all the nodes whose precursor is $a$: 
\begin{equation}
\mathfrak{s}(a)=\{b\in \mathrm{Q}_0 \,|\, \mathfrak{p}(b)=a\}\,.    
\end{equation}

\medskip

Recall that for a gauge node $a\in \mathrm{Q}_0$, the associated vector-multiplet scalars are $\{\varphi^{(a)},\sigma^{(a)}\}$. 
For an arrow $I\in \textrm{Q}_1\cup\{I^{\mathtt{f}\to\mathtt{r}}\}$, note that in a tree-type quiver each arrow is specified by its target, therefore we can label the hypermultiplet scalars associated to the arrow $I$ as 
\begin{equation}\label{eq:XY_by_target}
\{X^{(t(I))}:=X_I\, ,\, Y^{(t(I))}:=Y_I\}\,,    
\end{equation}
which are in the representation $(R,\bar{R})$ of $U(N^{(t(I))})\times U(N^{(s(I))})$.
In particular,
\begin{equation}
X^{(\mathtt{r})}:=X_{I^{\mathtt{f}\rightarrow \mathtt{r}}}
\qquad\textrm{and}\qquad
Y^{(\mathtt{r})}:=Y_{I^{\mathtt{f}\rightarrow \mathtt{r}}} \,.
\end{equation}
Then, since every node $a\in \mathrm{Q}_0$ in a tree-type quiver is equivalently described by a unique path from $\mathtt{f}$ to $a$, we can also define an alternative set of variables for the hypermultiplet scalars:
\begin{equation}\label{eq:XYtilde}
\{\tilde{X}^{(a)}\, ,\, \tilde{Y}^{(a)}\}\,,    
\end{equation}
by
\begin{equation}\label{eq:X_nest}
\begin{aligned}
\tilde{X}^{(\mathtt{r})}:=X^{(\mathtt{r})}\,, &\qquad \tilde{X}^{(a)}:=X^{(a)}\tilde{X}^{(\mathfrak{p}(a))}\,;\\
\tilde{Y}^{(\mathtt{r})}:=Y^{(\mathtt{r})}\,, &\qquad \tilde{Y}^{(a)}:=\tilde{Y}^{(\mathfrak{p}(a))}Y^{(a)}\,.
\end{aligned}    
\end{equation}
Namely, each new variable is a successive product of the old ones along the path. 
The index structures of $\tilde{X}^{(a)}$ and $\tilde{Y}^{(a)}$ are  $(\tilde{X}^{(a)})^{p}_{\ i}$ and $(\tilde{Y}^{(a)})^{i}_{\ p}\,,$ respectively, where 
$p$ is the gauge index for $U(N^{(a)})$ and $i$ is the flavor index for $U(N_\mathtt{f})$.

Once we obtain the solution in terms of  $\{\tilde{X}^{(a)},\tilde{Y}^{(a)}\}$, we can recover $\{X^{(a)},Y^{(a)}\}$ by solving the linear equations
\begin{equation}
X^{(a)}\tilde{X}^{(\mathfrak{p}(a))}=\tilde{X}^{(a)}\,,\qquad \tilde{Y}^{(\mathfrak{p}(a))}Y^{(a)}=\tilde{Y}^{(a)}\,.
\end{equation}
Note that for each $\{X^{(a)}\, ,\, Y^{(a)}\}$, we have $N^{(a)}\times N^{(\mathfrak{p}(a))}$ variables but have $N^{(a)}\times N_{\mathtt{f}}$ equations. 
With the condition \eqref{eq:non-increasing} (see later), we have more equations than variables, so generically we will not lose information translating between $\{X^{(a)}\, ,\, Y^{(a)}\}$ and $\{\tilde{X}^{(a)}\, ,\, \tilde{Y}^{(a)}\}$.

\medskip

Finally,  we also impose the following conditions on the tree-type quiver gauge theory:
\begin{enumerate}
\item The ranks of the gauge factors are non-increasing along each path:
\begin{equation}\label{eq:non-increasing}
 N^{(a)}  \leq N^{(\mathfrak{p}(a))} \,,
\end{equation} 
where $N^{(\mathfrak{p}(\mathtt{r}))}:=N_{\mathtt{f}}$.
This will guarantee that the isolated massive vacuum solutions exist, see \eqref{eq:NaNf} and \eqref{eq:Nanon-increase}.
\item 
The complex masses $\{m_1,m_2,\dots,m_{N_{\mathtt{f}}}\}$ are generic; the real FI parameters $\{t_1,t_2,\dots,t_{|\mathrm{Q}_0|}\}$ are chosen to be negative, without loss of generality, see \eqref{eq:tRnegative}. 

\end{enumerate}

\subsection{``Good-bad-ugly" theories of tree type}

Following the terminology defined in \cite{Gaiotto_2009}, the 3D $\mathcal{N}=4$ gauge theories can be divided into three classes: the ``good", the ``bad", and the ``ugly" theories, according to the IR behaviors of the $\frac{1}{2}$-BPS monopole operators. 
\begin{itemize}
\item In a ``good" theory, the IR scaling dimensions of the monopoles obey the unitary bound, and therefore in the IR the theory flows to an SCFT whose R-symmetry can be directly read off from the UV description. 
\item In a ``bad" theory, the scaling dimensions of certain monopoles violate the unitary bound, and therefore in the IR the theory flows to an SCFT whose R-symmetry differs from the one in the UV. 
\item Finally, an ``ugly" theory lies between the ``good" and the ``bad" ones, where some monopole operators could saturate the unitary bound, and in the IR the theory flows to an SCFT whose R-symmetry can also be seen in the UV. 
\end{itemize}

For quiver gauge theories, the information of being ``good", ``ugly" or ``bad" is encoded in the quiver diagram, in particular in the ranks of the gauge and flavor group factors.  
For each gauge node $a\in\textrm{Q}_0$, its \textit{excess}\footnote{It is also termed ``\textit{balance}" in some references, see e.g. \cite{Hanany_2024}.} $\mathfrak{b}^{(a)}$ is defined by \cite{Gaiotto_2009}: 
\begin{equation}\label{eq:excess}
\mathfrak{b}^{(a)}=-2N^{(a)}+\sum_{c\in \mathfrak{j}(a)}|a\leftrightarrow c|  \, N^{(c)}\,,
\end{equation}
where $\mathfrak{j}(a)$ is the set of all the nodes (both gauge and flavor) that are adjacent to $a$, and $|a\leftrightarrow c|$ denotes the number of arrows between $a$ and $c$. 
The good-bad-ugly classification of a quiver gauge theory depends on the values of $\mathfrak{b}^{(a)}$'s:
\begin{itemize}
\item If $\mathfrak{b}^{(a)}\geq 0$ for all the nodes $a\in \mathrm{Q}_0$, then the theory is ``good".

\item If the minimal value of $\mathfrak{b}^{(a)}$ ($a\in \textrm{Q}_0$) is $-1$, then the theory is ``ugly".

\item If there exists at least one $a\in \textrm{Q}_0$ such that $\mathfrak{b}^{(a)}\leq -2$, then the theory is ``bad".
\end{itemize}

In Sec.~\ref{ssec:tree-type_quiver} we have imposed various restrictions on our tree-type theories. 
In particular we require that the ranks of the (unitary) flavor and gauge group factors are non-increasing along each path starting from the flavor node:
\begin{equation}\label{eq:non-increasingFull}
N^{(b)}\leq N^{(a)}\,,\quad b\in \mathfrak{s}(a)\,,\quad a\in \textrm{Q}_0\cup\{\mathtt{f}\}\,.
\end{equation}
In App.~\ref{appssec:GoodBadUgly}, we prove that for any tree-type quiver (with a finite number of  nodes), there exists a choice of the ranks of the unitary gauge and flavor groups satisfying the condition \eqref{eq:non-increasingFull} and that the corresponding 3D $\mathcal{N}=4$ quiver gauge theory is ``good".
Similarly, we can easily make a tree-type theory ``ugly" or ``bad". 
Note that, while it is easy to render the tree-type quiver ``good", ``ugly" or ``bad", it's hard (or even impossible) to keep it \textit{balanced}, that is, to keep $\mathfrak{b}^{(a)}=0$ for every $a\in\textrm{Q}_0$, without breaking our assumptions.\footnote{For example, we have required that there is only one flavor node in our extended quiver. If we allow more flavor nodes it will be easier to make the nodes balanced.} 
Indeed, as pointed out in \cite{Gaiotto_2009}, a balanced tree-type quiver must be the Dynkin quiver of ADE type or extended ADE type. 
Moreover, the non-increasing condition \eqref{eq:non-increasingFull} is also violated. 
For example, an extended ADE-type Dynkin quiver can be balanced only when $N_{\mathtt{f}}=0$, i.e.\ there is no flavor symmetry.

\subsection{Vortices in tree-type quiver gauge theories}
\subsubsection{Vacua}\label{sssec:vacua}

We start by analyzing the vacua of the tree-type theories. 

For a general 3D $\mathcal{N}=4$ gauge theory, its supersymmetric vacuum equations are summarized in \eqref{eq:SUSY-vacua}.
Specializing to quiver gauge theories, the set of vacuum equations \eqref{eq:SUSY-vacua} become as follows. For each node $a\in \textrm{Q}_0$:
\begin{equation}
\begin{aligned}
[\varphi^{(a)},\varphi^{(a)\dag}]=[\varphi^{(a)},\sigma^{(a)}]&=0\,,\\
\sum_{\substack{I\in\textrm{Q}_1 \\ t(I)=a}}
(\bar{X}_{I}T^{(a)}X_I-Y_I T^{(a)}\bar{Y}_I)+\sum_{\substack{I\in\textrm{Q}_1 \\ s(I)=a}}(\bar{Y}_I T^{(a)}Y_I-X_{I}T^{(a)}\bar{X}_I)+t_{\mathbb{R}}^{(a)}&=0\,,\\
\sum_{\substack{I\in\textrm{Q}_1 \\ t(I)=a}}
Y_I T^{(a)}X_I+\sum_{\substack{I\in\textrm{Q}_1 \\ s(I)=a}}X_I T^{(a)}Y_I+t_{\mathbb{C}}^{(a)}&=0\,,\\
\end{aligned}
\end{equation}
and for each arrow in the extended quiver $I\in\textrm{Q}_1\cup\{I^{\mathtt{f}\to\mathtt{r}}\}$,
\begin{equation}
\begin{aligned}
(\varphi^{(t(I))}+\varphi^{(s(I))})X_I&=(\sigma^{(t(I))}+\sigma^{(s(I))})X_I=0\,,\\
(\varphi^{(t(I))}+\varphi^{(s(I))})Y_I&=(\sigma^{(t(I))}+\sigma^{(s(I))})Y_I=0\,,\\
\end{aligned}
\end{equation}
where we have defined $\varphi^{(\mathtt{f})}:=m_{\mathbb{C}}$ and $\sigma^{(\mathtt{f})}:=m_{\mathbb{R}}$. 
We will specialize to the case \eqref{eq:Mixed}, namely, setting $m_\mathbb{R}=t_{\mathbb{C}}=0$ from now on.

\medskip

Now let us focus on the 3D $\mathcal{N}=4$ quiver gauge theories whose quivers are of tree type. 
First of all, using the field redefinition \eqref{eq:X_nest}, we can rewrite the set of vacuum equations such that they are all labeled by the nodes of the quiver $\mathrm{Q}$:
\begin{align}
[\varphi^{(a)},\varphi^{(a)\dag}]=[\varphi^{(a)},\sigma^{(a)}]&=0\,,\label{eq:varphi-sigma}\\
\overline{\tilde{X}^{(a)}}T^{(a)}_n \tilde{X}^{(a)}-\tilde{Y}^{(a)} T^{(a)}_n \overline{\tilde{Y}^{(a)}}&=-t^{(a)}_{\mathbb{R}}\delta_{n,0}\,,\label{eq:vacuum1}\\
\tilde{Y}^{(a)} T^{(a)}_n \tilde{X}^{(a)}&=0\,,\label{eq:vacuum2}\\
\varphi^{(a)}\tilde{X}^{(a)}+\tilde{X}^{(a)}m_{\mathbb{C}}=0\,,\quad\tilde{Y}^{(a)}\varphi^{(a)}+m_{\mathbb{C}}\tilde{Y}^{(a)}&=0\,,\label{eq:vacuum_XY_matrix}\\
\sigma^{(a)}\tilde{X}^{(a)}=0\,,\quad\quad\sigma^{(a)}\tilde{Y}^{(a)}&=0\,,\label{eq:sigma-XY}
\end{align}
where $T^{(a)}_n$ with $n=0,1,\dots,(N^{(a)})^2-1$ are the generators of the gauge group $U(N^{(a)})$ that corresponds to the node $a\in \mathrm{Q}_0$.
Note that the equations \eqref{eq:vacuum_XY_matrix}, \eqref{eq:sigma-XY} are matrix equations, where $\tilde{X}^{(a)}$ (resp.\ $\tilde{Y}^{(a)}$) is a $N^{(a)}\times N_{\mathtt{f}}$ (resp.\ $N_{\mathtt{f}}\times N^{(a)}$) matrix, and
\begin{equation}
m_{\mathbb{C}}=\text{diag}(m_1,m_2,\dots,m_{N_\mathtt{f}})\quad\quad\,\in \mathfrak{t}_{\mathbb{C}}^{(H)}\,.
\end{equation}

\medskip
In App.~\ref{appssec:VacuumEq} we explain in detail how to solve this set of vacuum equations.
First, they can be simplified into
\begin{equation}
\begin{aligned}
\sum_{i=1}^{N_\mathtt{f}}(\overline{\tilde{X}^{(a)}})^i_{\ p} (\tilde{X}^{(a)})^p_{\ i}-(\tilde{Y}^{(a)})^i_{\ p} (\overline{\tilde{Y}^{(a)}})^p_{\ i}&=-\frac{t^{(a)}_\mathbb{R}}{N^{(a)}}\,,    \\
\sum_{i=1}^{N_\mathtt{f}}(\tilde{Y}^{(a)})^i_{\ p}  (\tilde{X}^{(a)})^p_{\ i}&=0\,, 
\end{aligned}    
\end{equation}
for $p=1,2,\dots,N^{(a)}$, and 
\begin{align}
(\varphi^{(a)}_p+m_i)(\tilde{X}^{(a)})^p_{\ i}=(\varphi^{(a)}_p+m_i)(\tilde{Y}^{(a)})^i_{\ p}&=0\,,
\\
\sigma^{(a)}_p(\tilde{X}^{(a)})^p_{\ i}=\sigma^{(a)}_p(\tilde{Y}^{(a)})^i_{\ p}&=0\,,
\end{align}
where $p=1,2,\dots,N^{(a)}$, $i=1,2,\dots,N_{\mathtt{f}}$ are not summed. 

\medskip
Solving these equations (for details see App.~\ref{appssec:VacuumEq}), we obtain that the gauge-invariant label of a supersymmetric vacuum is a collection of length-$N^{(a)}$ subsets of $\{1,2,\dots,N_\mathtt{f}\}$: 
\begin{equation}\label{eq:vacua_index_sets}
\mathcal{I}^{(a)}=\{\bar{i}(1),\bar{i}(2),\dots,\bar{i}(N^{(a)})\} 
\ \subset \
\mathcal{I}_{\mathtt{f}}:=\{1,2,\dots,N_\mathtt{f}\}\,,
\end{equation}
for each $a\in \mathrm{Q}_0$.
The vacuum solution that corresponds to the index sets $\{\mathcal{I}^{(a)}\}$, denoted by $\nu_{\{\mathcal{I}^{(a)}\}}$, is given by 
\begin{equation}\label{eq:vacuaSol}  
(\tilde{X}^{(a)})^p_{\ i}=\sqrt{\frac{-t^{(a)}_\mathbb{R}}{N^{(a)}}}\delta_{i,\bar{i}(p)}\,, \qquad (\tilde{Y}^{(a)})^i_{\ p}=0\,,
\end{equation}
with 
\begin{equation}\label{eq:vacuaSolconstraint} 
\varphi^{(a)}_p=-m_{\bar{i}(p)}\,, \qquad  \sigma^{(a)}=0\,,
\end{equation}
up to shuffling of $\bar{i}(p)$ for different $p$'s. 

\medskip

The vacuum configuration of $\tilde{X}^{(a)}$ satisfies the following gauge-invariant conditions:
\begin{equation}\label{eq:vacuum_condition}
\text{det}(\tilde{X}^{(a)}_{J^{(a)}})\propto \left(\frac{-t^{(a)}_{\mathbb{R}}}{N^{(a)}}\right)^{\frac{N^{(a)}}{2}}\delta_{J^{(a)},\ \mathcal{I}^{(a)}}\,, \qquad a\in \textrm{Q}_0\,,
\end{equation}
where $J^{(a)}$ is an arbitrary length-$N^{(a)}$ subset of $\mathcal{I}_{\mathtt{f}}$, $\text{det}(\tilde{X}^{(a)}_{J^{(a)}})$ is the determinant of the $N^{(a)}\times N^{(a)}$ minor of $\tilde{X}^{(a)}$ with columns in $J^{(a)}$, and $\delta_{J^{(a)},\ \mathcal{I}^{(a)}}$ is the delta function on sets.
The matrix $\tilde{X}^{(a)}$ has maximal rank:
\begin{equation}
\text{Rank}(\tilde{X}^{(a)})=N^{(a)} \leq N_{\mathtt{f}}\,,\qquad a\in Q_0\,,
\end{equation}
and if we view the rows of $\tilde{X}^{(a)}$ as $N_{\mathtt{f}}$-dimensional vectors, these vectors are orthogonal to each other
\begin{equation}\langle (\tilde{X}^{(a)})^p|(\tilde{X}^{(a)})^q\rangle\sim\delta^{pq}\,.
\end{equation}
Due to the relation $\tilde{X}^{(a)}=X^{(a)}\tilde{X}^{(\mathfrak{p}(a))}$ (see \eqref{eq:X_nest}), we need the ranks of the gauge groups to be non-increasing along the path:
\begin{equation}\label{eq:Nanon-increase}
N^{(a)}
\ \leq \ 
N^{(\mathfrak{p}(a))} 
\ \leq \ 
N_{\mathtt{f}}\,.    
\end{equation}
The index sets must be nested along each path:
\begin{equation}
\mathcal{I}^{(a)}
\ \subset \
\mathcal{I}^{(\mathfrak{p}(a))}
\ \subset \
\mathcal{I}_{\mathtt{f}} \,.
\end{equation}

\subsubsection{Vortex states and vortex Hilbert space}\label{sssec:vortex_Hilbert_space}

Now we consider the Hilbert space of the vortices that approach the vacuum $\nu_{\{\mathcal{I}^{(a)}\}}$ at spatial infinity, which we denote by $\mathcal{H}_\nu$.
As reviewed in Sec.~\ref{sssec:vortex_moduli}, these configurations correspond to principal $G$-bundles over $\mathbb{CP}^1$ with holomorphic sections $(X,Y)$ of the associated bundles in the representation $(R,\bar{R})$.

For quiver gauge theories, they are described by the following equations.
\begin{enumerate}
\item First, eq.~\eqref{eq:vortex-t-indep} and \eqref{eq:vortex_constant} mean that the vortices are time-independent and the scalars $\varphi,\sigma$ are constant and live in the Cartan subalgebra of $\mathfrak{g}$.

\item Then for each $a\in \text{Q}_0$:
\begin{equation}\label{eq:vortex_eq_quiver_a}
\begin{aligned}
\sum_{\substack{I\in\textrm{Q}_1 \\ t(I)=a}}
(\bar{X}_I T^{(a)}X_I-Y_I T^{(a)}\bar{Y}_I)+\sum_{\substack{I\in\textrm{Q}_1 \\ s(I)=a}}(\bar{Y}_I T^{(a)}Y_I-X_{I}T^{(a)}\bar{X}_I)+t_{\mathbb{R}}^{(a)}&=4[D^{(a)}_z,D^{(a)}_{\bar{z}}]\,,\\
\sum_{\substack{I\in\textrm{Q}_1 \\ t(I)=a}}
Y_I T^{(a)}X_I+\sum_{\substack{I\in\textrm{Q}_1 \\ s(I)=a}}X_I T^{(a)}Y_I+t_{\mathbb{C}}^{(a)}&=0\,.\\
\end{aligned}
\end{equation}

\item And for each $I\in\text{Q}_1\cup\{I^{\mathtt{f}\to\mathtt{r}}\}$:
\begin{equation}\label{eq:vortex_eq_quiver_I}
\begin{aligned}
\partial_t X_I= &D_{\bar{z}}X_I=0\,,\quad &&(\varphi^{(t(I))}+\varphi^{(s(I))})X_I=(\sigma^{(t(I))}+\sigma^{(s(I))})X_I=0\,,\\
\partial_t Y_I= &D_{\bar{z}}Y_I\ =0\,,\quad &&(\varphi^{(t(I))}+\varphi^{(s(I))})Y_I\ =(\sigma^{(t(I))}+\sigma^{(s(I))})Y_I\ =0\,.\\
\end{aligned}
\end{equation}

\item Topologically, the vortices are classified by $|\mathrm{Q}_0|$ integers, the vortex numbers:
\begin{equation}\label{eq:na}
\nn^{(a)}=\frac{1}{2\pi}\int_{\mathbb{CP}^1} \mathrm{Tr}(F^{(a)})\,,\qquad a\in \mathrm{Q}_0\,,
\end{equation}
which are essentially the winding numbers of the transition functions. 
\end{enumerate}

\medskip

Let us now solve the vortex equations in the mixed branch \eqref{eq:Mixed} with the $\Omega$-deformation turned on.
Recall that we can equivalently drop the real conditions in \eqref{eq:vortex_eq_quiver_a} and at the same time complexify the gauge group, see \eqref{eq:MnuC}; similar to the vacuum solutions, let us solve the equations in terms of the  variables $\{\tilde{X}^{(a)},\tilde{Y}^{(a)}\}$:
\begin{align}
\partial_t\tilde{X}^{(a)}=0\,, \qquad D_{\bar{z}}\tilde{X}^{(a)}&=0\,,\label{eq:vortex_X_holomorphic}\\
\partial_t\tilde{Y}^{(a)}=0\,, \qquad
D_{\bar{z}}\tilde{Y}^{(a)}&=0\,, \label{eq:vortex_Y_holomorphic} \\
\tilde{Y}^{(a)} T^{(a)}_n \tilde{X}^{(a)}&=0\,,\label{eq:vortex1}
\end{align}
and 
\begin{align}
\varphi^{(a)}\tilde{X}^{(a)}+\tilde{X}^{(a)}m_{\mathbb{C}}+(\epsilon z\partial_z+\frac{\eps}{2})\tilde{X}^{(a)}&=0\,,\label{eq:vortex_X_matrix}\\
\tilde{Y}^{(a)}\varphi^{(a)}+m_{\mathbb{C}}\tilde{Y}^{(a)}+(\epsilon z\partial_z+\frac{\eps}{2})\tilde{Y}^{(a)}&=0\,,\label{eq:vortex_Y_matrix}\\
\sigma^{(a)}_p(\tilde{X}^{(a)})^p_{\ i}&=0\,,\label{eq:vortex_sigma_X1}\\
\sigma^{(a)}_p(\tilde{Y}^{(a)})^i_{\ p}&=0\,.\label{eq:vortex_sigma_Y1}
\end{align}
Among these equations, 
\eqref{eq:vortex_X_holomorphic} -- \eqref{eq:vortex1} define the vortex moduli space $\mathcal{M}_{\nu}$, which is the space of time-independent (and not necessarily supersymmetric) vortex configurations that approach the vacuum $\nu$ as $z\rightarrow \infty$; 
whereas the constraint equations \eqref{eq:vortex_X_matrix} --  \eqref{eq:vortex_sigma_Y1} select $\frac{1}{2}$-BPS  vortex configurations among $\mathcal{M}_{\nu}$, which are also the fixed points of the moduli space under $T_{H}\times U(1)_{\epsilon}$ where $T_{H}$ is the maximal torus of the flavor symmetry of the Higgs branch, generated by $m_{\mathbb{C}}\in\mathfrak{t}^{(H)}_{\mathbb{C}} $, and $U(1)_{\epsilon}$ is the diagonal combination of $U(1)_H$ and $U(1)_E$, parameterized by $\epsilon$.

\medskip

We want to find the solution of 
\eqref{eq:vortex_X_holomorphic} -- \eqref{eq:vortex_sigma_Y1} that satisfies \eqref{eq:na} and approaches the vacuum solution \eqref{eq:vacuaSol} at $z\rightarrow \infty$.
Let us solve these equations step by step.
\begin{enumerate}
\item Using a complex gauge transformation one can set $A_{\bar{z}}=0$, and then \eqref{eq:vortex_X_holomorphic} and \eqref{eq:vortex_Y_holomorphic} simply state that $\tilde{X}^{(a)},\tilde{Y}^{(a)}$ are time-independent and holomorphic:
\begin{equation}\label{eq:XYholo}
\tilde{X}^{(a)}=\tilde{X}^{(a)}(z)\,,\qquad \tilde{Y}^{(a)}=\tilde{Y}^{(a)}(z)\,. 
\end{equation}
\item For a vortex configuration over $\mathbb{CP}^1$, the vortex number \eqref{eq:na} gives the winding number of the transition function between the southern hemisphere (containing $z=0$) and the northern hemisphere (containing $z=\infty$). 
We want to write down the solutions on the southern hemisphere, namely, on the $z$-plane. 

Consider the GL$_{\mathbb{C}}(N^{(a)})$-bundle, characterized by the following singular transition function:
\begin{equation}\label{eq:transition}
g^{(a)}(z)=i \ \text{diag}(z^{-k^{(a)}_1},{z^{-k^{(a)}_2}},\dots,{z^{-k^{(a)}_{N^{(a)}}}})\in \text{GL}_{\mathbb{C}}(N^{(a)})\,,
\end{equation}
modulo regular gauge transformations, where $k^{(a)}_1,k^{(a)}_2,\dots,k^{(a)}_{N^{(a)}}$ are integers satisfying the following constraint: 
\begin{equation}
\begin{aligned}
\nn^{(a)}=\frac{1}{2\pi}\int_{\mathbb{CP}^1}\mathrm{Tr}(F^{(a)})&=\oint_{\textrm{equator}}\frac{\mathrm{d}z}{2\pi \textrm{i}}\mathrm{Tr}\left(ig^{(a)}(z)^{-1}\partial_zg^{(a)}(z)\right)
=\sum_{p=1}^{N^{(a)}}k^{(a)}_p\,.
\end{aligned}
\end{equation}

\item Next we solve the sections $(\tilde{X}^{(a)},\tilde{Y}^{(a)})$ of the associated bundles in the representation $(R,\bar{R})$, where $R$ is the fundamental representation of $G$. 
In order for $\tilde{X}^{(a)}(z)$ to approach the vacuum $\nu_{\{\mathcal{I}^{(a)}\}}$ at $z=\infty$, a necessary condition is that $\tilde{X}^{(a)}(z)$ must have maximal rank. 
Imposing this condition, we now solve \eqref{eq:vortex_X_matrix}. 
Following the procedures that we used when solving the vacua (see App.~\ref{appssec:VacuumEq}), we obtain:
\begin{equation}\label{eq:vortex_X}
\left(\tilde{X}^{(a)}(z)\right)^p_{\ i}=\sqrt{\frac{-t_{\mathbb{R}}^{(a)}}{N^{(a)}}} \delta_{i,\bar{i}(p)}\ z^{k^{(a)}_p}\,, 
\end{equation}
with
\begin{equation}\label{eq:vortex_X_varphi}
\varphi^{(a)}_p=-m_{\bar{i}(p)}-(k^{(a)}_p+\frac{1}{2})\epsilon\,,
\end{equation}
for $p=1,2,\dots,N^{(a)}$, where $\bar{i}(p)$ are elements of $\mathcal{I}^{(a)}$, see \eqref{eq:vacua_index_sets}, and the coefficient $\sqrt{\frac{-t_{\mathbb{R}}^{(a)}}{N^{(a)}}} $ is chosen such that $\tilde{X}^{(a)}$ tends to the vacuum $\nu_{\{\mathcal{I}^{(a)}\}}$ as $z\to\infty $. 
Moreover, since $\tilde{X}^{(a)}(z)$ are holomorphic functions, we also have the following consistency conditions:
\begin{equation}
    k^{(a)}_p\geq 0\,,\ \text{for\ }\forall \ (a,p)\,,
\end{equation}
and consequently $\nn^{(a)}\geq 0$.

\item Similar to the vacuum solution, since the matrix $\tilde{X}^{(a)}(z)$ has the maximal rank, the solution of \eqref{eq:vortex_sigma_X1} is simply:
\begin{equation}\label{eq:vortex_sigma}
\sigma^{(a)}_p=0\,.
\end{equation}

\item There are several ways to determine the solutions for $\tilde{Y}^{(a)}(z)$. 
One is to solve $\eqref{eq:vortex_Y_matrix}$, with $\varphi^{(a)}_p$ given by \eqref{eq:vortex_X}, and then impose \eqref{eq:vortex1}, with the  result being \begin{equation}\label{eq:Yz=0}
\tilde{Y}^{(a)}(z)=0 \,.   
\end{equation}
An easier way to obtain this result is to use the fact that $\tilde{Y}^{(a)}$ is a holomorphic section but lives in the representation $\bar{R}$ of GL$_{\mathbb{C}}(N^{(a)})$. Since $k^{(a)}_p\geq 0$, such a section must be trivial. 
And since $\tilde{Y}^{(a)}=0$ the equation \eqref{eq:vortex_sigma_Y1} is again trivially solved. 

\end{enumerate}

\medskip

In summary, we have the following solutions for vortex states in $\mathcal{H}_\nu$:
\begin{equation}\label{eq:vortex_state}
\begin{aligned}
&\left(\tilde{X}^{(a)}(z)\right)^p_{\ i}=\sqrt{\frac{-t_{\mathbb{R}}^{(a)}}{N^{(a)}}} \delta_{i,\bar{i}(p)}\ z^{k^{(a)}_p}\,, \qquad \left(\tilde{Y}^{(a)}(z)\right)^i_{\ p}=0\,, 
\end{aligned}
\end{equation}
together with 
\begin{equation}\label{eq:vortex_state_phisigma}
\begin{aligned}
&\varphi^{(a)}_p=-m_{\bar{i}(p)}-(k^{(a)}_p+\frac{1}{2})\epsilon\,,\qquad \qquad
\sigma^{(a)}_p=0\,,
\end{aligned}
\end{equation}
where $\bar{i}(p)$'s are elements of the index set $\mathcal{I}^{(a)}$ that specifies the boundary condition $\nu$, see \eqref{eq:vacua_index_sets}.
Within $\mathcal{H}_{\nu}$, each vortex state is  specified by 
$|\mathrm{Q}_0|$ vectors:
\begin{equation}
\vec{k}^{(a)}\,,\qquad a\in \mathrm{Q}_0\,,
\end{equation}
where each $\vec{k}^{(a)}$ is a length-$N^{(a)}$ \textit{decomposition} of the vortex number $\nn^{(a)}$, namely $\vec{k}^{(a)}$ is an $N^{(a)}$-dimensional vector of non-negative integers that satisfy
\begin{equation}\label{eq:kn}
\sum_{p=1}^{N^{(a)}}k^{(a)}_p=\nn^{(a)}\,.
\end{equation}
In addition, the decomposition $\vec{k}^{(a)}$ for the node $a$ and the one for its precursor $\mathfrak{p}(a)$, $\vec{k}^{(\mathfrak{p}(a))}$,  are not completely independent but related in the following way.
First, for each $p\in\{1,2,\dots,N^{(a)}\}$, there exists a unique $\tilde{p}\in \{1,2,\dots,N^{(\mathfrak{p}(a))}\}$ such that \begin{equation}
\mathcal{I}^{(a)}_{p}=\mathcal{I}^{(\mathfrak{p}(a))}_{\tilde{p}}\,.    
\end{equation} 
Then from the definition of $\tilde{X}^{(a)}(z)$, since $X^{(a)}(z)$ is holomorphic, see \eqref{eq:XYholo}, $(\tilde{X}^{(a)}(z))^{p}_{\ i}$ should have higher degree than the corresponding term of $\tilde{X}^{(\mathfrak{p}(a))}$, namely 
\begin{equation}\label{eq:tree-codi}
k^{(a)}_{p}\geq k^{(\mathfrak{p}(a))}_{\tilde{p}}\,, \qquad p\in \{1,2,..,N^{(a)}\}\,. 
\end{equation}
From now on we will denote the states with $\{\vec{k}^{(a)}\}$ satisfying the conditions above in $\mathcal{H}_\nu$ simply as $|k)_{\nu}$, and we will often drop the subscript $\nu$.

\subsubsection{Vortex moduli space}

The vortex moduli space $\mathcal{M}_{\nu}$ is given by \eqref{eq:vortex_X_holomorphic} -- \eqref{eq:vortex1}, whose solutions are  time-independent (but not necessarily supersymmetric) vortex configurations that approach the vacuum $\nu$ as $z\rightarrow\infty$.
The $\frac{1}{2}$-BPS vortex configurations \eqref{eq:vortex_state} with \eqref{eq:vortex_state_phisigma}
are fixed points of $\mathcal{M}_{\nu}$ under $T_{H}\times U(1)_{\epsilon}$, and they are labeled by $k:=\{\vec{k}^{(a)}\}$ with $\vec{k}^{(a)}$ being the collection of integers $k^{(a)}_p$'s satisfying \eqref{eq:kn} and \eqref{eq:tree-codi}.
Accordingly, the vortex $\mathcal{M}_\nu$ has the cell decomposition\footnote{In \cite{Bullimore:2016hdc} there are two ways to decompose the moduli space: the cell decomposition and the decomposition into strata, for which the parametrizations of the moduli space are different.
In this paper we adopt the cell decomposition.}
\begin{equation}
\mathcal{M}_\nu=\bigcup_{\nn} \mathcal{M}^{\nn}_\nu
\qquad \textrm{with} \qquad
\mathcal{M}^\nn_\nu=\bigcup_k \mathcal{M}^{\nn,k}_\nu\,,
\end{equation}
where each cell $\mathcal{M}^{\nn,k}_\nu$ contains one $\frac{1}{2}$-BPS vortex configuration, denoted by $|k)$.
The vortex Hilbert space $\mathcal{H}_{\nu}$ consists of all such $|k)$.

\medskip

Now let us study the cell $\mathcal{M}^{\nn,k}_\nu$ using the moduli matrix approach \cite{Eto:2005yh,Eto:2006pg}.
It is given by the deformation of the $(\tilde{X}^{(a)}, \tilde{Y}^{(a)})$ around the fixed point \eqref{eq:vortex_state} with \eqref{eq:vortex_state_phisigma}, subject to \eqref{eq:vortex_X_holomorphic} -- \eqref{eq:vortex1} and the boundary condition that they should approach the vacuum $\nu$ given by \eqref{eq:vacuaSol}
as $z\rightarrow \infty$, up to gauge transformations.

First of all, away from the fixed point, $\tilde{Y}^{(a)}(z)=0$ still holds.
The reason is that  $\tilde{Y}^{(a)}(z)$ is a holomorphic section but lives in the representation $\bar{R}$ of GL$_{\mathbb{C}}(N^{(a)})$, and since $k^{(a)}_p\geq 0$, such a section must be trivial ---otherwise there will be poles, since the components of $\tilde{Y}^{(a)}$ have non-positive degrees.
Therefore we only need to consider the deformation of $\tilde{X}^{(a)}(z)$ around \eqref{eq:vortex_state} with \eqref{eq:vortex_state_phisigma}, still subject to \eqref{eq:vortex_X_holomorphic} but not \eqref{eq:vortex_X_matrix} and \eqref{eq:vortex_sigma_X1}.
\begin{enumerate}
\item First, in the gauge $A_{\bar{z}}=0$,  \eqref{eq:vortex_X_holomorphic} still implies that $\tilde{X}^{(a)}(z)$ are time-independent and holomorphic, and for a given $\vec{k}^{(a)}$, the $(p,i)$ elements of $\tilde{X}^{(a)}(z)$ are polynomials of $z$ with degree $k^{(a)}_p$, see \eqref{eq:vortex_state}.
\item The deformation $\tilde{X}^{(a)}(z)$ should approach the vacuum $\nu$ given by \eqref{eq:vacuaSol}
as $z\rightarrow \infty$, up to a gauge transformation.
Namely, $\tilde{X}^{(a)}(z)$ at $z=\infty$ sits in the $G_{\mathbb{C}}$-orbit of the vacuum $\nu_{\{\mathcal{I}^{(a)}\}}$, and as a result
\begin{equation}\label{eq:vacuum_condition_z}
\text{deg(}\text{det}(\tilde{X}^{(a)}_{J^{(a)}}(z)))=
\left\{
\begin{array}{lc}
\nn^{(a)}\,,   & J^{(a)}=\mathcal{I}^{(a)}\,, \\
\leq \nn^{(a)}-1,  & J^{(a)}\neq\mathcal{I}^{(a)}\,,
\end{array}\right.
\end{equation}
where $J^{(a)}$ is an arbitrary length-$N^{(a)}$ subset of $\mathcal{I}_{\mathtt{f}}$, c.f.\ the similar condition at the vacuum \eqref{eq:vacuum_condition}.
\item We have solved for the vacuum solutions \eqref{eq:vacuaSol} and the vortex solutions \eqref{eq:vortex_state} row by row, but to describe the deformations around the fixed points,  we need to express $\tilde{X}^{(a)}(z)$ column by column. 
Consider the inverse map $\bar{p}$ of the map $\bar{i}:\{1,2,\dots,N^{(a)}\}\to\mathcal{I}^{(a)}$ that is defined in \eqref{eq:vacua_index_sets}, namely
\begin{equation}
\bar{p}:\mathcal{I}^{(a)}\to\{1,2,\dots,N^{(a)}\} 
\quad 
\textrm{such that } 
i=\bar{i}(\bar{p}(i))\,,\quad i\in \mathcal{I}^{(a)}\,,
\end{equation}
using which the vortex solution $\tilde{X}^{(a)}(z)$ in \eqref{eq:vortex_state} can be rewritten as 
\begin{equation}\label{eq:vortex_state_column}
(\tilde{X}^{(a)}(z))^p_{\ i}=\sqrt{\frac{-t^{(a)}_{\mathbb{R}}}{N^{(a)}}}\delta_{p,\bar{p}(i)}z^{k^{(a)}_{\bar{p}(i)}} \,.
\end{equation}
\item 
The deformation $\tilde{X}^{(a)}(z)$ around the fixed point \eqref{eq:vortex_state_column} (up to a gauge transformation) and subject to the boundary condition \eqref{eq:vacuum_condition_z} can be divided into three cases, according to whether the flavor index $i$ is in $ \mathcal{I}^{(a)}$, $ \mathcal{I}^{(\mathfrak{p}(a))}\backslash \mathcal{I}^{(a)}$, or $ \mathcal{I}_{\mathtt{f}}\backslash\mathcal{I}^{(\mathfrak{p}(a))}$.

\begin{enumerate}
\item For $i\in \mathcal{I}^{(a)}$, the minor $(\tilde{X}^{(a)})^p_{\ i}$ with $i\in \mathcal{I}^{(a)}$ is a square matrix.
Also, recall the condition 
\begin{equation}\label{eq:tree-codi_2}
k^{(a)}_{p}\geq k^{(\mathfrak{p}(a))}_{\tilde{p}}\,, \qquad p\in \{1,2,..,N^{(a)}\}\,,
\end{equation}
where for each $p$, $\tilde{p}$ is gauge index for $U(N^{(\mathfrak{p}(a))})$ such that $\mathcal{I}^{(a)}_{p}=\mathcal{I}^{(\mathfrak{p}(a))}_{\tilde{p}}$.
Therefore, the deformation of $(\tilde{X}^{(a)})^p_{\ i}$ for $i\in\mathcal{I}^{(a)}$ is given by:
\begin{equation}\label{eq:vortex_deformation1}
(\tilde{X}^{(a)}(z))^p_{\ i}=\sqrt{\frac{-t^{(a)}_{\mathbb{R}}}{N^{(a)}}}\delta_{p,\bar{p}(i)}z^{k^{(a)}_{\bar{p}(i)}}+\sum_{n=k^{(\mathfrak{p}(a))}_{\widetilde{\bar{p}(i)}}}^{k^{(a)}_{\bar{p}(i)}-1}(\delta x^{(a)})^p_{\ i,n}\ z^n\,,
\quad \text{for } 
i\in \mathcal{I}^{(a)}\,
\end{equation}
and $p=1,2,\dots,N^{(a)}$. 
Note that since the gauge transformations are row-operations for $\tilde{X}^{(a)}(z)$, in the $i^{\textrm{th}}$ column one can use the gauge freedom to eliminate the terms whose powers are higher than $(k^{(a)}_{\bar{p}(i)}-1)$ in all the $(p,i)$-components with $p\neq \bar{p}(i)$. 
Hence in \eqref{eq:vortex_deformation1} the summation has an upper bound $(k^{(a)}_{\bar{p}(i)}-1)$.

\item For $i\in \mathcal{I}^{(\mathfrak{p}(a))}\backslash \mathcal{I}^{(a)}$, we have
\begin{equation}\label{eq:vortex_deformation2}
(\tilde{X}^{(a)})^p_{\ i}= \sum_{n=k^{(\mathfrak{p}(a))}_{\bar{p}(i)}}^{k^{(a)}_p-1} (\delta x^{(a)})^p_{\ i,n}\ z^n\,,\qquad \text{for }
i\in \mathcal{I}^{(\mathfrak{p}(a))}\backslash \mathcal{I}^{(a)}\,
\end{equation}
and $p=1,2,\dots,N^{(a)}$. 
The upper bound of the summation in \eqref{eq:vortex_deformation2} is such that $\tilde{X}^{(a)}(z)$ preserves the boundary condition, i.e.\ approaches the vacuum $\nu_{\{\mathcal{I}^{(a)}\}}$ as $z\to\infty$. 
Indeed, with the transition function given in \eqref{eq:transition}, we easily find $(g^{(a)}\tilde{X}^{(a)}(z))^p_{\ i}|_{z\to\infty}=0$ for $i\in \mathcal{I}^{(\mathfrak{p}(a))}\backslash \mathcal{I}^{(a)}$. 
On the other hand, the lower bound in \eqref{eq:vortex_deformation2} again comes from \eqref{eq:tree-codi}.

\item For $i\notin \mathcal{I}^{(\mathfrak{p}(a))}$ we have simply:
\begin{equation}\label{eq:vortex_deformation3}
(\tilde{X}^{(a)})^p_{\ i}=0\,,\qquad 
\text{for }
i\notin \mathcal{I}^{(\mathfrak{p}(a))}\,
\end{equation}
and $p=1,2,\dots,N^{(a)}$.
\end{enumerate}
These deformation coefficients $\delta x^{(a)}$'s are the coordinates of the normal directions at the fixed point $|k)$ within $\mathcal{M}^{\nn,k}$. 
\end{enumerate}

\medskip

We can now compute the equivariant weight of the state $|k)\in\mathcal{H}_\nu$. Under the $G\times T_H\times U(1)_\eps$ transformation
\begin{equation}
\left(X^{(a)}(z)\right)^p_{\ i}\longrightarrow (\varphi^{(a)}_p+m_i+\eps z\partial_z+\frac{\eps}{2})\left(X^{(a)}(z)\right)^p_{\ i}\,,
\end{equation}
where $\varphi^{(a)}_p$ is given in \eqref{eq:vortex_state_phisigma}, each coordinate $(\delta x^{(a)})^p_{\ i,n}$ in \eqref{eq:vortex_deformation1} -- \eqref{eq:vortex_deformation3} contributes a factor
\begin{equation}
(m_i-m_{\mathcal{I}^{(a)}_p}+(n-k^{(a)}_p)\eps)
\end{equation}
to the equivariant weight. Multiplying the contributions from all the normal directions, we obtain the following expression of the equivariant weight of the state $|k)$:
\begin{equation}\label{eq:omegak}
\omega_k=\prod_{a\in \mathrm{Q}_0}\omega^{(a)}_{k}\,,
\end{equation}
where the factor $\omega^{(a)}_{k}$ is
{\footnotesize
\begin{equation}
\omega^{(a)}_{k}=\prod_{p=1}^{N^{(a)}}\frac{\prod_{s=1}^{k^{(a)}_{p}}\left[\prod_{q=1}^{N^{(a)}}(m_{\mathcal{I}^{(a)}_{p}}-m_{\mathcal{I}^{(a)}_{q}}+(s-1-k^{(a)}_{q})\epsilon) \prod_{q\in \mathcal{I}^{(\mathfrak{p}(a))}/\mathcal{I}^{(a)}}(m_q-m_{\mathcal{I}^{(a)}_{p}}-s\epsilon)  \right]}{\prod_{q=1}^{N^{(\mathfrak{p}(a))}}\prod_{s=1}^{k^{(\mathfrak{p}(a))}_{q}}\left[m_{\mathcal{I}^{(\mathfrak{p}(a))}_{q}}-m_{\mathcal{I}^{(a)}_{p}}+(s-1-k^{(a)}_{p})\epsilon\right]}\,.
\end{equation}}
In addition we set $\vec{k}^{(\mathfrak{p}(\mathtt{r}))}=\vec{k}_{\mathtt{f}}=0$.

\subsection{Monopoles}\label{ssec:monopole_operator}

\subsubsection{Action of monopoles on vortices}\label{sssec:monopole_action}

The (first fundamental) minuscule monopole operators $\{\hat{v}^{(a)\pm}_{p}\}$ act on the vortex states as singular gauge transformations \cite{Bullimore:2016hdc}, which change the vortex number by one unit. 

\medskip

Let us first consider the action of the raising operators $\hat{v}^{(a)+}_p$, with  $p=1,2,\dots,N^{(a)}$, on the state $|k)\in\mathcal{H}_\nu$. 
These monopole operators correspond to the (Weyl orbit of the) first fundamental minuscule cocharacters
\begin{equation}
A^{(a)}_p=(0,0,\dots, \stackunder{$1$}{$p$},\dots,0)\quad \in \mathbb{Z}^{N^{(a)}}\left(= \mathrm{Hom}(U(1),U(N^{(a)}))\right)\,.
\end{equation}
The action of $\hat{v}^{(a)+}_p$ on $|k)$ is realized by applying the singular gauge transformation 
\begin{equation}
g^{(a)+}_{p}(z)=\textrm{diag}(1,1,\dots, \stackunder{$z$}{$p$},\dots,1)\quad \in \text{GL}_{\mathbb{C}}(N^{(a)})
\end{equation}
on the matrix $\tilde{X}^{(a)}(z)$ within the cell $\mathcal{M}^{\nn, k}$, given by \eqref{eq:vortex_deformation1} -- \eqref{eq:vortex_deformation3}. 
(Henceforth we will often refer to $\hat{v}^{(a)\pm}_p$ simply as monopole operators, dropping ``first fundamental minuscule".)

First of all, we see that
multiplying $\tilde{X}^{(a)}(z)$ by $g^{(a)+}_{p}(z)$ from the left corresponds to embedding the moduli space $\mathcal{M}_\nu^{\nn,k}$ as a subspace into $ \mathcal{M}_\nu^{\nn+1,k+\delta^{(a)}_{p}} $ 
where $+\delta^{(a)}_{p}$ means adding 1 to the $p^{\textrm{th}}$ component of $\vec{k}^{(a)}$.
However, in order that such an embedding makes sense, certain directions of $\mathcal{M}_\nu^{\nn,k}$ and $\mathcal{M}_\nu^{\nn+1,k+\delta^{(a)}_{p}}$ must be excluded. 
This can be explained in terms of the correspondence space reviewed in Sec.~\ref{sssec:algebra_from_action}. 
Taking the cell decompositions into consideration, we have the following diagram
\begin{equation}\label{eq:monopole_correspondence}
\xymatrix{
  &\quad\quad\mathcal{M}^{(\nn,k),(\nn+1,k+\delta^{(a)}_p)}_\nu \ar[dr]^{\pi'} \ar@{->}[dl]_-{\pi} \\
  \mathcal{M}^{\nn,k}_\nu \ar[rr]^{{\hat{v}^{(a)+}_p}} & &\mathcal{M}^{\nn+1,k+\delta^{(a)}_p}_\nu
}\,
\end{equation}
and the excluded directions are those that are not covered by the images of the projections $\pi$ and $\pi'$.

Let us consider them case by case. 
To distinguish the coordinates of the two moduli spaces, we denote the coordinates of $ \mathcal{M}_\nu^{\nn,k} $ by $\{(\delta x^{(a)})^p_{\ i,r}\}$ and those of $\mathcal{M}_\nu^{\nn+1,k+\delta^{(a)}_{p}}$ by $\{(\delta {x'}^{(a)})^p_{\ i,r}\}$.
\begin{itemize}
\item In the $p^{\textrm{th}}$ row of the matrices $\tilde{X}^{(a)}(z)$ and $(g^{(a)+}_p\tilde{X}^{(a)})(z)$, the components all have upper bounds for their degrees, see \eqref{eq:vortex_deformation1} -- \eqref{eq:vortex_deformation3}. 
The multiplication by $g^{(a)+}_p(z)$ must not break the upper bounds.
The danger comes from the $(p,i)$-components of the matrix $\tilde{X}^{(a)}(z)$ with $i\in\mathcal{I}^{(a)}\backslash\{\bar{i}(p)\}$. 
These polynomials have upper bounds for their degrees
\begin{equation}\label{eq:vortex_upperbound}
\mathrm{deg}((\tilde{X}^{(a)}(z))^p_{\ i})\leq k^{(a)}_{\bar{p}(i)}-1\,,\qquad i\in\mathcal{I}^{(a)}\backslash\{\bar{i}(p)\}\,.
\end{equation}
Note that since $\bar{p}(i)\neq p$ in \eqref{eq:vortex_upperbound}, $k^{(a)}_{\bar{p}(i)}$ takes the same value in both $\mathcal{M}_\nu^{\nn,k}$ and $\mathcal{M}_\nu^{\nn+1,k+\delta^{(a)}_{p}}$. 
Due to the multiplication by $g^{(a)+}_p(z)$, the degrees of the $(p,i)$-components of $\tilde{X}^{(a)}(z)$ (where $i\in\mathcal{I}^{(a)}\backslash\{\bar{i}(p)\}$) must not saturate their upper bounds, such that in the matrix $(g^{(a)+}_p\tilde{X}^{(a)})(z)$ the corresponding components will also not break their upper bounds.
Namely, we require that
\begin{equation}\label{eq:monopole_action_excluded1}
(\delta x^{(a)})^p_{\ i,k^{(a)}_{\bar{p}(i)}-1}=0\,,\qquad i\in\mathcal{I}^{(a)}\backslash\{\bar{i}(p)\}\,.
\end{equation} 
Hence we see the projection $\pi$ in \eqref{eq:monopole_correspondence} is not surjective, since the directions in \eqref{eq:monopole_action_excluded1} of the moduli space $\mathcal{M}^{\nn,k}_{\nu}$ are not covered by the image of $\pi$. 
The equivariant weight of these directions is
\begin{equation}\label{eq:excluded_weight1}
\omega^{\textrm{excluded}}_{k}=\prod_{q\neq p}^{N^{(a)}}(m_{\mathcal{I}^{(a)}_{q}}-m_{\mathcal{I}^{(a)}_{p}}+(k^{(a)}_{q}-k^{(a)}_{p}-1)\epsilon)\,.
\end{equation}
\item The excluded directions in {\small $\mathcal{M}_\nu^{\nn+1,k+\delta^{(a)}_p}$} come from the $(p,i)$-components of the matrix $(g^{(a)+}_p\tilde{X}^{(a)})(z)$ with $i\in\mathcal{I}^{(\mathfrak{p}(a))}$. Note that these components have lower bounds for the degrees of the allowed monomials, see \eqref{eq:vortex_deformation1} -- \eqref{eq:vortex_deformation3}. 
However due to the multiplication by $g^{(a)+}_p(z)$, these lower bounds are no longer saturated. 
Namely the image $(g^{(a)+}_p\tilde{X}^{(a)})(z)$ does not occupy all the directions of $\mathcal{M}_\nu^{\nn+1,k+\delta^{(a)}_{p}}$. 
    
To be precise, in the $(p,i)$-components of the matrix $(\tilde{X}^{(a)})(z)$ with $i\in \mathcal{I}^{(\mathfrak{p}(a))}$, the degrees of the monomials have lower bounds given by $\vec{k}^{(\mathfrak{p}(a))}$:
\begin{equation}
\text{Degree of monomial in } (\tilde{X}^{(a)}(z))^p_{\ i}\geq k^{(\mathfrak{p}(a))}_{\bar{p}(i)}\,,\qquad i\in \mathcal{I}^{(\mathfrak{p}(a))}\,,
\end{equation}
with $\bar{p}:\mathcal{I}^{(\mathfrak{p}(a))}\to\{1,2,\dots,N^{(\mathfrak{p}(a))}\}$. 
Due to the multiplication by $g^{(a)+}_p(z)$, in the matrix $(g^{(a)+}_p\tilde{X}^{(a)})(z)$ the corresponding lower bounds are no longer saturated. 
Namely, in the matrix $(g^{(a)+}_p\tilde{X}^{(a)})(z)$
\begin{equation}\label{eq:monopole_action_excluded2}
(\delta x'^{(a)})^p_{\ i,k^{(\mathfrak{p}(a))}_{\bar{p}(i)}}=0\,,\qquad i\in \mathcal{I}^{(\mathfrak{p}(a))}\,.
\end{equation}
We see that the projection $\pi'$ in \eqref{eq:monopole_correspondence} is also not surjective, and the directions in \eqref{eq:monopole_action_excluded2} of the moduli space {\small $\mathcal{M}_\nu^{\nn+1,k+\delta^{(a)}_{p}}$} are not covered by the image of $\pi'$, hence are excluded from the embedding. 
The equivariant weight of these directions is
\begin{equation}\label{eq:excluded_weight2}
\omega^{\textrm{excluded}}_{k+\delta^{(a)}_p}=\prod_{q=1}^{N^{(\mathfrak{p}(a))}}(m_{\mathcal{I}^{(\mathfrak{p}(a))}_{q}}-m_{\mathcal{I}^{(a)}_{p}}+(k^{(\mathfrak{p}(a))}_{q}-k^{(a)}_{p}-1)\epsilon)\,.
\end{equation}
\end{itemize}

We see that there are directions in both $\mathcal{M}_\nu^{\nn,k}$ and $\mathcal{M}_\nu^{\nn+1,k+\delta^{(a)}_{p}}$ that should be excluded from the embedding. 
Similar to \cite{Bullimore:2016hdc}, we must introduce the ``equivariant delta functions" to account for these directions. 
These delta functions are normalized by the equivariant weight of the directions they fix. 
For example, the vortex state $|k)$, up to a normalization factor, is essentially an equivariant delta function on the moduli space $\mathcal{M}_\nu$, supported on the fixed point which we label by $\mathsf{k}$ 
\begin{equation}\label{eq:vortex_as_delta}
|k)=\frac{1}{\omega_k}\boldsymbol{\delta}_{\mathsf{k}}\,,
\end{equation}
where $\omega_k$ is the equivariant weight. 
The equivariant delta functions are normalized by 
\begin{equation}
(\boldsymbol{\delta}_{\mathsf{k}'}|\boldsymbol{\delta}_{\mathsf{k}})=\omega_k \delta_{k',k}\,.
\end{equation}
Note that the normalization in \eqref{eq:vortex_as_delta} is consistent with \eqref{eq:vortex_normal}.

\medskip

Hence the action of $\hat{v}^{(a)+}_p$ can be described as follows: it acts on a state $|k)$ multiplied by an equivariant delta function corresponding to the directions in \eqref{eq:monopole_action_excluded1}, and gives a state $|k+\delta^{(a)}_p)$ multiplied by an equivariant delta function corresponding to the directions in \eqref{eq:monopole_action_excluded2}.
This introduces normalization factors that count the equivariant weights of the excluded directions, see \eqref{eq:excluded_weight1} and \eqref{eq:excluded_weight2}: 
{\small
\begin{equation}
\begin{aligned}
 \hat{v}^{(a)+}_{p}\left(\prod_{q\neq p}^{N^{(a)}}(m_{\mathcal{I}^{(a)}_{q}}-m_{\mathcal{I}^{(a)}_{p}}+(k^{(a)}_{q}-k^{(a)}_{p}-1)\epsilon)\right)|k)&\\
 =\prod_{q=1}^{N^{(\mathfrak{p}(a))}}(m_{\mathcal{I}^{(\mathfrak{p}(a))}_{q}}-m_{\mathcal{I}^{(a)}_{p}}+(k^{(\mathfrak{p}(a))}_{q}-&k^{(a)}_{p}-1)\epsilon)|k+\delta^{(a)}_{p})\,,
\end{aligned}
\end{equation}}
from which we obtain the action of $\hat{v}^{(a)+}_p$:
\begin{equation}\label{eq:v+_action}
\hat{v}^{(a)+}_{p}|k)
 =\frac{Q^{(\mathfrak{p}(a))}(\hat{\varphi}^{(a)}_p)}{\prod_{q\neq p}^{N^{(a)}}(\hat{\varphi}^{(a)}_p-\hat{\varphi}^{(a)}_q)}|k+\delta^{(a)}_{p})\,,
\end{equation}
where the operators $\hat{\varphi}^{(a)}_p$ are the Cartan operators, see below. 
The state $|k+\delta^{(a)}_{p})$ is specified by the decompositions
\begin{equation}
\{\vec{k}^{(b)}+\delta_{ba}\vec{\delta}_p|b\in \textrm{Q}_0\}\,,
\end{equation}
and these new decompositions must satisfy \eqref{eq:tree-codi}, otherwise $\hat{v}^{(a)+}_{p}|k)=0$. 
The polynomial $Q^{(a)}(x)$ is defined as
\begin{equation}\label{eq:Qpolynomial}
Q^{(a)}(x):=\prod_{p=1}^{N^{(a)}}(x-\hat{\varphi}^{(a)}_{p})\,.
\end{equation}

We can analyze the action of the lowering operators $\hat{v}^{(a)-}_p$ ($p=1,2,\dots,N^{(a)}$) in a completely analogous manner. 
The corresponding singular gauge transformation is
\begin{equation}
g^{(a)-}_{p}(z)=\textrm{diag}(1,1,\dots, \stackunder{$z^{-1}$}{$p$},\dots,1)\quad \in \text{GL}_{\mathbb{C}}(N^{(a)})\,,
\end{equation}
which embeds the moduli space $\mathcal{M}^{\nn,k}_\nu$ into $\mathcal{M}_\nu^{\nn-1,k-\delta^{(a)}_{p}}$.
Once again, there are directions of $\mathcal{M}^{\nn,k}_\nu$ that must be dropped, parameterized by the coefficients (saturating the upper bounds) from the $(q,\bar{i}(p))$-components with $q\neq p$ of $\tilde{X}^{(a)}$; there are also directions of  $\mathcal{M}_\nu^{\nn-1,k-\delta^{(a)}_{p}}$ that are not covered by the image of the embedding, parameterized by the coefficients (saturating the lower bounds) from the components at the $\bar{i}(p)$-th column of $\tilde{X}^{(b)}$ ($b\in\mathfrak{s}(a)$). 
Analogous computations lead to the following action of the lowering operator $\hat{v}^{(a)-}_{p}$:
\begin{equation}\label{eq:v-_action}
\hat{v}^{(a)-}_{p}|k)=
\left\{
\begin{array}{cl}
\frac{\prod_{b\in \mathfrak{s}(a)}(-1)^{N^{(b)}} Q^{(b)}(\hat{\varphi}^{(a)}_p)}{\prod_{q\neq p}^{N^{(a)}}(\hat{\varphi}^{(a)}_q-\hat{\varphi}^{(a)}_p)}
|k-\delta^{(a)}_{p})\,, \ & k^{(a)}_p>k^{(\mathfrak{p}(a))}_{\tilde{p}}\,, \\
0\,, & k^{(a)}_p=k^{(\mathfrak{p}(a))}_{\tilde{p}}\,.
\end{array}\right.
\end{equation}

Finally, the complex scalars in the vector-multiplet, $\hat{\varphi}^{(a)}_p$, act  as Cartan operators with vortices being their eigenvectors:
\begin{equation}\label{eq:varphi_action}
\hat{\varphi}^{(a)}_{p} |k)=-(m_{\mathcal{I}^{(a)}_{p}}+k^{(a)}_{p}\epsilon+\frac{1}{2}\epsilon)|k)\,.
\end{equation}
This is derived simply from the vortex solution \eqref{eq:vortex_state_phisigma}.

As pointed out in \cite[Sec.~4]{Bullimore:2016hdc}, the vortex Hilbert space $\mathcal{H}_\nu$ furnishes a Verma module of the Coulomb branch algebra: it is a highest weight representation, freely generated by the raising operators $\{\hat{v}^{(a)+}_p\}$ from the highest weight state, the ground state $|0)$.

\medskip
As examples, let us apply the results above to the A- and D-type quiver gauge theories, shown in Fig.~\ref{fig:Atype_quiver} and \ref{fig:Dtype_quiver}, respectively. 
The former was already studied in \cite{Bullimore:2016hdc}, and below we will reproduce the known results.
We will give more examples later, including the E-type quiver gauge theory (App.~\ref{appssec:Etype}) and the $K$-star quiver gauge theory (Sec.~\ref{sec:kleaf_example}), where the latter involves the tree-type quiver with high valency.
\begin{figure}[h] 
\centering
\begin{subfigure}[b]{.30\textwidth}
\centering
\begin{tikzpicture}[scale=0.5]
\draw (-5,0.2) circle(0.5);
\node (a1) at (-5,0.2) {$1$}; 
\draw (-4.5,0.2) to (-3.5,0.2); 
\draw (-3,0.2) circle(0.5);
\node (b) at (-3,0.2) {$2$}; 
\draw (-2.5,0.2) to (-2,0.2); 
\node (c) at (-1.5,0.2) {$\dots$};
\draw (-1,0.2) to (-0.5,0.2); 
\draw (0,0.2) circle(0.5);
\node (d) at (0,0.2) {$L$}; 
\draw (0.5,0.2) to (1.8,0.2);
\draw (1.8,-0.3) rectangle (2.8,0.7);
\node (e) at (2.3,0.2) {$\mathtt{f}$};
\end{tikzpicture}
\caption{A-type}
\label{fig:Atype_quiver}
\end{subfigure}
\hspace{20mm}
\begin{subfigure}[b]{.30\textwidth}
\centering
\begin{tikzpicture}[scale=0.5]
\draw (-5,1) circle(0.5);
\node (a1) at (-5,1) {1}; 
\draw[-] (-4.5,1) to (-3.5,0.2); 
\draw (-5,-0.6) circle(0.5);
\node (a2) at (-5,-0.6) {2}; 
\draw[-] (-4.5,-0.6) to (-3.5,0.2); 
\draw (-3,0.2) circle(0.5);
\node (b) at (-3,0.2) {3}; 
\draw[-] (-2.5,0.2) to (-2,0.2); 
\node (c) at (-1.5,0.2) {$\dots$};
\draw[-] (-1,0.2) to (-0.5,0.2); 
\draw (0,0.2) circle(0.5);
\node (d) at (0,0.2) {$L$}; 
\draw[-] (0.5,0.2) to (2,0.2);
\draw (2,-0.3) rectangle (3,0.7);
\node (e) at (2.5,0.2) {$\mathtt{f}$};
\end{tikzpicture}
\caption{D-type}
\label{fig:Dtype_quiver}
\end{subfigure}
\caption{Quivers for A- and D-type quiver gauge theories.}
\end{figure}

\subsubsubsection{Example: A-type quiver gauge theory}\label{ssssec:Atype_vortex}
For the A-type theory, without loss of generality let us choose the vacuum $\nu_{\{\mathcal{I}^{(a)}\}}$ where the nested sets $\{\mathcal{I}^{(1)},\mathcal{I}^{(2)},\dots,\mathcal{I}^{(L)}\}$ are given by
\begin{equation}\label{eq:Atype_vac}
\mathcal{I}^{(a)}=\{1,2,\dots,N^{(a)}\}\,,\qquad a=1,2,\dots,L\,.
\end{equation}
This is always possible by a relabeling of the indices of the mass parameters. 
Consider the Hilbert space $\mathcal{H}_\nu$ of vortices that approach this vacuum at spatial infinity. Each state $|k)\in \mathcal{H}_\nu$ is specified by $L$ vectors 
\begin{equation}
    \{\vec{k}^{(1)},\vec{k}^{(2)},\dots,\vec{k}^{(L)}\}\,,
\end{equation}
where each $\vec{k}^{(a)}$ is $N^{(a)}$-dimensional and consists of non-negative integral components. 
Moreover, these vectors satisfy
\begin{equation}\label{eq:A_condi}
k^{(a)}_p\geq k^{(a+1)}_p\,, \quad p=1,2,\dots,N^{(a)}\,,\quad a=1,2,\dots,L-1\,.
\end{equation}
The vortex number of $|k)$ is given by
\begin{equation}
\nn=\{\sum_{p=1}^{N^{(1)}}k^{(1)}_p,\sum_{p=1}^{N^{(2)}}k^{(2)}_p,\dots,\sum_{p=1}^{N^{(L)}}k^{(L)}_p\}\in \pi_1(G)\,.    
\end{equation}

Let us write down the action of the Coulomb branch operators on the vortex Hilbert space $\mathcal{H}_\nu$. Applying the earlier results \eqref{eq:v+_action} and \eqref{eq:v-_action}, one obtains:
\begin{equation}
\begin{aligned}
\hat{v}_{p}^{(a)+} |k)&=\frac{Q^{(a+1)}(\hat{\varphi}^{(a)}_{p})}{\prod_{q\neq p}^{N^{(a)}}(\hat{\varphi}^{(a)}_{p}-\hat{\varphi}^{(a)}_{q})}|k+\delta^{(a)}_{p})\,,\\
\hat{v}_{p}^{(a)-} |k)&=\frac{(-1)^{N^{(a-1)}}Q^{(a-1)}(\hat{\varphi}^{(a)}_{p})}{\prod_{q\neq p}^{N^{(a)}}(\hat{\varphi}^{(a)}_{q}-\hat{\varphi}^{(a)}_{p})}|k-\delta^{(a)}_{p})\,,
\end{aligned}
\end{equation}
where $\{\hat{\varphi}^{(a)}_p\}$ are the Cartan operators:
\begin{equation}
    \hat{\varphi}^{(a)}_p|k)=-(m_p+k^{(a)}_p\eps+\frac{\eps}{2})|k)\,,
\end{equation}
and $\{Q^{(a)}\}$ are the polynomials defined in \eqref{eq:Qpolynomial}. 

\subsubsubsection{D-type quiver gauge theory}\label{ssssec:Dtype_vortex}
Next, we consider the D-type quiver gauge theory, shown in Fig.~\ref{fig:Dtype_quiver}, and work out the corresponding vortex Hilbert space and the action of the Coulomb branch operators.

Using the results from Sec.~\ref{sssec:vacua}, the supersymmetric vacua of the theory are now specified by two sequences of nested index sets:
\begin{equation}
\begin{aligned}
&\mathcal{I}^{(1)}\subset\mathcal{I}^{(3)}\subset \dots\subset \mathcal{I}^{(L)}\subset\{1,2,\dots,N_{\mathtt{f}}\}\,,\\
&\mathcal{I}^{(2)}\subset\mathcal{I}^{(3)}\subset \dots\subset \mathcal{I}^{(L)}\subset\{1,2,\dots,N_{\mathtt{f}}\}\,.\\
\end{aligned}
\end{equation}
Note the sets $\mathcal{I}^{(1)}$ and $\mathcal{I}^{(2)}$ are allowed to have overlap.
Let us choose the vacuum $\nu_{\{\mathcal{I}^{(a)}\}}$ to be specified by
\begin{equation}\label{eq:D_vac1}
\mathcal{I}^{(a)}=\{1,2,\dots,N^{(a)}\}\,,\qquad a=3,4,\dots,L\,,
\end{equation}
and
\begin{equation}\label{eq:D_vac2}
    \mathcal{I}^{(1)}=\{1,2,\dots,N^{{(1)}}\}\,,\qquad \mathcal{I}^{(2)}=\{l+1,l+2,\dots,l+N^{{(2)}}\}\,,
\end{equation}
with $l$ a positive integer satisfying 
\begin{equation}
l\leq N^{(1)}\,,\qquad l+N^{^{(2)}}\leq N^{(3)}\,.
\end{equation}
We see the overlap of $\mathcal{I}^{(1)}$ and $\mathcal{I}^{(2)}$ is characterized by the integer $l$: they will have overlap if and only if $l<N^{(1)}$, and the length of the overlap is $N^{(1)}-l$.

We consider the Hilbert space $\mathcal{H}_\nu$ of the vortices that approach this vacuum at spatial infinity. 
Applying the results from Sec.~\ref{sssec:vortex_Hilbert_space} to the D-type theory, we see that each state $|k)\in \mathcal{H}_\nu$ is labeled by $L$ decompositions $\{\vec{k}^{(1)},\vec{k}^{(2)},\dots,\vec{k}^{(L)}\}$ such that
\begin{equation}
\text{Dim}(\vec{k}^{(a)})=N^{(a)}\,,\qquad a=1,2,\dots,L\,,
\end{equation}
and
\begin{equation}\label{eq:D_condi1}
k^{(a)}_p\geq k^{(a+1)}_p\,,\quad p=1,2,\dots,N^{(a)}\,,\quad a=3,4,\dots,L-1\,,
\end{equation}
and in particular
\begin{equation}\label{eq:D_condi2}
\begin{aligned}
k^{(1)}_p\geq k^{(3)}_p&\,,\quad p=1,2,\dots,N^{(1)}\,,\\
k^{(2)}_q\geq k^{(3)}_{q+l}&\,,\quad q=1,2,\dots,N^{(2)}\,.
\end{aligned}
\end{equation}
The vortex number of the state $|k)$ is 
\begin{equation}
\nn=\{\sum_{p=1}^{N^{(1)}}k^{(1)}_p\,,\sum_{p=1}^{N^{(2)}}k^{(2)}_p\,,\dots\,,\sum_{p=1}^{N^{(L)}}k^{(L)}_p\}\in \pi_1(G)\,.
\end{equation}

The action of monopole operators can also be easily obtained, using the formulae \eqref{eq:v+_action} and \eqref{eq:v-_action}. 
We have, for the raising operators
\begin{equation}
\hat{v}^{(a)+}_p|k)=\frac{Q^{(a+1)}(\hat{\varphi}^{(a)}_p)}{\prod_{q\neq p}^{N^{(a)}}(\hat{\varphi}^{(a)}_p-\hat{\varphi}^{(a)}_q)}|k+\delta^{(a)}_p)\,,\quad p=1,2,\dots,N^{(a)}\,,\quad a=2,3,\dots,L\,,
\end{equation}
and 
\begin{equation}
\hat{v}^{(1)+}_p|k)=\frac{Q^{(3)}(\hat{\varphi}^{(1)}_p)}{\prod_{q\neq p}^{N^{(1)}}(\hat{\varphi}^{(1)}_p-\hat{\varphi}^{(1)}_q)}|k+\delta^{(1)}_p)\,,\quad p=1,2,\dots,N^{(1)}.
\end{equation}
On the other hand, for the lowering operators we have
\begin{equation}
\hat{v}^{(a)-}_p|k)=\frac{(-1)^{N^{(a-1)}}Q^{(a-1)}(\hat{\varphi}^{(a)}_p)}{\prod_{q\neq p}^{N^{(a)}}(\hat{\varphi}^{(a)}_q-\hat{\varphi}^{(a)}_p)}|k-\delta^{(a)}_p)\,,\quad p=1,2,\dots,N^{(a)}\,,\quad a=4,5,\dots,L\,,
\end{equation}
and in particular, for $a=1,2,3$:
\begin{equation}
\begin{aligned}
\hat{v}^{(1)-}_p|k)&=\frac{1}{\prod_{q\neq p}^{N^{(1)}}(\hat{\varphi}^{(1)}_q-\hat{\varphi}^{(1)}_p)}|k-\delta^{(1)}_p)\,,\quad p=1,2,\dots,N^{(1)}\,,\\
\hat{v}^{(2)-}_p|k)&=\frac{1}{\prod_{q\neq p}^{N^{(2)}}(\hat{\varphi}^{(2)}_q-\hat{\varphi}^{(2)}_p)}|k-\delta^{(2)}_p)\,,\quad p=1,2,\dots,N^{(2)}\,,\\
\hat{v}^{(3)-}_p|k)&=\frac{(-1)^{N^{(1)}+N^{(2)}}Q^{(1)}(\hat{\varphi}^{(3)}_p)Q^{(2)}(\hat{\varphi}^{(3)}_p)}{\prod_{q\neq p}^{N^{(3)}}(\hat{\varphi}^{(3)}_q-\hat{\varphi}^{(3)}_p)}|k-\delta^{(3)}_p)\,,\quad p=1,2,\dots,N^{(3)}\,.
\end{aligned}
\end{equation}

\subsubsection{Coulomb branch algebra}\label{sssec:coulomb_branch_algebra}

From the action of the monopole operators $\hat{v}^{(a)\pm}_p$ and the complex scalar in the vector-multiplet $\hat{\varphi}^{(a)}_p$ on the vortex states, see \eqref{eq:v+_action}, \eqref{eq:v-_action} and \eqref{eq:varphi_action}, one can deduce the algebraic relations of these operators that are valid on these vortex states. 
In order to obtain the Coulomb branch algebra this way, we need to show that the representation is faithful. 
This is indeed easy to see. 
For the monopole operators $\{\hat{v}^{(a)\pm}_p\}$, since $\hat{v}^{(a)\pm}_p|k)\sim |k\pm\delta^{(a)}_p)$, we have ($\alpha,\beta=\pm$)
\begin{equation}
\hat{v}^{(a)\alpha}_p|k)= \hat{v}^{(b)\beta}_q|k)\,,\ \  \forall|k)\in\mathcal H_\nu\,,\quad \Longleftrightarrow \quad \alpha=\beta\,,\ a=b\,,\ p=q\,.
\end{equation}
On the other hand, since we have required that all the mass parameters are generic (see Sec.~\ref{ssec:tree-type_quiver}), the representation for the Cartan operators is also faithful, see \eqref{eq:varphi_action}.

\subsubsubsection{$\hat{v}^{+}$-$\hat{v}^{+}$ and $\hat{v}^{-}$-$\hat{v}^{-}$ relations}

For the $\hat{v}^{(a)+}_p$-$\hat{v}^{(b)+}_q$ and $\hat{v}^{(a)-}_p$-$\hat{v}^{(b)-}_q$ relations, where $p=1,2,\dots,N^{(a)}$ and $q=1,2,\dots,\\N^{(b)}$, we need to distinguish among the following three cases:
\begin{enumerate}
\item For $a$ and $b$ not adjacent to each other in the tree diagram, we have
\begin{equation}\label{eq:vpmvpm_relation1}
[\hat{v}^{(a)+}_p \,,\, \hat{v}^{(b)+}_q]=[
\hat{v}^{(a)-}_p \,,\,
\hat{v}^{(b)-}_q]=0\,.
\end{equation}
\item For $a$ and $b$ adjacent, without loss of generality we can set $a=\mathfrak{p}(b)$. 
Then we have
\begin{equation} \label{eq:vpmvpm_relation2}
\begin{aligned}
\hat{v}^{(a)\pm}_p \hat{v}^{(b)\pm}_q&=\hat{v}^{(b)\pm}_q\hat{v}^{(a)\pm}_p\left(\frac{\hat{\varphi}^{(b)}_q-\hat{\varphi}^{(a)}_p-\eps}{\hat{\varphi}^{(b)}_q-\hat{\varphi}^{(a)}_p}\right)^{\pm 1}\,.\\ 
\end{aligned}
\end{equation}
\item Finally for $a=b$, the relation between $\hat{v}^{(a)\pm}_p$ and $\hat{v}^{(a)\pm}_q$ ($p,q=1,2,\dots,N^{(a)}$) is
\begin{equation}\label{eq:vpmvpm_relation3}
\hat{v}^{(a)\pm}_p \hat{v}^{(a)\pm}_q=\hat{v}^{(a)\pm}_q\hat{v}^{(a)\pm}_p\left(\frac{\hat{\varphi}^{(a)}_q-\hat{\varphi}^{(a)}_p+\eps}{\hat{\varphi}^{(a)}_q-\hat{\varphi}^{(a)}_p-\eps}\right)^{\pm1}\,,\qquad p\neq q\,.
\end{equation}
\end{enumerate}

\subsubsubsection{$\hat{v}^{+}$-$\hat{v}^{-}$ relations}

From the action of the monopole operators \eqref{eq:v+_action} and \eqref{eq:v-_action}, it is straightforward to work out the algebraic relation of the $\hat{v}^{(a)+}_p$ and $\hat{v}^{(b)-}_q$ operators 
\begin{equation}
[\hat{v}^{(a)+}_p,\hat{v}^{(b)-}_q]=\delta_{ab}\delta_{pq} C^{(a)}_p\,,
\end{equation}
with $C^{(a)}_p$ defined as a combination of Cartan operators:
\begin{equation}
C^{(a)}_p=C^{(a)+-}_p-C^{(a)-+}_p\,,    
\end{equation}
with
{\footnotesize
\begin{equation}\label{eq:C+C-}
\begin{aligned}
C^{(a)+-}_p&:=\frac{\prod_{c\in \mathfrak{s}(a)}(-1)^{N^{(c)}}Q^{(c)}(\hat{\varphi}^{(a)}_p+\epsilon)\cdot Q^{(\mathfrak{p}(a))}(\hat{\varphi}^{(a)}_p)}{\prod_{q\neq p}^{N^{(a)}}(\hat{\varphi}^{(a)}_p-\hat{\varphi}^{(a)}_q)(\hat{\varphi}^{(a)}_q-\hat{\varphi}^{(a)}_p-\epsilon)}\,,\\
C^{(a)-+}_p&:=\frac{\prod_{c\in \mathfrak{s}(a)}(-1)^{N^{(c)}}Q^{(c)}(\hat{\varphi}^{(a)}_p)\cdot Q^{(\mathfrak{p}(a))}(\hat{\varphi}^{(a)}_p-\eps)}{\prod_{q\neq p}^{N^{(a)}}(\hat{\varphi}^{(a)}_p-\hat{\varphi}^{(a)}_q)(\hat{\varphi}^{(a)}_q-\hat{\varphi}^{(a)}_p+\epsilon)}\,.
\end{aligned}
\end{equation}}
Note that when $(a,p)=(b,q)$, the two combinations $\hat{v}^{(a)+}_p\hat{v}^{(a)-}_p$ and $\hat{v}^{(a)-}_p\hat{v}^{(a)+}_p$ are both Cartan generators, and they obey stronger relations:
\begin{equation}\label{eq:v+v-_relation}
\hat{v}^{(a)+}_p \hat{v}^{(a)-}_p= C^{(a)^{+-}}_p
\qquad\textrm{and}\qquad
\hat{v}^{(a)-}_p \hat{v}^{(a)+}_p=\ C^{(a)^{-+}}_p\,.
\end{equation}

\subsubsubsection{$\hat{\varphi}$-$\hat{v}^{\pm}$ relations}

From the action of the complex scalar \eqref{eq:varphi_action} and of the monopole operators \eqref{eq:v+_action}, \eqref{eq:v-_action}, we have
\begin{equation}\label{eq:phiv+- relation}
[\hat{\varphi}^{(a)}_p, \hat{v}^{(b)\pm}_q]=\mp\epsilon\delta_{ab}\delta_{pq}\hat{v}^{(b)\pm}_q\,.
\end{equation}

\medskip

Finally, note that if we set $\epsilon=0$ in all the equations \eqref{eq:vpmvpm_relation1} -- \eqref{eq:phiv+- relation} above, we obtain a commutative algebra. 
For example, setting $\epsilon=0$ in \eqref{eq:C+C-} leads to $C^{(a)+-}_p=C^{(a)-+}_p$, and we have
\begin{equation}
\hat{v}^{(a)+}_p\hat{v}^{(b)-}_q=\hat{v}^{(b)-}_q\hat{v}^{(a)+}_p\,.
\end{equation}

\subsection{Vortex quantum mechanics and triple quiver}\label{ssec:VQM_and_triple}

As reviewed in Sec.~\ref{ssec:review_vortex_QM}, placing the 3D theory on the $\Omega$-background $\mathbb{R}\times \mathbb{R}^2_{\epsilon}$ effectively compactifies the 3D theory into the $\mathcal{N}=4$ quantum mechanics, and a vortex configuration in 3D corresponds to a supersymmetric ground state in the quantum mechanics. 

\medskip

Recall that the $\mathcal{N}=4$ quantum mechanics decomposes into an infinite number of components $\bigcup$QM$(\nn,\nu)$, where  each summand QM$(\nn,\nu)$ is a vortex quantum mechanics, with vortex number $\nn$ and 
boundary condition $\nu$ of the vortex configuration that is chosen as the ground state.
The Higgs branch of QM$(\nn,\nu)$ reproduces the component $\mathcal{M}^{\nn}_\nu$ of the vortex moduli space $\mathcal{M}_\nu$, see Sec.~\ref{ssec:review_vortex_QM}.

\medskip

QM$(\nn,\nu)$ is determined by the following information: a  quiver $\mathrm{Q}$,
the boundary condition $\nu$ (which includes the information on the gauge ranks $\{N^{(a)}\}$ and the flavor rank $N_{\mathtt{f}}$), and the vortex number $\{\nn^{(a)}\}$.
We propose that for the 3D $\mathcal{N}=4$ quiver gauge theory with quiver $\mathrm{Q}$, the gauge part of the vortex quantum mechanics QM$(\nn,\nu)$ is given by the $\mathcal{N}=4$ quiver quantum mechanics based on the triple quiver  $\widehat{Q}=(\widehat{Q}_0,\widehat{Q}_1; s,t)$, defined as follows:
\begin{equation}
	\widehat{Q}_0:=\mathrm{Q}_0\,, \qquad \widehat{Q}_1:=\mathrm{Q}_1\cup \,  {}^{\textrm{op}}\mathrm{Q}_1\cup
	\{\mathrm{self\ loops}\}\,,
\end{equation}
where
\begin{equation}
	{}^{\textrm{op}}\mathrm{Q}_1:=\{I^{b\rightarrow a}\, |\, I^{a\rightarrow b} \in \mathrm{Q}_1\}
	\qquad\textrm{and} \qquad
	\{\mathrm{self\ loops}\}:=\{I^{a\rightarrow a}\,|\, a\in \mathrm{Q}_0\}\,.
\end{equation}
Note that the definition of the triple quiver is the same as the one appearing in Sec.~\ref{sssec:IntoN2}. However, here it encodes the field content of the quiver quantum mechanics. 

In Sec.~\ref{ssec:CBA_CoHA_general}  we will consider general quivers. 
Now let us focus on the tree-type quiver $\mathrm{Q}$.
The translation from the triple quiver $\widehat{Q}$ to the field content of the vortex quantum mechanics is
\begin{itemize}
\item each node $a\in \widehat{Q}_0$ represents a gauge factor $U(\nn^{(a)})$ (the gauge group on the $\nn^{(a)}$ vortex world-lines), as well as a 1D $\mathcal{N}=4$ vector-multiplet $(A^{(a)}_t,\lambda^{(a)},\bar{\lambda}^{(a)},\\\phi_m)$ (where $m=1,2,3$), living in the adjoint representation of $U(\nn^{(a)})$;
\item each self-loop $I^{a\rightarrow a}$, denoted as $C_a$, corresponds to a 1D $\mathcal{N}=4$ chiral multiplet $(C_a,\psi_{C_a})$ in the adjoint representation of $U(\nn^{(a)})$;
\item each pair of arrows $\{I^{b\rightarrow a},I^{a\rightarrow b}\}$, denoted $A_a$ and $B_a$ respectively,\footnote{See also  \eqref{eq:triple_notation}.} correspond to two 1D $\mathcal{N}=4$ chiral multiplets: $(A_a,\psi_{A_a})$ in the bi-fundamental representation of $U(\nn^{(a)})\times\overline{U(\nn^{(b)})}$ and $(B_a,\psi_{B_a})$ in the bi-fundamental representation of $U(\nn^{(b)})\times\overline{U(\nn^{(a)})}$.
\end{itemize}
The superpotential is the canonical potential for the triple quiver:
\begin{equation}
\widehat{W}=\sum_{a\in \widehat{Q}_0\backslash \{\mathtt{r}\}}\mathrm{Tr}\,(A_aC_aB_a-C_{\mathfrak{p}(a)}A_aB_a)\,.
\end{equation} 
The triple quiver $\widehat{Q}$ together with the superpotential $\widehat{W}$ encodes the gauge part of the field content of the vortex quantum mechanics QM$(\nn,\nu)$. 
Note that the gauge content only depends on the tree-type quiver $\mathrm{Q}$ and the vortex number $\nn$.

\medskip
The flavor part of the vortex quantum mechanics QM$(\nn,\nu)$ is determined by the boundary condition $\nu$ of the 3D vortex, which includes the information on the gauge ranks $\{N^{(a)}\}$ and the flavor rank $N_{\mathtt{f}}$; this flavor information is encoded in the \textit{framing} of the triple quiver, denoted by $\sharp$. 
We will draw the flavor groups as square nodes, and an arrow between a flavor node and a gauge node will represent a 1D $\mathcal{N}=4$ chiral multiplet in the bi-fundamental representation with respect to the corresponding flavor and gauge factors.
Finally, we will often use the arrow $I=A_a,B_a,C_a,q_a,\tilde{q}_a\dots$ to denote the corresponding field in the quiver quantum mechanics. 

\medskip

The gauge content of the vortex quantum mechanics QM$(\nn,\nu)$ can be straightforwardly written down given the quiver $\mathrm{Q}$ and the vortex number $\nn$.
On the other hand, the flavor content requires more computation. 
One can determine it by requiring that the supersymmetry ground state of the QM$(\nn,\nu)$ reproduces the corresponding 3D vortex configuration, as was done for the A-type theory in \cite{Bullimore:2016hdc}.
In this paper, we adopt a different approach. 

\medskip

In the next section, we will show that the vortex Hilbert space $\mathcal{H}_\nu$ of the tree-type quiver gauge theory with quiver $\mathrm{Q}$ furnishes a representation of the (shifted) quiver Yangian Y$(\widehat{Q},\widehat{W})$, with $\widehat{Q}$ the triple version of $\mathrm{Q}$ and $\widehat{W}$ the associated potential (see \eqref{eq:triple_notation} and \eqref{eq:tree_potential}). 
The representation is given by the framed triple quiver with potential $({}^{\sharp}\widehat{Q},{}^{\sharp}\widehat{W})$, where the framing $\sharp$ is determined by the boundary condition $\nu$ of the vortex, which contains the information on the gauge and flavor ranks, see Sec.~\ref{sssec:mono_as_QY}. Note that for the quiver Yangian, the framing is presented by a unique framing node $\infty$ as well as various framing arrows, and determines a quiver Yangian representation \cite{Li:2023zub} (see also Sec.~\ref{sssec:QY_rep}).

The $\mathcal{N}=4$ vortex quantum mechanics QM$(\nn,\nu)$ is thus a quiver quantum mechanics associated to $({}^\sharp\widehat{Q},{}^\sharp \widehat{W})$.
We will adopt the following translation between the framings in the two contexts, the quiver quantum mechanics and the quiver Yangian:
\begin{equation}\label{eq:translation_QM_QY}
	\text{QM v.s.\ QY:}\qquad
	\begin{tikzpicture}[baseline={([yshift=-.5ex]current bounding box.center)},vertex/.style={anchor=base,
			circle,fill=black!25,minimum size=18pt,inner sep=2pt}]
		\node (va) at (0,1) {$\bullet$};
		\node (r) at (0,-0.5) [rectangle,draw] {$r$};
		\draw[->] (r) edge (va);
	\end{tikzpicture}
	\Longleftrightarrow
	\begin{tikzpicture}[baseline={([yshift=-.5ex]current bounding box.center)},vertex/.style={anchor=base,circle,fill=black!25,minimum size=18pt,inner sep=2pt}]
		\node (va) at (0,1) {$\bullet$};
		\node (r) at (0,-0.5)  {$\infty$};
		\draw[->] (r) edge[bend left =50] (va);
		\draw[->] (r) edge[bend left =30] (va);
		\draw[->] (r) edge[bend right =30] (va);
		\draw[->] (r) edge[bend right =50] (va);
		\node at (0,0.35) {{\scriptsize $\dots$}};
		\node at (0,0.1) {{\small $r$}};
	\end{tikzpicture}\,\, ,
	\quad
	\begin{tikzpicture}[baseline={([yshift=-.5ex]current bounding box.center)},vertex/.style={anchor=base,circle,fill=black!25,minimum size=18pt,inner sep=2pt}]
		\node (va) at (0,1) {$\bullet$};
		\node (r) at (0,-0.5) [rectangle,draw] {$r$};
		\draw[->] (va) edge (r);
	\end{tikzpicture}
	\Longleftrightarrow
	\begin{tikzpicture}[baseline={([yshift=-.5ex]current bounding box.center)},vertex/.style={anchor=base,circle,fill=black!25,minimum size=18pt,inner sep=2pt}]
		\node (va) at (0,1) {$\bullet$};
		\node (r) at (0,-0.5)  {$\infty$};
		\draw[->] (va) edge[bend left =50] (r);
		\draw[->] (va) edge[bend left =30] (r);
		\draw[->] (va) edge[bend right =30] (r);
		\draw[->] (va) edge[bend right =50] (r);
		\node at (0,0.35) {{\scriptsize $\dots$}};
		\node at (0,0.1) {{\small $r$}};
	\end{tikzpicture}\,\,.
\end{equation}
This is to say, $r$ chiral multiplets in the fundamental (resp.\ anti-fundamental) representation of a flavor group $U(r)$ in the quiver quantum mechanics correspond to $r$ framing arrows to (resp.\ from) the framing node $\infty$ in the quiver Yangian, and vice versa. 
In the next section, we will specify the framing of the quiver Yangian which gives a representation isomorphic to the vortex Hilbert space $\mathcal{H}_\nu$. 
Then, using the translation above we can obtain the flavor groups of the vortex quantum mechanics.

\medskip

Let us now present the examples of A-type and D-type quiver gauge theories. 
The reader may refer to Sec.~\ref{sec:QYasMonopoleAlgebra} and App.~\ref{appsec:ADE_examples} for more details on the relevant quiver Yangians.

\subsubsection{Example: A-type}
As the first example, let us consider the vortex quantum mechanics of the A-type theory.
It was already studied in \cite{Bullimore:2016hdc} and we re-examine it here using our approach, for completeness and to compare with other examples.

The gauge part of the quantum mechanics QM$(\nn,\nu)$ of A-type is based on the triple quiver $\widehat{Q}$ of the A-type quiver $\mathrm{Q}$ and the framing $\sharp$ can be obtained by applying the translation \eqref{eq:translation_QM_QY} to the framed quiver of the A-type quiver Yangian (see App.~\ref{appssec:Atype_QY}). 
The result precisely reproduces the handsaw quiver\footnote{For the original study on the handsaw quiver, see \cite{Nakajima:2011yq}.} shown in Fig.~\ref{fig:handsaw}. 
\begin{figure}[h!]
\centering
\begin{tikzpicture}
 [->,auto=right, node distance=2cm,
shorten >=1pt, semithick,square/.style={
draw,
minimum width=width("#1"),
minimum height=width("#1")+2*\pgfshapeinnerysep,
node contents={#1}},scale=0.8]
\node (v1) at (-4,0.2)[circle,draw] { \small{$\nn^{(1)}$}};
\node (v2) at (-1.3,0.2)[circle,draw] {\small{$\nn^{(2)}$}};
\node (e) at (1.4,0.15) {$\dots$};
\node (vl) at (4,0.2)[circle,draw] {\small{$\nn^{(L)}$}};

\node (f1) at (-4,-1.5) [blue,square={\small{$\rho_{1}$}}];
\node (f2) at (-1.3,-1.5) [blue,square={\small{$\rho_{2}$}}];
\node (fl) at (4,-1.5) [blue,square={\small{$\rho_{_L}$}}];
\node (fl1) at (6,-1.5) [blue,square={\small{$\rho_{_{L+1}}$}}];
\node (fe) at (1.4,-1.2) {$\dots$};
			
\draw (v2) edge [bend right=15] node {{\tiny $A_1$}} (v1);
\draw (v1) edge [bend right=15] node {{\tiny $B_1$}} (v2);
\draw (v1) edge [in=110,out=70,loop] node {{\tiny $C_1$}} (v1);
		
\draw (e) edge [bend right=15] node {{\tiny $A_2$}} (v2);
\draw (v2) edge[bend right=15] node {{\tiny $B_2$}} (e);
\draw (v2) edge [in=110,out=70,loop] node {{\tiny $C_2$}} (v2);

\draw (vl) edge [bend right=15] node {{\tiny $A_{L-1}$}} (e);
\draw (e) edge [bend right=15] node {{\tiny $B_{L-1}$}} (vl);
\draw (vl) edge [in=110,out=70,loop] node {{\tiny $C_{L}$}} (vl);

\draw[->,blue] (f1) -- (v1) node [blue,pos=0.33,right,font=\footnotesize] {$q_1$};
\draw[->,blue] (v1) -- (f2) node [blue,pos=0.33,below,font=\footnotesize] {$\tilde{q}_1$};

\draw[->,blue] (f2) -- (v2) node [blue,pos=0.33,right,font=\footnotesize] {$q_2$};
\draw[->,blue] (v2) -- (fe) node [blue,pos=0.33,below,font=\footnotesize] {$\tilde{q}_2$};

\draw[->,blue] (fl) -- (vl) node [blue,pos=0.33,right,font=\footnotesize] {$q_{_L}$};
\draw[->,blue] (vl) -- (fl1) node [blue,pos=0.5,below,font=\footnotesize] {$\tilde{q}_{_L}$};
			
\draw[->,blue] (1.7,-0.7) to (fl);
\node[blue] (ex) at (2.5,-1.23) {{\footnotesize $\tilde{q}_{_{L-1}}$}};
\end{tikzpicture}
\caption{The handsaw quiver.}
\label{fig:handsaw}
\end{figure}
In the handsaw quiver $({}^{\sharp}\widehat{Q},{}^{\sharp}\widehat{W})$, each gauge node $a\in \widehat{Q}$ carries a gauge factor $U(\nn^{(a)})$, which is the gauge group on the world-lines of $\nn^{(a)}$ vortices. 
On the other hand, the ranks of the flavor groups can be translated from the framing of the quiver Yangian, which is specified by the vacuum $\nu$ (see \eqref{eq:Atype_vac}): 
\begin{equation}\label{eq:rhoDef}
\begin{aligned}
\rho_1&=|\mathcal{I}^{(1)}|=N^{(1)}\,,\qquad \rho_{L+1}=|\mathcal{I}_{\mathtt{f}}\backslash\mathcal{I}^{(L)}|=N_{\mathtt{f}}-N^{(L)}\,,\\
\rho_a&=|\mathcal{I}^{(a)}\backslash\mathcal{I}^{(a-1)}|=N^{(a)}-N^{(a-1)}\,,\quad (a=2,3,\dots,L)\,.
\end{aligned}
\end{equation}
The superpotential of the handsaw quiver quantum mechanics is given by 
\begin{equation}\label{eq:A_potential}
{}^{\sharp}\widehat{W}= \sum_{a=1}^{L-1} \mathrm{Tr}\ A_a(C_a B_a-B_a C_{a+1}+\tilde{q}_a q_{a+1})\,,
\end{equation}
see the superpotential of the corresponding quiver Yangian in App.~\ref{appssec:Atype_QY}. 

The handsaw quiver quantum mechanics QM$(\nn,\nu)$ can also be directly read off from the brane constructions of the $\frac{1}{2}$-BPS vortices \cite{Hanany:2003hp,Bullimore:2016hdc}, and it describes a 3D-1D coupled system \cite{Assel:2015oxa}. 
The vortex quantum mechanics lives on the D1-branes whereas the 3D $\mathcal{N}=4$ theory lives on the D3-branes which provide the framing of the vortex quantum mechanics.
As shown in \cite[Sec.~6]{Bullimore:2016hdc}, the $D$-term and $F$-term equations of the handsaw quiver quantum mechanics, together with the $T_H\times U(1)_\eps$ constraints from the complex masses and the $\Omega$-deformation, specify the vortex solutions $|k)$ with vortex number $\nn$ and boundary condition $\nu$. 
On the other hand, the monopole operators can be realized as matrix model degrees of freedom, inserted between two 1D theories QM$(\nn,\nu)$ and QM$(\nn',\nu)$.
All the results we have obtained in this section from the 3D perspective can be re-derived from this 1D formulation.

\subsubsection{Example: D-type}

As the second example, let us use the translation \eqref{eq:translation_QM_QY} to obtain the vortex quantum mechanics QM$(\nn,\nu)$ of the D-type theory, which is studied in Sec.~\ref{ssssec:Dtype_vortex}, and whose corresponding quiver Yangian Y$(\widehat{Q},\widehat{W})$ and framing $\sharp$ are given in Sec.~\ref{ssec:Dtype_theory}.
Translating the framing of the quiver Yangian into the flavor groups of QM$(\nn,\nu)$, we obtain the framed quiver shown in Fig.~\ref{fig:fig:Dtype_quiverver_QM}.
The gauge factors are the gauge groups on the world-lines of vortices, with their ranks given by the vortex number $\nn$. 
On the other hand, the ranks of the flavor groups are determined by $\nu$ (see \eqref{eq:D_vac1} and \eqref{eq:D_vac2}):
\begin{equation}
\begin{aligned}
&\rho_1=|\mathcal{I}^{(1)}\backslash\mathcal{I}^{(2)}|=l\,,
&&\rho_2=|\mathcal{I}^{(2)}\backslash\mathcal{I}^{(1)}|=N^{(2)}-N^{(1)}+l\,,\\
&\rho_3=|\mathcal{I}^{(3)}\backslash(\mathcal{I}^{(1)}\cup\mathcal{I}^{(2)})|=N^{(3)}-N^{(2)}-l\,,
&&\tilde{\rho}=|\mathcal{I}^{(1)}\cap\mathcal{I}^{(2)}|=N^{(1)}-l\,,
\end{aligned}
\end{equation}
and 
\begin{equation}
\rho_a=|\mathcal{I}^{(a)}\backslash\mathcal{I}^{(a-1)}|=N^{(a)}-N^{(a-1)}\,,\qquad a=4,5,\dots,L+1\,,
\end{equation}
with $N^{(L+1)}:=N_{\mathtt{f}}$. (Note that if $\mathcal{I}^{(1)}\cap\mathcal{I}^{(2)}=\emptyset$ i.e.\ $l=N^{(1)}$, the flavor node and the arrows that are drawn red in Fig.~\ref{fig:fig:Dtype_quiverver_QM} would be absent.)
The field content is completely analogous to that in the A-type case:
each arrow carries a 1D $\mathcal{N}=4$ chiral multiplet in the bi-fundamental representation, while each gauge node $a\in \widehat{Q}$ carries a 1D $\mathcal{N}=4$ vector-multiplet of $U(\nn^{(a)})$.
The superpotential for the theory $\textrm{QM}(\nn,\nu)$ is 
\begin{equation}
{}^{\sharp}\widehat{W}_{\textrm{D-type}}=\widehat{W}+ \widehat{W}_{\text{framing}}\,,
\end{equation}
with $\widehat{W}$ given by \eqref{eq:D_triple_superpotential} and 
\begin{equation}
\begin{aligned}
\widehat{W}_{\text{framing}}=&\mathrm{Tr}\,(q_1B_1A_2\tilde{q}_2+q_2B_2A_1\tilde{q}_1+q_3A_1\tilde{q}'_1-q_3A_2\tilde{q}'_2\\
&+s_2B_2\tilde{s}_3-s_1B_1\tilde{s}_3+\sum_{a=4}^{L}q_aA_{a-1}\tilde{q}_{a-1})\,.
\end{aligned}
\end{equation}
It is then straightforward to derive the $D$-terms and $F$-terms equations, which specify the vortex moduli space $\mathcal{M}^{\nn}_\nu$ of the D-type theory.
Turning on the complex masses and the $\Omega$-deformation will then introduce a $T_H\times U(1)_\eps$ action on the moduli space, which localizes the solution to the fixed points, i.e.\ all the states $|k)$ with vortex number $\nn$ and boundary condition $\nu$.
\begin{figure}[h]
\centering
\begin{tikzpicture}
[scale=0.9,->,auto=right, node distance=2cm,
shorten >=1pt, semithick,square/.style={
draw,
minimum width=width("#1"),
minimum height=width("#1")+2*\pgfshapeinnerysep,
node contents={#1}}]
\node (v1) at (-2,1.5)[circle,draw] {$\nn^{(1)}$};
\node (rho1) at (-4,1.5) [blue,square={$\rho_{1}$}];
\node (v2) at (-2,-1.5)[circle,draw] {$\nn^{(2)}$};
\node (rho2) at (-4,-1.5) [blue,square={$\rho_{2}$}];
\node (v3) at (0,0)[circle,draw] {$\nn^{(3)}$};
\node (rho3) at (0,-2) [blue,square={$\rho_{3}$}];
\node (rho3p) at (0,2.5) [red,square={$\tilde{\rho}$}];
\node (v4) at (2,0) [circle,draw] {$\nn^{(4)}$};
\node (e) at (4,-0.05) {$\dots$};
\node (vl) at (6,0)[circle,draw] {$\nn^{(L)}$};

\node (rho4) at (2,-2) [blue,square={$\rho_{4}$}];

\node (rhol) at (6,-2) [blue,square={$\rho_{L}$}];
\node (rholp) at (8,-2) [blue,square={\tiny{$\rho_{L+1}$}}];

\draw (v3) edge [bend right=15] node [pos=0.56,above,font=\tiny] {$A_1$} (v1);
\draw[->] (v1) -- (v3) node [pos=0.33,below,font=\tiny] {$B_1$};
\draw (v1) edge [in=110,out=70,loop] node {{\tiny $C_1$}} (v1);
			
\draw (v2) edge [bend right=15] node [pos=0.33,below] {{\tiny $B_2$}} (v3);
\draw (v3) edge node [pos=0.56,above] {{\tiny $A_2$}} (v2);
\draw (v2) edge [in=110,out=70,loop] node {{\tiny $C_2$}} (v2);

\draw (v3) edge [bend right=15] node {{\tiny $B_3$}} (v4);
\draw (v4) edge [bend right=15] node {{\tiny $A_3$}} (v3);
\draw (v3) edge [in=110,out=70,loop] node {{\tiny $C_3$}} (v3);

\draw (v4) edge [bend right=15] node {{\tiny $B_4$}} (e);
\draw (e) edge [bend right=15] node {{\tiny $A_4$}} (v4);
\draw (v4) edge [in=110,out=70,loop] node {{\tiny $C_4$}} (v4);

\draw (e) edge [bend right=15] node {{\tiny $B_{L-1}$}} (vl);
\draw (vl) edge [bend right=15] node {{\tiny $A_{L-1}$}} (e);
\draw (vl) edge [in=110,out=70,loop] node {{\tiny $C_{L}$}} (vl);

\draw[->,blue] (rho1) -- (v1) node [blue,pos=0.5,above,font=\footnotesize] {$q_1$};
\draw[->,blue] (v1) -- (rho2) node [blue,pos=0.33,above,font=\footnotesize] {$\tilde{q}_1$};
\draw[->,blue] (rho2) -- (v2) node [blue,pos=0.5,below,font=\footnotesize] {$q_2$};
\draw[->,blue] (v2) -- (rho1) node [blue,pos=0.33,below,font=\footnotesize] {$\tilde{q}_2$};

\draw[->,blue] (v1) -- (rho3) node [blue,pos=0.2,below,font=\footnotesize] {$\tilde{q}'_1$};
\draw[->,blue] (v2) -- (rho3) node [blue,pos=0.33,below,font=\footnotesize] {$\tilde{q}'_2$};
\draw[->,blue] (rho3) -- (v3) node [blue,pos=0.33,right,font=\footnotesize] {$q_3$};

\draw[->,blue] (v3) -- (rho4) node [blue,pos=0.33,below,font=\footnotesize] {$\tilde{q}_3$};
\draw[->,blue] (rho4) -- (v4) node [blue,pos=0.33,right,font=\footnotesize] {$q_4$};

\draw[->,blue] (v4) -- (3,-1) node [blue,pos=0.33,below,font=\footnotesize] {$\tilde{q}_4$};

\draw[->,blue] (5,-1) -- (rhol) node [blue,pos=0.1,below,font=\footnotesize] {$\tilde{q}_{L-1}$};
\draw[->,blue] (rhol) -- (vl) node [blue,pos=0.33,right,font=\footnotesize] {$q_L$};

\draw[->,blue] (vl) -- (rholp) node [blue,pos=0.33,below,font=\footnotesize] {$\tilde{q}_{L}$};

\draw[red] (v3) edge [bend right]  node [red,pos=0.5,right,font=\footnotesize] {$\tilde{s}_{3}$} (rho3p);
\draw[->,red] (rho3p) -- (v1) node [red,pos=0.33,above,font=\footnotesize] {$s_1$};
\draw[->,red] (rho3p) -- (v2) node [red,pos=0.65,right,font=\footnotesize] {$s_2$};

\end{tikzpicture}
\caption{The quiver quantum mechanics $\textrm{QM}(\nn,\nu)$ of D type.}
\label{fig:fig:Dtype_quiverver_QM}
\end{figure}

\section{Quiver Yangians as Coulomb branch algebras for tree-type theories}
\label{sec:QYasMonopoleAlgebra}

In this section, we will first reproduce the vortex Hilbert spaces $\mathcal{H}_{\nu}$ as representations of the quiver Yangian algebra of the triple quiver $\widehat{Q}$.
Then by comparing the action of the generators from the Coulomb branch algebra and those from the quiver Yangian on the same state, we will find the map between the two and show the isomorphism between the quantum Coulomb branch algebra and the (truncated shifted) quiver Yangian.
We will focus on tree-type quivers.

\subsection{Review: quiver Yangian}\label{ssec:QY_review}

We  give a lightning review on the quiver Yangian for general quivers and its representations for general quivers, for more details, see \cite{Li:2023zub}. 

\subsubsection{Quiver and potential}

Given a quiver $Q=\{Q_0,Q_1,s,t: Q_1\rightarrow Q_0\}$, we can associate with it a superpotential, or simply potential, which is a sum of cyclic monomials of arrows: 
\begin{equation}\label{eq:QY_potential}
	W=\sum_{d=1}^{F}\# M_d(\{I\})\,,
\end{equation}
where each monomial $M_d(\{I\})$ ($d=1,2,\dots,F$) is a closed loop of arrows:
\begin{equation}
	M_d(\{I\})=I_1\cdot I_2\cdot\dots\cdot I_n\,\text{ with }\ t(I_n)=s(I_1)\,,\qquad d=1,2,\dots,F\,.
\end{equation}
To each arrow of the quiver we assign a weight
\begin{equation}
	h_I\qquad \mathrm{for}\qquad  I\in Q_1\,.
\end{equation}
Then the potential \eqref{eq:QY_potential} enforces the \textit{loop constraints} among the weights $\{h_I\}$:
\begin{equation}\label{eq:loop_cons}
	\sum_{I\in M_d}h_I=0\,,\qquad d=1,2,\dots,F\,.
\end{equation}

We denote the number of arrows from node $a$ to $b$ by $|a\rightarrow b|$.
A quiver is said to be symmetric if $|a\rightarrow b|=|b\rightarrow a|$ for any $a,b\in Q_0$. 
The triple quivers that we will use to define the quiver Yangians in this paper are all symmetric. 

The quiver Yangian Y$(Q,W)$ is generated by $|Q_0|$ triplets of generators, one for each node $a$ in $Q_0$ \cite{Li:2020rij}:
\begin{equation}
	\begin{aligned}
		\mathrm{raising:}\qquad& e^{(a)}_n\,,\quad n\in \mathbb{N}_0\,,\\ 
		\mathrm{lowering:}\qquad& f^{(a)}_n\,,\quad n\in \mathbb{N}_0\,,\\ 
		\mathrm{Cartan:}\qquad& \psi^{(a)}_n\,,\quad n\in \mathbb{Z}\,.
	\end{aligned}
\end{equation}
We combine these three families of modes into three ``fields" by introducing   a spectral parameter $z$:
\begin{equation}\label{eq:QY_mode_expansion}
	e^{(a)}(z)\equiv \sum_{n=0}^{\infty}\frac{e^{(a)}_n}{z^{n+1}}\,,\qquad      f^{(a)}(z)\equiv \sum_{n=0}^{\infty}\frac{f^{(a)}_n}{z^{n+1}}\,,\qquad \psi^{(a)}(z)\equiv \sum_{n=0}^{\infty}\frac{\psi^{(a)}_n}{z^{n+1+\mathtt{s}^{(a)}}}\,, 
\end{equation}
where in the mode expansion of $\psi^{a}(z)$ we have already focused on the symmetric quivers. 
(For non-symmetric quivers, $n\in \mathbb{Z}$.)
The integers $\{\mathtt{s}^{(a)}\}$, called shifts, 
are introduced to adapt to the representations (see later). 
Strictly speaking, the quiver Yangian with non-zero shifts should be called the ``shifted quiver Yangian", but since in this paper generically they are  shifted, we will often drop the qualifier ``shifted".

The raising  and lowering operators, $e^{(a)}(z)$ and $f^{(a)}(z)$, can have Bose or Fermi statistics, depending on the number of the self-loops of the node $a$ \cite{Li:2020rij,Li:2023zub}. 
The Cartan operators $\psi^{(a)}(z)$ are always bosonic. 
In this paper, we focus on the quiver Yangians of the triple quivers,  where each node has one self-loop, and hence all the generators $\{e^{(a)}(z),f^{(a)}(z),\psi^{(a)}(z)\}$ will obey Bose statistics.

\subsubsection{Quadratic relations}
\label{sssec:Qurdratic}

Central to the definition of the quiver Yangian is the bonding factor $\varphi^{b\Leftarrow a}(z)$, defined as
\begin{equation}\label{eq:QY_bond}
	\varphi^{b\Leftarrow a}(z)=e^{i\pi t_{ab}}\frac{\prod_{I\in\{b\to a\}}(z+h_I)}{\prod_{I\in\{a\to b\}}(z-h_I)}\,,
\end{equation}
where $\{a\to b\}$ is the set of all the arrows from $a$ to $b$, and $t_{ab}$ is again the statistics factor \cite{Li:2023zub}. 
For the triple quiver we will choose $t_{ab}=0$. 
Finally, if there is no arrow between $a$ and $b$ we define $\varphi^{b\Leftarrow a}(z)=\varphi^{a\Leftarrow b}(z)=1$.

The quadratic relations of the quiver Yangian are \cite{Li:2020rij,Li:2023zub}
\begin{equation}\label{eq:QY_quadratic_relations}
	\begin{aligned}
		\psi^{(a)}(z)\psi^{(b)}(w)&=\psi^{(b)}(w)\psi^{(a)}(z)\,,\\
		\psi^{(a)}(z)e^{(b)}(w)&\simeq \varphi^{a\Leftarrow b}(z-w)e^{(b)}(w)\psi^{(a)}(z)\,,\\
		\psi^{(a)}(z)f^{(b)}(w)&\simeq \varphi^{a\Leftarrow b}(z-w)^{-1}f^{(b)}(w)\psi^{(a)}(z)\,,\\
		e^{(a)}(z)e^{(b)}(w)&\sim (-1)^{|a||b|}\varphi^{a\Leftarrow b}(z-w)e^{(b)}(w)e^{(a)}(z)\,,\\
		f^{(a)}(z)f^{(b)}(w)&\sim (-1)^{|a||b|}\varphi^{a\Leftarrow b}(z-w)^{-1}f^{(b)}(w)f^{(a)}(z)\,,\\
		[e^{(a)}(z), f^{(b)}(w)\}&\sim -\delta_{ab}\frac{\psi^{(a)}(z)-\psi^{(b)}(w)}{z-w}\,,
	\end{aligned}
\end{equation}
\noindent where ``$\simeq$" means equality up to $z^nw^{m\geq0}$ terms, while ``$\sim$" means equality up to $z^{n\geq 0}w^{m}$ and $z^nw^{m\geq0}$ terms. 
$|a|$ is the $\mathbb{Z}_2$-grading of node $a$, which determines the statistics of $e^{(a)}(z)$ and $f^{(a)}(z)$. 
For the triple quiver we can just set $|a|=0$ for every $a\in Q_0$, and $e^{(a)}(z)$ and $f^{(a)}(z)$ obey Bose statistics. 
$[\ ,\ \}$ is the bracket operation depending on the relative statistics of the operators. 
For the triple quivers it is just the commutator. 

The quadratic relations can also be expressed in terms of modes. 
In \eqref{eq:QY_quadratic_relations}, by moving the denominators of the bonding factors to the l.h.s., then plugging in the mode expansion \eqref{eq:QY_mode_expansion}, and finally expanding both sides and taking the singular terms, we obtain:
\small{
\begin{equation}\label{eq:QY_mode_relation}
\begin{aligned}
[\psi^{(a)}_m,\psi^{(b)}_n]&=0\,,\\
\sum_{k=0}^{|b\to a|}(-1)^{|b\to a|-k}\sigma^{b\to a}_{|b\to a|-k}[\psi^{(a)}_{n-\mathtt{s}^{(a)}}e^{(b)}_m]_k&=\sum_{k=0}^{|a\to b|}\sigma^{a\to b}_{|a\to b|-k}[e^{(b)}_m \psi^{(a)}_{n-\mathtt{s}^{(a)}}]^k\,,\\
\sum_{k=0}^{|a\to b|}\sigma^{a\to b}_{|a\to b|-k}[\psi^{(a)}_{n-\mathtt{s}^{(a)}}f^{(b)}_m]_k&=\sum_{k=0}^{|b\to a|}(-1)^{|b\to a|-k}\sigma^{b\to a}_{|b\to a|-k}[f^{(b)}_m \psi^{(a)}_{n-\mathtt{s}^{(a)}}]^k\,,\\
\sum_{k=0}^{|b\to a|}(-1)^{|b\to a|-k}\sigma^{b\to a}_{|b\to a|-k}[e^{(a)}_n e^{(b)}_m]_k&=(-1)^{|a||b|}\sum_{k=0}^{|a\to b|}\sigma^{a\to b}_{|a\to b|-k}[e^{(b)}_m e^{(a)}_n]^k\,,\\
\sum_{k=0}^{|a\to b|}\sigma^{a\to b}_{|a\to b|-k}[f^{(a)}_n f^{(b)}_m]_k&=(-1)^{|a||b|}\sum_{k=0}^{|b\to a|}(-1)^{|b\to a|-k}\sigma^{b\to a}_{|b\to a|-k}[f^{(b)}_m f^{(a)}_n]^k\,,\\
[e^{(a)}_n,f^{(b)}_m\}&=\delta_{ab}\psi^{(a)}_{n+m-\mathtt{s}^{(a)}}\,.
\end{aligned}
\end{equation}}

\noindent The indices of the $e$ and $f$ operators only run over $\mathbb{N}_0$; the indices of the Cartan operator run over $\mathbb{Z}$ for general quivers, but over $\mathbb{N}_0$ for symmetric quivers, which is the case for our triple quivers.
We have also used the following shorthand notations:
{\small
	\begin{equation}
		[A_n B_m]_k=\sum_{j=0}^k(-1)^{j}\binom{k}{j}A_{n+k-j}B_{m+j}\,,\quad
		[B_m A_n]^k=\sum_{j=0}^k(-1)^{j}\binom{k}{j}B_{m+j}A_{n+k-j}\,.
\end{equation}}

\noindent Finally $\sigma^{a\to b}_{k}$ is the $\text{$k^{\textrm{th}}$ symmetric sum of $\{h_I\}$ with }I\in \{a\to b\}$.
The relations in terms of modes are important for comparing the quiver Yangian with known algebras. 
For example, when the quiver is given by the triple quiver of the Dynkin diagram of a Lie algebra $\mathfrak{g}$, the mode relations of the quiver Yangian reproduce those of the original Yangian of $\mathfrak{g}$ (in the Chevalley basis).

\subsubsection{Quiver Yangian of the triple quiver for tree-type quiver}\label{ssec:QY_triple}

We will be interested in the quiver Yangian that is based on the triple version $\widehat{Q}$ of the quiver $\mathrm{Q}$ that specifies the 3D $\mathcal{N}=4$ gauge theory.
For a given quiver $\mathrm{Q}$, its triple quiver $\widehat{Q}=(\widehat{Q}_0,\widehat{Q}_1; s,t)$ is defined as follows:
\begin{equation}\label{eq:tripleQdef}
	\widehat{Q}_0:=\mathrm{Q}_0\,, \qquad \widehat{Q}_1:=\mathrm{Q}_1\cup \,  {}^{\textrm{op}}\mathrm{Q}_1\cup
	\{\mathrm{self\ loops}\}\,,
\end{equation}
where
\begin{equation}\label{eq:tripleQdef2}
	{}^{\textrm{op}}\mathrm{Q}_1:=\{I^{b\rightarrow a}\, |\, I^{a\rightarrow b} \in \mathrm{Q}_1\}
	\qquad\textrm{and} \qquad
	\{\mathrm{self\ loops}\}:=\{I^{a\rightarrow a}\,|\, a\in \mathrm{Q}_0\}\,.
\end{equation}
For a triple quiver, there exists a canonical potential $\widehat{W}$ that guarantees the equivalence between the CoHA of $(\widehat{Q},\widehat{W})$ and the pre-projective algebra $\Pi(\mathrm{Q})$, see Sec.~\ref{ssec:CBA_CoHA_general}.

\medskip

Let us now first consider the case when $\mathrm{Q}$ is of tree type, in which case the arrows in the triple quiver $\widehat{Q}$ can be labeled by 
\begin{equation}\label{eq:triple_notation}
\mathrm{Q}_1=\{A_a\}\,, \quad {}^{\textrm{op}}\mathrm{Q}_1=\{B_a\}\,, \quad a\in \mathrm{Q}_0\backslash\{\mathtt{r}\}\,,
\end{equation}
where $A_a:=I^{b\rightarrow a}\in\textrm{Q}_1$ is the arrow inherited from $\textrm{Q}$, specified by its target $a$, and $B_a:=I^{a\rightarrow b}$ is its reverse, and 
\begin{equation}
\{\mathrm{self\ loops}\}=\{C_a\}\,,\qquad a\in\textrm{Q}_0\,,
\end{equation}
where $C_a$ is the self-loop of the node $a$.
The canonical superpotential $\widehat{W}$ for the triple quiver $\widehat{Q}$ is:
\begin{equation}\label{eq:tree_potential}
\widehat{W}=\sum_{a\in \widehat{Q}_0\backslash \{\mathtt{r}\}}\mathrm{Tr}\,(A_aC_aB_a-C_{\mathfrak{p}(a)}A_aB_a)\,,
\end{equation}

We will be interested in the representations of the quiver Yangian Y$(\widehat{Q},\widehat{W})$, given by the ideals of the Jacobian algebra 
\begin{equation}
J(\widehat{Q},\widehat{W}) :=\frac{\mathbb{C}\widehat{Q}}{\{\partial_{I}\widehat{W}=0|I\in \widehat{Q}_1\}}   \,.
\end{equation}

Let us consider the assignment of the equivariant weights. 
Imposing the loop constraints from the potential \eqref{eq:tree_potential}, we have:
\begin{itemize}
\item the weight $h(C_a)$ takes the same value for every $a\in\widehat{Q}_0$, which we denote by
\begin{equation}
h(C_a)=\hbar\,,\qquad a\in \widehat{Q}_0\,;
\end{equation}
\item the weights $h(A_a)\,,h(B_a)$ for $a\in \widehat{Q}_0\backslash\{\mathtt{r}\}$ satisfy $h(A_a)+h(B_a)+h(C_a)=0$, which leads to the following weight assignment:
\begin{equation}\label{eq:weight_AB}
h(A_a)=h_a\,,\quad h(B_a)=-h_a-\hbar \,,\qquad a\in \widehat{Q}_0\backslash\{\mathtt{r}\}\,.
\end{equation}
\end{itemize}

In fact, the weight assignment \eqref{eq:weight_AB} still contains redundancies, which come from the gauge symmetries of the quiver nodes \cite{Li:2020rij}. 
We are allowed to perform the following spectral shifts:
\begin{equation}
h_{I^{a\to b}}\to h_{I^{a\to b}}+\varepsilon_a-\varepsilon_b\,,\qquad I^{a\to b}\in\widehat{Q}_1\,,
\end{equation}
with the shift parameters $\{\varepsilon_a|a\in\widehat{Q}_0\}$. 
Using appropriate spectral shifts, we can set $h(B_a)=0$ and hence ${h}_{a}=-\hbar$. 
The shift parameters are given as follows. 
We define $\textrm{path}(a):=\{a,\mathfrak{p}(a),\dots,\mathtt{r}\}$ as the set of nodes on the path from the root to $a$, and define $\textrm{path}'(a):=\textrm{path}(a)\backslash\{\mathtt{r}\}$. 
Then we have
\begin{equation}
\varepsilon_a=\sum_{b\in\textrm{path}'(a)} (h_b+\hbar)\,\ \text{ for }\ a\in\widehat{Q}_0\backslash\{\mathtt{r}\}\,  \quad\text{and }\ \varepsilon_{\mathtt{r}}=0\,.
\end{equation}
Note that $\hbar$, the weight of the self-loop, remains unchanged under the spectral shifts. 
In fact, it is the one and only one parameter of the quiver Yangian. In later sections, we will identify it with the $\Omega$-deformation parameter $\epsilon$ in the 3D theory, when we compare the action of the quiver Yangian generators with that of the Coulomb branch operators (see Sec.~\ref{sssec:translation_between_QY_monopole}). 
In summary, the weight assignment is given by:
\begin{equation}\label{eq:weight_assign_C}
h(C_a)=\hbar\,,\qquad a\in \widehat{Q}_0\,,
\end{equation}
and 
\begin{equation}\label{eq:weight_assign_AB}
h(A_a)=-\hbar\,,\quad h(B_a)=0 \,,\qquad a\in \widehat{Q}_0\backslash\{\mathtt{r}\}\,,
\end{equation}
and we denote the corresponding quiver Yangian by Y$_{\hbar}(\widehat{Q},\widehat{W})$, with the bonding factors given by (see \eqref{eq:QY_bond} with statistics factor $t_{ab}=0$):
\begin{equation}\label{eq:tree_bond}
\varphi^{a\Leftarrow b}(z)=\left\{
\begin{array}{cc}
\frac{z+\hbar}{z-\hbar} \,,  &a=b\,,  \\
\frac{z}{z+\hbar} \,, & b=\mathfrak{p}(a)\,,\\
\frac{z-\hbar}{z}\,,&a=\mathfrak{p}(b)\,,\\
1\,,&a,b\text{ not adjacent}\,.
\end{array}\right.
\end{equation}

\subsubsection{Representations of quiver Yangian}
\label{sssec:QY_rep}

Before we move on, let us briefly review the representations of the quiver Yangian for a general quiver, which are given by the framings of the quiver, for more details, 
see \cite{Li:2023zub}.

\subsubsubsection{Path algebra and Jacobian algebra}

A length-$\ell$ path in $Q$ is a sequence of arrows
\begin{equation}\label{eq:path}
	I_1\cdot I_2\cdot\dots\cdot I_\ell\qquad \mathrm{for}\qquad  I_1,I_2,\dots,I_\ell\in Q_1\,,
\end{equation}
satisfying $t(I_i)=s(I_{i+1})$. 
Note that there exist length-$0$ paths $I^{(a)}$ for each node $a\in Q_0$, satisfying $s(I^{(a)})=t(I^{(a)})=a$.
The path algebra $\mathbb{C}Q$ is an algebra whose underlying vector space has a basis given by all the paths of $Q$, with length $\ell \geq 0$, and the multiplication is defined by the concatenation of paths. 
The length-$0$ paths $I^{(a)}$ play the roles of projection operators:  $I^{(a)}\cdot\mathbb{C}Q$ consists of all the paths that start from node $a$.

The Jacobian algebra $J(Q,W)$ is defined as the path algebra modulo the relations from the potential:
\begin{equation}
	J(Q,W)=\mathbb{C}Q/R\,,
\end{equation}
where $R:=\{\partial_I W=0| I\in Q_1\}$.

\subsubsubsection{Representation of quiver Yangian}
\label{ssssec:QY_rep_and_framing}

A representation of the quiver Yangian is specified by a framing $\sharp$ of the quiver \cite{Galakhov:2021xum,Li:2023zub}.
For the toric Calabi-Yau quiver, the framing can be chosen arbitrarily \cite{Galakhov:2021xum}; whereas for a general quiver, the framing should be chosen such that the charge functions only have simple poles \cite{Li:2023zub}, see below.
As we will show in Sec.~\ref{sssec:vortex_representation}, all the framings that we will encounter in this paper satisfy the property that the charge functions for all the states in the representation have only simple poles.

We introduce a framing node which we denote by $\infty$, together with framing arrows between $\infty$ and quiver nodes $a_i,b_j\in Q_0$. 
The framed quiver $^\sharp Q=\{{}^{\sharp}Q_0,{}^{\sharp}Q_1\}$ with potential ${}^{\sharp}W$ is an extension of the quiver with potential $(Q,W)$:
\begin{equation}\label{eq:extendefig:Dtype_quiverver}
	\begin{aligned}
		{}^\sharp Q_0&=Q_0\cup\{\infty\}\,,\qquad ^\sharp Q_1=Q_1\cup\{\infty\to a_i\}\cup \{b_j\to \infty\}\,,\\
		{}^\sharp W&=W+\text{Monomials involving framing arrows}\,.
	\end{aligned}
\end{equation}
The framing arrows of the form $\infty\to a_i$ are said to be \textit{in-coming}, while those of the form $b_j\to \infty$ are said to be \textit{out-going}.

Let $\mathbb{C}^\sharp Q$ be the path algebra of the framed quiver ${}^\sharp Q$. 
The Jacobian algebra $J^\sharp(Q,W)$ for the framed quiver is defined by:
\begin{equation}\label{eq:jacob}
	J^\sharp(Q,W)=\left(\cup_{\{a_i\}}I^{\infty\to a_i}\cdot J(^\sharp Q,^\sharp W)\right) \backslash\left(\cup_{\{b_j\}}J(^\sharp Q,^\sharp W)\cdot I^{b_j \to \infty}\right)\,.
\end{equation}
A representation of the quiver Yangian is constructed using the basis in $J^\sharp(Q,W)$: we use the basis vectors in $\mathbb{C}Q$ (given by individual paths) and subject them to the path equivalence relations.  
A path ending at $a\in\textrm{Q}_0$ is called a path of color $a$, and a path of color $a$, up to path equivalence, is sometimes called an atom of color $a$.

A state $|\Pi\rangle$ in the representation is a set of linearly independent elements of $J^\sharp(Q,W)$:
\begin{equation}\label{eq:general_quiver_state}
	\Pi=\{p_1,p_2,\dots\}\,.
\end{equation}
Each state $\Pi$ is an eigenstate of the Cartan generator $\psi^{(a)}(z)$:
\begin{equation}
	\psi^{(a)}(z)|\Pi\rangle=\Psi_{\Pi}^{(a)}(z) |\Pi\rangle\,,
\end{equation}
where the eigenvalue $\Psi_{\Pi}^{(a)}(z)$ is called the charge function of the state and is of the form
\begin{equation}\label{eq:QY_charge_function}
	\Psi_{\Pi}^{(a)}(z)={}^\sharp\psi^{(a)}_0(z)\cdot\prod_{p\in \Pi}\varphi^{a\Leftarrow t(p)}(z-h(p))\,.
\end{equation}
Each element in \eqref{eq:general_quiver_state} contributes a bonding factor $\varphi$, with the starting color given by $t(p)$ (the target of the path $p$) and shifted by $h(p)$, which is the equivariant weight of the path, defined as
\begin{equation}
	h(p)=\sum_{I\in p}h_I\,.
\end{equation}
In addition, the ground state of the representation, which is an empty set but contains the information of the framing, also contributes a factor, the ground state charge function ${}^\sharp\psi^{(a)}_0(z)$, and it is specified by the framing:
\begin{equation}\label{eq:ground_charge}
	{}^\sharp\psi^{(a)}_0(z)=\frac{\prod_{I\in\{a\to\infty\}}(z+h_I)}{\prod_{I\in\{\infty\to a\}}(z-h_I)}\,.
\end{equation}

For a given state $\Pi$, the action of the $e$ operators (resp.\ $f$ operators) is to create a new state with one more (resp.\ less) atom.
For a given $a\in Q_0$, the poles of the charge function $\Psi_{\Pi}^{(a)}(z)$ determine the action of $e^{(a)}(z)$ and $f^{(a)}(z)$ on $|\Pi\rangle$. 
Denote the set of poles by
\begin{equation}
	S^{(a)}_\Pi:=\{\text{Poles of }\ \Psi_{\Pi}^{(a)}(z)\}\,.
\end{equation}
The set of paths of color $a$ that can be  removed by $f^{(a)}(z)$ is given by
\begin{equation}    
	\mathcal{R}^{(a)}_\Pi:=\{p|\ p\in\Pi\,,t(p)=a\,,\ h(p)\in S^{(a)}_\Pi\}\,, 
\end{equation}
and we denote the set of removing poles (weights of the paths in $\mathcal{R}^{(a)}_\Pi$) by:
\begin{equation}
	R^{(a)}_\Pi:=\{h(p)|\ p\in \mathcal{R}^{(a)}_\Pi\}\,.
\end{equation}
Then the set of adding poles is defined as the complement of $R^{(a)}_\Pi$ within $S^{(a)}_\Pi$:
\begin{equation}
	A^{(a)}_\Pi=S^{(a)}_\Pi\backslash R^{(a)}_\Pi\,.
\end{equation}
Finally, the set of paths of color $a$ that can be added by the action of $e^{(a)}(z)$ is given by
\begin{equation}
\mathcal{A}^{(a)}_\Pi=\bigcup_{p\in \Pi}\bigl\{ p\cdot I|\ I\in \{t(p)\to a\}\,,\  h(p\cdot I)\in A^{(a)}_\Pi\bigr\}\,.
\end{equation}

All the framings that we will encounter in this paper satisfy the property  that the charge functions for all the states in the representation have only simple poles.
For such representations,\footnote{Otherwise one has to first resolve the higher-order poles, see \cite[App.~C]{Li:2023zub}.} the action of $e^{(a)}(z)$ and $f^{(a)}(z)$ is given by
\begin{align}
e^{(a)}(z)|\Pi\rangle&=\sum_{p\in \mathcal{A}^{(a)}_\Pi}\frac{e^{\frac{i\pi}{4}|a\to a|}}{z-h(p)}\left(\Res_{u=h(p)}\Psi_{\Pi}^{(a)}(u)\right)^{\frac{1}{2}}|\Pi\cup\{p\}\rangle\,,\label{eq:QY_e_action}\\
	f^{(a)}(z)|\Pi\rangle&=\sum_{p\in \mathcal{R}^{(a)}_\Pi}\frac{e^{\frac{i\pi}{4}|a\to a|}}{z-h(p)}\left(\Res_{u=h(p)}\Psi_{\Pi}^{(a)}(u)\right)^{\frac{1}{2}}|\Pi\backslash\{p\}\rangle\,,\label{eq:QY_f_action}
\end{align}
where $|a\to a|$ is the number of self-loops of the node $a$. 
For triple quivers we have $|a\to a|=1$ for every $a\in Q_0$.

\medskip

Once the framing is given, one can build the representation by applying the raising operators $e^{(a)}(z)$ iteratively, starting from the ground state, 
which satisfies:
\begin{equation}
	\Pi_0=\emptyset\,,
	\qquad \psi^{(a)}(z) |\Pi_0\rangle= {}^\sharp\psi^{(a)}_0(z) |\Pi_0\rangle\,,
	\qquad f^{(a)}(z) |\Pi_0\rangle=0\,.
\end{equation}
The resulting representation is a highest weight representation. 
Note that the information of the entire representation is encoded in the ground state charge functions $\{{}^\sharp\psi^{(a)}_0(z)\}$.

\medskip

In the mode expansion of the Cartan generators $\{\psi^{(a)}(z)\}$ in \eqref{eq:QY_mode_expansion}, we have introduced the ``shifts" $\{\mathtt{s}^{(a)}\}$ in order to adapt to general representations, and they are given by the difference of the number of poles and the number of zeros (with multiplicity) in the ground state charge function, or equivalently,  the difference of the numbers of the in-coming and out-going framing arrows: 
\begin{equation}\label{eq:QY_shift}
\mathtt{s}^{(a)}=|\infty\to a|-|a\to\infty|\,.
\end{equation}
The quiver Yangian with non-zero shifts is also sometimes called ``shifted quiver Yangian" to emphasize the shift. 

Furthermore, for a general representation $\mathcal{R}$, the zeros in the ground state charge function \eqref{eq:ground_charge}, corresponding to the arrows going back to the framing node,  ``truncate" the representation by canceling certain potential adding poles corresponding to some states higher up. 
As a result, such a representation is not a Verma module w.r.t.\ the shifted quiver Yangian, and the effect of this cancellation can also be implemented by introducing certain algebraic relations, in addition to the defining relations in \eqref{eq:QY_quadratic_relations}.  
By quotienting out these additional relations from the original shifted quiver Yangian,
one can thus define a ``truncated shifted quiver Yangian", w.r.t.\ which the representation $\mathcal{R}$ becomes a Verma module.

\subsection{Vortex Hilbert space as quiver Yangian representation}
\label{ssec:vortex_as_QY_rep}

We will now  show that the vortex Hilbert space $\mathcal{H}_\nu$ is a representation of the shifted quiver Yangian Y$_{\hbar}(\widehat{Q},\widehat{W})$ of the triple quiver, and in fact a Verma module of the corresponding truncated shifted quiver Yangian.

\subsubsection{Ground state charge function  and framing of triple quiver}
\label{sssec:mono_as_QY}

Recall that a representation of a quiver Yangian is specified by the framing, or equivalently, the ground state charge functions (see Sec.~\ref{sssec:QY_rep}).

To proceed, let us first introduce some new tools and operations. 
To describe the ground state charge functions and the framing, we need to use the concept of a \textit{multi-set}, which is a set where we also keep track of the multiplicities of the items in the set.
We also need the following operations on multi-sets:
\begin{enumerate}
\item Union with multiplicity,  denoted by $\bigsqcup$. \\
For two multi-sets $\mathcal{S}_1$ and $\mathcal{S}_2$, if an element $p$ has multiplicity $n_1$ in $\mathcal{S}_1$ and $n_2$ in $\mathcal{S}_2$, then its multiplicity in $\mathcal{S}_1\bigsqcup \mathcal{S}_2$ is $n_1+n_2$.
\item  Difference (or subtraction) with multiplicity, denoted by $/$.\\ 
For two multi-sets $\mathcal{S}_1$ and $\mathcal{S}_2$, if an element $p$ has multiplicity $n_1$ in $\mathcal{S}_1$ and $n_2$ in $\mathcal{S}_2$, then its multiplicity in $\mathcal{S}_1 / \mathcal{S}_2$ is $\mathrm{Max}\{n_1-n_2,0\}$.
\end{enumerate}
    
For a general tree-type theory, we propose that the charge functions of the ground state in the representation are 
\begin{equation}\label{eq:GSchargeFunc}
^\sharp\psi^{(a)}_0(z)=\frac{\prod_{p\in \mathcal{I}^{(\mathfrak{p}(a))}/\mathcal{I}^{(a)}} (z+\hbar-m_p) \prod_{q\in \tilde{\mathcal{I}}^{(\mathfrak{s}(a))}/\mathcal{I}^{(a)} }(z-m_q)}{\prod_{r\in \mathcal{I}^{(a)}/\mathcal{I}^{(\mathfrak{s}(a))} }(z-m_r)}\,, \qquad a\in \widehat{Q}_0\,,
\end{equation}
where 
\begin{equation}
\mathcal{I}^{(\mathfrak{s}(a))}:=\bigcup_{b\in \mathfrak{s}(a)}\mathcal{I}^{(b)}
\qquad\textrm{and} \qquad
\tilde{\mathcal{I}}^{(\mathfrak{s}(a))}:=\bigsqcup_{b\in \mathfrak{s}(a)}\mathcal{I}^{(b)}\,.    
\end{equation}
The numerator factor  $\prod_{q\in \tilde{\mathcal{I}}^{(\mathfrak{s}(a))}/\mathcal{I}^{(a)}}(z-m_q)$ in \eqref{eq:GSchargeFunc} would not appear when the index sets $\mathcal{I}^{(b_i)}$ for different $b_i\in\mathfrak{s}(a)$ do not intersect, in which case  $\tilde{\mathcal{I}}^{(\mathfrak{s}(a))}=\mathcal{I}^{(\mathfrak{s}(a))}\subset \mathcal{I}^{(a)}$ and therefore $\tilde{\mathcal{I}}^{(\mathfrak{s}(a))}/\mathcal{I}^{(a)}=\emptyset$.
(For A-type theory, this is always the case.)
The representation given by \eqref{eq:GSchargeFunc} will be called the vortex representation and denoted by $\mathcal{R}_\nu$.

Comparing \eqref{eq:GSchargeFunc} with \eqref{eq:ground_charge}, we  see that the weights of the framing arrows are specified by the mass parameters $\{m_i\}$ of the 3D $\mathcal{N}=4$ theory and the weight parameter $\hbar$ of the quiver Yangian, and there are various loop constraints among these framing arrows.
Note also that the ground state charge function \eqref{eq:GSchargeFunc} depends on the index sets $\{\mathcal{I}^{(a)}\}$ that specify the boundary condition $\nu$ of the vortices. 

\medskip

The framing $\sharp$ that corresponds to the ground state charge functions \eqref{eq:GSchargeFunc} is as follows.
We introduce the framing node $\infty$. 
As before, a framing arrow is said to be \textsl{in-coming} if it is from $\infty$ to some $a\in \widehat{Q}_0$, and \textsl{out-going} otherwise. 
The framing is as follows.
For each node $a\in\widehat{Q}_0$, there are:
\begin{tcolorbox}[breakable]
\begin{enumerate}
\item $|\mathcal{I}^{(a)}/\mathcal{I}^{(\mathfrak{s}(a))}|$ in-coming arrows $\infty\rightarrow a$, with weights in the set
\begin{equation}\label{eq:Incoming}
\{m_p\, |\, p\in \mathcal{I}^{(a)}/\mathcal{I}^{(\mathfrak{s}(a))}\}\,.  
\end{equation}
\item $|\mathcal{I}^{(\mathfrak{p}(a))}/\mathcal{I}^{(a)}|$ out-going arrows $a\rightarrow \infty$, with weights in the set
\begin{equation}\label{eq:Outgoing1}
\{\hbar-m_p\, |\, p\in \mathcal{I}^{(\mathfrak{p}(a))}/\mathcal{I}^{(a)}\}\,.    
\end{equation}
\item $|\tilde{\mathcal{I}}^{(\mathfrak{s}(a))}/\mathcal{I}^{(a)}|$ out-going arrows $a\rightarrow \infty$ with weights in the multi-set
\begin{equation}\label{eq:Outgoing2}
\{-m_p\, |\, p\in \tilde{\mathcal{I}}^{(\mathfrak{s}(a))}/\mathcal{I}^{(a)}\}\,.    
\end{equation}
\end{enumerate}
\end{tcolorbox}
\noindent The superpotential involving the framing arrows are encoded in the constraints satisfied by their weight assignment: the arrows whose equivariant weights add up to zero form a closed loop in the framed quiver, and contribute a monomial to the superpotential.
As before, when the index sets $\mathcal{I}^{(b_i)}$ for different $b_i\in\mathfrak{s}(a)$ do not intersect, the third type of arrows will not appear.  

The framing above (or equivalently, the ground state charge functions \eqref{eq:GSchargeFunc}) completely specifies a representation of the shifted quiver Yangian with the shift given by \eqref{eq:QY_shift}.
We can now show that this representation, denoted by $\mathcal{R}_\nu$, is isomorphic to the vortex Hilbert space $\mathcal{H}_{\nu}$ of the 3D theory.

\medskip

As a consistency check, let us first show that the shift $\mathtt{s}^{(a)}$ is independent of the vacuum $\nu$, but only depends on the ranks of the gauge and flavor groups.
Consider an element $p$ in the multi-set $\tilde{\mathcal{I}}^{(\mathfrak{s}(a))}/\mathcal{I}^{(a)}$, with multiplicity $n_p > 0$. 
Such an element appears in $(n_p+1)$ index sets of $\{\mathcal{I}^{(b)}|b\in \mathfrak{s}(a)\}$. 
However, in the union $\mathcal{I}^{(\mathfrak{s}(a))}=\cup_{b\in \mathfrak{s}(a)}\mathcal{I}^{(b)}$ only one copy of $p$ is included. 
Namely, the length of $\mathcal{I}^{(\mathfrak{s}(a))}$ is given by:
\begin{equation}
    |\mathcal{I}^{(\mathfrak{s}(a))}|=\sum_{b\in\mathfrak{s}(a)}N^{(b)}-\sum_p n_p\,,
\end{equation}
where the second summation is over all the elements in $\tilde{\mathcal{I}}^{(\mathfrak{s}(a))}/\mathcal{I}^{(a)}$. 
Hence, the number of in-coming arrows is:
\begin{equation}
\#^{(a)}_{\text{in}}=|\mathcal{I}^{(a)}/\mathcal{I}^{(\mathfrak{s}(a))}|=N^{(a)}-\sum_{b\in\mathfrak{s}(a)}N^{(b)}+\sum_p n_p\,.
\end{equation}
On the other hand, the number of out-going arrows is
\begin{equation}
\#^{(a)}_{\text{out}}=|\mathcal{I}^{(\mathfrak{p}(a))}/\mathcal{I}^{(a)}|+|\tilde{\mathcal{I}}^{(\mathfrak{s}(a))}/\mathcal{I}^{(a)}|=N^{(\mathfrak{p}(a))}-N^{(a)}+\sum_p n_p\,.
\end{equation}
Note that in the trivial case, i.e.\ when the index sets $\mathcal{I}^{(b)}$ for different $b\in \mathfrak{s}(a)$ do not intersect, all the $n_p$ in the equations above are 0.
It is then straightforward to compute the shift $\mathtt{s}^{(a)}$
\begin{equation}
\mathtt{s}^{(a)}=\#^{(a)}_{\text{in}}-\#^{(a)}_{\text{out}}=2N^{(a)}-N^{(\mathfrak{p}(a))}-\sum_{b\in\mathfrak{s}(a)}N^{(b)}=-\mathfrak{b}^{(a)}\,,
\end{equation}
which is nothing but the reverse of the excess $\mathfrak{b}^{(a)}$ (see \eqref{eq:excess}).

\subsubsection{Enumerating states in the vortex representations}\label{sssec:vortex_representation}

Let us first enumerate the states in the vortex representation given by the ground state charge function \eqref{eq:GSchargeFunc}.    
We introduce $|\widehat{Q}_0|$ vectors 
\begin{equation}
\{\vec{k}^{(1)},\vec{k}^{(2)},\dots,\vec{k}^{(|\widehat{Q}_0|)}\}    
\end{equation}
to label a state in the representation, with each $\vec{k}^{(a)}$ a vector of dimension $N^{(a)}$. 
The component $k^{(a)}_{p}$ ($p=1,2,\dots,N^{(a)}$) is the number of atoms (i.e.\ paths) of color $a$ that start with the framing arrow of weight $m_{\mathcal{I}^{(a)}_{p}}$. 
We label the state by $|k\rangle$
\begin{equation}
|k\rangle\triangleq |\{k^{(a)}_{p}\}\rangle\,,
\end{equation}
and the ground state is $|0\rangle$. 

By construction, each such state is an eigenstate of the Cartan operator $\psi^{(a)}(z)$
\begin{equation}\label{eq:psi-action}
\psi^{(a)}(z)|k\rangle=\Psi^{(a)}_k(z)|k\rangle\,,
\end{equation}
and we now compute the charge function $\Psi^{(a)}_k(z)$. By definition \eqref{eq:QY_charge_function}: 
\begin{equation}
    \Psi^{(a)}_k(z)={}^{\sharp}\psi^{(a)}_0(z)\cdot \psi^{(\mathfrak{p}(a))}_k(z)\cdot \psi^{(\mathfrak{s}(a))}_k(z)\cdot \psi^{(a)}_k(z)\,,
\end{equation}
where ${}^{\sharp}\psi^{(a)}_0(z)$ is the ground state charge function given by \eqref{eq:GSchargeFunc}, and $\psi^{(\mathfrak{p}(a))}_k(z)\,, \psi^{(\mathfrak{s}(a))}_k(z)$, $\psi^{(a)}_k(z)$ are the contributions from the precursor of $a$, the successors of $a$, and $a$ itself, respectively. Using the bonding factors \eqref{eq:tree_bond} we obtain:
{\small
\begin{equation}
\psi^{(\mathfrak{p}(a))}_k(z)=\prod_{q=1}^{N^{(\mathfrak{p}(a))}}\prod_{s=0}^{k^{(\mathfrak{p}(a))}_q-1}\frac{z-m_{\mathcal{I}^{(\mathfrak{p}(a))}_q}-s\hbar}{z-m_{\mathcal{I}^{(\mathfrak{p}(a))}_q}-(s-1)\hbar}=\prod_{q=1}^{N^{(\mathfrak{p}(a))}}\frac{z-m_{\mathcal{I}^{(\mathfrak{p}(a))}_q}-(k^{(\mathfrak{p}(a))}_q-1)\hbar}{z-m_{\mathcal{I}^{(\mathfrak{p}(a))}_q}+\hbar}\,,
\end{equation}
}
and similarly
{\small
\begin{align}
\psi^{(\mathfrak{s}(a))}_k(z)&=\prod_{b\in\mathfrak{s}(a)}\prod_{q=1}^{N^{(b)}}\frac{z-m_{\mathcal{I}^{(b)}_q}-k^{(b)}_q\hbar}{z-m_{\mathcal{I}^{(b)}_q}}\,,\\ \psi^{(a)}_k(z)&=\prod_{p=1}^{N^{(a)}}\frac{(z-m_{\mathcal{I}^{(a)}_p})(z-m_{\mathcal{I}^{(a)}_p}+\hbar)}{(z-m_{\mathcal{I}^{(a)}_p}-k^{(a)}_p\hbar)(z-m_{\mathcal{I}^{(a)}_p}-(k^{(a)}_p-1)\hbar)}\,.
\end{align}}

\noindent Multiplying all these terms together and after canceling various factors in the numerator and denominator, we finally obtain:
\begin{tcolorbox}
{\small
\begin{equation}\label{eq:tree-char}
\Psi^{(a)}_k(z)=\frac{\prod_{b\in \mathfrak{s}(a)}\prod_{p=1}^{N^{(b)}}(z-m_{\mathcal{I}^{(b)}_{p}}-k^{(b)}_p\hbar)\prod_{q=1}^{N^{(\mathfrak{p}(a))}}(z-m_{\mathcal{I}^{(\mathfrak{p}(a))}_{q}}-(k^{(\mathfrak{p}(a))}_{q}-1)\hbar)}{\prod_{p=1}^{N^{(a)}}(z-m_{\mathcal{I}^{(a)}_{p}}-(k^{(a)}_{p}-1)\hbar)(z-m_{\mathcal{I}^{(a)}_{p}}-k^{(a)}_{p}\hbar)}\,.
\end{equation}}
\end{tcolorbox}
\noindent 
Note that when the mass parameters $\{m_1,m_2,\dots ,m_{N_{\texttt{f}}}\}$ are generic, all the poles of $\Psi^{(a)}_k(z)$ are simple poles, guaranteeing that the action \eqref{eq:e-action} and \eqref{eq:f-action} are consistent with the algebraic relations \eqref{eq:QY_quadratic_relations}.

\medskip

Given a state $|k\rangle$, applying $e^{(a)}(z)$ on it would create a new state $|k+\delta^{(a)}_{p}\rangle$ if
\begin{equation}
z=m_{\mathcal{I}^{(a)}_{p}}+k^{(a)}_{p}\hbar\,
\end{equation} is an adding pole of $\Psi^{(a)}_k(z)$, and in this case we say $k^{(a)}_{p}$ is increasable. 
For $b\in \mathfrak{s}(a)$ if $\mathcal{I}^{(a)}_{p}\in \mathcal{I}^{(b)}$, we denote the index of $\mathcal{I}^{(a)}_{p}$ in $\mathcal{I}^{(b)}$ by $p^{a\to b}$:
\begin{equation}
\mathcal{I}^{(a)}_{p}=\mathcal{I}^{(b)}_{p^{a\to b}}\,,\qquad p^{a\to b}\in\{1,2,\dots,N^{(b)}\}.
\end{equation}
From the charge function \eqref{eq:tree-char} we can see that, for a state $|k\rangle$ with $\{\vec{k}^{(1)},\vec{k}^{(2)},\dots,\vec{k}^{(|\mathrm{Q}_0|)}\}$ satisfying condition \eqref{eq:tree-codi}, $k^{(a)}_{p}$ is increasable if and only if 
\begin{equation}
k^{(a)}_{p}<k^{(b)}_{p^{a\to b}}\,,\ \qquad\forall b \in \mathfrak{s}(a)\ \text{such that}\ \mathcal{I}^{(a)}_{p}\in \mathcal{I}^{(b)}\,,
\end{equation}
where $p^{a\to b}$ is the index of $\mathcal{I}^{(a)}_{p}$ in $\mathcal{I}^{(b)}$. 
If any one of the inequalities above becomes an equality, the adding pole $z=m_{\mathcal{I}^{(a)}_{p}}+k^{(a)}_{p}\hbar$ is canceled and $k^{(a)}_p$ is no longer increasable. 
We conclude that the condition \eqref{eq:tree-codi}, which is satisfied by the vortices, is also preserved by the states in $\mathcal{R}_\nu$. 
Thus, by applying the $e$ operators on the ground state $|0\rangle$ recursively, one constructs a representation $\mathcal{R}_\nu$ of the shifted quiver Yangian that is isomorphic to the vortex Hilbert space $\mathcal{H}_{\nu}$: there exists a natural correspondence
\begin{equation}\label{eq:|k>|k)}
|k\rangle\longleftrightarrow|k)\,,
\end{equation}
where the vectors $k=\{\vec{k}^{(1)},\vec{k}^{(2)},\dots,\vec{k}^{(|\textrm{Q}_0|)}\}$ take all possible values subject to the condition \eqref{eq:tree-codi}.
Following the construction in \cite{Li:2023zub}, the action of the raising and lowering operators on the state $|k\rangle$ is given by:
\begin{align}
e^{(a)}(z)|k\rangle&=\sum_{p=1}^{N^{(a)}}\frac{e^{i\frac{\pi}{4}}}{z-m_{\mathcal{I}^{(a)}_{p}}-k^{(a)}_p \hbar}\left(\mathrm{Res}_{u=(m_{\mathcal{I}^{(a)}_{p}}+k^{(a)}_{p}\hbar)}\Psi^{(a)}_k(u)\right)^{\frac{1}{2}}|k+\delta^{(a)}_{p}\rangle\,,\label{eq:e-action}\\
f^{(a)}(z)|k\rangle&=\sum_{p=1}^{N^{(a)}}\frac{e^{i\frac{\pi}{4}}}{z-m_{\mathcal{I}^{(a)}_{p}}-(k^{(a)}_{p}-1)\hbar}\left(\mathrm{Res}_{u=(m_{\mathcal{I}^{(a)}_{p}}+(k^{(a)}_{p}-1)\hbar)}\Psi^{(a)}_k(u)\right)^{\frac{1}{2}}|k-\delta^{(a)}_{p}\rangle\,,\label{eq:f-action}
\end{align}  
where $\Psi^{(a)}_k(z)$ is the charge function of $|k\rangle$ (see \eqref{eq:tree-char}).

\medskip

Finally, note that the representation $\mathcal{R}_\nu$ is not a Verma module w.r.t.\ the entire shifted quiver Yangian. 
This is due to the zeros in the ground state charge function $\psi^{(a)}_0(z)$ \eqref{eq:GSchargeFunc}, which are designed to cancel the corresponding ``would-be" adding poles and thus ``truncate" the representation -- the growth of the representation along the paths that correspond to the ``would-be" adding poles is stopped. 
Accordingly, certain combinations of the modes would annihilate the states. 
We will discuss in Sec.~\ref{sssec:truncated_QY} how to translate this truncation of the representation into the truncation of the algebra:  we will define the \textit{truncated} shifted quiver Yangian Y$^{\text{trun.}}_{\hbar}(\widehat{Q},\widehat{W})$, which is a quotient algebra of Y$_{\hbar}(\widehat{Q},\widehat{W})$ such that the truncated representation $\mathcal{R}_\nu$, as a module of Y$^{\text{trun.}}_{\hbar}(\widehat{Q},\widehat{W})$, is a Verma module.

\subsection{Relating quiver Yangian and monopole operators}\label{ssec:relate_QY_and_monopole}

We have constructed the vortex representation $\mathcal{R}_\nu$ of the quiver Yangian, which is isomorphic to the vortex Hilbert space $\mathcal{H}_\nu$. 
It is then natural to compare the action of the quiver Yangian generators and that of the Coulomb branch operators on the same states, and to relate the two sets of operators.

Below we will determine the map between the Coulomb branch operators and the quiver Yangian generators:
\begin{equation}
\{\hat{v}^{(a)\pm}_{p}\,, \hat{\varphi}^{(a)}_{p}  \} 
\quad \longleftrightarrow \quad
\{e^{(a)}(z),f^{(a)}(z),\psi^{(a)}(z)\}\,,
\end{equation}   
by comparing their action on the state $|k)$ in the vortex Hilbert space $\mathcal{H}_{\nu}$.
We then use this map to show the quadratic relations among these two sets of operators are equivalent.

\subsubsection{Translating between quiver Yangian generators and Coulomb branch operators}\label{sssec:translation_between_QY_monopole}

In order to study the action of $\{e^{(a)}(z),f^{(a)}(z),\psi^{(a)}(z)\}$ on a state $|k)\in\mathcal{H}_\nu$, we first need to fix the relative normalization between the two states $|k)\in\mathcal{H}_{\nu}$ and $|k\rangle\in\mathcal{R}_{\nu}$. 
It turns out that one convenient choice is 
\begin{equation}\label{eq:normalization}
|k\rangle=\sqrt{\omega_k}|k)\,,
\end{equation}
where $\omega_k$ is the equivariant weight of the state $|k)$ (see \eqref{eq:omegak}).

With the relative normalization of the states fixed, we can now compare the action of these two sets of operators. 
Recall the action \eqref{eq:e-action} and \eqref{eq:f-action}, with the charge function given by \eqref{eq:tree-char}. 
Taking the relative normalization \eqref{eq:normalization} into consideration, one finds
{\footnotesize
\begin{equation}
    \begin{aligned}
e^{(a)}(z)|k)=\sum_{p=1}^{N^{(a)}}&\frac{e^{i\frac{\pi}{4}}}{z-m_{\mathcal{I}^{(a)}_{p}}-k^{(a)}_p \hbar}\left(\mathrm{Res}_{u=(m_{\mathcal{I}^{(a)}_{p}}+k^{(a)}_{p}\hbar)}\Psi^{(a)}_k(u)\right)^{\frac{1}{2}}\sqrt{\frac{\omega_{k+\delta^{(a)}_{p}}}{\omega_k}}|k+\delta^{(a)}_{p})\,,\\
f^{(a)}(z)|k)=\sum_{p=1}^{N^{(a)}}&\frac{e^{i\frac{\pi}{4}}}{z-m_{\mathcal{I}^{(a)}_{p}}-(k^{(a)}_{p}-1)\hbar}\left(\mathrm{Res}_{u=(m_{\mathcal{I}^{(a)}_{p}}+(k^{(a)}_{p}-1)\hbar)}\Psi^{(a)}_k(u)\right)^{\frac{1}{2}}\sqrt{\frac{\omega_{k-\delta^{(a)}_{p}}}{\omega_k}}|k-\delta^{(a)}_{p})\,.
\end{aligned}
\end{equation}}
We have not yet specified the value of the parameter $\hbar$ of the quiver Yangian. 
To reproduce the Coulomb branch algebra (see Sec.~\ref{sssec:coulomb_branch_algebra}), we need the identification: 
\begin{equation}\label{eq:hbar=eps}
\hbar=\eps\,,
\end{equation}
where $\eps$ is the parameter of the $\Omega$-deformation in the 3D theory. 
Hence we see how the parameters of the 3D $\mathcal{N}=4$ theory manifest themselves in the corresponding quiver Yangian: 
\begin{itemize}
\item the $\Omega$-deformation parameter arises as the weight of the self-loops of the quiver Yangian;
\item the mass parameters arise in the weights of the framing arrows, which determine the desired vortex representation $\mathcal{R}_\nu$.
\end{itemize}
Then, with the explicit expressions for the equivariant weights and the charge functions, \eqref{eq:omegak} and \eqref{eq:tree-char}, and using \eqref{eq:hbar=eps}, we have
{\small
\begin{equation}\label{eq:e_on_vortices}
    \begin{aligned}
e^{(a)}(z)|k)=&(\frac{(-1)^{N^{(a)}+N^{(\mathfrak{p}(a))}-1}}{\epsilon})^{\frac{1}{2}}\sum_{p=1}^{N^{(a)}}\frac{e^{i\frac{\pi}{4}}}{z-m_{\mathcal{I}^{(a)}_{p}}-k^{(a)}_p \epsilon}\\
 &\times \frac{\prod_{q=1}^{N^{(\mathfrak{p}(a))}}(m_{\mathcal{I}^{(\mathfrak{p}(a))}_{q}}-m_{\mathcal{I}^{(a)}_{p}}+(k^{(\mathfrak{p}(a))}_{q}-k^{(a)}_{p}-1)\epsilon)}{\prod_{q\neq p}^{N^{(a)}}(m_{\mathcal{I}^{(a)}_{q}}-m_{\mathcal{I}^{(a)}_{p}}+(k^{(a)}_{q}-k^{(a)}_{p}-1)\epsilon)}|k+\delta^{(a)}_{p})\\
 =&\varepsilon^+(a)\sum_{p=1}^{N^{(a)}}\frac{1}{z-m_{\mathcal{I}^{(a)}_{p}}-k^{(a)}_p \epsilon}\frac{Q^{(\mathfrak{p}(a))}(\hat{\varphi}^{(a)}_p)}{\prod_{q\neq p}^{N^{(a)}}(\hat{\varphi}^{(a)}_p-\hat{\varphi}^{(a)}_q)} |k+\delta^{(a)}_{p})\,,
\end{aligned}
\end{equation}}
and
{\small
\begin{equation}\label{eq:f_on_vortices}
\begin{aligned}
f^{(a)}(z)|k)=&(\frac{(-1)^{N^{(a)}+N^{(\mathfrak{p}(a))}}}{\epsilon})^{\frac{1}{2}}\sum_{p=1}^{N^{(a)}}\frac{e^{i\frac{\pi}{4}}}{z-m_{\mathcal{I}^{(a)}_{p}}-(k^{(a)}_p-1) \epsilon}\\
&\times \frac{\prod_{b\in \mathfrak{s}(a)}\prod_{q=1}^{N^{(b)}}(m_{\mathcal{I}^{(a)}_{p}}-m_{\mathcal{I}^{(b)}_{q}}+(k^{(a)}_{p}-k^{(b)}_{q}-1)\epsilon)}{\prod_{q\neq p}^{N^{(a)}}(m_{\mathcal{I}^{(a)}_{p}}-m_{\mathcal{I}^{(a)}_{q}}+(k^{(a)}_{p}-k^{(a)}_{q}-1)\epsilon)}|k-\delta^{(a)}_{p})\\
=&\varepsilon^-(a)\sum_{p=1}^{N^{(a)}}\frac{1}{z-m_{\mathcal{I}^{(a)}_{p}}-(k^{(a)}_p-1) \epsilon}  \frac{\prod_{b\in \mathfrak{s}(a)}(-1)^{N^{(b)}}Q^{(b)}(\hat{\varphi}^{(a)}_p)}{\prod_{q\neq p}^{N^{(a)}}(\hat{\varphi}^{(a)}_q-\hat{\varphi}^{(a)}_p)}|k-\delta^{(a)}_{p})\,,
\end{aligned}
\end{equation}}
where we have used the following shorthand notations:
\begin{equation}\label{eq:epsilon+-}
\varepsilon^\pm(a):=\frac{e^{i\pi(\frac{N^{(a)}+N^{(\mathfrak{p}(a))}}{2}\mp\frac{1}{4})}}{\epsilon^{\frac{1}{2}}}\,.
\end{equation}
Note also that, when evaluating the square roots in \eqref{eq:e_on_vortices} and \eqref{eq:f_on_vortices} we have picked certain branches such that
{\small
\begin{equation}
(\frac{(-1)^{N^{(a)}+N^{(\mathfrak{p}(a))}-1}}{\epsilon})^{\frac{1}{2}}=\frac{e^{i\pi(\frac{N^{(a)}+N^{(\mathfrak{p}(a))}-1}{2})}}{\epsilon^{\frac{1}{2}}}\,,\quad  (\frac{(-1)^{N^{(a)}+N^{(\mathfrak{p}(a))}}}{\epsilon})^{\frac{1}{2}}=\frac{e^{i\pi(\frac{N^{(a)}+N^{(\mathfrak{p}(a))}}{2})}}{\epsilon^{\frac{1}{2}}}\,.
\end{equation}}

\noindent Finally, since the states $|k)\in\mathcal{H}_{\nu}$ and $|k \rangle\in\mathcal{R}_\nu$ are related by a normalization, it is easy to see that the action of $\psi^{(a)}(z)$ on $|k)$ is
\begin{equation}\label{eq:psi_on_vortices}
\psi^{(a)}(z)|k)=\Psi^{(a)}_k|k)\,,
\end{equation}
where the charge function $\Psi^{(a)}_k$ is still given by \eqref{eq:tree-char}.

Comparing \eqref{eq:e_on_vortices}, \eqref{eq:f_on_vortices}, \eqref{eq:psi_on_vortices} with the earlier results \eqref{eq:v+_action}, \eqref{eq:v-_action} and \eqref{eq:varphi_action} (which are now on the same set of states $|k)\in \mathcal{H}_\nu$), we obtain the following correspondence between the operators $\{\hat{v}^{(a)\pm}_{p}\,, \hat{\varphi}^{(a)}_{p}  \}$ and $\{e^{(a)}(z),f^{(a)}(z),\psi^{(a)}(z)\}$.

\begin{itemize}
\item Raising operators.
Comparing \eqref{eq:e_on_vortices} with the action of the raising operators in the Coulomb branch algebra (see \eqref{eq:v+_action}), we have
\begin{equation}
\begin{aligned}
e^{(a)}(z)&=\varepsilon^+(a)\sum_{p=1}^{N^{(a)}} \hat{v}_{p}^{(a)+} \sum_{k^{(a)}_p=0}^{\infty} \frac{\delta(\hat{\varphi}^{(a)}_{p}|-m_{\mathcal{I}^{(a)}_{p}}-k^{(a)}_{p}\epsilon-\frac{\epsilon}{2})}{z-m_{\mathcal{I}^{(a)}_{p}}-k^{(a)}_{p}\epsilon}\,,\\  
&=\varepsilon^+(a)\sum_{p=1}^{N^{(a)}} \hat{v}^{(a)+}_{p}\frac{1}{z+\hat{\varphi}^{(a)}_{p}+\frac{\epsilon}{2}}\,,
\end{aligned}
\end{equation}
where $\delta(a|b)$ is another notation for the Kronecker delta  $\delta_{ab}$, and we have summed over $k^{(a)}_p$ to reach the second line.

\item Lowering operators. 
Similarly, comparing \eqref{eq:f_on_vortices} with the action of the lowering operators in the Coulomb branch algebra (see \eqref{eq:v-_action}), we have
\begin{equation}
\begin{aligned}
f^{(a)}(z)&=\varepsilon^-(a)\sum_{p=1}^{N^{(a)}}\hat{v}^{(a)-}_{p}\sum_{k^{(a)}_{p}=0}^{\infty}\frac{\delta(\hat{\varphi}^{(a)}_{p}|-m_{\mathcal{I}^{(a)}_{p}}-k^{(a)}_{p}\epsilon-\frac{\epsilon}{2})}
{z-m_{\mathcal{I}^{(a)}_{p}}-(k^{(a)}_p-1)\epsilon}\,,\\
&=\varepsilon^-(a)\sum_{p=1}^{N^{(a)}}\hat{v}^{(a)-}_{p}\frac{1}{z+\hat{\varphi}^{(a)}_{p}+\frac{3\epsilon}{2}}\,.
\end{aligned}
\end{equation}

\item Cartan operators. 
Finally, comparing \eqref{eq:psi_on_vortices} with the action of the Cartan operators in the Coulomb branch algebra (see \eqref{eq:varphi_action}), we have
\begin{equation}
\begin{aligned}
\psi^{(a)}(z)&=\frac{\prod_{b\in \mathfrak{s}(a)}\prod_{q=1}^{N^{(b)}}(z+\hat{\varphi}^{(b)}_{q}+\frac{\epsilon}{2})\cdot\prod_{q=1}^{N^{(\mathfrak{p}(a))}}(z+\hat{\varphi}^{(\mathfrak{p}(a))}_{q}+\frac{3\epsilon}{2})}{\prod_{p=1}^{N^{(a)}}(z+\hat{\varphi}^{(a)}_{p}+\frac{\epsilon}{2})(z+\hat{\varphi}^{(a)}_{p}+\frac{3\epsilon}{2})}\,.\\
\end{aligned}
\end{equation}
\end{itemize}

In summary, the map from the quiver Yangian operators $
\{e^{(a)}(z),f^{(a)}(z),\psi^{(a)}(z)\}$ to the Coulomb branch operators $\{\hat{v}^{(a)\pm}_{p}\,, \hat{\varphi}^{(a)}_{p}  \}$  is:
{\small
\begin{tcolorbox}[breakable,ams equation]
\label{eq:tree_QYmono1}
\begin{aligned}
e^{(a)}(z)
&=\varepsilon^+(a)\sum_{p=1}^{N^{(a)}} \hat{v}^{(a)+}_{p}\frac{1}{z+\hat{\varphi}^{(a)}_{p}+\frac{\epsilon}{2}}\,,\\
f^{(a)}(z)
&=\varepsilon^-(a)\sum_{p=1}^{N^{(a)}}\hat{v}^{(a)-}_p \frac{1}{z+\hat{\varphi}^{(a)}_{p}+\frac{3\epsilon}{2}}\,,\\
\psi^{(a)}(z)&=\frac{\prod_{b\in \mathfrak{s}(a)}\prod_{q=1}^{N^{(b)}}(z+\hat{\varphi}^{(b)}_{q}+\frac{\epsilon}{2})\cdot\prod_{q=1}^{N^{(\mathfrak{p}(a))}}(z+\hat{\varphi}^{(\mathfrak{p}(a))}_{q}+\frac{3\epsilon}{2})}{\prod_{p=1}^{N^{(a)}}(z+\hat{\varphi}^{(a)}_{p}+\frac{\epsilon}{2})(z+\hat{\varphi}^{(a)}_{p}+\frac{3\epsilon}{2})}\,.
\end{aligned}  
\end{tcolorbox}}
\noindent
Plugging in the mode expansion \eqref{eq:QY_mode_expansion} we have the map in terms of the modes:
{\footnotesize
\begin{equation}\label{eq:mono_QY_mode}
\begin{aligned}
e^{(a)}_n&=\varepsilon^{+}(a)\sum_{p=1}^{N^{(a)}}\hat{v}^{(a)+}_p (-\hat{\varphi}^{(a)}_p-\frac{\eps}{2})^n\,,\qquad f^{(a)}_n=\varepsilon^{-}(a)\sum_{p=1}^{N^{(a)}}\hat{v}^{(a)-}_p (-\hat{\varphi}^{(a)}_p-\frac{3\eps}{2})^n\,,\\
\psi^{(a)}_n&=\varepsilon^{+}(a)\varepsilon^{-}(a)\sum_{p=1}^{N^{(a)}}\left((-\hat{\varphi}^{(a)}_p-\frac{3\eps}{2})^{n+\mathtt{s}^{(a)}}C^{(a)+-}_p-(-\hat{\varphi}^{(a)}_p-\frac{\eps}{2})^{n+\mathtt{s}^{(a)}}C^{(a)-+}_p\right)\,,
\end{aligned}
\end{equation}}
where $C^{(a)+-}_p$ and $C^{(a)-+}_p$ are defined in \eqref{eq:C+C-}. 
One can see that unlike the monopole operators $\hat{v}^{(a)\pm}_p$ and the complex scalars $\hat{\varphi}^{(a)}_p$, the quiver Yangian operators $\{e^{(a)}_n, \psi^{(a)}_n,f^{(a)}_n\}$ are automatically Weyl-invariant. 

\medskip

Conversely, we can also express the Coulomb branch operators in terms of the quiver Yangian generators:
\begin{equation}\label{eq:tree_QYmono2}
\begin{aligned}
\hat{v}^{(a)+}_{p}&=\frac{1}{\varepsilon^+(a)}\sum_{n\geq 0}\oint_{m_{\mathcal{I}^{(a)}_{p}}+n\epsilon}\frac{dz}{2\pi \text{i}}e^{(a)}(z)\,,\\
\hat{v}^{(a)-}_{p}&=\frac{1}{\varepsilon^-(a)}\sum_{n\geq0}\oint_{m_{\mathcal{I}^{(a)}_{p}}+(n-1)\epsilon}\frac{dz}{2\pi \text{i}}f^{(a)}(z)\,,\\
\hat{\varphi}^{(a)}_{p}&=-m_{\mathcal{I}^{(a)}_{p}}-\mathrm{Max}( K^{(a)}_{p})\ \epsilon-\frac{\epsilon}{2}\,, 
\end{aligned} 
\end{equation}
\noindent where the integration is performed on an infinitesimal circle enclosing the pole, and $K^{(a)}_p$ is a set-valued operator defined as
\begin{equation}
K^{(a)}_{p} :=\{n\in \mathbb{Z}|\mathrm{Res}_{z=(m_{\mathcal{I}^{(a)}_{p}}+n\epsilon)}\psi^{(a)}(z)\neq 0\}\,.
\end{equation}

\subsubsection{Truncated shifted quiver Yangian}\label{sssec:truncated_QY}

Recall that the vortex representation $\mathcal{R}_\nu$, constructed in Sec.~\ref{ssec:vortex_as_QY_rep}, is not a Verma module w.r.t.\ the shifted quiver Yangian Y$_{\hbar}(\widehat{Q},\widehat{W})$, due to the zeros in the ground state charge function $\psi^{(a)}_0(z)$ \eqref{eq:GSchargeFunc} that truncate the representation.
One can translate this truncation of the representation into a truncation of the algebra: 
namely, we can define the truncated version of the shifted quiver Yangian, Y$^{\text{trun.}}_{\hbar}(\widehat{Q},\widehat{W})$, such that the vortex representation $\mathcal{R}_\nu$ is a Verma module of Y$^{\text{trun.}}_{\hbar}(\widehat{Q},\widehat{W})$.

We define the truncated shifted quiver Yangian Y$^{\text{trun.}}_{\hbar}(\widehat{Q},\widehat{W})$ using the  mode relations \eqref{eq:mono_QY_mode}. 
These relations define a map from the modes $\{e^{(a)}_n,f^{(a)}_n,\psi^{(a)}_n\}$ of the shifted quiver Yangian generators to the Coulomb branch algebra $\mathbb{C}_\eps[\mathcal{M}_C]$, which we denote by $\mathcal{F}$:
\begin{equation}\label{eq:map_QY_to_mono}
\mathcal{F}:\ \{e^{(a)}_n,f^{(a)}_n,\psi^{(a)}_n\}\to \{\hat{v}^{(a)\pm}_p,\hat{\varphi}^{(a)}_p\}\,, \quad \text{with the correspondence given by \eqref{eq:mono_QY_mode}}\,,
\end{equation}
and we define the truncated shifted quiver Yangian Y$^{\text{trun.}}_{\hbar}(\widehat{Q},\widehat{W})$ as the pre-image of this map:
\begin{equation}
\mathrm{Y}^{\text{trun.}}_{\hbar}(\widehat{Q},\widehat{W}):=\mathcal{F}^{-1}(\mathbb{C}_\eps[\mathcal{M}_C])\,.
\end{equation}
Then the vortex representation $\mathcal{R}_\nu$ is a Verma module of the truncated shifted quiver Yangian Y$^{\text{trun.}}_{\hbar}(\widehat{Q},\widehat{W})$ by construction. 

One might try to define the truncated algebra more explicitly by quotienting out the additional relations that one can deduce from the disappearance of the ``would-be" adding poles, caused by zeros in $\psi^{(a)}_0(z)$.
This is similar to how one can define the $\mathcal{W}_{N}$ algebra from the $\mathcal{W}_{\infty}$ algebra, except that the situation here is much more complicated, partially due to the interference of the set of relations from different zeros. 

\subsubsection{Matching the quadratic relations}
Using \eqref{eq:tree_QYmono1} and \eqref{eq:tree_QYmono2}, one can translate the algebraic relations of the quiver Yangian into those of the Coulomb branch operators, and vice versa. 
We will use the quadratic relations of the quiver Yangian to reproduce the Coulomb branch algebra.

\subsubsubsection{From $e$-$f$ relations to $\hat{v}^+$-$\hat{v}^-$ relations}

We now show that the $\hat{v}^+$-$\hat{v}^-$ relation \eqref{eq:v+v-_relation} is equivalent to the $e$-$f$ relation of the quiver Yangian
\begin{equation}\label{eq:QY_ef}
[e^{(a)}(z),f^{(b)}(w)]\simeq -\delta_{ab}\frac{\psi^{(a)}(z)-\psi^{(b)}(w)}{z-w}\,.
\end{equation}
Plugging the expressions of $\hat{v}^{(a)+}_p$ and $\hat{v}^{(b)-}_q$ in terms of the $e$ and $f$ generators from \eqref{eq:tree_QYmono2}, we have 
\begin{equation}
\begin{aligned}
[\hat{v}^{(a)+}_p,\hat{v}^{(b)-}_q]=&\frac{1}{\varepsilon^+(a)\varepsilon^-(b)} \sum_{n_1,n_2\geq 0} \oint_{m_{\mathcal{I}^{(a)}_{p}}+n_1\epsilon}\frac{dz}{2\pi \text{i}} \oint_{m_{\mathcal{I}^{(b)}_{q}}+(n_2-1)\epsilon}\frac{dw}{2\pi \text{i}}[e^{(a)}(z),f^{(b)}(w)]\,,\\
=&\frac{1}{\varepsilon^+(a)\varepsilon^-(b)} \sum_{n_1,n_2\geq 0} \oint_{m_{\mathcal{I}^{(a)}_{p}}+n_1\epsilon}\frac{dz}{2\pi \text{i}} \oint_{m_{\mathcal{I}^{(b)}_{q}}+(n_2-1)\epsilon}\frac{dw}{2\pi \text{i}}\delta_{ab}\frac{\psi^{(b)}(w)-\psi^{(a)}(z)}{z-w}\,,
\end{aligned}
\end{equation}
where for the second equality we have used the $e$-$f$ relation \eqref{eq:QY_ef}. 
Then integrating over $w$,  we have
\begin{equation}
[\hat{v}^{(a)+}_p,\hat{v}^{(b)-}_q]=\frac{\delta_{ab}\epsilon}{(-1)^{N^{(a)}+N^{(\mathfrak{p}(a))}}}\sum_{n_1,n_2\geq 0}\oint_{m_{\mathcal{I}^{(a)}_{p}}+n_1\epsilon}\frac{dz}{2\pi \text{i}}\frac{\text{Res}_{w=(m_{\mathcal{I}^{(a)}_{q}}+(n_2-1)\epsilon)}\ \psi^{(a)}(w)}{z-m_{\mathcal{I}^{(a)}_{q}}-(n_2-1)\epsilon}\,,
\end{equation}
where we have plugged in the explicit expression of $\varepsilon^\pm(a)$, see \eqref{eq:epsilon+-}. Then integrating over $z$ gives $\delta_{pq}\delta_{n_1,n_2-1}$, and we thus obtain
\begin{equation}\label{eq:v+v-_proof}
[\hat{v}^{(a)+}_p,\hat{v}^{(b)-}_q]=\frac{\delta_{ab}\delta_{pq}\epsilon}{(-1)^{N^{(a)}+N^{(\mathfrak{p}(a))}}}\sum_{n\geq 0} \text{Res}_{w=(m_{\mathcal{I}^{(a)}_{p}}+n\epsilon)}\ \psi^{(a)}(w)\,.
\end{equation}
As the final step, we translate $\psi^{(a)}(w)$ back into $\hat{\varphi}^{(a)}_p$'s, using the third line of \eqref{eq:tree_QYmono1}. Since $\hat{\varphi}^{(a)}_p$ takes values of the form $-m_{\mathcal{I}^{(a)}_{p}}-\#\epsilon-\frac{\epsilon}{2}$, the summation over $n$ guarantees that the two poles $w=-\hat{\varphi}^{(a)}_p-\frac{\epsilon}{2}$ and $w=-\hat{\varphi}^{(a)}_p-\frac{3\epsilon}{2}$ are covered, and we obtain
\begin{equation}
\begin{aligned}
&\sum_{n\geq0}\text{Res}_{w=(m_{\mathcal{I}^{(a)}_{p}}+n\epsilon)}\ \psi^{(a)}(w)\\
=&
\frac{1}{\epsilon}\left(\frac{\prod_{c\in \mathfrak{s}(a)}\prod_{q=1}^{N^{(c)}}(\hat{\varphi}^{(c)}_{q}-\hat{\varphi}^{(a)}_{p})\cdot\prod_{q=1}^{N^{(\mathfrak{p}(a))}}(\hat{\varphi}^{(\mathfrak{p}(a))}_{q}-\hat{\varphi}^{(a)}_{p}+\epsilon)}{\prod_{q\neq p}^{N^{(a)}}(\hat{\varphi}^{(a)}_{q}-\hat{\varphi}^{(a)}_{p})(\hat{\varphi}^{(a)}_{q}-\hat{\varphi}^{(a)}_{p}+\epsilon)}\right.\\
&-\left.\frac{\prod_{c\in \mathfrak{s}(a)}\prod_{q=1}^{N^{(c)}}(\hat{\varphi}^{(c)}_{q}-\hat{\varphi}^{(a)}_{p}-\eps)\cdot\prod_{q=1}^{N^{(\mathfrak{p}(a))}}(\hat{\varphi}^{(\mathfrak{p}(a))}_{q}-\hat{\varphi}^{(a)}_{p})}{\prod_{q\neq p}^{N^{(a)}}(\hat{\varphi}^{(a)}_{q}-\hat{\varphi}^{(a)}_{p})(\hat{\varphi}^{(a)}_{q}-\hat{\varphi}^{(a)}_{p}-\epsilon)}\right)
= \frac{(-1)^{N^{(a)}+N^{(\mathfrak{p}(a))}}}{\epsilon} C^{(a)}_p\,.
\end{aligned}
\end{equation}
Substituting this expression into \eqref{eq:v+v-_proof}, we finally reproduce \eqref{eq:v+v-_relation} from the $e$-$f$ relation \eqref{eq:QY_ef} of the quiver Yangian.

\subsubsubsection{$\hat{v}^+$-$\hat{v}^+$ and $\hat{v}^-$-$\hat{v}^-$ relations}

Recall that the $\hat{v}^{(a)+}_p$-$\hat{v}^{(b)+}_q$ and $\hat{v}^{(a)-}_p$-$\hat{v}^{(b)-}_q$ relations depend on the relative positions of the nodes $a$ and $b$ in the quiver $\mathrm{Q}$, and there are three possibilities, given in \eqref{eq:vpmvpm_relation1} -- \eqref{eq:vpmvpm_relation3}. 
We now show that they can be reproduced from the $e$-$e$ (resp.\ $f$-$f$) relation of the quiver Yangian:
\begin{equation}\label{eq:QY_ee}
\begin{aligned}
e^{(a)}(z)e^{(b)}(w)&\sim (-1)^{|a||b|}\varphi^{a\Leftarrow b}(z-w)e^{(b)}(w)e^{(a)}(z)\,,  \\
f^{(a)}(z)f^{(b)}(w)&\sim (-1)^{|a||b|}\varphi^{a\Leftarrow b}(z-w)^{-1}f^{(b)}(w)f^{(a)}(z)\,,
\end{aligned}
\end{equation}
where, for the triple quiver, we have $|a|=0$ for $a\in\widehat{Q}_0$.

We begin with 
\begin{equation}
\begin{aligned}
\hat{v}^{(a)+}_p\hat{v}^{(b)+}_q=&\frac{1}{\varepsilon^+(a)\varepsilon^+(b)}\sum_{n_1,n_2\geq 0} \oint_{m_{\mathcal{I}^{(a)}_{p}}+n_1\epsilon}\frac{dz}{2\pi \text{i}} \oint_{m_{\mathcal{I}^{(b)}_{q}}+n_2\epsilon}\frac{dw}{2\pi \text{i}}\ e^{(a)}(z)e^{(b)}(w)\,,\\
=&\frac{1}{\varepsilon^+(a)\varepsilon^+(b)}\sum_{n_1,n_2\geq 0} \oint_{m_{\mathcal{I}^{(a)}_{p}}+n_1\epsilon}\frac{dz}{2\pi \text{i}} \oint_{m_{\mathcal{I}^{(b)}_{q}}+n_2\epsilon}\frac{dw}{2\pi \text{i}}\ \varphi^{a\Leftarrow b}(z-w) e^{(b)}(w)e^{(a)}(z)\,,\\
\end{aligned}
\end{equation}
where we have used the $e$-$e$ relation \eqref{eq:QY_ee} to get the second line. 
Similarly, for the $\hat{v}^{(a)-}_p$ operators, we have
\begin{equation}
\begin{aligned}
\hat{v}^{(a)-}_p\hat{v}^{(b)-}_q =\frac{1}{\varepsilon^-(a)\varepsilon^-(b)}\sum_{n_1,n_2\geq 0} \oint_{m_{\mathcal{I}^{(a)}_{p}}+(n_1-1)\epsilon}\frac{dz}{2\pi \text{i}} \oint_{m_{\mathcal{I}^{(b)}_{q}}+(n_2-1)\epsilon}\frac{dw}{2\pi \text{i}}&\\
\times \varphi^{a\Leftarrow b}(z-w)^{-1} f^{(b)}(w)&f^{(a)}(z)\,,
\end{aligned}
\end{equation}
where we have used the second line of \eqref{eq:tree_QYmono2} and the $f$-$f$ relation in \eqref{eq:QY_ee}.

\medskip

Now we proceed with the three different cases in \eqref{eq:vpmvpm_relation1} -- \eqref{eq:vpmvpm_relation3}.
\begin{enumerate}
\item For $a,b$ not adjacent, the bonding factor $\varphi^{a\Leftarrow b}(z-w)$ is trivial and we immediately obtain \eqref{eq:vpmvpm_relation1}.

\item For $a=\mathfrak{p}(b)$, the bonding factor is $\varphi^{a\Leftarrow b}(z-w)=\frac{z-w-\epsilon}{z-w}$, and hence we have
{\small
    \begin{align}
\hat{v}^{(a)+}_p\hat{v}^{(b)+}_q=&\frac{1}{\varepsilon^+(a)\varepsilon^+(b)}\sum_{n_1,n_2\geq 0} \oint_{m_{\mathcal{I}^{(a)}_{p}}+n_1\epsilon}\frac{dz}{2\pi \text{i}} \oint_{m_{\mathcal{I}^{(b)}_{q}}+n_2\epsilon}\frac{dw}{2\pi \text{i}}\ \frac{z-w-\epsilon}{z-w} e^{(b)}(w)e^{(a)}(z)\,,\notag\\
=&\hat{v}^{(b)+}_q\frac{1}{\varepsilon^+(a)}\sum_{n_1\geq 0} \oint_{m_{\mathcal{I}^{(a)}_{p}}+n_1\epsilon}\frac{dz}{2\pi \text{i}} \  e^{(a)}(z)\frac{z+\hat{\varphi}^{(b)}_q-\frac{\epsilon}{2}}{z+\hat{\varphi}^{(b)}_q+\frac{\epsilon}{2}}\,,
\end{align}
}

\noindent where we have translated the quiver Yangian operators into the monopole operators and evaluated the contour integral over $w$, and similarly for $\hat{v}^{(a)-}_p\hat{v}^{(b)-}_q$.
Then evaluating the contour integral over $z$ reproduces \eqref{eq:vpmvpm_relation2}.

\item Finally, for $a=b$ the bonding factor is $\varphi^{a\Leftarrow a}(z-w)=\frac{z-w+\epsilon}{z-w-\eps}$, then we have
{\small
    \begin{align}
\hat{v}^{(a)+}_p\hat{v}^{(a)+}_q=&\frac{1}{(\varepsilon^+(a))^2}\sum_{n_1,n_2\geq 0} \oint_{m_{\mathcal{I}^{(a)}_{p}}+n_1\epsilon}\frac{dz}{2\pi \text{i}} \oint_{m_{\mathcal{I}^{(a)}_{q}}+n_2\epsilon}\frac{dw}{2\pi \text{i}}\ \frac{z-w+\epsilon}{z-w-\epsilon} e^{(a)}(w)e^{(a)}(z)\,,\notag\\
=&\hat{v}^{(a)+}_q\frac{1}{\varepsilon^+(a)} \sum_{n_1\geq 0} \oint_{m_{\mathcal{I}^{(a)}_{p}}+n_1\epsilon}\frac{dz}{2\pi \text{i}} \  e^{(a)}(z)\frac{z+\hat{\varphi}^{(a)}_q+\frac{3\epsilon}{2}}{z+\hat{\varphi}^{(a)}_q-\frac{\epsilon}{2}}\,,
\end{align}
}

\noindent and similarly for $\hat{v}^{(a)-}_p\hat{v}^{(a)-}_q$. 
They reproduce \eqref{eq:vpmvpm_relation3} after we evaluate the contour integral over $z$.
\end{enumerate}

\medskip

In summary, we have checked explicitly that the $\hat{v}^{+}$-$\hat{v}^{+}$  and $\hat{v}^{+}$-$\hat{v}^{-}$ relations, \eqref{eq:vpmvpm_relation1} --\eqref{eq:vpmvpm_relation3}, can be reproduced from the $e$-$e$ and $f$-$f$ relations of the quiver Yangian.

\subsubsubsection{$\hat{\varphi}$-$\hat{v}^\pm$ relations}
We now look at the relations between the $\hat{\varphi}^{(a)}_p$ and $\hat{v}^{(a)\pm}_{q}$ operators, see \eqref{eq:phiv+- relation}.
As an example we focus on the $\hat{\varphi}$-$\hat{v}^+$ relation.
We show that
\begin{equation}\label{eq:phiv+ relation}
[\hat{\varphi}^{(a)}_p, \hat{v}^{(b)+}_q]=-\epsilon\delta_{ab}\delta_{pq}\hat{v}^{(b)+}_q
\end{equation}
can be derived from the $\psi$-$e$ relation
of the quiver Yangian:
\begin{equation}\label{eq:QY_psi_e}
\psi^{(a)}(z)e^{(b)}(w)\sim \varphi^{a\Leftarrow b}(z-w)e^{(b)}(w)\psi^{(a)}(z)\,.
\end{equation}
We will focus on the non-trivial part of $\hat{\varphi}^{(a)}_p$, namely $\text{Max}( K^{(a)}_{p})$, and work out its commutator with $\hat{v}^{(b)+}_q$. We have
\begin{equation}
\begin{aligned}
&\text{Max}( K^{(a)}_{p}) \hat{v}^{(b)+}_q \\
=&\frac{1}{\varepsilon^+(b)}\sum_{n\geq 0}\oint_{m_{\mathcal{I}^{(b)}_{q}}+n\epsilon}\frac{dw}{2\pi \text{i}}
\text{Max}\{ k\in \mathbb{Z}|\text{Res}_{z=(m_{\mathcal{I}^{(a)}_{p}}+k\epsilon)}\psi^{(a)}(z)\neq 0\}\ \ e^{(b)}(w)\,,\\
=&\frac{1}{\varepsilon^+(b)}\sum_{n\geq 0}\oint_{m_{\mathcal{I}^{(b)}_{q}}+n\epsilon}\frac{dw}{2\pi \text{i}}e^{(b)}(w) \text{Max}\{ k\in \mathbb{Z}|\text{Res}_{z=(m_{\mathcal{I}^{(a)}_{p}}+k\epsilon)}\ \varphi^{a\Leftarrow b}(z-w)\psi^{(a)}(z)\neq 0\}\,,\\
=&\hat{v}^{(b)+}_q\sum_{n\geq 0} \delta(\hat{\varphi}^{(b)}_{q}|-m_{\mathcal{I}^{(b)}_{q}}-n\epsilon-\frac{\epsilon}{2})\\
&\qquad \qquad \times\text{Max}\{ k\in \mathbb{Z}|\text{Res}_{z=(m_{\mathcal{I}^{(a)}_{p}}+k\epsilon)}\ \varphi^{a\Leftarrow b}(z-m_{\mathcal{I}^{(b)}_{q}}-n\epsilon)\psi^{(a)}(z)\neq 0\}\,,\\
=&\hat{v}^{(b)+}_q \text{Max}\{ k\in \mathbb{Z}|\text{Res}_{z=(m_{\mathcal{I}^{(a)}_{p}}+k\epsilon)}\ \varphi^{a\Leftarrow b}(z+\hat{\varphi}^{(b)}_{q}+\frac{\epsilon}{2})\psi^{(a)}(z)\neq 0\}\,,
\end{aligned}
\end{equation}
where in the second line we have used the $\psi^{(a)}$-$e^{(b)}$ relation \eqref{eq:QY_psi_e}, namely $\psi^{(a)}(z)$ is to be multiplied by the bonding factor $\varphi^{a\Leftarrow b}(z-w)$ when passing through $e^{(b)}(w)$. 
In the next two lines we performed the contour integral to translate $e^{(b)}(w)$ back into $\hat{v}^{(b)+}_q$. 
To proceed, we need to distinguish between two different cases:
\begin{itemize}
\item For $a\neq b$, there are two possibilities: 
\begin{itemize}
\item When $a,b$ are not adjacent, the bonding factor is trivial.
\item When $a,b$ are adjacent to each other, then either $b=\mathfrak{p}(a)$ or $a=\mathfrak{p}(b)$. 
In both cases the pole of the bonding factor $\varphi^{a\Leftarrow b}(z+\hat{\varphi}^{(b)}_{q}+\frac{\epsilon}{2})$ is just canceled by the zero of $\psi^{(a)}(z)$, see \eqref{eq:tree_bond} and \eqref{eq:tree_QYmono1}.
\end{itemize}
In both cases, the pole structure of $\varphi^{a\Leftarrow b}(z+\hat{\varphi}^{(b)}_{q}+\frac{\epsilon}{2})\psi^{(a)}(z)$ is the same as that of $\psi^{(a)}(z)$, hence we have $\text{Max}( K^{(a)}_{p}) \hat{v}^{(b)+}_q= \hat{v}^{(b)+}_q\text{Max}( K^{(a)}_{p})$.
    
\item For $a=b$, the bonding factor is 
\begin{equation}\varphi^{a\Leftarrow a}(z+\hat{\varphi}^{(a)}_{q}+\frac{\epsilon}{2})=\frac{z+\hat{\varphi}^{(a)}_{q}+\frac{3\epsilon}{2}}{z+\hat{\varphi}^{(a)}_{q}-\frac{\epsilon}{2}}\,,\end{equation}
and it is straightforward to check \begin{equation}\text{Max}\{ k\in \mathbb{Z}|\text{Res}_{z=(m_{\mathcal{I}^{(a)}_{p}}+k\epsilon)}\ \varphi^{a\Leftarrow a}(z+\hat{\varphi}^{(a)}_{q}+\frac{\epsilon}{2})\psi^{(a)}(z)\neq 0\}=\text{Max}( K^{(a)}_{p})+\delta_{pq}\,.\end{equation}
It follows that $\text{Max}( K^{(a)}_{p}) \hat{v}^{(a)+}_q= \hat{v}^{(a)+}_q(\text{Max}( K^{(a)}_{p})+\delta_{pq})$.
\end{itemize}
To summarize, the commutator of $\text{Max}( K^{(a)}_{p})$ and $\hat{v}^{(b)+}_q$ can be written as
\begin{equation}
[\text{Max}( K^{(a)}_{p}),\hat{v}^{(b)+}_q]=\delta_{ab}\delta_{pq}\hat{v}^{(b)+}_q\,,
\end{equation}
and the relation \eqref{eq:phiv+ relation} follows.

\subsection{Characters of the representations}

For a tree-type quiver theory, by definition the character (or partition function) of the Hilbert space $\mathcal{H}_\nu$ is 
\begin{equation}\label{eq:character1}
\mathcal{Z}_{\mathcal{H}_\nu}=\sum_{\{\vec{k}^{(a)}\}}\prod_{a\in \mathrm{Q}_0}x_a^{\ \sum_{p=1}^{N^{(a)}}k^{(a)}_p}\,,
\end{equation} 
where the summation is over all the states in $\mathcal{H}_{\nu}$, i.e.\ over the decompositions $\{\vec{k}^{(a)}\}$ satisfying \eqref{eq:tree-codi}. 

The character \eqref{eq:character1} can be evaluated in closed form. 
For each node $a\in \mathrm{Q}_0$, define Tree$(a)$ as the set of the nodes in tree$(a)$, the sub-tree with root $a$.
Then the character \eqref{eq:character1} can be rewritten as
\begin{equation}\label{eq:character2}
\mathcal{Z}_{\mathcal{H}_{\nu}}=\prod_{a\in \mathrm{Q}_0}\prod_{p=1}^{N^{(a)}}\frac{1}{1-\prod_{b\in \textrm{Tree}(a)}x[b,\mathcal{I}^{(a)}_p]}\,,
\end{equation}
with
\begin{equation}
x[b,\mathcal{I}^{(a)}_p]=\left\{ 
\begin{array}{lll}
x_b&  , & \mathcal{I}^{(a)}_p \in \mathcal{I}^{(b)} \\
1&  , &  \mathcal{I}^{(a)}_p \notin \mathcal{I}^{(b)}
\end{array}
\right.\,.
\end{equation}

\medskip

Now we can check that the partition function \eqref{eq:character2} of the vortex Hilbert space $\mathcal{H}_{\nu}$ can be reproduced as the character of the corresponding vortex representation of the quiver Yangian.
We recall that the action of the quiver Yangian provides an efficient algorithm for computing the character of the representation, which is straightforward to implement in Mathematica; for this so-called quiver Yangian algorithm, see e.g.\ \cite[Sec.~3.2]{Li:2023zub}.
We applied this algorithm to the vortex representation $\mathcal{R}_\nu$ of the quiver Yangian for a few tree examples, and checked that the characters agree with the partition functions \eqref{eq:character2} to high orders.

\subsection{Example: D-type theory}\label{ssec:Dtype_theory}
In Sec.~\ref{ssssec:Dtype_vortex} we have worked out the supersymmetric vacua $\nu_{\{\mathcal{I}^{(a)}\}}$, the vortex Hilbert space $\mathcal{H}_\nu$ and the action of the Coulomb branch operators for the D-type quiver gauge theory.
Let us now study the corresponding quiver Yangian. 
The quiver Yangian that reproduces the Coulomb branch algebra of the D-type theory is constructed using the triple quiver shown in Fig.~\ref{fig:fig:Dtype_quiver_QY}, with the potential (see \eqref{eq:tree_potential})
\begin{equation}\label{eq:D_triple_superpotential}
\widehat{W}=\mathrm{Tr}\,(A_1C_1B_1-C_3A_1B_1)+\sum_{a=2}^{L-1}\mathrm{Tr}\,(A_aC_aB_a-C_{a+1}A_aB_a)\,,
\end{equation}
and the weight assignment (see \eqref{eq:weight_assign_C} and \eqref{eq:weight_assign_AB})
\begin{equation}
h(A_a)=-\epsilon\,,\qquad h(B_a)=0\,,\qquad h(C_a)=\epsilon\,,
\end{equation}
where we have used the parameter identification $\hbar=\epsilon$.
\begin{figure}[h]
\centering
\begin{tikzpicture}
[->,auto=right, node distance=2cm,
shorten >=1pt, semithick]
\node (v1) at (-0.5,1.5)[circle,draw] {{\tiny1}};
\node (v2) at (-2.5,0.2)[circle,draw] {{\tiny2}};
\node (v3) at (-0.5,0.2)[circle,draw] {{\tiny 3}};
\node (e) at (1.7,0.15) {$\dots$};
\node (vl) at (3.8,0.2)[circle,draw] {{\tiny $L$}};

\draw (v3) edge [bend right=15] node [pos=0.5,right] {{\tiny $A_1$}} (v1);
\draw (v1) edge[bend right=15] node [pos=0.5,left]  {{\tiny $B_1$}} (v3);
\draw (v1) edge [in=20,out=-20,loop] node {{\tiny $C_1$}} (v1);
			
\draw (v2) edge [bend right=15] node [pos=0.5,below] {{\tiny $B_2$}} (v3);
\draw (v3) edge [bend right=15] node [pos=0.5,above] {{\tiny $A_2$}} (v2);
\draw (v2) edge [in=110,out=70,loop] node {{\tiny $C_2$}} (v2);

\draw (v3) edge [bend right=15] node {{\tiny $B_3$}} (e);
\draw (e) edge [bend right=15] node {{\tiny $A_3$}} (v3);
\draw (v3) edge [in=-70,out=-110,loop] node {{\tiny $C_3$}} (v3);

\draw (e) edge [bend right=15] node {{\tiny $B_{L-1}$}} (vl);
\draw (vl) edge [bend right=15] node {{\tiny $A_{L-1}$}} (e);
\draw (vl) edge [in=110,out=70,loop] node {{\tiny $C_{L}$}} (vl);
\end{tikzpicture}
\caption{Triple quiver for the D-type shifted quiver Yangian.}
\label{fig:fig:Dtype_quiver_QY}
\end{figure}
We will now focus on the representation $\mathcal{R}_\nu$ of the quiver Yangian which reproduces the vortex Hilbert space $\mathcal{H}_\nu$. 
Applying our proposal of Sec.~\ref{ssec:vortex_as_QY_rep}, the charge functions of the ground state of the representation $\mathcal{R}_\nu$ are given by:
\begin{equation}\label{eq:D_ground_charge}
\begin{aligned}
^\sharp\psi^{(1)}_0(z)&=\frac{\prod_{p=N^{(1)}+1}^{N^{(3)}}(z-m_p+\epsilon)}{\prod_{p=1}^{N^{(1)}}(z-m_p)}\,,\quad ^\sharp\psi^{(2)}_0(z)=\frac{\prod_{p\in \mathcal{I}^{(3)}\backslash\mathcal{I}^{(2)}}(z-m_p+\epsilon)}{\prod_{p=1}^{N^{(2)}}(z-m_p)}\,,\\
^\sharp\psi^{(3)}_0(z)&=\frac{\prod_{p=N^{(3)}+1}^{N^{(4)}}(z-m_p+\epsilon)\prod_{p\in \mathcal{I}^{(1)}\cap \mathcal{I}^{(2)}}(z-m_p)}{\prod_{p\in \mathcal{I}^{(3)}\backslash(\mathcal{I}^{(1)}\cup \mathcal{I}^{(2)})}(z-m_p)}\,,\\
^\sharp\psi^{(a)}_0(z)&=\frac{\prod_{p=N^{(a)}+1}^{N^{(a+1)}}(z-m_p+\epsilon)}{\prod_{p=N^{(a-1)}+1}^{N^{(a)}}(z-m_p)}\,,\quad 4\leq a\leq L\,,
\end{aligned}
\end{equation}
from which we immediately obtain the following framing of the D-type triple quiver:
\begin{enumerate}
\item For node $a=1,2$, there are $N^{(a)}$ in-coming arrows with weights 
\begin{equation} \label{eq:Dtype_framing_first}
\{m_p|p\in \mathcal{I}^{(a)}\}\,,\end{equation}
as well as $(N^{(3)}-N^{(a)})$ out-going arrows with weights
\begin{equation}\{\epsilon-m_p|p\in\mathcal{I}^{(3)}\backslash \mathcal{I}^{(1)}\}\,.\end{equation}

\item For node $a\geq4$, there are $(N^{(a)}-N^{(a-1)})$ in-coming arrows with weights
\begin{equation}\{m_p|p\in\mathcal{I}^{(a)}\backslash \mathcal{I}^{(a-1)}\}\,,\end{equation}
as well as $(N^{(a+1)}-N^{(a)})$ out-going arrows with weights
\begin{equation}\{\epsilon-m_p|p\in\mathcal{I}^{(a+1)}\backslash \mathcal{I}^{(a)}\}\,.\end{equation}

\item For node 3, there are $(N^{(3)}-N^{(2)}-l)$ in-coming arrows with weights
\begin{equation}\{m_p|p\in \mathcal{I}^{(3)}\backslash (\mathcal{I}^{(1)}\cup\mathcal{I}^{(2)})\}\,,\end{equation}
as well as $(N^{(4)}-N^{(3)})+(N^{(1)}-l)$ out-going arrows with weights
\begin{equation}\label{eq:Dtype_framing_last}
\{\epsilon-m_p|p\in \mathcal{I}^{(4)}\backslash \mathcal{I}^{(3)}\}\bigcup\{-m_p|p\in \mathcal{I}^{(1)}\cap\mathcal{I}^{(2)}\}\,.
\end{equation}
Here we have used the fact $(\mathcal{I}^{(1)}\sqcup\mathcal{I}^{(2)})/\mathcal{I}^{(3)}=\mathcal{I}^{(1)}\cap\mathcal{I}^{(2)}$.
\end{enumerate} 
Note that the shifts $\{\mathtt{s}^{(a)}\}$ do not depend on the vacuum $\nu$, in particular, they are independent of $l$.
It is straightforward to check that the framing above specifies the vortex representation $\mathcal{R}_\nu$ of the (truncated shifted) quiver Yangian, which reproduces the vortex Hilbert space $\mathcal{H}_\nu$.
Consider a general state in $\mathcal{R}_\nu$. 
Let us still denote the number of paths of color $a\in\widehat{Q}$ which start with the framing arrow of weight $m_{\mathcal{I}^{(a)}_p}$ by $k^{(a)}_p$, and label the state by $|k\rangle=|\{k^{(a)}_p\}\rangle$.
With the ground state charge functions \eqref{eq:D_ground_charge} and the bonding factors \eqref{eq:tree_bond}, one readily computes the charge functions of the state $|k\rangle$:
{\footnotesize
\begin{equation}\label{eq:Dtype_charge}
\begin{aligned}
\Psi^{(1)}_k(z)&=\frac{\prod_{q=1}^{N^{(3)}}(z-m_{q}-(k^{(3)}_{q}-1)\eps)}{\prod_{p=1}^{N^{(1)}}(z-m_{p}-(k^{(1)}_{p}-1)\eps)(z-m_{p}-k^{(1)}_{p}\eps)}\,,\\
\Psi^{(2)}_k(z)&=\frac{\prod_{q=1}^{N^{(3)}}(z-m_{q}-(k^{(3)}_{q}-1)\eps)}{\prod_{p=1}^{N^{(2)}}(z-m_{p+l}-(k^{(2)}_{p}-1)\eps)(z-m_{p+l}-k^{(2)}_{p}\eps)}\,,\\
\Psi^{(3)}_k(z)&=\frac{\prod_{b=1,2}\prod_{q=1}^{N^{(b)}}(z-m_{\mathcal{I}^{(b)}_{q}}-k^{(b)}_q\eps)\prod_{q=1}^{N^{(4)}}(z-m_{q}-(k^{(4)}_{q}-1)\eps)}{\prod_{p=1}^{N^{(3)}}(z-m_{p}-(k^{(3)}_{p}-1)\eps)(z-m_{p}-k^{(3)}_{p}\eps)}\,,\\
\Psi^{(a)}_k(z)&=\frac{\prod_{q=1}^{N^{(a-1)}}(z-m_{q}-k^{(a-1)}_q\eps)\prod_{q=1}^{N^{(a+1)}}(z-m_{q}-(k^{(a+1)}_{q}-1)\eps)}{\prod_{p=1}^{N^{(a)}}(z-m_{p}-(k^{(a)}_{p}-1)\eps)(z-m_{p}-k^{(a)}_{p}\eps)}\,,\quad a=4,5,\dots,L\,.
\end{aligned}
\end{equation}}
From the pole structures of the charge functions above, we see that the conditions \eqref{eq:D_condi1} and \eqref{eq:D_condi2} are preserved by the states in $\mathcal{R}_\nu$. 
Consider a state in $\mathcal{R}_\nu$, which is generated by applying various $e$ operators successively on the ground state $|0\rangle$. In the state-generating process, all the adding poles that break the conditions \eqref{eq:D_condi1} and \eqref{eq:D_condi2} will be canceled. 
As a result of this cancellation, we obtain a representation $\mathcal{R}_\nu$ where each state $|k\rangle$ is labeled by the same set of vectors $\{\vec{k}^{(1)},\vec{k}^{(2)},\dots , \vec{k}^{(L)}\}$ as those labeling the states in the vortex Hilbert space $\mathcal{H}_\nu$: these vectors take all possible values subject to \eqref{eq:D_condi1} and \eqref{eq:D_condi2}.
This leads to the isomorphism between $\mathcal{H}_\nu$ and $\mathcal{R}_\nu$.

\section{\texorpdfstring{Example with high valency: $K$-star}{Example with high valency: K-star}}
\label{sec:kleaf_example}
We have proposed the following correspondence:
\begin{equation}\label{eq:correspondence}
\begin{aligned}
\text{Quantum Coulomb branch algebra}\quad&\longleftrightarrow\quad \text{Quiver Yangian}\,,\\
\text{Vortex Hilbert space $\mathcal{H}_{\nu}$}\quad&\longleftrightarrow \quad\text{Representation $\mathcal{R}_{\nu}$ of quiver Yangian}\,,
\end{aligned}
\end{equation}
and demonstrated this with the examples of the ADE-type quivers (see Sec.~\ref{sec:QYasMonopoleAlgebra} and App.~\ref{appsec:ADE_examples}).
For these ADE examples, the valency of the tree is $\leq 2$.
In this section, we consider an example with higher valency: the quiver with the shape of a starfish with $K$ ($\geq 3$) legs of length one,  called $K$-star for short. 
We will use it to demonstrate how the framings derived from the vortex Hilbert spaces truncate the representations in such a way as to evade the potential problem of the non-simple poles. (Note that this problem doesn't arise for trees with valency $\leq 2$.)

\subsection{\texorpdfstring{$K$-star quiver and its triple quiver}{K-star quiver and its triple quiver}}

The $K$-star quiver that we choose as the tree-type quiver for the 3D $\mathcal{N}=4$ theory has $K+1$ nodes, with the root $\mathtt{r}$ sitting in the center and $K$ nodes in the peripheral, each connected to $\mathtt{r}$ by one arrow.
For the triple quiver of the $K$-star quiver, using the convention of Sec.~\ref{ssec:QY_triple},  the arrow from $\mathtt{r}$ to $a$ for $a\in\{1,2,\dots,K\}$ is labeled by $A_a$, while its reverse is labeled by $B_a$, and 
the self-loop of node $a$  is denoted by $C_a$ for $a=1,2,\dots,K,\mathtt{r}$.
An example with $K=3$ is shown in Fig.~\ref{fig:k=3}, where the tree-type quiver is in fact the Dynkin quiver of $D_4$, and the position of the flavor node makes the quiver a 3-valent tree.
\begin{figure}[h]	
\centering
\begin{subfigure}[b]{0.30\textwidth}
\centering
\begin{tikzpicture}
[->,auto=right, node distance=2cm,
shorten >=1pt, semithick]
\node (v4) at (0,0) [circle,draw] {$\mathtt{r}$};
\node (v3) at (1.5,0)[circle,draw] {3};
\node (v2) at (0,1.5)[circle,draw] {2};
\node (v1) at (-1.5,0)[circle,draw] {1};
\node (f) at (0,-1)[rectangle,draw]{$\mathtt{f}$};
			
\draw (v4) edge (v1);
\draw (v4) edge (v2);
\draw (v4) edge (v3);
\draw (f) edge (v4);
			
\end{tikzpicture}
\caption{$D_4$ quiver gauge theory.}
\label{fig:k=3_quiver}
\end{subfigure}  
\hspace{20mm}
\begin{subfigure}[b]{.30\textwidth}
\centering
\begin{tikzpicture}
[->,auto=right, node distance=2cm,
shorten >=1pt, semithick]
\node (v4) at (0,0) [circle,draw] {$\mathtt{r}$};
\node (v1) at (-2,0)[circle,draw] {1};
\node (v2) at (0,1.5)[circle,draw] {2};
\node (v3) at (2,0)[circle,draw] {3};

\draw (v1) edge [bend right=15] node {{\tiny $B_1$}} (v4);
\draw (v2) edge [bend right=15] node {{\tiny $B_2$}} (v4);
\draw (v3) edge [bend right=15] node {{\tiny $B_3$}} (v4);
\draw (v4) edge [bend right=15] node {{\tiny $A_1$}} (v1);
\draw (v4) edge [bend right=15] node {{\tiny $A_2$}} (v2);
\draw (v4) edge [bend right=15] node {{\tiny $A_3$}} (v3);
		
\draw (v1) edge [in=110,out=70,loop] node {{\tiny $C_1$}} (v1);
\draw (v2) edge [in=110,out=70,loop] node {{\tiny $C_2$}} (v2);
\draw (v3) edge [in=110,out=70,loop] node {{\tiny $C_3$}} (v3);
\draw (v4) edge [in=290,out=250,loop] node {{\tiny $C_{\mathtt{r}}$}} (v4);
			
\end{tikzpicture}
\caption{Triple quiver of $D_4$.}
\label{fig:k=3_triple}
\end{subfigure}
\caption{$D_4$ quiver as a $3$-star.}
\label{fig:k=3}
\end{figure}

The quiver Yangian is constructed using this triple quiver, with the following potential (see \eqref{eq:tree_potential}): 
\begin{equation}
\widehat{W}=\sum_{a=1}^{K}\mathrm{Tr}\,(A_aC_aB_a-C_{\mathtt{r}}A_aB_a)\,.
\end{equation}
Applying the weight assignment \eqref{eq:weight_assign_C} and \eqref{eq:weight_assign_AB} to this case and using the identification $\hbar=\epsilon$, we have
\begin{equation}
\begin{aligned}
&h(A_a)=-\epsilon\,,\quad h(B_a)=0\,,\quad &&a=1,2,\dots,K\,,\\
&h(C_a)=\epsilon\,,&& a=1,2,\dots,K,\mathtt{r}\,.
\end{aligned}
\end{equation}

\subsection{The vortex Hilbert space}
We consider the 3D $\mathcal{N}=4$ quiver gauge theory of the $K$-star. 
For simplicity, we choose the following ranks for the gauge and flavor groups:
\begin{equation}
N^{(a)}=1\,\quad\text{for }\  1\leq a \leq K\,, \qquad
N^{(\mathtt{r})}=K+1\,,\quad N_{\mathtt{f}}=K+2\,.
\end{equation} 
Note that this theory is ``good", since every gauge node has non-negative excess. 
The $N_{\mathtt{f}}$ complex masses $\{m_1,m_2,\dots,m_{K+2}\}$ and $|\textrm{Q}_0|$ (negative) real FI parameters\\ $\{t_{\mathbb{R}}^{(1)},t_{\mathbb{R}}^{(2)},\dots,t_{\mathbb{R}}^{(K+1)}\}$ are also turned on.

\medskip

The solutions of the vacua are characterized by the index sets which are nested along the paths in the tree, i.e.\ $\mathcal{I}^{(a)}\subset \mathcal{I}^{(\mathtt{r})}\subset\mathcal{I}_{\mathtt{f}}=\{1,2,\dots,K+2\}$ for $a=1,2,\dots,K$. 
Since there are $K\geq3$ branches in the tree, the possible intersections of the index sets on different branches become very complicated. 
Let us consider the simplest case, where the $K$ branches have no overlap with each other:
\begin{equation}\label{eq:kleaf_vacuum}
\mathcal{I}^{(a)}=\{a\}\quad\text{for }\ a=1,2,\dots,K\,, \qquad 
\mathcal{I}^{(\mathtt{r})}=\{1,2,\dots,K+1\}\,.
\end{equation}
This simplest choice is already enough to illustrate the key ideas.

The vortex Hilbert space $\mathcal{H}_\nu$, where the vacuum $\nu$ is specified by \eqref{eq:kleaf_vacuum}, can be easily obtained using the results from Sec.~\ref{sssec:vortex_Hilbert_space}. 
We find that a state $|k)\in\mathcal{H}_\nu$ is specified by $K+1$ vectors 
\begin{equation}
\{\vec{k}^{(1)},\vec{k}^{(2)},\dots,\vec{k}^{(K)},\vec{k}^{(\mathtt{r})}\}\,,
\end{equation}
where $\vec{k}^{(a)}$ is a length-$N^{(a)}$ decomposition of the vortex number $\nn^{(a)}$.\footnote{Since $N^{(a)}=1$ for $a=1,\dots,K$, we have simply $k^{(a)}=\nn^{(a)}$ for $a=1,2,\dots,K$.} 
Moreover, they satisfy
\begin{equation} \label{eq:kleaf_condi}
k^{(a)}\geq k^{(\mathtt{r})}_a\,,\qquad a=1,2,\dots,K\,.
\end{equation}
As an example, when $K=3$ we can arrange the components of the decompositions $\{\vec{k}^{(1)},\vec{k}^{(2)},\vec{k}^{(3)},\vec{k}^{(\mathtt{r})}\}$ into the following matrix:
\begin{equation}
\left(
\begin{array}{cccc}
k^{(1)} &* &* & k^{(\mathtt{r})}_1\\
*&k^{(2)}&* &k^{(\mathtt{r})}_2\\
*&* &k^{(3)}&k^{(\mathtt{r})}_3\\
*&* &* &k^{(\mathtt{r})}_4
\end{array}\,\right),
\end{equation}
where the $a^{\textrm{th}}$ column is a decomposition of $\nn^{(a)}$ ($a=1,2,3,\mathtt{r}$), and in each row the components are non-increasing from the left to the right.
	
\subsection{Vortex Hilbert space  as Quiver Yangian representation}

After obtaining the vortex Hilbert space $\mathcal{H}_\nu$ of the theory, we will now construct a representation $\mathcal{R}_{\nu}$ of the quiver Yangian of the triple quiver that can reproduce this Hilbert space.

\subsubsection{From vortex states to states in quiver Yangian representation}

First, we demand that in the representation $\mathcal{R}_\nu$, the numbers of the various atoms precisely reproduce the corresponding vortex numbers. 
Let us again take $K=3$ as an example. 
A natural proposal is that all the states in the representation $\mathcal{R}_\nu$ are composed of the four clusters shown in Fig.~\ref{fig:k=3_paths}.
\begin{figure}[h] 
\centering
\begin{subfigure}[b]{0.25\textwidth}
\centering
\begin{tikzpicture}[scale=0.7]
[->,auto=right, node distance=2cm, shorten >=1pt, semithick]
\node (vf1) at (0.5,0) {\footnotesize{$\infty$}};
\node (v1) at (2,0) [circle,draw] {\tiny{1}};
\node (v4) at (3.5,0) [circle,draw] {\tiny{$\mathtt{r}$}};
\node (v11) at (2,-1) [circle,draw] {\tiny{1}};
\node (vd1) at (2,-1.9)  {\dots};
\node (v12) at (2,-2.8) [circle,draw] {\tiny{1}};
\node (v41) at (3.5,-1) [circle,draw] {\tiny{$\mathtt{r}$}};
\node (v5leaf) at (3.5,-1.9)  {\dots};
\node (v42) at (3.5,-2.8) [circle,draw] {\tiny{$\mathtt{r}$}};
				
\draw (vf1) edge (v1);
\draw (v1) edge (v4);
\draw (v1) edge (v11);
\draw (v11) edge (vd1);
\draw (vd1) edge (v12);
\draw (v4) edge (v41);
\draw (v41) edge (v5leaf);
\draw (v5leaf) edge (v42);
				
\end{tikzpicture}
\caption{}
\end{subfigure}  
\begin{subfigure}[b]{.25\textwidth}
\centering
\begin{tikzpicture}[scale=0.7]
[->,auto=right, node distance=2cm,
shorten >=1pt, semithick]
\node (vf1) at (0.5,0) {\footnotesize{$\infty$}};
\node (v1) at (2,0) [circle,draw] {\tiny{2}};
\node (v4) at (3.5,0) [circle,draw] {\tiny{$\mathtt{r}$}};
\node (v11) at (2,-1) [circle,draw] {\tiny{2}};
\node (vd1) at (2,-1.9)  {\dots};
\node (v12) at (2,-2.8) [circle,draw] {\tiny{2}};
\node (v41) at (3.5,-1) [circle,draw] {\tiny{$\mathtt{r}$}};
\node (v5leaf) at (3.5,-1.9)  {\dots};
\node (v42) at (3.5,-2.8) [circle,draw] {\tiny{$\mathtt{r}$}};
				
\draw (vf1) edge (v1);
\draw (v1) edge (v4);
\draw (v1) edge (v11);
\draw (v11) edge (vd1);
\draw (vd1) edge (v12);
\draw (v4) edge (v41);
\draw (v41) edge (v5leaf);
\draw (v5leaf) edge (v42);
				
\end{tikzpicture}
\caption{}
\end{subfigure}
\begin{subfigure}[b]{0.25\textwidth}
\centering
\begin{tikzpicture}[scale=0.7]
[->,auto=right, node distance=2cm,
shorten >=1pt, semithick]
\node (vf1) at (0.5,0) {\footnotesize{$\infty$}};
\node (v1) at (2,0) [circle,draw] {\tiny{3}};
\node (v4) at (3.5,0) [circle,draw] {\tiny{$\mathtt{r}$}};
\node (v11) at (2,-1) [circle,draw] {\tiny{3}};
\node (vd1) at (2,-1.9)  {\dots};
\node (v12) at (2,-2.8) [circle,draw] {\tiny{3}};
\node (v41) at (3.5,-1) [circle,draw] {\tiny{$\mathtt{r}$}};
\node (v5leaf) at (3.5,-1.9)  {\dots};
\node (v42) at (3.5,-2.8) [circle,draw] {\tiny{$\mathtt{r}$}};
				
\draw (vf1) edge (v1);
\draw (v1) edge (v4);
\draw (v1) edge (v11);
\draw (v11) edge (vd1);
\draw (vd1) edge (v12);
\draw (v4) edge (v41);
\draw (v41) edge (v5leaf);
\draw (v5leaf) edge (v42);
				
\end{tikzpicture}
\caption{}
\end{subfigure}
\begin{subfigure}[b]{.22\textwidth}
\centering
\begin{tikzpicture}[scale=0.7]
[->,auto=right, node distance=2cm,
shorten >=1pt, semithick]
\node (vf1) at (0.5,0) {\footnotesize{$\infty$}};
\node (v4) at (2,0) [circle,draw] {\tiny{$\mathtt{r}$}};
\node (v41) at (2,-1) [circle,draw] {\tiny{$\mathtt{r}$}};
\node (v5leaf) at (2,-1.9)  {\dots};
\node (v42) at (2,-2.8) [circle,draw] {\tiny{$\mathtt{r}$}};

\draw (vf1) edge (v4);
\draw (v4) edge (v41);
\draw (v41) edge (v5leaf);
\draw (v5leaf) edge (v42);
				
\end{tikzpicture}
\caption{}
\end{subfigure}
\caption{Four candidate clusters that make up the states in $\mathcal{R}_\nu$ (for $K=3$).}
\label{fig:k=3_paths}
\end{figure}

Consider a state in $\mathcal{R}_\nu$ that consists of the four clusters  in Fig.~\ref{fig:k=3_paths}, and let us denote the numbers of $1,2,3$-colored atoms by $k^{(1)},k^{(2)},k^{(3)}$, respectively. 
And for the atoms of color $\texttt{r}$, their numbers in the four clusters shown in Fig.~\ref{fig:k=3_paths} are denoted by $(k^{(\texttt{r})}_1,k^{(\texttt{r})}_2,k^{(\texttt{r})}_3,k^{(\texttt{r})}_4)$ respectively, and the total number is $\sum_{p=1}^{N^{(\texttt{r})}} k^{(\texttt{r})}_p=\nn^{(\texttt{r})}$. 

\medskip
	
A key feature of this representation is that all the paths are one-directional. 
When a path reaches a node $a\in \mathrm{Q}_0$, it flows either to the precursor of $a$ or to $a$ itself (through a self-loop). 
For example, there is no path like {\small $\infty\to \circled{1}\to \circled{\texttt{r}}\to\circled{1}$, $\infty\to \circled{2}\to \circled{\texttt{r}}\to\circled{1}$ }or {\small $\infty\to \circled{\texttt{r}}\to\circled{1}$} in our representation. 
Consequently, in the $1^{\textrm{st}}$ cluster, the adding poles of $\texttt{r}$-colored atoms are controlled solely by the 1-colored atoms in front of them (and also by themselves, of course), but are independent of the atoms of the other colors. 
Thus intuitively there should exist relation(s) between the atom numbers $k^{(1)}$ and $k^{(\texttt{r})}_1$.
Later we will see, the relation turns out to be $k^{(1)}\geq k^{(\texttt{r})}_1$ -- precisely what we want.
Similar arguments hold for the $2^{\textrm{nd}}$ and the $3^{\textrm{rd}}$ clusters and the conditions $k^{(a)}\geq k^{(\texttt{r})}_a$, $a=2,3$ (see \eqref{eq:kleaf_condi}). 
Finally, for the $4^{\textrm{th}}$ cluster, the adding poles of the $\texttt{r}$-colored atoms are independent of the 1,2,3-colored atoms, which means that there is no constraint on $k^{(\texttt{r})}_4$ (except for $k^{(\texttt{r})}_4\in\mathbb{N}_0$).

\medskip

The discussion above can be straightforwardly generalized to arbitrary $K\geq3$, for which  
we have $N^{(\texttt{r})}=K+1$ candidate clusters (shown in Fig.~\ref{fig:k_paths}), and every state in $\mathcal{R}_\nu$ is composed by them. 
The $a^{\textrm{th}}$ cluster (where $a=1,2,\dots,K$) consists of $k^{(a)}$ $a$-colored atoms and $k^{(\texttt{r})}_a$ $\mathtt{r}$-colored atoms, while the $(K+1)$-st cluster consists of $k^{(\mathtt{r})}_{K+1}$ $\mathtt{r}$-colored atoms. 
Moreover, the numbers of the various atoms should satisfy \eqref{eq:kleaf_condi}.
\begin{figure}[h] 
\centering
\begin{subfigure}[b]{.30\textwidth}
\centering
\begin{tikzpicture}[scale=0.7]
[->,auto=right, node distance=2cm,
shorten >=1pt, semithick]
\node (vf1) at (0.5,0) {\footnotesize{$\infty$}};
\node (v3) at (2,0) [circle,draw] {\tiny{$a$}};
\node (v4) at (3.5,0) [circle,draw] {\tiny{$\mathtt{r}$}};
\node (v31) at (2,-1) [circle,draw] {\tiny{$a$}};
\node (vd3) at (2,-1.9)  {\dots};
\node (v32) at (2,-2.8) [circle,draw] {\tiny{$a$}};
\node (v41) at (3.5,-1) [circle,draw] {\tiny{$\mathtt{r}$}};
\node (v5leaf) at (3.5,-1.9)  {\dots};
\node (v42) at (3.5,-2.8) [circle,draw] {\tiny{$\mathtt{r}$}};
				
\draw (vf1) edge (v3);
\draw (v3) edge (v4);
\draw (v3) edge (v31);
\draw (v31) edge (vd3);
\draw (vd3) edge (v32);
\draw (v4) edge (v41);
\draw (v41) edge (v5leaf);
\draw (v5leaf) edge (v42);
				
\end{tikzpicture}
\caption{The $a^{\textrm{th}}$ candidate.
($a=1,2,\dots,K$)}
\end{subfigure}
\hspace{20mm}
\begin{subfigure}[b]{.30\textwidth}
\centering
\begin{tikzpicture}[scale=0.7]
[->,auto=right, node distance=2cm,
shorten >=1pt, semithick]
\node (vf1) at (0.5,0) {\footnotesize{$\infty$}};
\node (v4) at (2,0) [circle,draw] {\tiny{$\mathtt{r}$}};
\node (v41) at (2,-1) [circle,draw] {\tiny{$\mathtt{r}$}};
\node (v5leaf) at (2,-1.9)  {\dots};
\node (v42) at (2,-2.8) [circle,draw] {\tiny{$\mathtt{r}$}};

\draw (vf1) edge (v4);
\draw (v4) edge (v41);
\draw (v41) edge (v5leaf);
\draw (v5leaf) edge (v42);
\end{tikzpicture}
\caption{The $(K+1)$-st candidate.}
\end{subfigure}
\caption{The $(K+1)$ candidate clusters that compose the states in $\mathcal{R}_\nu$ (for general $K\geq 3$).}
\label{fig:k_paths}
\end{figure}
\subsubsection{Deriving the framing}

Next, we apply the general proposal for the  framing in  Sec.~\ref{sssec:mono_as_QY} to this case, and show that the corresponding representation reproduces precisely the enumeration of the states in the representation $\mathcal{R}_{\nu}$.
Translating the framing prescription \eqref{eq:Incoming} -- \eqref{eq:Outgoing2} to the current case, we have
\begin{enumerate}
\item For each node $a=1,2,\dots,K,\mathtt{r}$, there is an in-coming arrow with weight $m_a$ (where we set $m_{\mathtt{r}}\equiv m_{K+1}$).
\item For each node $a=1,2,\dots,K$, there are $K$ out-going arrows with weights 
\begin{equation}
\{\epsilon-m_b|\ 1\leq b\leq K+1\  \text{ and } b\neq a \}\,.
\end{equation}

\item For the node $\mathtt{r}$, there is an out-going arrow with weight $\epsilon-m_{K+2}$.	
\end{enumerate}
The ground state charge functions that correspond to this framing are
{\small
\begin{equation}
{}^\sharp\psi^{(a)}_0(z)=\frac{\prod_{b=1,b\neq a}^{K+1}(z+\eps-m_b)}{z-m_a}\,\ \text{ for }\  \ a=1,2,\dots,K\,,\qquad {}^\sharp\psi^{(\mathtt{r})}_0(z)=\frac{z+\eps-m_{K+2}}{z-m_{K+1}}\,,
\end{equation}
}
from which one can apply the raising operators iteratively to generate all the states in $\mathcal{R}_{\nu}$.

\medskip

To demonstrate in more detail the mechanism of how the out-going arrows truncate the representation to its proper size, let us solve for the framing again, by demanding that the resulting representation can reproduce precisely all the states in the representation $\mathcal{R}_{\nu}$.\footnote{Note that it is non-trivial that such a solution exists, namely, that we can indeed construct a representation such that every state is made by the $K+1$ clusters shown in Fig.~\ref{fig:k_paths}, and the representation $\mathcal{R}_\nu$ is isomorphic to the vortex Hilbert space $\mathcal{H}_\nu$.}

Let us start by only introducing the in-coming arrows from step-1, then the corresponding ground state charge functions are:
\begin{equation}
{^\sharp}\psi^{(a)}_{0,\text{naive}}(z)=\frac{1}{z-m_a}\,,\qquad a=1,2,\dots,K,\mathtt{r}\,.
\end{equation}
One can then work out all the level-1 states in $\mathcal{R}_\nu$, and confirm that these states are consistent with our expectation. 
Namely, the representation reproduces the Hilbert space $\mathcal{H}_\nu$ to the first level. 
We have the following $K+1$ states at this level:
\begin{equation}
|k^{(a)}=1\rangle\,\ \text{ for }\  a=1,2,\dots,K\,\quad\text{and }\quad 
|k^{(\mathtt{r})}_{K+1}=1\rangle\,.
\end{equation}

\medskip

Let us then proceed by induction. 
Assuming we have a state $|k\rangle$, made up by the $K+1$ clusters shown in Fig.~\ref{fig:k_paths}, with the numbers of the various atoms satisfying the condition \eqref{eq:kleaf_condi},  
we now compute its charge functions. 
The results are
{\footnotesize
\begin{equation}
\Psi^{(a)}_{k,\text{naive}}(z)=\frac{z-m_a-(k^{(\mathtt{r})}_a-1)\epsilon}{(z-m_a-(k^{(a)}-1)\epsilon)(z-m_a-k^{(a)}\epsilon)}\prod_{\substack{b=1\\b\neq a}}^{K+1}\frac{z-m_b-(k^{(\mathtt{r})}_b-1)\epsilon}{z-m_b+\epsilon}\,,\quad a=1,2,\dots,K\,,
\end{equation}}
and
{\footnotesize
\begin{equation}
\Psi^{(\mathtt{r})}_{k,\text{naive}}(z)=\frac{(z-m_{\mathtt{r}}+\epsilon)}{(z-m_{\mathtt{r}}-(k^{(\mathtt{r})}_{K+1}-1)\epsilon)(z-m_{\mathtt{r}}-k^{(\mathtt{r})}_{K+1}\epsilon)}\prod_{a=1}^{K}\frac{(z-m_a+\epsilon)(z-m_a-k^{(a)}\epsilon)}{(z-m_a-(k^{(\mathtt{r})}_a-1)\epsilon)(z-m_a-k^{(\mathtt{r})}_a\epsilon)}\,.
\end{equation}}

One immediately realizes there are many unwanted poles in the charge functions. 
For example, the poles $m_a-\eps$ ($a=2,3,\dots,K,\mathtt{r}$) in $\Psi^{(1)}_{k,\text{naive}}(z)$ correspond to the paths 
\begin{equation}
\infty\to\circled{a}\to\circled{\texttt{r}}\to\circled{1}\ \text{ for }\ a=2,3,\dots,K\quad\text{ and }\quad \infty\to\circled{\texttt{r}}\to\circled{1}\,,
\end{equation}
which are unwanted. 
These poles should be canceled by introducing new zeros.\footnote{In fact, introducing new zeros will completely cut off the problematic sub-tree from the representation.} 
For the atoms of the other colors there are also unwanted poles. 
We see that we need to introduce zeros to cancel them.
Namely, we need to add out-going framing arrows whose weights are uniquely determined by the desired cancellation, and this gives the out-going arrows in step-2 and 3.\footnote{The arrow introduced in step-3 actually does not change the representation, since it doesn't lead to any cancellation, nor any new pole. However, it modifies the value of the charge function, which determines the action of the $e$ and $f$ operators, see \eqref{eq:QY_e_action} and \eqref{eq:QY_f_action}. 
This is crucial for the identification of the quiver Yangian generators and the monopole operators by comparing their action (see Sec.~\ref{sssec:translation_between_QY_monopole}).}

The charge functions of the state $|k\rangle$ now become:
\begin{equation}\label{eq:kleaf_charge1}
\Psi^{(a)}_k(z)=\frac{\prod_{b=1}^{K+1}(z-m_b-(k^{(\mathtt{r})}_b-1)\epsilon)}{(z-m_a-(k^{(a)}-1)\epsilon)(z-m_a-k^{(a)}\epsilon)}\,,\qquad a=1,2,\dots,K\,
\end{equation}
and 
\begin{equation}\label{eq:kleaf_charge2}
\Psi^{(\mathtt{r})}_k(z)=\frac{\prod_{a=1}^{N_{\mathtt{f}}}(z-m_a+\epsilon)\cdot\prod_{a=1}^K(z-m_a-k^{(a)}\epsilon)}{\prod_{a=1}^{N^{(\mathtt{r})}}(z-m_a-(k^{(\mathtt{r})}_a-1)\epsilon)(z-m_a-k^{(\mathtt{r})}_a\epsilon)}\,,
\end{equation}
which agree with the general result \eqref{eq:tree-char} specialized to this case. 
The pole structures are just as desired. 
The charge functions only have simple poles, which have clear physical interpretations:
\begin{itemize}
\item In each of the $\Psi^{(a)}_k(z)$ where $a=1,2,\dots,K$, there are two poles, corresponding to one removable atom and one addable atom in the $a^{\textrm{th}}$ cluster, respectively.
\item In $\Psi^{(\mathtt{r})}_k(z)$ there are $2N^{(\mathtt{r})}=2K+2$ poles, corresponding to $K+1$ removable atoms and $K+1$ addable atoms in all the $K+1$ clusters.
\end{itemize}
Most importantly, the condition \eqref{eq:kleaf_condi} is also satisfied: e.g.\ when $k^{(1)}=k^{(\mathtt{r})}_1$ the removing pole $z=m_1+k^{(1)}\epsilon$ in $\Psi^{(1)}_k(z)$ and the adding pole $z=m_1+k^{(\mathtt{r})}_1\epsilon$ in $\Psi^{(\mathtt{r})}_k(z)$ are both canceled, namely, $k^{(1)}$ cannot be decreased and $k^{(\mathtt{r})}_1$ cannot be increased. 
As a result, the condition $k^{(1)}\geq k^{(\mathtt{r})}_1$ is preserved.

\medskip

Thus we have checked that the general proposal for the  framing in  Sec.~\ref{sssec:mono_as_QY} is indeed correct for the $K$-star example, and moreover, the general charge function \eqref{eq:tree-char}, when applied to the current example (see \eqref{eq:kleaf_charge1} and \eqref{eq:kleaf_charge2}), has the right pole structure.

\medskip

Finally, note that the framing that we have derived has the property that, for each state in the corresponding representation, the charge function has \textit{only simple poles}, which is not true for a general framing.
(Recall that for a general quiver with potential, not all framings give rise to representations with only simple poles, see \cite{Li:2023zub}.)

For example, consider the simplest framing $\infty\to \circled{\texttt{r}}$, with the weight of the framing arrow equal to $0$. 
The corresponding ground state charge function has only a pole, and no zero.
To proceed, let us focus on the $3$-star.
The first state with a non-simple pole appears at level-$4$, shown in Fig.~\ref{fig:k=3_state}.
Using the prescription  \eqref{eq:QY_charge_function}, the charge function $\Psi^{(\mathtt{r})}(z)$ of this state is 
\begin{equation}
\Psi^{(\mathtt{r})}(z)=\frac{z^2}{(z-\epsilon)(z+\epsilon)^2}\,,
\end{equation}
which has a second-order pole at $z=-\eps$, which is not pure. 
As a result, the action of the quiver Yangian generators breaks down on this state. 
\begin{figure}[h]	
\centering
\begin{tikzpicture}[scale=0.7]
[->,auto=right, node distance=2cm,
shorten >=1pt, semithick]
\node (v4) at (0,0) [circle,draw] {\tiny{4}};
\node (v1) at (2,1)[circle,draw] {\tiny{1}};
\node (v2) at (2,0)[circle,draw] {\tiny{2}};
\node (v3) at (2,-1)[circle,draw] {\tiny{3}};
\node (f) at (-1.5,0) {\small{$\infty$}};

\draw (f) edge (v4);
\draw (v4) edge node[pos=0.7,below] {{\tiny $A_1$}} (v1);
\draw (v4) edge node[pos=0.7,below] {{\tiny $A_2$}} (v2);
\draw (v4) edge node[pos=0.7,below] {{\tiny $A_3$}} (v3);
		
\end{tikzpicture}
\caption{For a framing that does not correspond to any vortex Hilbert space, a state can have a non-simple pole.}
\label{fig:k=3_state}
\end{figure}

For the triple quivers of the tree-type quivers, the potential \eqref{eq:tree_potential} is not strong enough to guarantee that for an arbitrary framing, all the states in the corresponding representation have only simple poles in their charge functions. 
However, the framing that does correspond to a vortex Hilbert space has enough zeros in the ground state charge functions so that the corresponding representation is truncated to such an extent that no higher-order poles arise.

\section{Quantum Coulomb branch algebra for general quivers}\label{sec:Conjecture}

Our main conjecture is that 
for a 3D $\mathcal{N}=4$ quiver gauge theory based on the quiver $\mathrm{Q}$ and with unitary gauge group, the quantum Coulomb branch algebra can be formulated as a certain truncated shifted quiver Yangian 
based on the \textit{triple quiver} $\widehat{Q}$ of the original quiver $\mathrm{Q}$.
In previous sections, we have explicitly checked this conjecture when $\mathrm{Q}$ is a tree-type quiver.
In this section, we will give the chain of mathematical arguments leading to the conjecture. 
In the next section, we will study the non-simply-laced quivers and quivers with edge-loops, and show that the known results on quantum Coulomb branch algebra (for general parameter $\hbar$) on ABCDEFG-type quiver and Jordan quiver satisfy this conjecture \cite{Braverman:2016pwk,Nakajima:2019olw,Kodera:2016faj_jordan_quiver}.

\subsection{Coulomb branch algebra and CoHA of triple quivers}\label{ssec:CBA_CoHA_general}

\subsubsection{Pre-projective algebra}

Mathematically, the appearance of the triple quiver can be understood as first passing to the pre-projective algebra, then to the pre-projective CoHA,  and finally to the critical CoHA \cite{Ginzburg:2006fu,Yang_2014,Yang_2016,yang2017}.
The double quiver $\bar{Q}$ of the quiver $\mathrm{Q}$ is defined as 
\begin{equation}
\label{eq:doubleQuiver}
\bar{Q}_0:=\mathrm{Q}_0\,, \qquad \bar{Q}_1:=\mathrm{Q}_1\cup \,  {}^{\textrm{op}}\mathrm{Q}_1\,. 
\end{equation}
The pre-projective algebra of $\mathrm{Q}$ is defined as
\begin{equation}
\Pi(\mathrm{Q})=\mathbb{C}\bar{Q}/(R)\,, 
\end{equation}
where $(R)$ is the two-sided ideal generated by the quadratic element 
\begin{equation}
R:= \sum_{I^{a\to b}\in \mathrm{Q}_1}[I^{a\to b},I^{b\to a}]\,, 
\end{equation}
where the summation is over all the arrows $I^{a\to b}$ in the original quiver $\mathrm{Q}$.
The moduli stack of representations of the pre-projective algebra $\Pi(\mathrm{Q})$ is equal to the cotangent bundle of the moduli stack of representations of the original quiver $\mathrm{Q}$ \cite{Ginzburg:2006fu}.

\subsubsection{CoHA of triple quiver}

The pre-projective CoHA is defined to be the cohomological Hall algebra 
(CoHA) that is associated to the CY$_2$ category of representations of the pre-projective algebra $\Pi(\mathrm{Q})$ \cite{Yang_2014}, and it was then shown to be isomorphic to the critical CoHA of \cite{Kontsevich:2010px} for the triple quiver H$(\widehat{Q},\widehat{W})$ \cite{Yang_2016}.\footnote{Sometimes the pre-projective CoHA is called 2D CoHA whereas the critical CoHA of the triple quiver is called 3D CoHA.}

\medskip

The underlying vector space of the critical CoHA H$(\widehat{Q},\widehat{W})$ is given by the space of equivariant cohomology of semi-stable quiver representations of $\widehat{Q}$ subject to $\partial \widehat{W}=0$ \cite{Kontsevich:2010px}. 
The multiplication $\mathfrak{m}_{\textrm{KS}}$ of CoHA is a convolution product defined via the standard pull-backs and push-forwards of cohomology classes along certain correspondences, defined in terms of the moduli spaces of extensions of quiver representations \cite{Kontsevich:2010px}, in a similar way to the multiplication in the Coulomb branch algebra, see Sec.~\ref{ssec:CBalgebra}. 

We will be interested in the critical CoHA of the triple quiver $(\widehat{Q},\widehat{W})$,  defined in \eqref{eq:tripleQdef} with \eqref{eq:tripleQdef2}.
The canonical superpotential of the triple quiver is given by \cite{Ginzburg:2006fu}:
\begin{equation}\label{eq:triple_canonical_potential}
\widehat{W}=\sum_{I^{a\to b}\in\mathrm{Q}_1} \text{Tr}(I^{a\to b}I^{b\to b}I^{b\to a}-I^{a\to a}I^{a\to b}I^{b\to a})\,.
\end{equation}
With this potential, the critical CoHA of the triple quiver  H$(\widehat{Q},\widehat{W})$ is isomorphic to the  pre-projective  CoHA \cite{Yang_2016}.
From now on, we will refer to the critical CoHA simply as CoHA.

\subsubsection{Relation between Coulomb branch algebra and CoHA of triple quivers}

During the previous three  sections, we have shown  that for a given tree-type quiver $\mathrm{Q}_{\textrm{tree}}$, the quantum Coulomb branch algebra is 
\begin{equation}
\mathcal{A}_{\hbar}(\mathrm{Q}_{\textrm{tree}}, \{N^{(a)}\}) = \textrm{truncated shifted Y}_{\hbar} (\widehat{Q},\widehat{W})   \,,
\end{equation}
where Y$_{\hbar}(\widehat{Q},\widehat{W})$ only depends on the quiver $\mathrm{Q}_{\textrm{tree}}$ whereas the 
information of the truncation and shift is contained in the ranks of the gauge group and flavor group factors $\{N^{(a)}\}$, with $a\in \mathrm{Q}_0\cup \{\mathtt{f}\}$.
For a general quiver $\mathrm{Q}$, we propose that this factorization of the information still holds: namely
\begin{equation}
\mathcal{A}_{\hbar}(\mathrm{Q},\{N^{(a)}\})=\textrm{truncated shifted }\tilde{\mathcal{A}}_{\hbar}(\mathrm{Q})   
\end{equation}
where there exists a ``universal" Coulomb branch algebra $\tilde{\mathcal{A}}_{\hbar}(\mathrm{Q})$ that only depends on $\mathrm{Q}$ and for the same $\mathrm{Q}$, $\mathcal{A}_{\hbar}(\mathrm{Q},\{N^{(a)}\})$ with different ranks $\{N^{(a)}\}$ differ only in the truncation and shift.
The truncation can be defined via the pre-image of the homomorphism 
\begin{equation}
\tilde{\mathcal{A}}_{\hbar}(\mathrm{Q}) 
\ \rightarrow \
\mathcal{A}_{\hbar}(\mathrm{Q},\{N^{(a)}\})\,.
\end{equation}

\medskip

The results of \cite[Theorem 6.6]{Nakajima:2015txa} imply that there is a natural action of the equivariant CoHA of the triple quiver on the Coulomb branch, from which we can deduce\footnote{We thank Gufang Zhao for explaining this point to us. See \cite{Rapcak:2020ueh} for the proof of this statement for the Jordan quiver.} 
the homomorphism from the spherical equivariant CoHA to the positive subalgebra of the quantum Coulomb branch algebra $\mathcal{A}_{\hbar}(\mathrm{Q},\{N^{(a)}\})$:
\begin{equation}
\mathrm{sH}(\widehat{Q},\widehat{W})^{\textrm{eq}} 
\ \rightarrow \
\mathcal{A}_{\hbar}^{+} (\mathrm{Q},\{N^{(a)}\})\end{equation}
with non-trivial kernel that corresponds to the truncation.
Based on the result on tree-type quivers, we conjecture the isomorphism for general $\mathrm{Q}$:
\begin{equation}
\mathrm{sH}(\widehat{Q},\widehat{W})^{\textrm{eq}} 
\ \simeq \
\tilde{\mathcal{A}}_{\hbar}^{+} (\mathrm{Q})  \,.
\end{equation}

\subsection{Spherical equivariant CoHA and Shuffle algebra}

For general quiver with potential $(Q,W)$, the CoHA is difficult to study.
However, the spherical subalgebra of the equivariant CoHA $\textrm{sH}(\widehat{Q},\widehat{W})^{\textrm{eq}}$  has a much easier presentation in terms of the spherical equivariant shuffle algebra \cite{Kontsevich:2010px}:
\begin{equation}
\textrm{sH}(\widehat{Q},\widehat{W})^{\textrm{eq}}\simeq \mathbf{sSh}(\widehat{Q})^{\textrm{eq}}\,,	
\end{equation}
where on the r.h.s.\ the information of the potential $\widehat{W}$ only appears as the loop constraints among $\{h_I\}$.

\medskip

Let us briefly review the definition of the (equivariant) shuffle algebra ${\bf Sh}(\widehat{Q})^{\textrm{eq}}$ and its subalgebra ${\bf sSh}(\widehat{Q})^{\textrm{eq}}$.
The underlying vector space of the shuffle algebra ${\bf Sh}(\widehat{Q})^{\textrm{eq}}$ is graded:
\begin{equation}
{\bf Sh}(\widehat{Q})^{\textrm{eq}}	=\bigoplus_{\vec{d}} {\bf Sh}_{\vec{d}}(\widehat{Q})^{\textrm{eq}}\,,
\end{equation}
where $\vec{d}=\{d_1,\dots,d_{|\mathrm{Q}_0|}\}$ corresponds to the dimensional-vector of the quiver representation and  ${\bf Sh}_{\vec{d}}(\widehat{Q})^{\textrm{eq}}$ is the space of  (anti)symmetric polynomials of the collection of variables $\{x^{(a)}_{i}\}$, with $a\in \mathrm{Q}_0$ and $i=1,\ldots, d_a$:
\begin{equation}
{\bf Sh}_{\vec{d}}(\widehat{Q})^{\textrm{eq}} =\mathbb{Q}[\{x^{(a)}_{1},\dots x^{(a)}_{d_a}\}]^{\prod_a S_{d_a}}\,.
\end{equation}
We denote such an (anti)symmetric polynomial by $[p(x)]_{\vec{d}}$.
(Note that these polynomials are (anti)symmetric upon the exchange of variables that are associated with
the same quiver node $a$, $x^{(a)}_i$ and $x^{(a)}_j$.)

The equivariant shuffle algebra ${\bf Sh}(\widehat{Q})^{\textrm{eq}}$ is given by the shuffle product \cite{Kontsevich:2010px,yang2017,Galakhov:2021vbo}
\begin{equation}
\shuffle:\quad {\bf Sh}_{\vec{d}}(\widehat{Q})^{\textrm{eq}}_{}\times  {\bf Sh}_{\vec{d}'}(\widehat{Q})^{\textrm{eq}}_{}\longrightarrow  {\bf Sh}_{\vec{d}+\vec{d}'}(\widehat{Q})^{\textrm{eq}}_{}\;,
\end{equation}
where the shuffle product $\shuffle$ between  two (anti)symmetric polynomials $[f(x)]_{\vec{d}}$ and $[g(y)]_{\vec{d}'}$ is given by 
\begin{equation}
[f(x)]_{\vec{d}}
\ \shuffle \
[g(y)]_{\vec{d}'} :=\left[\sum_{\textrm{shuffle }\vec \sigma} \epsilon(\vec\sigma) \,(\mathrm{fac}\times f \times \,g)( \vec\sigma(x\cup y))\right]_{\vec{d} + \vec{d}'}\,.
\end{equation}
First of all, the information of the triple quiver with potential $(\widehat{Q},\widehat{W})$ is contained in the rational factor $\textrm{fac}$:
\begin{equation}\label{eq:facDef}
\textrm{fac}(x,y) := 
\textrm{sgn}
\prod_{{a,b}} 
\prod_{i,j} \lambda_{b,a}(y^{(b)}_j, x^{(a)}_i)\,, 
\qquad \textrm{with} \quad 
\lambda_{a,b}(z,w)\myequiv\frac{\prod\lm_{I\in\{b\to a\}}\left(z-w-h_I\right)}{\left(z-w\right)^{\delta_{b,a}}}\;,
\end{equation}
where $\{h_I\}$ are the equivariant weights of the arrows $\{I\in \widehat{Q}_1\}$ and they satisfy the loop constraint imposed by the $F$-term condition $\partial \widehat{W}=0$.
The sum is over all possible shuffles $\vec \sigma=\{\sigma_1,\dots,\sigma_{|\mathrm{Q}_0|}\}$, where each element 
$\sigma_a \in (S_{d_a} \times  S_{d'_a})\backslash S_{d_a+d'_a}$ 
combines the two sets of the variables $x$ and $y$ together into $x\cup y$, shuffles them, and then divides them  into two new sets $x^{(a)}_{i}$ and $y^{(a)}_{i}$, on which $\mathrm{fac}(x,y)f(x)g(y)$ is evaluated. 
The total number of shuffles is $\prod\lm_{a\in Q_0}\binom{d_a+d_a' }{d_a}$.
The sgn and $\epsilon(\vec{\sigma})$ are sign factors that ensure the symmetry or anti-symmetry, which we omit here.

\medskip

The spherical shuffle algebra  $\mathbf{sSh}(\widehat{Q})^{\textrm{eq}}$ is defined as
\begin{equation}
\mathbf{sSh}(\widehat{Q})^{\textrm{eq}}
:\quad 
\textrm{subalgebra of } \mathbf{Sh}(\widehat{Q})^{\textrm{eq}} 
\textrm{ generated by monomials } 
\{ (x^{(a)})^m \}
\end{equation}
where $\{ (x^{(a)})^m \}$ correspond to dim-1 quiver representations.

\medskip

\subsection{Spherical equivariant CoHA v.s. quiver Yangian}

There exists a homomorphism from the positive subalgebra of the quiver Yangian to the spherical equivariant shuffle algebra \cite{Galakhov:2021vbo}:
\begin{equation}\label{eq:shuffle_homo}
\xi:\quad  \textrm{Y}^+(\widehat{Q},\widehat{W})\longrightarrow {\bf sSh}(\widehat{Q})^{\textrm{eq}}\;,
\end{equation}
with
\begin{equation}
\xi: \quad e^{(a)}_{n}\longrightarrow (x^{(a)})^n\,.
\end{equation}
It is straightforward to check \cite{Galakhov:2021vbo}:
\begin{equation}
\begin{split}
\xi\left[\lambda_{a,b}(z,w)\hat\mye^{(a)}(z)\hat\mye^{(b)}(w)\right]&=\lambda_{a,b}(z,w)\,\xi\left[\hat\mye^{(a)}(z)\right]\shuffle\; \xi\left[\hat\mye^{(b)}(w)\right]\\
&=\xi\left[\lambda_{b,a}(w,z)\hat\mye^{(b)}(w)\hat\mye^{(a)}(z)\right]\,,
\end{split}
\end{equation}
which obeys the quadratic $e$-$e$ relation in \eqref{eq:QY_quadratic_relations}.
The other relations from \eqref{eq:QY_quadratic_relations} can be checked similarly.
Then the extension to the whole algebra is guaranteed by the shuffle product associativity. 
We conjecture that for the triple quiver $(\widehat{Q},\widehat{W})$, the homomorphism \eqref{eq:shuffle_homo} can be promoted to the algebra isomorphism:
\begin{equation}
 \textrm{Y}^+(\widehat{Q},\widehat{W})\simeq {\bf sSh}(\widehat{Q})^{\textrm{eq}}\;.
\end{equation}
For more studies on triple quivers see e.g.\ \cite{feigin1995vectorbundlesellipticcurve,enriquez1998correlationfunctionsdrinfeldcurrents,Rapcak:2018nsl,Rapcak:2020ueh,2013arXiv1302.6202N}.

\subsection{Quiver Yangians as Coulomb branch algebras}

Putting everything together leads to the conjecture on the isomorphism between the positive nilpotent subalgebra of the ``universal" quantum Coulomb branch algebra $\tilde{\mathcal{A}}_{\hbar}(\mathrm{Q})$ and that of the quiver Yangian of the triple quiver
\begin{equation}
\tilde{\mathcal{A}}^{+}_{\hbar}(\mathrm{Q})
\ \simeq \
\textrm{sH}(\widehat{Q},\widehat{W})^{\textrm{eq}}
\ \simeq \
\mathbf{sSh}(\widehat{Q})^{\textrm{eq}} 
\ \simeq \
\mathrm{Y}^{+}(\widehat{Q},\widehat{W})\,.
\end{equation}
Taking the Drinfeld double, we then have the conjectured isomorphism 
\begin{equation}
\tilde{\mathcal{A}_{\hbar}}(\mathrm{Q})
\ \simeq \
\mathrm{Y}_{\hbar} (\widehat{Q},\widehat{W})
\end{equation}
and therefore
\begin{equation}
\mathcal{A}_{\hbar}(\mathrm{Q},\{N^{(a)}\})
\ \simeq \
\textrm{truncated shifted }
\mathrm{Y}_{\hbar}(\widehat{Q},\widehat{W}) \,.
\end{equation}
(Note that although the quiver Yangian might have more than one parameter, only one, denoted by $\hbar$, corresponds to the quantization parameter, which equals to the $\Omega$-deformation parameter $\epsilon$.)
We have checked the isomorphism on the universal algebra $\tilde{\mathcal{A}_{\hbar}}(\mathrm{Q})$ for the tree-type quivers.
In the next section we will explain the non-simply-laced quivers and quivers with edge-loops.

\medskip

We make a final comment.
The pre-projective CoHA is in fact spherical, namely the spherical subalgebra is the same as the full algebra \cite{negut2023}. 
Together with  the isomorphism between the pre-projective CoHA and the critical CoHA \cite{Yang_2016}, this shows that the critical CoHA of triple quiver H$(\widehat{Q},\widehat{W}
)$ is spherical.
(For related results on the shuffle algebras see \cite{negut2022}.) 
This helps explain why the (first fundamental) minuscule monopole operators (together with vector-multiplet scalars) generate the full quantum Coulomb branch algebra (with $\hbar\neq 0$), not just the spherical subalgebra. 

\section{Non-simply-laced quivers and quivers with edge-loops}
\label{sec:BeyondSimple}
In this section, we consider the non-simply-laced quivers and quivers with edge-loops.

\subsection{Non-simply-laced quivers
}\label{ssec:effective_quiver_yangian}

Consider the quiver Yangian Y$(\widehat{Q},\widehat{W})$ associated to the triple quiver of a non-simply-laced $\mathrm{Q}$ without edge-loop, where   $\widehat{Q}$ and $\widehat{W}$ are given by \eqref{eq:tripleQdef} and \eqref{eq:triple_canonical_potential}. 
Previously for the triple quivers of the tree-type quivers, we have considered the simplest torus action that rescales all the arrows by the same factor. 
(Recall that the torus action on the superpotential gives the constraint among the equivariant weights of the arrows.)
We can do the same for the non-simply-laced quivers, and the resulting quiver Yangian Y$(\widehat{Q},\widehat{W})$ would in general have more than one independent parameter.

On the other hand, recall that one can define a Yangian for a quiver without edge-loop, as follows.
Let $\mathfrak{g}_{\mathrm{Q}}$ be the symmetric Kac-Moody Lie algebra associated to a quiver $\mathrm{Q}$ without edge-loop, where the Cartan matrix of $\mathfrak{g}_{\mathrm{Q}}$ is 
$C=2-A-A^T$ with $A$ being the adjacency matrix of Q.
The quadratic relations of the Yangian Y$_\hbar(\mathfrak{g}_{\mathrm{Q}})$ are given as follows \cite{Yang_2014,yang2017}:\footnote{Note that $\hbar^{\textrm{here}}=-\hbar^{\textrm{there}}$.}
{\small
\begin{equation}\label{eq:Y(g_Q)}
\begin{aligned}
[\psi^{(a)}(z),\psi^{(b)}(w)]&=0\,,\\
\psi^{(a)}(z)e^{(b)}(w)&\simeq \frac{z-w+C_{ba}\frac{\hbar}{2}}{z-w-C_{ba}\frac{\hbar}{2}}e^{(b)}(w)\psi^{(a)}(z)\,,\\
\psi^{(a)}(z)f^{(b)}(w)&\simeq \frac{z-w-C_{ba}\frac{\hbar}{2}}{z-w+C_{ba}\frac{\hbar}{2}}f^{(b)}(w)\psi^{(a)}(z)\,,\\
e^{(a)}(z)e^{(b)}(w)&\sim \frac{z-w+C_{ab}\frac{\hbar}{2}}{z-w-C_{ab}\frac{\hbar}{2}}e^{(b)}(w)e^{(a)}(z)\,,\\
f^{(a)}(z)f^{(b)}(w)&\sim \frac{z-w-C_{ba}\frac{\hbar}{2}}{z-w+C_{ba}\frac{\hbar}{2}}f^{(b)}(w)f^{(a)}(z)\,,\\
[e^{(a)}(z),f^{(b)}(w)]&\sim-\delta_{ab}\frac{\psi^{(a)}(z)-\psi^{(b)}(w)}{z-w}\,,
\end{aligned}    
\end{equation}
}

\noindent and the relations in terms of the modes $\{\psi^{(a)}_n,e^{(a)}_n,f^{(a)}_n\}$ can be obtained from \eqref{eq:Y(g_Q)} as in Sec.~\ref{sssec:Qurdratic}.
There is only one parameter $\hbar$ in Y$_\hbar(\mathfrak{g}_{\mathrm{Q}})$.

\subsubsection{Effective triple quiver and potential}

We will now show that by choosing a different torus action on Y$(\widehat{Q},\widehat{W})$, we can reproduce Y$_\hbar(\mathfrak{g}_{\mathrm{Q}})$.
Given a pair of nodes $a\neq b\in\mathrm{Q}_0$, we denote the arrows from $a$ to $b$ in the original $\mathrm{Q}$ by $\{I^{a\rightarrow b}_i|i=1,2,\dots, |a\to b|\} $, and the reverses of these arrows in ${}^{\textrm{op}}\mathrm{Q}$ by $\{I^{b\rightarrow a}_i|I^{a\rightarrow b}_i\in \{a\to b\}\}$. 
Following \cite{Yang_2014}, we modify the torus action on $\widehat{Q}$ such that the loop constraints (induced by the torus action on $\widehat{W}$) on the equivariant weights of the arrows $I\in \widehat{Q}_1$ become:
\begin{equation}\label{eq:loop_constraints_assumption}
\frac{h(I^{a\rightarrow b}_i)}{m^{a\rightarrow b}_i }+\frac{h(I^{b\rightarrow a}_i) }{m^{b\rightarrow a}_i }=-h(I^{a\rightarrow a}) =-h(I^{b\rightarrow b})
\,, \quad a\neq b \in \mathrm{Q}_0\,, \quad 
i=1,\dots, |a\rightarrow b|\,.
\end{equation}

To compare with Yangian of $\mathfrak{g}_{\mathrm{Q}}$, the integers $\{m_i^{a\to b},m_{i}^{b\to a}\}$ can be chosen as:
\begin{equation}\label{eq:canonical_choice}
m^{a\rightarrow b}_{i}=|a\to b|+2-2i\,,\quad 
m^{b\rightarrow a}_i=-|a\to b|+2i\,,
\end{equation}
see also \cite{Yang_2014}.
The solution of the equivariant weights of arrows is then:
\begin{equation}\label{eq:canonical_weight}
h(I^{a\rightarrow b}_i)=(-|a\to b|-2+2i)\frac{\hbar}{2}\,, \quad h(I^{b\rightarrow a}_i)=(|a\to b|-2i)\frac{\hbar}{2}\,,
\quad 
h(I^{a\rightarrow a})=\hbar\,.
\end{equation}

With this choice of the equivariant weights, there are lots of  cancellations among the contributions of these arrows to the bonding factor $\varphi^{b\Leftarrow a}(z)$:
\begin{equation}\label{eq:BFtriple}
\varphi^{b\Leftarrow a}(z)=\frac{\prod_{s=2-|a\to b|}^{|a\to b|}(z-s\frac{\hbar}{2})}{\prod_{s=-|a\to b|}^{|a\to b|-2}(z-s\frac{\hbar}{2})}=\frac{z-|a\to b|\frac{\hbar}{2}}{z+|a\to b|\frac{\hbar}{2}}\,.
\end{equation}
This suggests that for the triple quiver $\widehat{Q}$ of a general quiver $\mathrm{Q}$ and with the equivariant weights \eqref{eq:canonical_weight}, the associated quiver Yangian can be rewritten in terms of a smaller effective triple quiver $(\widehat{Q}^{\text{eff}},\widehat{W}^{\text{eff}})$.
The multiple pairs of opposite arrows between $a,b$ in $\widehat{Q}$ are reduced to a \textit{single} pair of opposite arrows in $\widehat{Q}^{\text{eff}}$, which we denote by $\{I^{a\to b},I^{b\to a}\}$; the self-loops $I^{a\rightarrow a}$ do not change.
The weights of the arrows $I\in \widehat{Q}^{\text{eff}}_1$ are: 
\begin{equation}\label{eq:weightsEff}
h(I^{a\rightarrow b})=-|a\to b|\frac{\hbar}{2}
\,, \quad h(I^{b\rightarrow a})=-|a\to b|\frac{\hbar}{2}
\,,\quad
h({I^{a\to a}})=\hbar\,.
\end{equation}
The bonding factors for all $a,b\in\mathrm{Q}_0$ can be written 
 as \begin{equation}\label{eq:non-tree_bond}
\varphi^{b\Leftarrow a}(z)=\frac{z+C_{ab}\frac{\hbar}{2}}{z-C_{ab}\frac{\hbar}{2}}\,,\qquad a,b\in\mathrm{Q}_0\,,
\end{equation} 
(Note that in the previous sections on the tree-type quiver (with $|a\rightarrow b|=1$), we have performed spectral shifts such that $h(I^{b\rightarrow a})=0$, and consequently $h(I^{a\rightarrow b})=-|a\to b|\hbar$, while the weight of the self-loop remains unchanged, see \eqref{eq:weight_assign_C} and \eqref{eq:weight_assign_AB}.)

Accordingly, the canonical potential $\widehat{W}$ for the original triple quiver $\widehat{Q}$ (see \eqref{eq:triple_canonical_potential}) needs to be modified into an effective potential $\widehat{W}^{\text{eff}}$ in terms of the arrows $I\in \widehat{Q}^{\text{eff}}_1$. For a given node $a\in\mathrm{Q}_0$, the relevant term undergoes the following change:
\begin{equation}
\sum_{i=1,\dots, |a\to b|}I_i^{a\to b}I_i^{b\to a}I^{a\to a}\in \widehat{W}\Longrightarrow I^{a\to b}I^{b\to a} (I^{a\to a})^{|C_{ab}|} \in \widehat{W}^{\text{eff}}\,,
\end{equation}
where we have used $|C_{ab}|=|a\to b|$ for adjacent $a,b\in \mathrm{Q}_0$. 
One can check that with the simplest torus action where $I\rightarrow \lambda I$ for all $I\in \widehat{Q}^{\text{eff}}_1$, the potential $\widehat{W}^{\text{eff}}$ imposes the loop constraint
\begin{equation}
h(I^{a\to b})+h(I^{b\to a} )+|C_{ab}|h(I^{a\to a})=0\,,  
\end{equation}
satisfied by the choice of the equivariant weights \eqref{eq:weightsEff}. 
It is straightforward to check that the quiver Yangian Y$(\widehat{Q}^{\textrm{eff}},\widehat{W}^{\textrm{eff}})$, with  quadratic relations \eqref{eq:QY_quadratic_relations} and the bonding factor \eqref{eq:non-tree_bond}, precisely reproduces the quadratic relations of Y$_{\hbar}(\mathfrak{g}_Q)$, given in \eqref{eq:Y(g_Q)}.

\medskip

Since once the set of nodes $\widehat{Q}_0$ is given, the quiver Yangian only depends on the bonding factors, and Y$(\widehat{Q},\widehat{W})$ with the new torus action and Y$_{\hbar}(\widehat{Q}^{\textrm{eff}},\widehat{W}^{\textrm{eff}})$ share the same bonding factor \eqref{eq:non-tree_bond},
we conclude that 
\begin{equation}
\textrm{Y}(\widehat{Q},\widehat{W}) |_{T^{*}_{\textrm{modified}}}
\ \simeq \  
\textrm{Y}_{\hbar}(\widehat{Q}^{\textrm{eff}},\widehat{W}^{\textrm{eff}}) 
\ \simeq \ 
\textrm{Y}_{\hbar}(\mathfrak{g}_Q) 
\,.    
\end{equation}

Assuming the main conjecture in Sec.~\ref{sec:Conjecture}, it is natural to propose that for general quivers without edge-loop, the Coulomb branch algebra can be given by the (truncated shifted) quiver Yangian based on the (naive) triple quiver and potential $(\widehat{Q},\widehat{W})$ (see \eqref{eq:tripleQdef} and \eqref{eq:triple_canonical_potential}) but with the loop constraints given by the modified torus action, or equivalently, the one based on the effective triple quiver and potential $(\widehat{Q}^{\text{eff}},\widehat{W}^{\text{eff}})$ with the original loop constraints.

\subsubsection{Examples: non-simply-laced Dynkin quivers}

As examples of non-simply laced tree-type quivers without edge-loop, we present  the Dynkin quivers of non-simply-laced Lie algebras and construct the corresponding effective triple quivers and potentials, on which the quiver Yangian is based.

For the BCFG-type theory, the Dynkin quivers Q are given by
\begin{equation}
\begin{aligned}
&B_L:\quad 
\begin{tikzpicture}[baseline={([yshift=0ex]current bounding box.center)},vertex/.style={anchor=base, circle,fill=black!25,minimum size=18pt,inner sep=2pt}]
\node (v1) at (-3,0.2)[circle,draw] {};
\node (v2) at (-2,0.2)[circle,draw] {};
\node (e) at (-1,0.15) {$\dots$};
\node (vl1) at (0,0.2)[circle,draw] {};
\node (vl) at (1,0.2)[circle,draw] {};
\node (em) at (0.5,0.2) {};
\draw (v1) edge (v2);
\draw (e) edge (vl1);
\draw (v2) edge (e);
\draw[double] (em.west) -- (vl);
\draw[-{Classical TikZ Rightarrow[length=1mm,]},double,] (vl1) -- (em.east);
\end{tikzpicture}
\qquad
&&C_L:\quad 
\begin{tikzpicture}[baseline={([yshift=0ex]current bounding box.center)},vertex/.style={anchor=base,circle,fill=black!25,minimum size=18pt,inner sep=2pt}]
\node (v1) at (-3,0.2)[circle,draw] {};
\node (v2) at (-2,0.2)[circle,draw] {};
\node (e) at (-1,0.15) {$\dots$};
\node (vl1) at (0,0.2)[circle,draw] {};
\node (vl) at (1,0.2)[circle,draw] {};
\node (em) at (-2.5,0.2) {};
\draw (vl1) edge (vl);
\draw (v2) edge (e);
\draw (e) edge (vl1);
\draw[double] (em.west) -- (v2);
\draw[-{Classical TikZ Rightarrow[length=1mm,]},double,] (v1) -- (em.east);
\end{tikzpicture}
\\
&F_4 :\quad 
\begin{tikzpicture}[baseline={([yshift=0ex]current bounding box.center)},vertex/.style={anchor=base,circle,fill=black!25,minimum size=18pt,inner sep=2pt}]
\node (v1) at (-3,0.2)[circle,draw] {};
\node (v2) at (-2,0.2)[circle,draw] {};
\node (v3) at (-1,0.2)[circle,draw] {};
\node (v4) at (0,0.2)[circle,draw] {};
\node (em) at (-1.5,0.2) {};
\draw (v1) edge (v2);
\draw (v3) edge (v4);
\draw[double] (em.west) -- (v3);
\draw[-{Classical TikZ Rightarrow[length=1mm,]},double,] (v2) -- (em.east);
\end{tikzpicture}
\qquad
&&G_2 :\quad 
\begin{tikzpicture}[baseline={([yshift=0ex]current bounding box.center)},vertex/.style={anchor=base,
circle,fill=black!25,minimum size=18pt,inner sep=2pt}]
\node (v1) at (-3,0.2)[circle,draw] {};
\node (v2) at (-2,0.2)[circle,draw] {};
\node (em) at (-2.5,0.2) {};
\draw (em.west) -- (v2);
\draw[-{Classical TikZ Rightarrow[length=1mm,]},] (v1) -- (em.east);
\draw (-2.8,0.24) -- (-2.2,0.24);
\draw (-2.8,0.16) -- (-2.2,0.16);
\end{tikzpicture}
\end{aligned}
\end{equation}
Recall that in the triple quiver $\widehat{Q}^{\text{eff}}$, the adjacent nodes are linked by a single pair of opposite arrows,
and information of the multiplicity of arrows is instead encoded in the potential $\widehat{W}^{\text{eff}}$:
\begin{equation}
\widehat{W}^{\text{eff}}=\sum\text{Tr}(I^{a\to b}I^{b\to a}(I^{a\to a})^{|C_{ab}|}-I^{b\to a}I^{a\to b}(I^{b\to b})^{|C_{ba}|})\,.
\end{equation}
Hence, we obtain the following pairs of $(\widehat{Q}^{\text{eff}},\widehat{W}^{\text{eff}})$  for the BCFG-type theories:
{\small
\begin{equation}\label{eq:triple_B}
\begin{aligned}
B_L:\ & \widehat{Q}^{\text{eff}}_B=\begin{tikzpicture}[baseline={([yshift=-3ex]current bounding box.center)},vertex/.style={anchor=base,circle,fill=black!25,minimum size=18pt,inner sep=2pt}]
\node (v1) at (-3,0.2)[circle,draw] {1};
\node (v2) at (-1.5,0.2)[circle,draw] {2};
\node (e) at (0,0.15) {$\dots$};
\node (vl) at (1.5,0.2)[circle,draw] {$L$};

\draw[->] (v2) edge [bend right=15] node[above] {{\tiny $A_1$}} (v1);
\draw[->] (v1) edge [bend right=15] node[below] {{\tiny $B_1$}} (v2);
\draw (v1) edge [in=110,out=70,loop] node [above]{{\tiny $C_1$}} (v1);
			
\draw[->] (v2) edge [bend right=15] node[below] {{\tiny $B_2$}} (e);
\draw[->] (e) edge [bend right=15] node[above] {{\tiny $A_2$}} (v2);
\draw (v2) edge [in=110,out=70,loop] node[above] {{\tiny $C_2$}} (v2);
			
\draw[->] (e) edge [bend right=15] node[below] {{\tiny $B_{L-1}$}} (vl);
\draw[->] (vl) edge [bend right=15] node[above] {{\tiny $A_{L-1}$}} (e);
\draw (vl) edge [in=110,out=70,loop] node[above] {{\tiny $C_{L}$}} (vl);
	\end{tikzpicture}\,,\\
&\widehat{W}_B^{\text{eff}}=\sum_{a=1}^{L-2}\text{Tr}(B_aA_aC_a-A_aB_aC_{a+1})+\text{Tr}(B_{L-1}A_{L-1}C_{L-1}-A_{L-1}B_{L-1}C_L^2)\,,
\end{aligned}
\end{equation}}
{\small
\begin{equation}\label{eq:triple_C}
\begin{aligned}
C_L:\ &\widehat{Q}^{\text{eff}}_C=\begin{tikzpicture}[baseline={([yshift=-3ex]current bounding box.center)},vertex/.style={anchor=base,
			circle,fill=black!25,minimum size=18pt,inner sep=2pt}]
\node (v1) at (-3,0.2)[circle,draw] {1};
\node (v2) at (-1.5,0.2)[circle,draw] {2};
\node (e) at (0,0.15) {$\dots$};
\node (vl) at (1.5,0.2)[circle,draw] {$L$};

\draw[->] (v2) edge [bend right=15] node[above] {{\tiny $A_1$}} (v1);
\draw[->] (v1) edge [bend right=15] node[below] {{\tiny $B_1$}} (v2);
\draw (v1) edge [in=110,out=70,loop] node [above]{{\tiny $C_1$}} (v1);
			
\draw[->] (v2) edge [bend right=15] node[below] {{\tiny $B_2$}} (e);
\draw[->] (e) edge [bend right=15] node[above] {{\tiny $A_2$}} (v2);
\draw (v2) edge [in=110,out=70,loop] node[above] {{\tiny $C_2$}} (v2);
			
\draw[->] (e) edge [bend right=15] node[below] {{\tiny $B_{L-1}$}} (vl);
\draw[->] (vl) edge [bend right=15] node[above] {{\tiny $A_{L-1}$}} (e);
\draw (vl) edge [in=110,out=70,loop] node[above] {{\tiny $C_{L}$}} (vl);
	\end{tikzpicture}\,,\\
&\widehat{W}_C^{\text{eff}}=\sum_{a=2}^{L-1}\text{Tr}(B_aA_aC_a-A_aB_aC_{a+1})+\text{Tr}(B_{1}A_{1}C_{1}-A_{1}B_{1}C_2^2)\,,
\end{aligned}
\end{equation}}
{\small
\begin{equation}\label{eq:triple_F}
\begin{aligned}
F_4:\ &\widehat{Q}^{\text{eff}}_F=\begin{tikzpicture}[baseline={([yshift=-3ex]current bounding box.center)},vertex/.style={anchor=base,
			circle,fill=black!25,minimum size=18pt,inner sep=2pt}]
\node (v1) at (-3,0.2)[circle,draw] {1};
\node (v2) at (-1.5,0.2)[circle,draw] {2};
\node (v3) at (0,0.15) [circle,draw] {3};
\node (v4) at (1.5,0.2)[circle,draw] {4};

\draw[->] (v2) edge [bend right=15] node[above] {{\tiny $A_1$}} (v1);
\draw[->] (v1) edge [bend right=15] node[below] {{\tiny $B_1$}} (v2);
\draw (v1) edge [in=110,out=70,loop] node [above]{{\tiny $C_1$}} (v1);
			
\draw[->] (v2) edge [bend right=15] node[below] {{\tiny $B_2$}} (v3);
\draw[->] (v3) edge [bend right=15] node[above] {{\tiny $A_2$}} (v2);
\draw (v2) edge [in=110,out=70,loop] node[above] {{\tiny $C_2$}} (v2);

\draw[->] (v3) edge [bend right=15] node[below] {{\tiny $B_3$}} (v4);
\draw[->] (v4) edge [bend right=15] node[above] {{\tiny $A_3$}} (v3);
\draw (v3) edge [in=110,out=70,loop] node[above] {{\tiny $C_3$}} (v3);

\draw (v4) edge [in=110,out=70,loop] node[above] {{\tiny $C_4$}} (v4);
\end{tikzpicture}\,,\\
&\widehat{W}_F^{\text{eff}}=\sum_{a=1,3}\text{Tr}(B_aA_aC_a-A_aB_aC_{a+1})+\text{Tr}(B_{2}A_{2}C_{2}-A_{2}B_{2}C_3^2)\,,
\end{aligned}
\end{equation}}
and finally
{\small
\begin{equation}\label{eq:triple_G}
G_2:\ \widehat{Q}^{\text{eff}}_G=\begin{tikzpicture}[baseline={([yshift=-3ex]current bounding box.center)},vertex/.style={anchor=base,circle,fill=black!25,minimum size=18pt,inner sep=2pt}]
\node (v1) at (-3,0.2)[circle,draw] {1};
\node (v2) at (-1.5,0.2)[circle,draw] {2};

\draw[->] (v2) edge [bend right=15] node[above] {{\tiny $A_1$}} (v1);
\draw[->] (v1) edge [bend right=15] node[below] {{\tiny $B_1$}} (v2);
\draw (v1) edge [in=110,out=70,loop] node [above]{{\tiny $C_1$}} (v1);

\draw (v2) edge [in=110,out=70,loop] node [above]{{\tiny $C_2$}} (v2);
\end{tikzpicture}\,,\qquad
\widehat{W}_G^{\text{eff}}=\text{Tr}(B_{1}A_{1}C_{1}-A_{1}B_{1}C_2^3)\,.
\end{equation}
}

\noindent
Note that all these BCFG-type effective quivers $\widehat{Q}^{\text{eff}}$ take the forms of triple quivers of the linear type, while the information of the multiplicities of arrows is encoded in the superpotential $\widehat{W}^{\text{eff}}$, from which one can solve for the equivariant weights of the arrows.

The quiver Yangian Y$_{\hbar}(\widehat{Q}^{\text{eff}},\widehat{W}^{\text{eff}})$ of the BCFG type is then given by \eqref{eq:QY_quadratic_relations} with the bonding factors \eqref{eq:non-tree_bond} and it reproduces the Yangian Y$_{\hbar}(\mathfrak{g})$ with $\mathfrak{g}$ being the BCFG-type Lie algebra; according to our proposal, this is the quantum Coulomb branch algebra for the 3D $\mathcal{N}=4$ quiver gauge theories with BCFG-type quivers.
This then reproduces the result of \cite{Nakajima:2019olw}, derived using a different method.

\subsection{\texorpdfstring{Jordan quiver and affine Yangian of $\mathfrak{gl}_1$}{Jordan quiver and affine Yangian of gl1}}

As the last example, let us consider the simplest case of the quiver with edge-loop(s), the Jordan quiver, which has  one node with one edge-loop.
The corresponding triple quiver is 
\begin{equation}
\widehat{Q}=\begin{tikzpicture}[baseline={([yshift=-2ex]current bounding box.center)},vertex/.style={anchor=base, circle,fill=black!25,minimum size=18pt,inner sep=2pt}]
\node (v0) at (0,0) [circle, draw] {$0$};
\draw (v0) edge [in=110,out=70,loop] node [above] {{\tiny $C$}} (v0);
\draw (v0) edge [in=230,out=190,loop] node [left] {{\tiny $A$}} (v0);
\draw (v0) edge [in=350,out=310,loop] node [right] {{\tiny $B$}} (v0);
\end{tikzpicture}\,,\quad
\widehat{W}=\text{Tr}\ C(AB-BA)\,.
\end{equation}
Let us denote the equivariant weights of the arrows $A,B,C$ in the triple quiver by $h_1,h_2,h_3$, which satisfy the loop constraint $h_1+h_2+h_3=0$. 
The bonding factor is 
\begin{equation}\label{eq:BFagl1}
\varphi^{0\Leftarrow0}(z)=\frac{\prod^{3}_{i=1}(z+h_i)}{\prod^{3}_{i=1}(z-h_i)}\,,
\end{equation}
As shown in \cite{Li:2020rij}, the quiver Yangian Y$(\widehat{Q},\widehat{W})$ with \eqref{eq:BFagl1} reproduces the affine Yangian of $\mathfrak{gl}_1$,
studied in \cite{Tsymbaliuk:2014fvq,Prochazka:2015deb,Gaberdiel:2017dbk}.

On the other hand, the quantum Coulomb branch algebra for the Jordan quiver was shown to be isomorphic to the (truncated shifted) Yangian of affine $\mathfrak{gl}_1$ \cite{Kodera:2016faj_jordan_quiver, Rapcak:2020ueh}, which is exactly the quiver Yangian associated to the triple quiver of the Jordan quiver \cite{Li:2020rij}.
Note that in \cite{Kodera:2016faj_jordan_quiver}, the two independent parameters of the affine Yangian of $\mathfrak{gl}_1$ are denoted as $\hbar$ and $\mathbf{t}$, where $\hbar$ is still the quantization parameter while $\mathbf{t}$ is an extra parameter that already appears in the classical Coulomb branch algebra of the Jordan quiver.

\section{Discussion}
\label{sec:Discussion}

We have already summarized our main results in Sec.~\ref{sec:Intro}, and now we discuss some implications of our results and interesting future directions.

\medskip

\noindent
\textbf{Bootstrap.}
In the first half of this paper, our approach of constructing the quantum Coulomb branch algebra for the tree-type quivers, following \cite{Bullimore:2016hdc}, is reminiscent of how the quiver Yangian was first defined as  the BPS algebra for the $\frac{1}{2}$-BPS D-brane bound states system in type IIA string theory on a generic toric Calabi-Yau threefold \cite{Li:2020rij}. 

Recall that the idea of the BPS algebra was first proposed by Harvey and Moore in \cite{Harvey:1996gc} as the algebra of BPS states where the underlying vector space is the BPS Hilbert space and the multiplication of the algebra is given by  (some properly defined) ``fusion" of the BPS states \cite{Harvey:1996gc}.
For example, the cohomological Hall algebra (CoHA) \cite{Kontsevich:2008fj,Kontsevich:2010px,Rapcak:2018nsl} defined for a quiver with potential can be viewed as a concrete formulation of the BPS algebra, where the BPS Hilbert space is given by the equivariant cohomology of the quiver representation space and the multiplication is given by a certain convolution product  $\mathfrak{m}_{\textrm{KS}}$ on them.\footnote{In the context of 4D $\mathcal{N}=2$ theories where the $\frac{1}{2}$-BPS sectors are given by $\mathcal{N}=4$ quiver quantum mechanics with $2$-acyclic quivers and generic potential, the CoHA as the BPS algebra was recently studied in \cite{Gaiotto:2024fso}; see also \cite[App.~B]{Gaiotto:2024fso} for the physical underpinning of the CoHA.
}

However, one can still hope for a reformulation of the BPS algebra in terms of explicit relations among generators. 
For the $\frac{1}{2}$-BPS D-brane bound state systems in type IIA string theory on a generic toric Calabi-Yau threefold, one can first determine how the  generators of the BPS algebra act on the BPS states, using their description in terms of certain 3D colored crystals \cite{Ooguri:2008yb} and obeying certain basic properties of the BPS algebra. Then one can deduce the relations among the generators and thus define the quiver Yangian from this action \cite{Li:2020rij}.\footnote{For a computation in terms of $\mathcal{N}=4$ quiver quantum mechanics see \cite{Galakhov:2020vyb}, for generalization to non-vacuum representations and the definition of shifted quiver Yangian see \cite{Galakhov:2021xum}, and for generalizations to K-theory and elliptic cohomologies see \cite{Galakhov:2021vbo}.}
Namely, the quiver Yangian as  the BPS algebra was ``bootstrapped" from its action on the BPS states.
(Note that the resulting quiver Yangian is the Drinfeld double of CoHA, which only contains the raising operators. )

Our approach of constructing the quantum Coulomb branch algebra for the tree-type quivers can be viewed as a bootstrap procedure in the same spirit of how the quiver Yangian was defined as the BPS algebra for the type IIA string theory on toric Calabi-Yau threefolds, except that here the quiver Yangian serves as the algebra of monopoles (together with the vector-multiplet scalar).
Note that the reason that our bootstrap approach using only the (first fundamental) minuscule monopoles (for the theories with unitary gauge groups) works is related to the fact that the CoHA for the triple quiver is spherical. 

The resulting quantum Coulomb branch algebra for a 3D $\mathcal{N}=4$ unitary quiver gauge theory with tree-type quiver $\mathrm{Q}$ is given by the truncated shifted Yangian  Y$_{\hbar}(\widehat{Q},\widehat{W})$ based on the triple quiver $\widehat{Q}$ of $\mathrm{Q}$. 
As a consistency check, 
the information of the truncation and shift is contained in the ranks of the gauge and flavor groups but does not depend on the vortex Hilbert space $\mathcal{H}_{\nu}$, which appears in the intermediate steps of the computation.
Based on the results on the tree-type quivers and a series of mathematical arguments, we then generalize the main result (as a conjecture) to arbitrary quivers and check this conjecture against earlier known results.

\medskip

\noindent
\textbf{Absence of higher-order poles.}
For a given quiver $\mathrm{Q}$, the Hilbert space of vortices approaching the vacuum $\nu$ at spatial infinity furnishes the vortex representation $\mathcal{R}_{\nu}$ of the shifted quiver Yangian Y$_{\hbar}(\widehat{Q},\widehat{W})$.
The representation $\mathcal{R}_{\nu}$ of the quiver Yangian is given by the framing of the triple quiver $\widehat{Q}$ and the information is entirely captured by the eigenvalue functions of the Cartan generators on the ground state of the representation, which are rational functions of the equivariant weights of the framing arrows. 

For the quiver with potential $(Q,W)$ from the toric Calabi-Yau threefolds, for any highest weight representation given by the framing of $Q$, the eigenvalue functions of the Cartan generators (the so-called charge functions) of any states have only simple poles, corresponding to the positions where the raising (or lowering) operators can add (or remove) an atom, which is an element in the Jacobian algebra $J(Q,W)$, i.e.\ a path (in the path algebra  $\mathbb{C}Q$ up to path equivalence).
In contrast, for a general quiver with potential, there can be representations where the charge functions can have multiple poles \cite{Li:2023zub}.

In this paper, the framing of the triple quiver $\widehat{Q}$ was determined in such a manner that the states in the resulting quiver Yangian representation $\mathcal{R}_{\nu}$ match the states in the corresponding vortex Hilbert space $\mathcal{H}_{\nu}$. 
We do not need to impose the ``simple pole" constraint, but  the charge functions of all the states  turn out to have only simple poles, in any $\mathcal{R}_{\nu}$ and for any tree-type quiver $\mathrm{Q}$. 
It is very reassuring, given that the states in a \textit{general} representation of the quiver Yangian for a general triple quiver could have charge functions with higher-order poles (see \cite[App.~C]{Li:2023zub}) -- namely those thorny representations simply do not appear in this physical setup.

\medskip

\medskip

Let us end with some interesting problems for future investigations.

\begin{itemize}
\item We would like to check our conjecture more explicitly for 3D $\mathcal{N}=4$ unitary quiver gauge theories with general quivers, using the approach of the Sec.~\ref{sec:vortex_tree_typeQGT} and \ref{sec:QYasMonopoleAlgebra}, in particular the non-simply-laced quivers and quivers with multi-edge-loops. 
\item It would be interesting to generalize to 3D $\mathcal{N}=4$ quiver gauge theories whose gauge factors are not $U(N)$ groups, such as $SO(N)$ or $Sp(N)$.
This is interesting because we can check whether the quantum Coulomb branch algebra is still spherical.
\item We would also like to explore features of the quiver Yangians beyond the quadratic relations, such as the Serre relations and the co-product structures, and their physical manifestation. 
\end{itemize}
We hope to report on these in future work.

\section*{Acknowledgments}

We would like to thank Davide Gaiotto, Matthias Gaberdiel, Xinyu Zhang, and in particular Gufang Zhao and Hiraku Nakajima for very helpful discussions.
WL is supported by NSFC  No.\ 12275334, No.\ 12447101 and is grateful for the hospitality of the Kavli Institute for Theoretical Physics at Santa Barbara, where part of this work was carried out. 

\appendix

\section{\texorpdfstring{Review on 3D $\mathcal{N}=4$ gauge theories}{Review on 3D N=4 gauge theories}}
\label{appsec:Rev-3D-N=4}

\subsection{\texorpdfstring{3D $\mathcal{N}=4$ supersymmetry}{3D N=4 supersymmetry}}
\label{ssec:3D_N=4_SUSY}
	
6D $\mathcal{N}=1$, 4D $\mathcal{N}=2$ and 3D $\mathcal{N}=4$ supersymmetries all have 8 supercharges, and are related by dimensional reduction. 
The 6D $\mathcal{N}=1$ R-symmetry is $SU(2)$,\footnote{On a 6D spinor one cannot impose Weyl and Majorana conditions at the same time. 
However, as shown in \cite{Kugo:1982bn} one can impose a $SU(2)$-reality condition: $\bar{\psi}_{i\dot{\alpha}}=\epsilon_{ij}B_{\dot{\alpha}}^{\beta}\psi^j_{\beta}$, where the unitary matrix $B$ satisfies $B^*B=-1$, $i,j=1,2$ are $SU(2)$ indices, and $\alpha,\dot{\alpha}=1,2,3,4$ are 6D spinor indices. 
The 6D $\mathcal{N}=1$ supercharges are described by a Weyl spinor or equivalently, by two Weyl and $SU(2)$-Majorana spinors $\mathcal{Q}^i_\alpha$, satisfying the superalgebra $\{\mathcal{Q}^i_\alpha,\mathcal{Q}^j_\beta\}=\epsilon^{ij} P_\mu \Gamma^\mu_{\alpha\beta}$ with $\Gamma^\mu$ being 6D Pauli matrices. 
In this formulation, the $SU(2)$ R-symmetry is manifest.} which after dimensional reduction becomes the $SU(2)_H$ factor of R-symmetry in the 3D $\mathcal{N}=4$ theory. 
The rotations in $x^{3,4,5}$-directions reduce to the $SU(2)_C$ factor. 
Altogether they form the R-symmetry $SU(2)_{H}\times SU(2)_C\sim SO(4)$ of the 3D $\mathcal{N}=4$ theories. 
The 6D $\mathcal{N}=1$ supercharges become the 3D $\mathcal{N}=4$ supercharges $\mathcal{Q}^{a\dot{a}}_{\alpha}$, where $a,\dot{a}$ are the indices of $SU(2)_{H}\times SU(2)_C$ and $\alpha$ is the  spinor index of the 3D (Euclidean) Lorentz group $SO(3)_E\sim SU(2)_E$.
The 3D $\mathcal{N}=4$ superalgebra is 
\begin{equation}\label{eq:SUSY3DN4}
\{\mathcal{Q}_\alpha^{a\dot{a}},\mathcal{Q}_\beta^{b\dot{b}}\}
=-2\epsilon^{ab}\epsilon^{\dot{a}\dot{b}}\sigma^\mu_{\alpha\beta}P_\mu+2\epsilon_{\alpha\beta}(\epsilon^{ab}Z^{\dot{a}\dot{b}}+\epsilon^{\dot{a}\dot{b}}Z^{ab})\,,
\end{equation}
where $(\sigma^{\mu})^{\alpha}_{\ \beta}$ are the Pauli matrices, $\epsilon^{12}=\epsilon_{21}=1$.
The central charges $Z^{ab}$ and $Z^{\dot{a}\dot{b}}$ act as combinations of the flavor and the gauge transformations, see later.

\medskip
 
The 3D $\mathcal{N}=4$ multiplets can be obtained by dimensional reduction from the 6D $\mathcal{N}=1$ multiplets \cite{Seiberg:1996nz}: 
\begin{itemize}
\item  Consider a 6D $\mathcal{N}=1$ vector-multiplet $(\Psi^{\textrm{6D}}_{\textrm{v.m.}}, \textrm{A}^{6D}_\mu)$, where $\Psi^{\textrm{6D}} $ is the $4$-dimensional Weyl spinor and $\textrm{A}^{6D}_\mu$ is the gauge boson. 
Upon dimensional reduction to 3D, it becomes a 3D $\mathcal{N}=4$ vector-multiplet $(\vec{\phi}, \Psi^{a\dot{a}}_{\textrm{v.m.}},A_{\mu})$, where $\vec{\phi}=A^{\textrm{6D}}_{3,4,5}$ is a $SU(2)_C$ triplet, and the four $SU(2)_E$ spinors $\Psi^{a\dot{a}}_{\textrm{v.m.}}$ form the representation $(\mathbf{2}, \mathbf{2})$ of $SU(2)_{H}\times SU(2)_C$. $A_{\mu}$ is the gauge field in 3D, which is dual to a scalar $\phi$ (the dual photon) through $\mathrm{d}\phi=*\mathrm{d}A$; 
the dual photon $\phi$ is periodic $\phi\sim \phi+2\pi g^2$, where $g$ is the gauge coupling of $A_\mu$. 
All these fields are in the adjoint representation of the gauge group $G$.
		
\item The 6D $\mathcal{N}=1$ hypermultiplet $(\Phi^{1,2}, \Psi^{\textrm{6D}}_{\textrm{h.m.}})$ consists of 2 complex scalars and a Weyl spinor. 
After dimensional reduction, it becomes a 3D $\mathcal{N}=4$ hypermultiplet $(\Phi^{1,2}, \Psi^{\dot{1},\dot{2}}_{\textrm{h.m.}})$, transforming in a unitary representation $R$ of $G$. 
\end{itemize}

\medskip

The $\mathcal{N}=4$ supersymmetry can be described in $\mathcal{N}=2$ language. 
We choose an $\mathcal{N}=2$ subalgebra with generators:\footnote{We require that the generators $\mathcal{Q},\bar{\mathcal{Q}}$ of the $\mathcal{N}=2$ superalgebra are invariant under the $U(1)_H\times U(1)_C$, and this prevents us from using linear combination of $\mathcal{Q}^{a\dot{a}}$. 
Moreover, $\{\mathcal{Q},\bar{\mathcal{Q}}\}\sim P$ requires that $\mathcal{Q}$ and $\mathcal{Q}$ have reverse charges, i.e.\ the only possible candidates for $\{\mathcal{Q},\bar{\mathcal{Q}}\}$ are $\{\mathcal{Q}^{1\dot{1}},\mathcal{Q}^{2\dot{2}}\}$ and $\{\mathcal{Q}^{2\dot{1}},\mathcal{Q}^{1\dot{2}}\}$, and they are related by R-symmetry.}
\begin{equation}\label{eq:N=2SUSY}
\mathcal{Q}_{\alpha}\equiv  \mathcal{Q}^{2\dot{2}}_{\alpha}  
\qquad \textrm{and}\qquad
\bar{\mathcal{Q}}_{\alpha}\equiv \mathcal{Q}^{1\dot{1}}_{\alpha}\,,
\end{equation}
whose superalgebra follows from \eqref{eq:SUSY3DN4}:
\begin{equation}
\{\mathcal{Q}_\alpha,\bar{\mathcal{Q}}_\beta\}=-2\sigma^\mu_{\alpha\beta}P_\mu+2\epsilon_{\alpha\beta}Z \,,\qquad\{\mathcal{Q}_\alpha,\mathcal{Q}_\beta\}=\{\bar{\mathcal{Q}}_\alpha,\bar{\mathcal{Q}}_\beta\}=0\,,
\end{equation}
with $Z$ the real central charge $Z:=Z^{12}+Z^{\dot{1}\dot{2}}$.
The 3D $\mathcal{N}=2$ supersymmetry \eqref{eq:N=2SUSY} only preserves the $U(1)_H\times U(1)_C \subset SU(2)_H\times SU(2)_C$ R-symmetry.
	
A 3D $\mathcal{N}=4$ vector-multiplet is decomposed into a 3D $\mathcal{N}=2$ vector-multiplet together with a 3D $\mathcal{N}=2$ chiral multiplet
\begin{equation}
\begin{aligned}
&\textrm{3D $\mathcal{N}=4$ v.m.} \quad (\vec{\phi}, \Psi^{a\dot{a}}_{\textrm{v.m.}},A_{\mu}) \\
=\quad &\textrm{3D $\mathcal{N}=2$ v.m.} \quad (\sigma, \lambda,A_{\mu}) \quad
\oplus \quad \textrm{3D $\mathcal{N}=2$ c.m.} \quad (\varphi, \eta)\,,
\end{aligned}    
\end{equation}
where $\sigma$ and $\varphi$ are real and complex scalars, transforming as a triplet of $SU(2)_C$; $\lambda$ and $\eta$ are complex spinors defined as
\begin{equation}
\lambda=\Psi^{1\dot{1}}_{\textrm{v.m.}}\,,\quad \bar{\lambda}=\Psi^{2\dot{2}}_{\textrm{v.m.}}\,,\qquad \eta=\Psi^{2\dot{1}}_{\textrm{v.m.}}\,,\quad \bar{\eta}=\Psi^{1\dot{2}}_{\textrm{v.m.}}\,,
\end{equation}
and $(\lambda,\eta)$ transform as a doublet under $SU(2)_H$.
	
On the other hand, a 3D $\mathcal{N}=4$ hypermultiplet is decomposed into a 3D $\mathcal{N}=2$ chiral multiplet together with an anti-chiral multiplet
\begin{equation}
\begin{aligned}
&\textrm{3D $\mathcal{N}=4$ h.m.} \quad (\Phi^{a}, \Psi^{\dot{a}}_{\textrm{h.m.}})  \\
=\quad &\textrm{3D $\mathcal{N}=2$ c.m.} \quad (X, \psi^{X}) 
\quad
\oplus \quad 
\textrm{3D $\mathcal{N}=2$ a.c.m.} \quad (\bar{Y}, \bar{\psi}^{Y}) \,,
\end{aligned}    
\end{equation}
where $X$ and $\bar{Y}$ transform as a doublet under $SU(2)_H$, and the fermionic fields are defined by
\begin{equation}
\psi^X=\Psi^{\dot{2}}_{\textrm{h.m.}}\,,\quad   \bar{\psi}^Y=\Psi^{\dot{1}}_{\textrm{h.m.}}\,,
\end{equation}
and $(\psi^X,\bar{\psi}^{Y})$ transforms as a doublet under $SU(2)_C$. 
In $\mathcal{N}=2$ language, the chiral multiplets $(X,\psi^X)\oplus (Y,\psi^Y)$ live in the symplectic representation $ R\oplus \bar{R}$.

Finally, the indices $a=1,2$ (resp.\ $\dot{a}=\dot{1},\dot{2}$) correspond to charges $\frac{1}{2},-\frac{1}{2}$ under the maximal torus $U(1)_H$ (resp.\ $U(1)_C$). 
For completeness, we list the $U(1)_H\times U(1)_C$ charges of the various objects \cite{Bullimore:2016nji}:
\begin{equation}\label{tab:charges}
\begin{array}{c|cccccccccc}
&A_\mu &\sigma&\varphi&\lambda_{\pm}&\bar{\lambda}_{\pm}&\eta_{\pm}&\bar{\eta}_{\pm} & X,Y &\psi^X_{\pm},\psi^Y_{\pm}&\bar{\psi}^X_{\pm},\bar{\psi}^Y_{\pm} \\
\hline   U(1)_H&0&0&0&\frac{1}{2}&-\frac{1}{2}&-\frac{1}{2}&\frac{1}{2}&\frac{1}{2}&0&0\\
U(1)_C&0&0&1&\frac{1}{2}&-\frac{1}{2}&\frac{1}{2}&-\frac{1}{2}&0&-\frac{1}{2}&\frac{1}{2}
\end{array}\,.
\end{equation}

\medskip
	
We will write the 3D $\mathcal{N}=4$ theory in terms of the 3D $\mathcal{N}=2$ superfields. 
The $\mathcal{N}=2$ superspace has the coordinates $(x,\theta,\bar{\theta})$ where $\theta$ is a two-component spinor of $U(1)_H\times U(1)_C$ charge $(\frac{1}{2},\frac{1}{2})$. 
A general supersymmetry transformation with parameter $(\zeta,\bar{\zeta})$ can be written as $e^{i(\zeta \mathtt{Q}+\bar{\zeta}\bar{\mathtt{Q}})}$, with the supercharge representation
\begin{equation}
\mathtt{Q}_{\alpha}:=-i\frac{\partial}{\partial\theta^\alpha}-\sigma^\mu_{\alpha\beta}\bar{\theta}^\beta\partial_\mu\,,
\qquad \bar{\mathtt{Q}}_{\alpha}:=-i\frac{\partial}{\partial\bar{\theta}^\alpha}-\theta^\beta\sigma^\mu_{\beta\alpha}\partial_\mu \,,   
\end{equation}
and the $\mathcal{N}=2$ super-derivatives 
\begin{equation}
\mathtt{D}_{\alpha}:=\frac{\partial}{\partial\theta^\alpha}+i\sigma^\mu_{\alpha\beta}\bar{\theta}^\beta\partial_\mu\,,
\qquad \bar{\mathtt{D}}_{\alpha}:=\frac{\partial}{\partial\bar{\theta}^\alpha}+i\theta^\beta\sigma^\mu_{\beta\alpha}\partial_\mu \,.    
\end{equation}

The 3D $\mathcal{N}=2$ chiral multiplet $(\varphi,\eta)$ sits in the chiral superfield
\begin{equation}
\mathtt{\Phi}_c(x,\theta,\bar{\theta})=\varphi(x)+\sqrt{2}\theta\eta(x)+i\theta\sigma^{\mu}\bar{\theta}\partial_\mu\varphi(x)-\theta^2 F(x)-\frac{i}{\sqrt{2}}\theta^2\partial_\mu \eta(x) \sigma^{\mu}\bar{\theta}-\frac{1}{4}\theta^2\bar{\theta}^2\partial^2\varphi(x) \,,
\end{equation}
where $F(x)$ is an auxiliary field. 
It satisfies $\bar{\mathtt{D}}_a\mathtt{\Phi}_c(x,\theta,\bar{\theta})=0$. 
The 3D  $\mathcal{N}=2$ chiral multiplet 
$(X,\psi^X)$ (together with an auxiliary field $F_X$)
sits in the chiral superfield:
\begin{equation}
\begin{aligned}
\mathtt{\Phi}_X(x,\theta,\bar{\theta})=X(x)+\sqrt{2}\theta\psi^X(x)+i\theta\sigma^{\mu}\bar{\theta}&\partial_\mu X(x)-\theta^2 F_X(x)\\
&-\frac{i}{\sqrt{2}}\theta^2\partial_\mu \psi^X(x) \sigma^{\mu}\bar{\theta}-\frac{1}{4}\theta^2\bar{\theta}^2\partial^2 X(x) \,,
\end{aligned}
\end{equation}
and likewise for the $\mathcal{N}=2$ chiral multiplet $\mathtt{\Phi}_Y(x,\theta,\bar{\theta})$ which contains $(Y,\psi^Y,F_Y)$. 
The $U(1)_H\times U(1)_C$ charges of $\mathtt{\Phi}_c$, $\mathtt{\Phi}_X$, $\mathtt{\Phi}_Y$ are $(0,1),(\frac{1}{2},0),(\frac{1}{2},0)$, respectively. 
On the other hand, the 3D $\mathcal{N}=2$ vector-multiplet $(\sigma,\lambda,A_\mu)$ belongs to the real superfield $\mathtt{V}(x,\theta,\bar{\theta})$:
\begin{equation}\label{eq:VDef}
\mathtt{V}(x,\theta,\bar{\theta})=\theta\sigma^{\mu}\bar{\theta}A_\mu(x)+i\theta^2\bar{\theta}\bar{\lambda}(x)-i\bar{\theta}^2\theta\lambda(x)+i\theta\bar{\theta}\sigma(x)+\frac{1}{2}\theta^2\bar{\theta}^2D(x)\,,
\end{equation}
written in the Wess-Zumino gauge, where $D(x)$ is an auxiliary field.\footnote{All these expressions can be obtained directly by dimensional reduction from the 4D $\mathcal{N}=1$ theory.} The $U(1)_H\times U(1)_C$ charge of $\mathtt{V}$ is $(0,0)$. 
Finally, the superfields $\mathtt{\Phi}_{c}$ and $\mathtt{V}$ transform in the adjoint representation of the gauge group $G$, whereas $(\mathtt{\Phi}_X,\mathtt{\Phi}_Y)$ transforms in the representation $(R,\bar{R})$ of the gauge group $G$.

\subsection{\texorpdfstring{3D $\mathcal{N}=4$ gauge theory}{3D N=4 gauge theory}}\label{ssec:3D_SYM}
A 3D $\mathcal{N}=4$ gauge theory is specified by a gauge group $G$ and a symplectic representation $(R, \bar{R})$, in which the hypermultiplets transform.

The theory also has global symmetries, decomposed as $G_H\times G_C$. 
$G_H$ is the flavor symmetry, which rotates the hypermultiplets. 
It is taken to be the centralizer of $G$ within the unitary symmetry group $U(R)$ of the representation $R$.\footnote{More generally, the flavor symmetry $G_H$ can be taken to be the normalizer of $G$ within $U(R)$, modulo $G$ itself: $G_H=N_{U(R)}(G)/G$; we will not consider this more general situation in this paper.} 
On the other hand, $G_C$ is the large gauge symmetry corresponding to the periodicity of the dual photon $\phi$ and hence $G_C \approx U(1)^{\mathrm{\#\ of\ }U(1)\ \mathrm{factors\ in}\ G}$, which defines the $U(1)$ rotation of monopole operators \cite{Bullimore:2016hdc}.

\subsubsection{\texorpdfstring{Lagrangian of 3D $\mathcal{N}=4$ gauge theory and deformations}{Lagrangian of 3D N=4 gauge theory and deformations}}
\label{appssec:Lagrangian}

In $\mathcal{N}=2$ language, the action of a 3D $\mathcal{N}=4$ gauge theory is
\begin{equation}
\mathcal{S}_{\mathcal{N}=4}
=  \int \mathrm{d}^3x\, \mathcal{L}_V
+\int \mathrm{d}^3x\,\mathcal{L}_H\,.
\end{equation}
The Lagrangian for the vector-multiplet is 
\begin{equation}\label{eq:vec-L} 
\mathcal{L}_V=
\frac{1}{g^2}\int\mathrm{d}^4\theta \,
\mathrm{Tr}\left(\frac{1}{4}\mathtt{\Sigma}\mathtt{\Sigma^\dag}+ \mathtt{\Phi}_c^\dag e^{-2\mathtt{V}}\mathtt{\Phi}_c e^{2\mathtt{V}}\right)\,, 
\end{equation}
where the $\mathcal{N}=2$ linear multiplet $\mathtt{\Sigma}$ in the kinetic term is defined in terms of the real superfield $\mathtt{V}$ (see \eqref{eq:VDef}) \cite{Aharony_1997}:
\begin{equation}
\mathtt{\Sigma}=\epsilon^{\alpha\beta}\bar{\mathtt{D}}_\alpha \mathtt{D}_\beta \mathtt{V}\,,
\end{equation}
whose leading component is the real scalar $\sigma$. 
The Lagrangian for the hypermultiplet is
\begin{equation}\label{eq:hyper-L}
\mathcal{L}_H=\int\mathrm{d}^4\theta \left(\mathtt{\Phi}_X^\dag e^{2\mathtt{V}} \mathtt{\Phi}_X+ \mathtt{\Phi}_Y e^{-2\mathtt{V}} \mathtt{\Phi}_Y^\dag\right)
+ \left(\int\mathrm{d}^2 \theta \, W + h.c.\right)\,,
\end{equation} 
with the superpotential 
\begin{equation}\label{eq:WDef}
W=  \mathtt{\Phi}_Y\mathtt{\Phi}_c\mathtt{\Phi}_X\,.
\end{equation}
	
\medskip
The action of the supercharges on the various fields can be easily derived from dimensional reduction \cite{Aharony_1997,Closset_2019}. 
As an example, let us write down the transformations under $\mathcal{Q}^{1\dot{1}}$ for various fermionic fields:
\begin{equation}\label{eq:Q11VM}
\text{Vector-multiplet:}\left\{
\begin{array}{l}
\{\mathcal{Q}^{1\dot{1}}_\alpha,\lambda_\beta\} =0\,,   \\
\{\mathcal{Q}^{1\dot{1}}_\alpha,\bar{\lambda}_\beta\}=-i\epsilon_{\alpha\beta}D+i\frac{1}{2}\sigma^{\mu}_{\alpha\beta}(D_\mu\sigma-\frac{1}{2}\epsilon_{\mu\nu\rho}F^{\nu\rho})\,,\\
\{\mathcal{Q}^{1\dot{1}}_\alpha,\eta_\beta\}=-\sqrt{2}\epsilon_{\alpha\beta}[\sigma,\varphi]-\sqrt{2}\sigma^\mu_{\alpha\beta}D_\mu\varphi\,,\\
\{\mathcal{Q}^{1\dot{1}}_\alpha,\bar{\eta}_\beta\}=-i\sqrt{2}\epsilon_{\alpha\beta}\bar{F}\,,
\end{array}
\right.
\end{equation}
and
\begin{equation}\label{eq:Q11HM}
\text{Hypermultiplet:}\left\{
\begin{array}{l}
\{\mathcal{Q}^{1\dot{1}}_\alpha,\psi^X_\beta\}=-\sqrt{2}\epsilon_{\alpha\beta}\sigma X-\sqrt{2}\sigma^\mu_{\alpha\beta}D_\mu X\,,\\
\{\mathcal{Q}^{1\dot{1}}_\alpha,\bar{\psi}^X_\beta\}=-i\sqrt{2}\epsilon_{\alpha\beta}\bar{F}_X\,,\\
\{\mathcal{Q}^{1\dot{1}}_\alpha,\psi^Y_\beta\}=-\sqrt{2}\epsilon_{\alpha\beta}Y\sigma -\sqrt{2}\sigma^\mu_{\alpha\beta}D_\mu Y\,,\\
\{\mathcal{Q}^{1\dot{1}}_\alpha,\bar{\psi}^Y_\beta\}=-i\sqrt{2}\epsilon_{\alpha\beta}\bar{F}_Y\,,
\end{array}\right.
\end{equation}
where $D_\mu=\partial_\mu-iA_\mu$ is the covariant derivative, $F_{\mu\nu}=i[D_\mu,D_\nu]$ is the field strength of the gauge field, and the Levi-Civita symbol $\epsilon_{\mu\nu\rho}$ satisfies $\epsilon_{123}=1$. 
In \eqref{eq:Q11VM} and \eqref{eq:Q11HM}, the auxiliary fields $F$ and $D$ are to be replaced by their on-shell values, which can be read off from the 3D $\mathcal{N}=4$ Lagrangian \eqref{eq:hyper-L} and \eqref{eq:vec-L}, or via dimensional reduction from the 4D $\mathcal{N}=2$ theory. 
For the $F$'s we have
\begin{equation}
\bar{F}_X=\frac{\partial W}{\partial X}=Y\varphi \,,
\qquad 
\bar{F}_Y=\frac{\partial W}{\partial Y}=\varphi X\,,
\qquad
\bar{F}=\frac{\partial W}{\partial \varphi}=\mu_{\mathbb{C}}(T)
\,,
\end{equation}
where $W$ is the superpotential \eqref{eq:WDef}, and $T$ denotes the generators of the gauge group $G$. 
For the auxiliary field $D$ we have
\begin{equation}
\begin{aligned}
D=[\varphi,\varphi^\dag]-g^2\mu_{\mathbb{R}}(T)\,.
\end{aligned}
\end{equation}
Here $\mu_{\mathbb{R}},\mu_{\mathbb{C}}\in \mathfrak{g}^*$ are the real and complex momentum maps of the $G$-action on $(X,Y)$, given by
\begin{equation}
\mu_{\mathbb{R}}(T)=\bar{X}TX-YT\bar{Y}\,,\qquad \mu_{\mathbb{C}}(T)= YT X\,.
\end{equation}
Using the $R$-symmetry $SU(2)_H\times SU(2)_C$ and the charges of the various fields listed in \eqref{tab:charges}, one can obtain the action of the other supercharges. 
	
\medskip
	
The theory allows for deformations in terms of masses and FI-parameters, as follows. 
The real and complex masses, $m_{\mathbb{R}}$ and $m_{\mathbb{C}}$, are the constant background values of $\mathtt{V}(x,\theta,\bar{\theta})$ and $\mathtt{\Phi}_c(x,\theta,\bar{\theta})$, taking values in the Cartan subalgebra $\mathfrak{t}^{(H)}$ of the flavor symmetry $\mathfrak{g}^{(H)}$ and its complexified version $\mathfrak{t}^{(H)}_{\mathbb{C}}$, respectively. 
These mass deformations can be implemented by the shifts 
\begin{equation}
\begin{aligned}
&\sigma\to\sigma+m_{\mathbb{R}}
\,,\qquad m_{\mathbb{R}}\in \mathfrak{t}^{(H)}\,,\\
&\varphi\to\varphi+m_{\mathbb{C}}\,,\qquad m_{\mathbb{C}}\in \mathfrak{t}^{(H)}_{\mathbb{C}}\,.
\end{aligned}
\end{equation}
On the other hand, the real and complex FI-parameters, $t_{\mathbb{R}}$ and $t_{\mathbb{C}}$, are introduced by adding the following terms to the Lagrangian:
\begin{equation}\label{eq:FI_para}
t^A_{\mathbb{R}}\int\mathrm{d}^4 \theta \, \mathtt{V}^A
\qquad \text{and}\qquad t^A_{\mathbb{C}}\int\mathrm{d}^2\theta \, \mathtt{\Phi}_c^A+h.c.\,,
\end{equation}
where $A$ counts the abelian factors of the gauge group. (For example, $A=1,2,\dots,|\mathrm{Q}_0|$ for the quiver gauge theory.) 
This effectively modifies $\bar{F}$ and $D$ to
\begin{equation}
\bar{F}^A= \mu_{\mathbb{C}}(T^A)+t^A_{\mathbb{C}}\,, \qquad
D^A= -[\varphi,\varphi^\dag]^A+g^2(\mu_{\mathbb{R}}(T^A)+t^A_{\mathbb{R}})\,.
\end{equation}
The FI-parameters can also be viewed as background values of fields, if we introduce a twisted vector-multiplet. 
The field content of the twisted vector-multiplet is the same as that of the vector-multiplet, but with the $U(1)_H$ and the $U(1)_C$ charges exchanged. 
In $\mathcal{N}=2$ language, the $\mathcal{N}=4$  twisted vector-multiplet consists of an $\mathcal{N}=2$ 
vector-multiplet $\tilde{\mathtt{V}}(x,\theta,\bar{\theta})$ and an $\mathcal{N}=2$ chiral multiplet $\tilde{\mathtt{\Phi}}_c(x,\theta,\bar{\theta})$, with $U(1)_H\times U(1)_C$ charges $(0,0)$ and $(1,0)$, respectively. 
The coupling between the  $\mathcal{N}=4$ twisted vector-multiplet and the $\mathcal{N}=4$ vector-multiplet is given by the following two terms:
\begin{equation}\label{eq:coupleTV-V}
\int \mathrm{d}^4\theta \,\mathrm{Tr}(\tilde{\mathtt{\Sigma}}\mathtt{V})
\qquad \text{and}\qquad 
\int \mathrm{d}^2\theta \, \mathrm{Tr}(\tilde{\mathtt{\Phi}}_c\mathtt{\Phi}_c)+h.c.\,,
\end{equation}
where the first term is the $\mathcal{N}=2$ version of the mixed Chern-Simons term $\tilde{A}\mathrm{d}A$.
Comparing with \eqref{eq:FI_para}, one can then identify the FI-parameters with the VEV's of the scalars of the twisted vector-multiplet, which take values in the Cartan subalgebra $\mathfrak{t}^{(C)}$ of the large gauge symmetry $\mathfrak{g}^{(C)}$ and its complexified version, respectively:
\begin{equation}
\langle\tilde{\sigma}\rangle\rightarrow t_{\mathbb{R}}\in \mathfrak{t}^{(C)}\,,\qquad \langle\tilde{\varphi}\rangle\rightarrow t_{\mathbb{C}}\in\mathfrak{t}^{(C)}_{\mathbb{C}}\,.
\end{equation}

\medskip 
	
Finally let us discuss the central charges of the 3D $\mathcal{N}=4$ theories. 
The central charges $Z^{ab}$ and $Z^{\dot{a}\dot{b}}$ are symmetric rank-2 tensor (triplet) of $SU(2)_H$ and $SU(2)_C$, respectively. 
By direct computations of  $\{\mathcal{Q}^{1\dot{1}},\mathcal{Q}^{2\dot{1}}\}$ and etc (after introducing the twisted vector-multiplets), one can see that the central charges $Z^{\dot{a}\dot{b}}$ are proportional to the $SU(2)_C$ triplet of vector-multiplet scalars
\begin{equation}
Z^{\dot{1}\dot{1}}=(Z^{\dot{2}\dot{2}})^\dag\propto \varphi+m_{\mathbb{C}}\,,\qquad Z^{\dot{1}\dot{2}}\propto \sigma+m_{\mathbb{R}}\,.
\end{equation}
On the other hand, $Z^{ab}$ is proportional to the scalars of the twisted vector-multiplet
\begin{equation}
Z^{11}=(Z^{22})^\dag\propto t_{\mathbb{C}}\,,\qquad Z^{12}\propto t_{\mathbb{R}}\,.
\end{equation}
	
\subsubsection{BPS equations: moduli space, vortices and monopoles}

In this paper we focus on the algebra of monopole operators in 3D $\mathcal{N}=4$ theories, which is obtained by considering the action of monopoles on the vortices. 
In Sec.~\ref{ssec:BPS_in_3D_N4} we have reviewed aspects of three types of BPS solutions of the 3D $\mathcal{N}=4$ theory: the supersymmetric vacua, the $\frac{1}{2}$-BPS vortices, and the $\frac{1}{2}$-BPS monopoles. 
The supercharges they preserve are  \cite{Bullimore:2016hdc}:
\begin{equation}\label{eq:vor_mono_SUSY}
\begin{aligned}
\text{Vacua:}&\quad \{\mathcal{Q}^{1\dot{1}}_{\pm},\mathcal{Q}^{1\dot{2}}_{\pm},\mathcal{Q}^{2\dot{1}}_{\pm},\mathcal{Q}^{2\dot{2}}_{\pm}\}\,,\\
\text{Vortices:}&\quad \{\mathcal{Q}^{1\dot{1}}_{-},\mathcal{Q}^{1\dot{2}}_{-},\mathcal{Q}^{2\dot{1}}_{+},\mathcal{Q}^{2\dot{2}}_{+}\}\,,\\
\text{Monopoles:}&\quad \{\mathcal{Q}^{1\dot{1}}_{-},\mathcal{Q}^{2\dot{1}}_{-},\mathcal{Q}^{1\dot{1}}_{+},\mathcal{Q}^{2\dot{1}}_{+}\}\,.\\
\end{aligned}
\end{equation} 
Together they preserve a quarter of supersymmetry, $\mathcal{Q}^{1\dot{1}}_{-}$ and $\mathcal{Q}^{2\dot{1}}_{+}$.

\medskip

The action of $\mathcal{Q}^{1\dot{1}}_\alpha$ on the various fields was given in \eqref{eq:Q11VM} and \eqref{eq:Q11HM}. 
Applying the $SU(2)_H$ transformation on them then gives the action of $\mathcal{Q}^{2\dot{1}}_\alpha$:
\begin{equation}\label{eq:Q21VM}
\text{Vector-multiplet:}\left\{
\begin{array}{l}
\{\mathcal{Q}^{2\dot{1}}_\alpha,\lambda_\beta\} =\sqrt{2}\epsilon_{\alpha\beta}[\sigma,\varphi]+\sqrt{2}\sigma^\mu_{\alpha\beta}D_\mu\varphi\,,   \\
\{\mathcal{Q}^{2\dot{1}}_\alpha,\bar{\lambda}_\beta\}=i\sqrt{2}\epsilon_{\alpha\beta}F\,,\\
\{\mathcal{Q}^{2\dot{1}}_\alpha,\eta_\beta\}=0\,,\\
\{\mathcal{Q}^{2\dot{1}}_\alpha,\bar{\eta}_\beta\}=i\epsilon_{\alpha\beta}D^{'}-i\frac{1}{2}\sigma^{\mu}_{\alpha\beta}(D_\mu\sigma-\frac{1}{2}\epsilon_{\mu\nu\rho}F^{\nu\rho})\,,
\end{array}
\right.
\end{equation}
and
\begin{equation}\label{eq:Q21HM}
\text{Hypermultiplet:}\left\{
\begin{array}{l}
\{\mathcal{Q}^{2\dot{1}}_\alpha,\psi^X_\beta\}=-\sqrt{2}\epsilon_{\alpha\beta}\sigma \bar{Y}-\sqrt{2}\sigma^\mu_{\alpha\beta}D_\mu \bar{Y}\,,\\
\{\mathcal{Q}^{2\dot{1}}_\alpha,\bar{\psi}^X_\beta\}=i\sqrt{2}\epsilon_{\alpha\beta}F_Y\,,\\
\{\mathcal{Q}^{2\dot{1}}_\alpha,\psi^Y_\beta\}=\sqrt{2}\epsilon_{\alpha\beta}\bar{X}\sigma +\sqrt{2}\sigma^\mu_{\alpha\beta}D_\mu \bar{X}\,,\\
\{\mathcal{Q}^{2\dot{1}}_\alpha,\bar{\psi}^Y_\beta\}=-i\sqrt{2}\epsilon_{\alpha\beta}F_X\,,
\end{array}\right.
\end{equation}
where the auxiliary fields are
\begin{equation}
D'=[\varphi,\varphi^\dag]+g^2\mu_{\mathbb{R}}(T)\,,\quad F=\bar{X}T\bar{Y}\,,\quad F_X=\varphi \bar{Y}\,,\quad F_Y=\bar{X}\varphi\,.
\end{equation}

\medskip

It is straightforward to work out the action of the other supercharges, and then the BPS equations follow. 
For later convenience, we introduce the complex coordinates $z:=x^1+ix^2$ and $\bar{z}:=x^1-ix^2$, and define correspondingly $D_z:=\frac{1}{2}(D_1-iD_2)$ and $D_{\bar{z}}:=\frac{1}{2}(D_1+iD_2)$.
In addition, we define $
\mathcal{D}_t:=D_t-(\sigma+m_{\mathbb{R}})$.
(Here $D_1,D_2,D_t$ are components of the covariant derivative $D=\mathrm{d}-iA$.) 

\medskip 

Note that the three types of BPS objects all obey the $\frac{1}{4}$-BPS conditions imposed by the invariance under $\mathcal{Q}^{1\dot{1}}_{-}$ and $\mathcal{Q}^{2\dot{1}}_{+}$, which are \cite{Bullimore:2016hdc}:
\begin{itemize}
\item For the hypermultiplet $(X,Y,\psi^X,\psi^Y)$,
\begin{equation}\label{eq:14BSPhyper}
D_{\bar{z}}X=D_{\bar{z}}Y=\mathcal{D}_t X=\mathcal{D}_t Y=0\,.
\end{equation}
 
\item For the vector-multiplet $(A_\mu,\eta,\lambda,\varphi,\sigma)$, the equations are
\begin{equation}\label{eq:14BPSvector}
\begin{aligned}
&[D_{\bar{z}},\mathcal{D}_t]=0\,,\quad \quad [D_z,\varphi]=[D_{\bar{z}},\varphi]=[D_t,\varphi]=0\,,\\
&4[D_z, D_{\bar{z}}]-[\mathcal{D}_t,\mathcal{D}^\dag_t]=\mu_{\mathbb{R}}(T)+t_{\mathbb{R}}\,,\quad\mu_{\mathbb{C}}(T)+t_{\mathbb{C}}=0\,,
\end{aligned}
\end{equation}
where the conjugate of $\mathcal{D}_t$ is defined as $\mathcal{D}_t^\dag:=D_t+(\sigma+m_{\mathbb{R}})$. 
\item Since $\{\mathcal{Q}_{+}^{2\dot{1}},\mathcal{Q}_{-}^{1\dot{1}}\}=-2Z^{1\dot{1}}\sim \varphi+m_{\mathbb{C}}$ vanishes on any field, we have
\begin{equation}\label{eq:14BPSremaining}
[\varphi,\sigma]=[\varphi,\varphi^\dag]=0\,,\qquad (\varphi+m_{\mathbb{C}})X=Y(\varphi+m_{\mathbb{C}})=0\,.
\end{equation}
\end{itemize}
These equations can also be derived via dimensional reduction of the 4D $\mathcal{N}=2$ gauge theory. 

\section{Computational details}\label{appsec:details}

\subsection{The good, the bad, and the ugly}
\label{appssec:GoodBadUgly}
In this subsection we will prove that for any tree-type quiver (with a finite number of  nodes), there exists a choice of the ranks of the unitary gauge and flavor groups satisfying the condition \eqref{eq:non-increasingFull} and that the corresponding 3D $\mathcal{N}=4$ quiver gauge theory is ``good".

\medskip

We prove this by induction. 
We need a few definitions before we start.
We first define the height of the node $a\in \mathrm{Q_0}$ to be
\begin{equation}
\textrm{hei}(a)=\# \textrm{ of arrows from $\mathtt{f}$ to $a$}
\,,\qquad
\textrm{hei}(\mathtt{f})=0\,,
\end{equation}
and the height of the tree to be
\begin{equation}
\textrm{hei}(\mathrm{Q})= \textrm{max}(\textrm{hei}(a)) \ \textrm{for all } a\in \mathrm{Q}_0   \,.
\end{equation}
Next, let us denote a sub-tree (of the full tree-type quiver) with the root $a$ by $\textrm{tree}(a)$, and define a ``good sub-tree" rooted at $a\in \mathrm{Q}_0$ to be a sub-tree that is rooted at $a$ and that satisfies the following two conditions:
\begin{equation}
\mathfrak{b}^{(c)}\geq 0 \qquad \textrm{for any }\ c\in \{ \textrm{nodes on  \textrm{tree}$(a)$} \textit{ except for $a$}\} 
\end{equation}
and 
\begin{equation}
N^{(d)}\leq N^{(c)}    
\qquad \textrm{for any }\ c\in \{\textrm{nodes on \textrm{tree}}(a) \} \textrm{ and }
d\in \{\textrm{nodes on \textrm{tree}}(c)\}\,.
\end{equation}

We will show that we can choose the ranks of the nodes step by step, starting from those with the maximal height and then moving towards the root of the tree, such that the whole tree $\textrm{Q}$ is a ``good" quiver.
The key point is that once all the nodes with height $n$ are the roots of good sub-trees, we can make all the nodes at height $(n-1)$ roots of good sub-trees by choosing their ranks, namely we can make all the nodes of height $n$ to have non-negative excesses while preserving the non-increasing condition \eqref{eq:non-increasingFull}.

\begin{enumerate}
\item We start with the nodes with the maximal height, $\textrm{hei}(Q)$. 
We can choose arbitrary positive ranks for them, and this satisfies the good sub-tree conditions trivially. 

\item Assume that we have specified the ranks of all the nodes at height $\geq n$ ($n\in\{1,2,\dots,\textrm{hei(Q)}\}$), such that all the nodes at height $n$ are roots of good sub-trees.
We now show that we can choose the ranks for the nodes with height $(n-1)$, such that these new nodes are also roots of good sub-trees.

Consider a node $a\in \textrm{Q}_0\cup\{\mathtt{f}\}$ at height $(n-1)$, we denote the set of its successors by
\begin{equation}
\mathfrak{s}(a)=\{c_1,c_2,\dots,c_m\}\,,
\end{equation}
where $m\in\mathbb{N}$ is finite (possibly zero).
Each $c_i$ is the root of a good sub-tree and has the excess
\begin{equation}
\mathfrak{b}^{(c_i)}
=-2N^{(c_i)}+N^{(a)}+\sum_{d\in \mathfrak{s}(c_i)}N^{(d)}\,,
\end{equation}
which is positively related to $N^{(a)}$. 
Hence, we can always set $N^{(a)}$ large enough such that
\begin{equation}
N^{(a)}\geq N^{(c_i)}
\qquad\textrm{and}\qquad \mathfrak{b}^{(c_i)}\geq 0 
\end{equation}
for all $i=1,2,\dots,m$.
Then, together with the condition that all the $c_i$ are roots of good sub-trees, we have made the node $a$ also the root of a good sub-tree.

We can repeat this for all the nodes at height $(n-1)$, and render all of them the roots of good sub-trees.

\item Repeating this procedure iteratively then reaches the flavor node $\mathtt{f}$: we can make $\mathtt{f}$ the root of a good sub-tree, which is the entire tree.

\end{enumerate}
We conclude that by choosing the ranks of the nodes height by height, moving from the maximal height to the flavor node (with minimal height $0$), we have generated a good tree-type quiver gauge theory while preserving the non-increasing condition \eqref{eq:non-increasingFull}. 
The tree-structure guarantees that the induction will always terminate.\footnote{Note that the tree structure guarantees that there is no arrow between nodes of the same height. 
}

Similarly, we can easily make a tree-type theory ``ugly" or ``bad": we can simply adjust the ranks of the nodes at height $(n-1)$ such that some of their successors have excess $-1$ or $\leq-2$. 

\subsection{Solving the vacuum equations}
\label{appssec:VacuumEq}

In this subsection we give the details for how to solve the vacuum equations \eqref{eq:varphi-sigma} -- \eqref{eq:sigma-XY}.

Let us first simplify these equations before we solve them. 
\begin{enumerate}
\item Given that $(\varphi^{(a)}\, , \, \sigma^{(a)})$ live in the Lie algebra $\mathfrak{g}^{(a)}$ of the gauge factor $U(N^{(a)})$, \eqref{eq:varphi-sigma} simply implies
\begin{equation}
\varphi^{(a)}\in \mathfrak{t}^{(a)}_{\mathbb{C}},\quad\sigma^{(a)}\in \mathfrak{t}^{(a)}\,.
\end{equation}
Below we will denote their components by:
\begin{equation}
    \begin{aligned}
        \varphi^{(a)}&=\text{diag}(\varphi^{(a)}_1,\varphi^{(a)}_2,\dots,\varphi^{(a)}_{N^{(a)}}) \quad\in \mathfrak{t}^{(a)}_{\mathbb{C}}\,,\\
        \sigma^{(a)}&=\text{diag}(\sigma^{(a)}_1,\sigma^{(a)}_2,\dots,\sigma^{(a)}_{N^{(a)}}) \quad\in \mathfrak{t}^{(a)}\,.
    \end{aligned}
\end{equation}
\item Let us choose $T^{(a)}_0\propto \mathbb{I}$, and then $T^{(a)}_n$ with $n=1,2,\dots (N^{(a)})^2-1$ span all the traceless $N^{(a)}\times N^{(a)}$ matrices; therefore \eqref{eq:vacuum1} can be equivalently written into the following two equations:
\begin{align}
\sum_{i=1}^{N_\mathtt{f}}\sum^{N^{(a)}}_{p=1}(\overline{\tilde{X}^{(a)}})^i_{\ p} (\tilde{X}^{(a)})^p_{\ i}-(\tilde{Y}^{(a)})^i_{\ p} (\overline{\tilde{Y}^{(a)}})^p_{\ i}&=-t^{(a)}_{\mathbb{R}}\,,\label{eq:vacuum_XY1}\\
\sum_{i=1}^{N_\mathtt{f}}\sum^{N^{(a)}}_{p,q=1}(\overline{\tilde{X}^{(a)}})^i_{\ p} (M^{(a)})^p_{\ q} (\tilde{X}^{(a)})^q_{\ i}-(\tilde{Y}^{(a)})^i_{\ p} (M^{(a)})^p_{\ q}(\overline{\tilde{Y}^{(a)}})^q_{\ i}&=0\,, 
\label{eq:vacuum_XYM2}
\end{align}
where $M^{(a)}$ is a traceless $N^{(a)}\times N^{(a)}$  matrix.
We can choose $M^{(a)}$ to take the following form:
\begin{equation}
M^{(a)}=\textrm{diag}(0,\dots,0, \stackunder{$1$}{$p$},0,\dots, 0,\stackunder{$-1$}{$q$},0,\dots,0)  \,,
\end{equation}
and then \eqref{eq:vacuum_XYM2} gives 
\begin{equation}\label{eq:vacuum_traceless}
\sum_{i=1}^{N_\mathtt{f}}(\overline{\tilde{X}^{(a)}})^i_{\ p} (\tilde{X}^{(a)})^p_{\ i}-(\tilde{Y}^{(a)})^i_{\ p} (\overline{\tilde{Y}^{(a)}})^p_{\ i}=\sum_{i=1}^{N_\mathtt{f}}(\overline{\tilde{X}^{(a)}})^i_{\ q} (\tilde{X}^{(a)})^q_{\ i}-(\tilde{Y}^{(a)})^i_{\ q} (\overline{\tilde{Y}^{(a)}})^q_{\ i}\,,
\end{equation}
for $p, q = 1,2,\dots, N^{(a)}$.
Plugging \eqref{eq:vacuum_traceless} into \eqref{eq:vacuum_XY1} then gives
\begin{equation}\label{eq:vacuum_XY4}
\sum_{i=1}^{N_\mathtt{f}}(\overline{\tilde{X}^{(a)}})^i_{\ p} (\tilde{X}^{(a)})^p_{\ i}-(\tilde{Y}^{(a)})^i_{\ p} (\overline{\tilde{Y}^{(a)}})^p_{\ i}=-\frac{t^{(a)}_\mathbb{R}}{N^{(a)}}\,, 
\end{equation}
for $p=1,2,\dots,N^{(a)}$.

\item Similarly, the equation \eqref{eq:vacuum2} is equivalent to 
\begin{equation}\label{eq:vacuum_XY5}
\sum_{i=1}^{N_\mathtt{f}}\sum^{N^{(a)}}_{p,q=1}(\tilde{Y}^{(a)})^i_{\ p} (\tilde{M}^{(a)})^p_{\ q} (\tilde{X}^{(a)})^q_{\ i}=0\,,
\end{equation}
where $\tilde{M}^{(a)}$ is any $N^{(a)}\times N^{(a)}$ matrix. 
It can be chosen as
\begin{equation}
\tilde{M}^{(a)}=\textrm{diag}(0,\dots,0, \stackunder{$1$}{$p$},0,\dots,0)  \,,  
\end{equation}
and then \eqref{eq:vacuum_XY5} gives
\begin{equation}\label{eq:vacuum_xy3}
\sum_{i=1}^{N_\mathtt{f}}(\tilde{Y}^{(a)})^i_{\ p}  (\tilde{X}^{(a)})^p_{\ i}=0\,,   
\end{equation}
for $p=1,2,\dots,N^{(a)}$.

\item For the matrix equations \eqref{eq:vacuum_XY_matrix} and \eqref{eq:sigma-XY}, their $(p,i)$ components are:
\begin{align}
(\varphi^{(a)}_p+m_i)(\tilde{X}^{(a)})^p_{\ i}=(\varphi^{(a)}_p+m_i)(\tilde{Y}^{(a)})^i_{\ p}&=0\,,
\label{eq:vacuum_XY2}\\
\sigma^{(a)}_p(\tilde{X}^{(a)})^p_{\ i}=\sigma^{(a)}_p(\tilde{Y}^{(a)})^i_{\ p}&=0\,,\label{eq:vacuum_sigma_XY1}
\end{align}
where $p=1,2,\dots,N^{(a)}$, $i=1,2,\dots,N_{\mathtt{f}}$ are not summed.

\end{enumerate}

The set of vacuum equations is then \eqref{eq:vacuum_XY4}, \eqref{eq:vacuum_xy3}, \eqref{eq:vacuum_XY2} and \eqref{eq:vacuum_sigma_XY1}, which we can now solve step by step.

\begin{enumerate}
\item Let us first consider \eqref{eq:vacuum_XY2}. 
For each $p=1,2,\dots,N^{(a)}$, if $\varphi^{(a)}_p$ is not equal to the minus of any $m_i$ ($i=1,2,\dots,N_{\mathtt{f}}$), then \eqref{eq:vacuum_XY2} would require
\begin{equation}
(\tilde{X}^{(a)})^p_{\ i}=(\tilde{Y}^{(a)})^i_{\ p}=0\,,\qquad\textrm{for }\,  \forall \, i\,,
\end{equation}
which contradicts with \eqref{eq:vacuum_XY4}; 
therefore, we need $\varphi^{(a)}_p=-m_{\bar{i}(p)}$ for some 
$\bar{i}(p)\in\{1,2,\dots,N_{\mathtt{f}}\}$, so that $(\tilde{X}^{(a)})^p_{\ \bar{i}(p)}$ and $(\tilde{Y}^{(a)})^p_{\ \bar{i}(p)}$ can be non-zero.  
Namely, for each $p=1,2,\dots,N^{(a)}$, we have the following solution of \eqref{eq:vacuum_XY2}:
\begin{equation}\label{eq:solution1}
\varphi^{(a)}_p=-m_{\bar{i}(p)}\,,\qquad  
(\tilde{X}^{(a)})^p_{\ i}=\alpha^{p}_{\ \bar{i}(p)}\, \delta_{i,\bar{i}(p)}, \qquad 
(\tilde{Y}^{(a)})^i_{\ p}=\beta^{\bar{i}(p)}_{\ p} \delta_{i,\bar{i}(p)}\,,
\end{equation}
where $\alpha^{p}_{\ \bar{i}(p)},\beta^{\bar{i}(p)}_{\ p}$ are to be fixed using the remaining equations.

\item Equation \eqref{eq:vacuum_xy3} is equivalent to
\begin{equation}\label{eq:alphabeta}
\alpha^{p}_{\ \bar{i}(p)}\beta^{\bar{i}(p)}_{\ p}=0\,.
\end{equation}

\item Then plugging \eqref{eq:solution1} and \eqref{eq:alphabeta} into \eqref{eq:vacuum_XY4} gives the equations on $\alpha^{p}_{\ \bar{i}(p)},\beta^{\bar{i}(p)}_{\ p}$, and we see that the solutions are actually independent of $p$ and they  
depend on the sign of $t_{\mathbb{R}}$:
\begin{equation}
\begin{aligned}
\alpha^{p}_{\ \bar{i}(p)}=0\,, \qquad 
\beta^{\bar{i}(p)}_{\ p}=\sqrt{\frac{t^{(a)}_\mathbb{R}}{N^{(a)}}}\,,&\qquad \qquad t_{\mathbb{R}}> 0 \,, \\
\alpha^{p}_{\ \bar{i}(p)}=\sqrt{\frac{-t^{(a)}_\mathbb{R}}{N^{(a)}}}\,, \qquad 
\beta^{\bar{i}(p)}_{\ p}=0\,,&\qquad \qquad t_{\mathbb{R}}< 0\,.
\end{aligned}
\end{equation}
Without loss of generality, we choose 
\begin{equation}\label{eq:tRnegative}
t_{\mathbb{R}}< 0\,,    
\end{equation}
namely
\begin{equation}\label{eq:XYsol}
\begin{aligned}
\varphi^{(a)}_p=-m_{\bar{i}(p)}
\,, \qquad
(\tilde{X}^{(a)})^p_{\ i}=\sqrt{\frac{-t^{(a)}_\mathbb{R}}{N^{(a)}}}\delta_{i,\bar{i}(p)}
\,, \qquad 
(\tilde{Y}^{(a)})^i_{\ p}=0\,. 
\end{aligned}
\end{equation}
\item Now we show that when $p$ runs over $\{1,2,\dots,N^{(a)}\}$, all the $\bar{i}(p)$'s  are different.
Plugging \eqref{eq:XYsol} into \eqref{eq:vacuum_XYM2} gives
\begin{equation}
\sum_{p,q=1}^{N^{(a)}}\sum_{i=1}^{N_\mathtt{f}}\delta_{i,\bar{i}(p)} (M^{(a)})^p_{\ q} \delta_{i,\bar{i}(q)} =0\,,    
\end{equation}
where $M^{(a)}$ is traceless. 
Choosing $M^{(a)}$ to be the off-diagonal matrix units  $M^{(a)}=E^p_{\ q}$ ($p\neq q$), which is a matrix whose $(p,q)$-component is 1 with all other components vanishing, we obtain
\begin{equation}
\sum_{i=1}^{N_\mathtt{f}}
\delta_{i,\bar{i}(p)}\delta_{i,\bar{i}(q)} =0\,,   
\qquad \forall \ p\neq q\,.
\end{equation}
This requires that  
\begin{equation}\label{eq:baripq}
p\neq q 
\qquad \Longrightarrow \qquad
\bar{i}(p) \neq \bar{i}(q) \,,
\end{equation}
which necessarily requires
\begin{equation}\label{eq:NaNf}
N^{(a)}\leq N_{\mathtt{f}}\,.  
\end{equation}

\item Since the matrix $\tilde{X}^{(a)}$ has maximal rank, the solution of \eqref{eq:vacuum_sigma_XY1} is simply:
\begin{equation}
\sigma^{(a)}=0\,.
\end{equation}

\item Finally, since the gauge group $U(N^{(a)})$ acts on $\tilde{X}^{(a)}$ as a row operation, the gauge transformations shuffle the set of $\bar{i}(p)$ with $p=1,2,\dots, N^{(a)}$, where $\bar{i}(p)$'s are all different. 

\end{enumerate}

\section{Examples: A- and E-type theories}
\label{appsec:ADE_examples}
In Sec.~\ref{sec:vortex_tree_typeQGT} and \ref{sec:QYasMonopoleAlgebra} we have studied the D-type quiver gauge theory as an example to illustrate our proposals. 
In this section, we present more examples of finite ADE-type Dynkin quiver gauge theories.

\subsection{A-type}\label{appssec:Atype_QY}

In Sec.~\ref{ssssec:Atype_vortex}, we have studied the vortex Hilbert space $\mathcal{H}_\nu$ and the action of the monopole operators for the A-type quiver gauge theory.
For completeness, we will also give the corresponding (truncated shifted) quiver Yangian description.

The triple quiver for the quiver Yangian is shown in Fig.~\ref{fig:A_qui_QY}, where for clarity we have drawn many copies of the framing node $\infty$ which should be identified. 
Here the symbol $\{\rho\}$ beside a multi-arrow indicates the multiplicity $\rho$ of the corresponding arrows.
Note that this quiver is equivalent to the handsaw quiver (Fig.~\ref{fig:handsaw}), in the sense of Sec.~\ref{ssec:VQM_and_triple}.
\begin{figure}[h]
\centering
\begin{tikzpicture}
[->,auto=right, node distance=2cm,
			shorten >=1pt, semithick]
\node (v1) at (-3,0.2)[circle,draw] {1};
\node (v2) at (-0.5,0.2)[circle,draw] {2};
\node (e) at (1.75,0.15) {$\dots$};
\node (vl) at (4,0.2)[circle,draw] {$L$};

\draw (v2) edge [bend right=15] node {{\tiny $A_1$}} (v1);
\draw (v1) edge [bend right=15] node {{\tiny $B_1$}} (v2);
\draw (v1) edge [in=110,out=70,loop] node {{\tiny $C_1$}} (v1);
			
\draw (v2) edge [bend right=15] node {{\tiny $B_2$}} (e);
\draw (e) edge [bend right=15] node {{\tiny $A_2$}} (v2);
\draw (v2) edge [in=110,out=70,loop] node {{\tiny $C_2$}} (v2);

\draw (e) edge [bend right=15] node {{\tiny $B_{L-1}$}} (vl);
\draw (vl) edge [bend right=15] node {{\tiny $A_{L-1}$}} (e);
\draw (vl) edge [in=110,out=70,loop] node {{\tiny $C_{L}$}} (vl);
			
\node (inf1) at (-3,-1.3) {$\infty$};
\node (inf2) at (-0.5,-1.3) {$\infty$};
\node (infl) at (4,-1.3) {$\infty$};

\draw[->>] (inf1) edge node[left] {{\scriptsize $\{\rho_1\}$}} (v1);
\draw[->>] (v1) edge node[pos=0.4,below] {{\scriptsize $\{\rho_2\}$}} (inf2);

\draw[->>] (inf2) edge node[pos=0.5,left] {{\scriptsize $\{\rho_2\}$}} (v2);
\draw[->>] (v2) edge node[pos=0.4,below] {{\scriptsize $\{\rho_3\}$}} (0.9,-0.7);

\node (infe) at (1.75,-0.8) {$\dots$};

\draw[->>] (infl) edge node[left] {{\scriptsize $\{\rho_L\}$}} (vl);
\draw[->>] (2.4,-0.6) -- (infl);
\node (rholm) at (2.7,-1.05) {{\scriptsize $\{\rho_L\}$}};

\node (infl1) at (6,-1.3) {$\infty$};
\draw[->>] (vl) edge  (infl1);
\node (rholp) at (5.4,-0.3) {{\scriptsize $\{\rho_{L+1}\}$}};

\end{tikzpicture}
\caption{Triple quiver with framing for the A-type shifted quiver Yangian.}
\label{fig:A_qui_QY}
\end{figure}

The associated superpotential is given by (see \eqref{eq:tree_potential}):
{\small
\begin{equation}\label{eq:A_triple_potential}
\widehat{W}=\sum_{a\geq1}^{L-1}\mathrm{Tr}\,(A_aC_aB_a-C_{a+1}A_aB_a)\,.
\end{equation}}
The assignment of the equivariant weights is:
\begin{align}
h(A_a)=-\epsilon\,, \qquad h(B_a)=0\,,\qquad h(C_a)=\epsilon\,,
\end{align}
where we have applied \eqref{eq:weight_assign_C} and \eqref{eq:weight_assign_AB} and used the parameter identification $\hbar=\epsilon$, and the framing of the quiver is given as follows. For each node $a=1,2,\dots,L$, there are:
\begin{itemize}
\item $\rho_a=(N^{(a)}-N^{(a-1)})$ in-coming arrows with weights
\begin{equation}\{m_p|p\in \mathcal{I}^{(a)}/\mathcal{I}^{(a-1)} \}=\{m_{N^{(a-1)}+1},m_{N^{(a-1)}+2},\dots,m_{N^{(a)}}\}\,,\end{equation} where we define $\mathcal{I}^{(0)}=\emptyset$ and $N^{(0)}=0$.
		
\item $\rho_{a+1}=(N^{(a+1)}-N^{(a)})$ out-going arrows with weights
\begin{equation}\{\epsilon-m_p|p\in \mathcal{I}^{(a+1)}/\mathcal{I}^{(a)}\}=\{\epsilon-m_{N^{(a)}+1},\epsilon-m_{N^{(a)}+2},\dots,\epsilon-m_{N^{(a+1)}}\}\,,\end{equation} and we set $\mathcal{I}^{(L+1)}=\mathcal{I}_{\mathtt{f}}$.
\end{itemize}
The superpotential involving the framing arrows are encoded in the weight assignment above. 
This framing determines a representation for the shifted quiver Yangian which is isomorphic to the vortex Hilbert space $\mathcal{H}_\nu$ of the 3D theory. 
The charge functions of the ground state are:
\begin{equation}\label{eq:A_groundstate}
^\sharp\psi^{(a)}_0(z)=\frac{\prod_{q=N^{(a)}+1}^{N^{(a+1)}}(z-m_q+\epsilon)}{\prod_{p=N^{(a-1)}+1}^{N^{(a)}}(z-m_p)}\,,\qquad a\in\{1,2,\dots,L\}\,.
\end{equation}
As before, for a state we let $k^{(a)}_{p}$ ($a=1,2,\dots,L$ and $p=1,2,\dots,N^{(a)}$) be the number of atoms of color $a$ that start with the framing arrow of weight $m_{\mathcal{I}^{(a)}_{p}}=m_{p}$, and we denote the state by $|k\rangle$.
	
We can now prove by induction that this representation is isomorphic to $\mathcal{H}_\nu$. 
For a state $|k\rangle$ with $\{\vec{k}^{(a)}\}$ satisfying the condition \eqref{eq:tree-codi}, to determine whether or not $k^{(a)}_{p}$ is increasable, we have the following adding rules:
\begin{itemize}
\item If $p\leq N^{(a)}$ then
\begin{equation}
k^{(a)}_p \ \text{is increasable}  \ \Longleftrightarrow k^{(a)}_{p}<k^{(a-1)}_{p}\,.
\end{equation}
		
\item If $p> N^{(a-1)}$ then $k^{(a)}_{p}$ is always increasable.
\end{itemize}
These rules can be directly read off from the following charge function $\psi^{(\alpha)}(z)|k\rangle=\Psi^{(\alpha)}_{k}(z) |k\rangle$, where using \eqref{eq:A_groundstate} and \eqref{eq:tree_bond} we have:
\begin{equation}
\Psi^{(a)}_{k}(z)=\frac{\prod_{q=1}^{N^{(a-1)}}(z-m_q-k^{(a-1)}_{q}\epsilon)\prod_{q=1}^{N^{(a+1)}}(z-m_q-(k^{(a+1)}_{q}-1)\epsilon)}{\prod_{p=1}^{N^{(a)}}(z-m_{p}-(k^{(a)}_{p}-1)\epsilon)(z-m_{p}-k^{(a)}_{p}\epsilon)}\,.
\end{equation}
With the adding rules, it's clear that by applying $e$ operators successively on the ground state $|0\rangle$, we obtain a representation of the quiver Yangian which reproduces the vortex Hilbert space $\mathcal{H}_\nu$. 
It is also straightforward to check that there exists a correspondence between the quiver Yangian and the Coulomb branch algebra. 
The correspondence is given by \eqref{eq:tree_QYmono1} and \eqref{eq:tree_QYmono2} with $\mathfrak{p}(a)=a+1$ and $\mathfrak{s}(a)=\{a-1\}$.

The character of the vortex Hilbert space $\mathcal{H}_\nu$ in the A-type theory can be easily obtained (see \eqref{eq:character2}):
\begin{equation}
\mathcal{Z}_{\textrm{A-type}} =\prod_{a=1}^L\left(\frac{1}{(1-x_a)(1-x_a x_{a+1})\dots(1-x_a x_{a+1}\dots x_L)}\right)^{\rho_a}\,,
\end{equation}
with $\rho_a$ ($a=1,2,\dots,L$) defined in \eqref{eq:rhoDef}.

\subsection{E-type}\label{appssec:Etype}
\subsubsection{Vortices and action of monopoles}
As one last example, let us consider the 3D $\mathcal{N}=4$ quiver gauge theory specified by the quiver shown in Fig.~\ref{E_qui}. 
When $L=6,7,8$ this quiver corresponds to the Dynkin diagrams of $E_6,E_7,E_8$.
\begin{figure}[h]
\centering
\begin{tikzpicture}[scale=0.5]
\draw (-7,0.2) circle(0.5);
\draw[-] (-6.5,0.2) to (-5.5,0.2);
\node (a1) at (-7,0.2) {$1$};
\draw (-3,2) circle(0.5);
\node (a3) at (-3,2) {$3$}; 
\draw[-] (-3,1.5) to (-3,0.7); 
\draw (-5,0.2) circle(0.5);
\node (a2) at (-5,0.2) {$2$}; 
\draw[-] (-4.5,0.2) to (-3.5,0.2); 
\draw (-3,0.2) circle(0.5);
\node (b) at (-3,0.2) {$4$}; 
\draw[-] (-2.5,0.2) to (-2,0.2); 
\node (c) at (-1.5,0.2) {$\dots$};
\draw[-] (-1,0.2) to (-0.5,0.2); 
\draw (0,0.2) circle(0.5);
\node (d) at (0,0.2) {$L$}; 
\draw[-] (0.5,0.2) to (1.5,0.2);
\draw (1.5,-0.3) rectangle (2.5,0.7);
\node (e) at (2,0.2) {$\mathtt{f}$};
\end{tikzpicture}
\caption{E-type quiver gauge theory.}
\label{E_qui}
\end{figure}

Similar to the D-type theory, in the E-type theory the vacua are described by two sequences of nested sets:
\begin{equation}
\begin{aligned}
&\mathcal{I}^{(3)}\subset\mathcal{I}^{(4)}\subset \dots\subset \mathcal{I}^{(L)}\subset\{1,2,\dots,N_{\mathtt{f}}\}\,,\\
&\mathcal{I}^{(1)}\subset\mathcal{I}^{(2)}\subset\mathcal{I}^{(4)}\subset \dots\subset \mathcal{I}^{(L)}\subset\{1,2,\dots,N_{\mathtt{f}}\}\,.\\
\end{aligned}
\end{equation}
The possible overlaps among different index sets now become more complicated. 
We need to distinguish among the following three cases:
\begin{enumerate}
\item $\mathcal{I}^{(2)}\cap\mathcal{I}^{(3)}=\emptyset\,;$
\item $\mathcal{I}^{(2)}\cap\mathcal{I}^{(3)}\neq\emptyset\,$ but $\mathcal{I}^{(1)}\cap\mathcal{I}^{(3)}=\emptyset\,;$
\item $\mathcal{I}^{(2)}\cap\mathcal{I}^{(3)}\neq\emptyset\,$ and $\mathcal{I}^{(1)}\cap\mathcal{I}^{(3)}\neq\emptyset\,$.
\end{enumerate}

Choosing a certain vacuum $\nu_{\{\mathcal{I}^{(a)}\}}$, let us next consider the Hilbert space $\mathcal{H}_\nu$ of vortices that approach $\nu_{\{\mathcal{I}^{(a)}\}}$ at spatial infinity. 
Applying the results from Sec.~\ref{sssec:vortex_Hilbert_space}, we find each state $|k)\in\mathcal{H}_\nu$ is specified by $L$ decompositions $\{\vec{k}^{(1)},\vec{k}^{(2)},\dots,\vec{k}^{(L)}\}$, with $\vec{k}^{(a)}$ of dimension $N^{(a)}$,
and they obey $k^{(a)}_p\geq k^{(\mathfrak{p}(a))}_{\tilde{p}}$ for $p=1,2,\dots,N^{(a)}$, $\tilde{p}\in\{1,2,\dots,N^{(\mathfrak{p}(a))}\}$ such that $\mathcal{I}^{(a)}_p=\mathcal{I}^{(\mathfrak{p}(a))}_{\tilde{p}}$. 
As before, the vortex number of the state $|k)$ is 
\begin{equation}
 \nn=\{\sum_{p=1}^{N^{(1)}}k^{(1)}_p\,,\sum_{p=1}^{N^{(2)}}k^{(2)}_p\,,\dots\,,\sum_{p=1}^{N^{(L)}}k^{(L)}_p\}\in \pi_1(G)\,.
\end{equation}

The action of the various monopole operators on the state $|k)$ can be obtained using the results from Sec.~\ref{sssec:monopole_action}. 
For the raising operators, one finds
\begin{equation}
    \hat{v}^{(a)+}_p|k)=\frac{Q^{(a+1)}(\hat{\varphi}^{(a)}_p)}{\prod_{q\neq p}^{N^{(a)}}(\hat{\varphi}^{(a)}_p-\hat{\varphi}^{(a)}_q)}|k+\delta^{(a)}_p)\,,\quad p=1,2,\dots,N^{(a)}\,,\quad a=1,3,4,\dots,L\,,
\end{equation}
and in particular for $a=2$:
\begin{equation}
    \hat{v}^{(2)+}_p|k)=\frac{Q^{(4)}(\hat{\varphi}^{(2)}_p)}{\prod_{q\neq p}^{N^{(2)}}(\hat{\varphi}^{(2)}_p-\hat{\varphi}^{(2)}_q)}|k+\delta^{(2)}_p)\,,\quad p=1,2,\dots,N^{(2)}.
\end{equation}
On the other hand, for the lowering operators one finds
\begin{equation}
     \hat{v}^{(a)-}_p|k)=\frac{(-1)^{N^{(a-1)}}Q^{(a-1)}(\hat{\varphi}^{(a)}_p)}{\prod_{q\neq p}^{N^{(a)}}(\hat{\varphi}^{(a)}_q-\hat{\varphi}^{(a)}_p)}|k-\delta^{(a)}_p)\,,\quad p=1,2,\dots,N^{(a)}\,,\quad a=2,5,6,\dots,L\,,
\end{equation}
and in particular for $a=1,3,4$:
\begin{equation}
\begin{aligned}
\hat{v}^{(1)-}_p|k)&=\frac{1}{\prod_{q\neq p}^{N^{(1)}}(\hat{\varphi}^{(1)}_q-\hat{\varphi}^{(1)}_p)}|k-\delta^{(1)}_p)\,,\quad p=1,2,\dots,N^{(1)}\,,\\
\hat{v}^{(3)-}_p|k)&=\frac{1}{\prod_{q\neq p}^{N^{(3)}}(\hat{\varphi}^{(3)}_q-\hat{\varphi}^{(3)}_p)}|k-\delta^{(3)}_p)\,,\quad p=1,2,\dots,N^{(3)}\,,\\
\hat{v}^{(4)-}_p|k)&=\frac{(-1)^{N^{(2)}+N^{(3)}}Q^{(2)}(\hat{\varphi}^{(4)}_p)Q^{(3)}(\hat{\varphi}^{(4)}_p)}{\prod_{q\neq p}^{N^{(4)}}(\hat{\varphi}^{(4)}_q-\hat{\varphi}^{(4)}_p)}|k-\delta^{(4)}_p)\,,\quad p=1,2,\dots,N^{(4)}\,.
\end{aligned}
\end{equation}

\subsubsection{The corresponding quiver Yangian}

The corresponding triple quiver for the shifted quiver Yangian is shown in Fig.~\ref{E_qui_QY}. 
The potential is given by (see \eqref{eq:tree_potential}):
{\small
\begin{equation}\label{eq:E_triple_potential}
\widehat{W}=\mathrm{Tr}\,(A_1C_1B_1-C_2A_1B_1)+\mathrm{Tr}\,(A_2C_2B_2-C_4A_2B_2)+\sum_{a\geq3}^{L-1}\mathrm{Tr}\,(A_aC_aB_a-C_{a+1}A_aB_a)\,,
\end{equation}}
and the weight assignment is (see \eqref{eq:weight_assign_C} and \eqref{eq:weight_assign_AB}):
\begin{equation}
h(A_a)=-\epsilon\,,\qquad h(B_a)=0\,,\qquad h(C_a)=\epsilon\,,
\end{equation}
where we have used the parameter identification $\hbar=\epsilon$.

For the vortex representation $\mathcal{R}_\nu$ in the E-type theory, we choose the framing as follows (see Sec.~\ref{sssec:mono_as_QY}):
\begin{enumerate}
\item For node $a=1,3$, there are $N^{(a)}$ in-coming arrows with weights
\begin{equation}\{m_p|p\in\mathcal{I}^{(a)}\}\,,\end{equation}
and $N^{(a+1)}-N^{(a)}$ out-going arrows with weights
\begin{equation}
\{\epsilon-m_p|p\in\mathcal{I}^{(a+1)}\backslash\mathcal{I}^{(a)} \}\,.
\end{equation}

\item For node $2$, there are $N^{(2)}-N^{(1)}$ in-coming arrows with weights \begin{equation}\{m_p|p\in\mathcal{I}^{(2)}\backslash\mathcal{I}^{(1)}\}\,,\end{equation} and $N^{(4)}-N^{(2)}$ out-going arrows with weights
\begin{equation}\{\epsilon-m_p|p\in \mathcal{I}^{(4)}\backslash\mathcal{I}^{(2)}\}\,.\end{equation}

\item For node $4$, there are $|\mathcal{I}^{(4)}/(\mathcal{I}^{(2)}\cup \mathcal{I}^{(3)})|$ in-coming arrows with weights
\begin{equation}\{m_p|p\in \mathcal{I}^{(4)}\backslash(\mathcal{I}^{(2)}\cup \mathcal{I}^{(3)})\}\,,\end{equation}
and $N^{(5)}-N^{(4)}+|\mathcal{I}^{(2)}\cap\mathcal{I}^{(3)}|$ out-going arrows with weights
\begin{equation}\{\epsilon-m_p|p\in \mathcal{I}^{(5)}\backslash\mathcal{I}^{(4)} \}\bigcup\{-m_p|p\in \mathcal{I}^{(2)}\cap\mathcal{I}^{(3)}\}\,,\end{equation}where we have used $(\mathcal{I}^{(2)}\sqcup\mathcal{I}^{(3)})/\mathcal{I}^{(4)}=\mathcal{I}^{(2)}\cap\mathcal{I}^{(3)}\,$.
    
\item Finally, for node $5\leq a\leq L$, there are $N^{(a)}-N^{(a-1)}$ in-coming arrows with weights \begin{equation}\{m_p|p\in \mathcal{I}^{(a)}\backslash\mathcal{I}^{(a-1)}\}\,,\end{equation}
and $N^{(a+1)}-N^{(a)}$ out-going arrows with weights
\begin{equation}\{\eps-m_p|p\in \mathcal{I}^{(a+1)}\backslash\mathcal{I}^{(a)}\}\,.\end{equation}
\end{enumerate}
\begin{figure}[h]
\centering
\begin{tikzpicture}
[scale=0.8,->,auto=right, node distance=2cm,
shorten >=1pt, semithick]
\node (v1) at (-7,0.2)[circle,draw] {1};
\node (v2) at (-5,0.2)[circle,draw] {2};
\node (v3) at (-3,2)[circle,draw] {3};
\node (v4) at (-3,0.2)[circle,draw] {4};
\node (v5) at (-0.5,0.2)[circle,draw] {5};
\node (e) at (1.4,0.15) {$\dots$};
\node (vl) at (4,0.2)[circle,draw] {$L$};

\draw (v1) edge [bend right=15] node {{\tiny $B_1$}} (v2);
\draw (v2) edge [bend right=15] node {{\tiny $A_1$}} (v1);
\draw (v1) edge [loop above] node {{\tiny $C_1$}} (v1);

\draw (v2) edge [bend right=15] node {{\tiny $B_2$}} (v4);
\draw (v4) edge [bend right=15] node {{\tiny $A_2$}} (v2);
\draw (v2) edge [loop above] node {{\tiny $C_2$}} (v2);

\draw (v4) edge [bend right=15]node {{\tiny $B_4$}} (v5);
\draw (v5) edge [bend right=15] node {{\tiny $A_4$}} (v4);
\draw (v4) edge [loop below] node {{\tiny $C_4$}} (v4);

\draw (v3) edge [bend right=15] node {{\tiny $B_3$}} (v4);
\draw (v4) edge [bend right=15] node {{\tiny $A_3$}} (v3);
\draw (v3) edge [loop above] node {{\tiny $C_3$}} (v3);

\draw (v5) edge [bend right=15] node {{\tiny $B_5$}} (e);
\draw (e) edge [bend right=15] node {{\tiny $A_5$}} (v5);
\draw (v5) edge [loop above] node {{\tiny $C_5$}} (v5);

\draw (e) edge [bend right=15] node {{\tiny $B_{L-1}$}} (vl);
\draw (vl) edge [bend right=15] node {{\tiny $A_{L-1}$}} (e);
\draw (vl) edge [loop above] node {{\tiny $C_{L}$}} (vl);

\end{tikzpicture}
\caption{Triple quiver for the E-type shifted quiver Yangian.}
\label{E_qui_QY}
\end{figure}

The framing gives the following ground state charge functions:
\begin{align}
^\sharp\psi^{(1)}_0(z)&=\frac{\prod_{p\in\mathcal{I}^{(2)}\backslash\mathcal{I}^{(1)}}(z-m_p+\epsilon)}{\prod_{p\in\mathcal{I}^{(1)}}(z-m_p)}\,,\\
^\sharp\psi^{(2)}_0(z)&=\frac{\prod_{p\in \mathcal{I}^{(4)}\backslash\mathcal{I}^{(2)}}(z-m_p+\epsilon)}{\prod_{p\in \mathcal{I}^{(2)}\backslash\mathcal{I}^{(1)}}(z-m_p)}\,,\\
^\sharp\psi^{(3)}_0(z)&=\frac{\prod_{p\in\mathcal{I}^{(4)}\backslash\mathcal{I}^{(3)}}(z-m_p+\epsilon)}{\prod_{p\in\mathcal{I}^{(3)}}(z-m_p)}\,,\\
^\sharp\psi^{(4)}_0(z)&=\frac{\prod_{p\in\mathcal{I}^{(5)}\backslash\mathcal{I}^{(4)}}(z-m_p+\epsilon)\prod_{p\in \mathcal{I}^{(2)}\cap \mathcal{I}^{(3)}}(z-m_p)}{\prod_{p\in \mathcal{I}^{(4)}\backslash(\mathcal{I}^{(2)}\cup \mathcal{I}^{(3)})}(z-m_p)}\,,\\
^\sharp\psi^{(a)}_0(z)&=\frac{\prod_{p\in\mathcal{I}^{(a+1)}\backslash\mathcal{I}^{(a)}}(z-m_p+\epsilon)}{\prod_{p\in \mathcal{I}^{(a)}\backslash\mathcal{I}^{(a-1)}}(z-m_p)}\,,\quad 5\leq a\leq L\,.
\end{align}
It is then straightforward to construct the entire representation $\mathcal{R}_\nu$, which reproduces the vortex Hilbert space $\mathcal{H}_\nu$, similar to earlier examples in Sec.~\ref{sec:QYasMonopoleAlgebra}.

The framing contains the complete information of the vortex quantum mechanics $\textrm{QM}(\nn,\nu)$, see Sec.~\ref{ssec:VQM_and_triple}. 
For the three different types of vacua, we have the following results: 
\begin{enumerate}
\item When $\mathcal{I}^{(2)}\cap\mathcal{I}^{(3)} =\emptyset$, the vortex quantum mechanics $\textrm{QM}(\nn,\nu)$ is described by the triple quiver shown in Fig.~\ref{1D_E}. 
For clarity we have represented the triple quiver by its skeleton, and focus on the framing. 
The ranks of the gauge groups are specified by the vortex number $\nn$, while those of the flavor groups are determined by the vacuum $\nu$:
\begin{equation}
\begin{aligned}
&\rho_1=|\mathcal{I}^{(1)}|\,,
&&\rho_2=|\mathcal{I}^{(2)}\backslash \mathcal{I}^{(1)}|\,,\\
&\rho_3=|\mathcal{I}^{(3)}|\,,
&&\rho_4=|\mathcal{I}^{(4)}\backslash(\mathcal{I}^{(2)}\cup\mathcal{I}^{(3)})|\,,
\end{aligned}
\end{equation}
and 
\begin{equation}\label{eq:E_rank_a5}
\rho_a=|\mathcal{I}^{(a)}\backslash \mathcal{I}^{(a-1)}|\,,\qquad a=5,6,\dots,L+1\,,
\end{equation}
with $\mathcal{I}^{(L+1)}:=\{1,2,\dots,N_\mathtt{f}\}$. 
The full superpotential of the theory is
\begin{equation}
\begin{aligned}
{}^\sharp \widehat{W}=&\widehat{W}+\mathrm{Tr}\,(q_1B_1B_2A_3\tilde{q}'_3+q_3B_3A_2\tilde{q}'_2\\
&+q_2A_1\tilde{q}_1-q_2B_2A_3\tilde{q}''_3+q_4A_2\tilde{q}_2-q_4A_3\tilde{q}_3+\sum_{a=5}^{L}q_aA_{a-1}\tilde{q}_{a-1})\,,
\end{aligned}
\end{equation}
with $\widehat{W}$ given in \eqref{eq:E_triple_potential}.

\item When $\mathcal{I}^{(2)}\cap\mathcal{I}^{(3)} \neq\emptyset$ but $\mathcal{I}^{(1)}\cap\mathcal{I}^{(3)} =\emptyset$, the vortex quantum mechanics $\textrm{QM}(\nn,\nu)$ is described by the triple quiver shown in Fig.~\ref{1D_E_2}. 
The ranks of the flavor groups are determined by the vacuum $\nu$:
\begin{equation}
\begin{aligned}
&\tilde{\rho}=|\mathcal{I}^{(2)}\cap\mathcal{I}^{(3)}|\,,
&&\rho_1=|\mathcal{I}^{(1)}|\,,\\
&\rho_2=|\mathcal{I}^{(2)}\backslash (\mathcal{I}^{(1)}\cup\mathcal{I}^{(3)})|\,,
&&\rho_3=|\mathcal{I}^{(3)}\backslash\mathcal{I}^{(2)}|\,,
&& \rho_4=|\mathcal{I}^{(4)}\backslash(\mathcal{I}^{(2)}\cup\mathcal{I}^{(3)})|\,,
\end{aligned}
\end{equation}
while $\rho_a$ for $a=5,6,\dots,L+1$ is the same as before (see \eqref{eq:E_rank_a5}). 
The full superpotential of the theory is
\begin{equation}
\begin{aligned}
{}^\sharp \widehat{W}=&\widehat{W}+\mathrm{Tr}\,(q_1B_1B_2A_3\tilde{q}'_3+q_3B_3A_2\tilde{q}'_2+q_2A_1\tilde{q}_1-s_2A_1\tilde{s}_1\\
&-q_2B_2A_3\tilde{q}''_3+q_4A_2\tilde{q}_2-q_4A_3\tilde{q}_3+s_2B_2\tilde{s}_4-s_3B_3\tilde{s}_4+\sum_{a=5}^{L}q_aA_{a-1}\tilde{q}_{a-1})\,,
\end{aligned}
\end{equation}
with $\widehat{W}$ given in \eqref{eq:E_triple_potential}.

\item Finally, when $\mathcal{I}^{(1)}\cap \mathcal{I}^{(3)}\neq \emptyset$, we have $\mathcal{I}^{(2)}/\mathcal{I}^{(1)}\subset \mathcal{I}^{(2)}\cap \mathcal{I}^{(3)}$. 
In fact, we have the following relation:
\begin{equation}
\mathcal{I}^{(2)}\cap \mathcal{I}^{(3)}=(\mathcal{I}^{(2)}/\mathcal{I}^{(1)})\cup(\mathcal{I}^{(1)}\cap \mathcal{I}^{(3)})\,.
\end{equation}
The vortex quantum mechanics $\textrm{QM}(\nn,\nu)$ in this case is described by the triple quiver shown in Fig.~\ref{1D_E_3}. 
The ranks of the flavor groups are
\begin{equation}
\begin{aligned}
&\tilde{\rho}=|\mathcal{I}^{(1)}\cap\mathcal{I}^{(3)}|\,,
&&\rho_1=|\mathcal{I}^{(1)}\backslash\mathcal{I}^{(3)}|\,,\\
&\rho_2=|\mathcal{I}^{(2)}\backslash \mathcal{I}^{(1)}|\,,
&&\rho_3=|\mathcal{I}^{(3)}\backslash\mathcal{I}^{(2)}|\,,
&&\rho_4=|\mathcal{I}^{(4)}\backslash(\mathcal{I}^{(2)}\cup\mathcal{I}^{(3)})|\,,\\
\end{aligned}
\end{equation}
and $\rho_a$ for $a=5,6,\dots,L+1$ is still given by \eqref{eq:E_rank_a5}. 
The full superpotential of the theory is
\begin{equation}
\begin{aligned}
{}^\sharp \widehat{W}=&\widehat{W}+\mathrm{Tr}\,(q_1B_1B_2A_3\tilde{q}'_3+q_3B_3A_2\tilde{q}'_2+s_1B_1B_2\tilde{s}_4-s_3B_3\tilde{s}_4\\
&+q_4A_2\tilde{q}_2-q_4A_3\tilde{q}_3+q_2A_1\tilde{q}_1-q_2B_2\tilde{q}'_4+q'_3B_3\tilde{q}'_4+\sum_{a=5}^{L}q_aA_{a-1}\tilde{q}_{a-1})\,,
\end{aligned}
\end{equation}
where, as before, $\widehat{W}$ is given in \eqref{eq:E_triple_potential}.
\begin{figure}[h!]
\centering
\begin{tikzpicture}
[->,auto=right, node distance=2cm,
shorten >=1pt, semithick,square/.style={
draw,
minimum width=width("#1"),
minimum height=width("#1")+2*\pgfshapeinnerysep,
node contents={#1}}]
\node (v1) at (-4,0)[circle,draw] {$\nn^{(1)}$};
\node (rho1) at (-4,2) [blue,square={$\rho_{1}$}];
\node (v2) at (-2,0)[circle,draw] {$\nn^{(2)}$};
\node (rho2) at (-2,-2) [blue,square={$\rho_{2}$}];
\node (v3) at (0,2)[circle,draw] {$\nn^{(3)}$};
\node (rho3) at (-2,2) [blue,square={$\rho_{3}$}];
\node (v4) at (0,0) [circle,draw] {$\nn^{(4)}$};
\node (rho4) at (0,-2) [blue,square={$\rho_{4}$}];
\node (v5) at (2,0) [circle,draw] {$\nn^{(5)}$};
\node (rho5) at (2,-2) [blue,square={$\rho_{5}$}];
\node (e) at (4,-0.05) {$\dots$};
\node (vl) at (6,0)[circle,draw] {$\nn^{(L)}$};
\node (rhol) at (6,-2) [blue,square={$\rho_{L}$}];
\node (rholp) at (8,-2) [blue,square={\tiny{$\rho_{L+1}$}}];

\draw[-] (v1) to (v2);
\draw[-] (v2) to (v4);
\draw[-] (v3) to (v4);
\draw[-] (v4) to (v5);
\draw[-] (v5) to (e);
\draw[-] (e) to (vl);

 \draw[blue] (rho1) edge  node [blue,pos=0.5,left,font=\tiny] {$q_1$} (v1);
\draw[->,blue] (v1) -- (rho2) node [pos=0.5,below,font=\tiny] {$\tilde{q}_1$};
\draw[blue] (rho2) edge node [blue,pos=0.56,left,font=\tiny] {$q_2$} (v2);
\draw[blue] (v2) edge node [blue,pos=0.56,left,font=\tiny] {$\tilde{q}'_2$} (rho3);
\draw[blue] (rho3) edge node [blue,pos=0.33,above,font=\tiny] {$q_3$} (v3);
\draw[blue] (v3) edge[bend right=20] node [blue,pos=0.56,above,font=\tiny] {$\tilde{q}'_3$} (rho1);
\draw[blue] (v3) edge[bend left] node [blue,pos=0.45,right,font=\tiny] {$\tilde{q}_3$} (rho4);
\draw[blue] (v3) edge node [blue,pos=0.33,right,font=\tiny] {$\tilde{q}''_3$} (rho2);
\draw[blue] (v2) edge node [blue,pos=0.56,below,font=\tiny] {$\tilde{q}_2$} (rho4);
\draw[blue] (rho4) edge node [blue,pos=0.56,right,font=\tiny] {$q_4$} (v4);
\draw[blue] (v4) edge node [blue,pos=0.56,below,font=\tiny] {$\tilde{q}_4$} (rho5);
\draw[blue] (rho5) edge node [blue,pos=0.56,right,font=\tiny] {$q_5$} (v5);
\draw[blue] (v5) edge node [blue,pos=0.33,below,font=\tiny] {$\tilde{q}_5$} (3,-1);

\draw[->,blue] (5,-1) -- (rhol) node [blue,pos=0.1,below,font=\tiny] {$\tilde{q}_{L-1}$};
\draw[blue] (rhol) edge node [blue,pos=0.56,right,font=\tiny] {$q_L$} (vl);
\draw[blue] (vl) edge node [blue,pos=0.33,below,font=\tiny] {$\tilde{q}_L$} (rholp);

\end{tikzpicture}
\caption{Vortex quantum mechanics $\textrm{QM}(\nn,\nu)$ for the E-type theory with $\mathcal{I}^{(2)}\cap\mathcal{I}^{(3)} =\emptyset\,$. For clarity, the triple quiver is represented by its skeleton.}
\label{1D_E}
\end{figure}

 \begin{figure}[h]
\centering
\begin{tikzpicture}
[->,auto=right, node distance=2cm,
shorten >=1pt, semithick,square/.style={
draw,
minimum width=width("#1"),
minimum height=width("#1")+2*\pgfshapeinnerysep,
node contents={#1}}]
\node (v1) at (-4,0)[circle,draw] {$\nn^{(1)}$};
\node (rho1) at (-4,2) [blue,square={$\rho_{1}$}];
\node (v2) at (-2,0)[circle,draw] {$\nn^{(2)}$};
\node (rho2) at (-2,-2) [blue,square={$\rho_{2}$}];
\node (v3) at (0,2)[circle,draw] {$\nn^{(3)}$};
\node (rho3) at (-2,2) [blue,square={$\rho_{3}$}];
\node (v4) at (0,0) [circle,draw] {$\nn^{(4)}$};
\node (rho4) at (0,-2) [blue,square={$\rho_{4}$}];
\node (v5) at (2,0) [circle,draw] {$\nn^{(5)}$};
\node (rho5) at (2,-2) [blue,square={$\rho_{5}$}];
\node (e) at (4,-0.05) {$\dots$};
\node (vl) at (6,0)[circle,draw] {$\nn^{(L)}$};
\node (rhol) at (6,-2) [blue,square={$\rho_{L}$}];
\node (rholp) at (8,-2) [blue,square={\tiny{$\rho_{L+1}$}}];
\node (rhot) at (-1.3,1) [red,square={\tiny{$\tilde{\rho}$}}];

\draw[-] (v1) to (v2);
\draw[-] (v2) to (v4);
\draw[-] (v3) to (v4);
\draw[-] (v4) to (v5);
\draw[-] (v5) to (e);
\draw[-] (e) to (vl);

 \draw[blue] (rho1) edge  node [blue,pos=0.5,left,font=\tiny] {$q_1$} (v1);
\draw[->,blue] (v1) -- (rho2) node [pos=0.5,below,font=\tiny] {$\tilde{q}_1$};
\draw[blue] (rho2) edge node [blue,pos=0.56,left,font=\tiny] {$q_2$} (v2);
\draw[blue] (v2) edge node [blue,pos=0.56,left,font=\tiny] {$\tilde{q}'_2$} (rho3);
\draw[blue] (rho3) edge node [blue,pos=0.33,above,font=\tiny] {$q_3$} (v3);
\draw[blue] (v3) edge[bend right=20] node [blue,pos=0.56,above,font=\tiny] {$\tilde{q}'_3$} (rho1);
\draw[blue] (v3) edge[bend left] node [blue,pos=0.45,right,font=\tiny] {$\tilde{q}_3$} (rho4);
\draw[blue] (v3) edge [bend left=15] node [blue,pos=0.6,below,font=\tiny] {$\tilde{q}''_3$} (rho2);
\draw[blue] (v2) edge node [blue,pos=0.56,below,font=\tiny] {$\tilde{q}_2$} (rho4);
\draw[blue] (rho4) edge node [blue,pos=0.56,right,font=\tiny] {$q_4$} (v4);
\draw[blue] (v4) edge node [blue,pos=0.56,below,font=\tiny] {$\tilde{q}_4$} (rho5);
\draw[blue] (rho5) edge node [blue,pos=0.56,right,font=\tiny] {$q_5$} (v5);
\draw[blue] (v5) edge node [blue,pos=0.33,below,font=\tiny] {$\tilde{q}_5$} (3,-1);

\draw[->,blue] (5,-1) -- (rhol) node [blue,pos=0.1,below,font=\tiny] {$\tilde{q}_{L-1}$};
\draw[blue] (rhol) edge node [blue,pos=0.56,right,font=\tiny] {$q_L$} (vl);
\draw[blue] (vl) edge node [blue,pos=0.33,below,font=\tiny] {$\tilde{q}_L$} (rholp);

\draw[red] (v4) edge node [red,pos=0.5,above,font=\tiny] {$\tilde{s}_4$} (rhot);
\draw[red] (v1) edge node [red,pos=0.33,above,font=\tiny] {$\tilde{s}_1$} (rhot);
\draw[red] (rhot) edge node [red,pos=0.33,above,font=\tiny] {$s_3$} (v3);
\draw[red] (rhot) edge node [red,pos=0.6,right,font=\tiny] {$s_2$} (v2);

\end{tikzpicture}
\caption{Vortex quantum mechanics $\textrm{QM}(\nn,\nu)$ for the E-type theory with $\mathcal{I}^{(2)}\cap\mathcal{I}^{(3)} \neq\emptyset\,$ but $\mathcal{I}^{(1)}\cap\mathcal{I}^{(3)} =\emptyset\,$.}
\label{1D_E_2}
\end{figure}

 \begin{figure}[h]
\centering
\begin{tikzpicture}
[->,auto=right, node distance=2cm,
shorten >=1pt, semithick,square/.style={
draw,
minimum width=width("#1"),
minimum height=width("#1")+2*\pgfshapeinnerysep,
node contents={#1}}]
\node (v1) at (-4,0)[circle,draw] {$\nn^{(1)}$};
\node (rho1) at (-4,2) [blue,square={$\rho_{1}$}];
\node (v2) at (-2,0)[circle,draw] {$\nn^{(2)}$};
\node (rho2) at (-2,-2) [blue,square={$\rho_{2}$}];
\node (v3) at (0,2)[circle,draw] {$\nn^{(3)}$};
\node (rho3) at (-2,2) [blue,square={$\rho_{3}$}];
\node (v4) at (0,0) [circle,draw] {$\nn^{(4)}$};
\node (rho4) at (0,-2) [blue,square={$\rho_{4}$}];
\node (v5) at (2,0) [circle,draw] {$\nn^{(5)}$};
\node (rho5) at (2,-2) [blue,square={$\rho_{5}$}];
\node (e) at (4,-0.05) {$\dots$};
\node (vl) at (6,0)[circle,draw] {$\nn^{(L)}$};
\node (rhol) at (6,-2) [blue,square={$\rho_{L}$}];
\node (rholp) at (8,-2) [blue,square={\tiny{$\rho_{L+1}$}}];
\node (rhot) at (-1.3,1) [red,square={\tiny{$\tilde{\rho}$}}];

\draw[-] (v1) to (v2);
\draw[-] (v2) to (v4);
\draw[-] (v3) to (v4);
\draw[-] (v4) to (v5);
\draw[-] (v5) to (e);
\draw[-] (e) to (vl);

 \draw[blue] (rho1) edge  node [blue,pos=0.5,left,font=\tiny] {$q_1$} (v1);
\draw[->,blue] (v1) -- (rho2) node [pos=0.5,below,font=\tiny] {$\tilde{q}_1$};
\draw[blue] (rho2) edge node [blue,pos=0.56,left,font=\tiny] {$q_2$} (v2);
\draw[blue] (v2) edge node [blue,pos=0.56,left,font=\tiny] {$\tilde{q}'_2$} (rho3);
\draw[blue] (rho3) edge node [blue,pos=0.33,above,font=\tiny] {$q_3$} (v3);
\draw[blue] (v3) edge[bend right=20] node [blue,pos=0.56,above,font=\tiny] {$\tilde{q}'_3$} (rho1);
\draw[blue] (v3) edge[bend left] node [blue,pos=0.45,right,font=\tiny] {$\tilde{q}_3$} (rho4);
\draw[blue] (rho2) edge [bend right=15] node [blue,pos=0.5,below,font=\tiny] {$q'_3$} (v3);
\draw[blue] (v2) edge node [blue,pos=0.2,below,font=\tiny] {$\tilde{q}_2$} (rho4);
\draw[blue] (rho4) edge node [blue,pos=0.56,right,font=\tiny] {$q_4$} (v4);
\draw[blue] (v4) edge node [blue,pos=0.56,below,font=\tiny] {$\tilde{q}_4$} (rho5);
\draw[blue] (rho5) edge node [blue,pos=0.56,right,font=\tiny] {$q_5$} (v5);
\draw[blue] (v5) edge node [blue,pos=0.33,below,font=\tiny] {$\tilde{q}_5$} (3,-1);

\draw[->,blue] (5,-1) -- (rhol) node [blue,pos=0.1,below,font=\tiny] {$\tilde{q}_{L-1}$};
\draw[blue] (rhol) edge node [blue,pos=0.56,right,font=\tiny] {$q_L$} (vl);
\draw[blue] (vl) edge node [blue,pos=0.33,below,font=\tiny] {$\tilde{q}_L$} (rholp);
\draw[blue] (v4) edge [bend left=10] node [blue,pos=0.6,below,font=\tiny] {$\tilde{q}'_4$} (rho2);

\draw[red] (v4) edge node [red,pos=0.5,above,font=\tiny] {$\tilde{s}_4$} (rhot);
\draw[red] (rhot) edge node [red,pos=0.66,above,font=\tiny] {$s_1$} (v1);
\draw[red] (rhot) edge node [red,pos=0.33,above,font=\tiny] {$s_3$} (v3);

\end{tikzpicture}
\caption{Vortex quantum mechanics $\textrm{QM}(\nn,\nu)$ for the E-type theory with $\mathcal{I}^{(2)}\cap\mathcal{I}^{(3)} \neq\emptyset\,$ and $\mathcal{I}^{(1)}\cap\mathcal{I}^{(3)} \neq\emptyset\,$.}
\label{1D_E_3}
\end{figure}

\end{enumerate}

\break
\bibliography{biblio}
\bibliographystyle{utphys}

\end{document}